\documentclass[a4paper,10pt,extrafontsizes]{memoir}
\pdfoutput=1

\usepackage[T1]{fontenc}
\usepackage{cmbright}
\usepackage[euler]{textgreek}
\usepackage{microtype}

\usepackage{graphicx}
\usepackage{amsmath,amssymb}
\usepackage{fancyvrb}
\usepackage{xcolor}
\usepackage{dsfont}
\usepackage{enumitem}
\usepackage{attachfile2}
\usepackage{xifthen}
\usepackage{siunitx}
\usepackage{pdfpages}
\usepackage{accsupp}
\usepackage{textcomp}
\usepackage{qrcode}
\usepackage{qcircuit}

\semiisopage[12]
\checkandfixthelayout

\newboolean{includesolutions}
\setboolean{includesolutions}{true}

\notepageref
\makepagenote
\let\oldpagenote\pagenote
\newcommand{\startsolutions}{\let\pagenote\oldpagenote}

\definecolor{livingcoral}{HTML}{FA7268}			
\definecolor{ultraviolet}{HTML}{5F4B8B}			
\definecolor{greenery}{HTML}{88B04B}			
\definecolor{radiantorchid}{HTML}{AD5E99}		
\definecolor{tangerinetango}{HTML}{DD4124}		

\definecolor{mathematicacolor}{rgb}{0,0,0.9}			
\definecolor{mathematicacommentcolor}{rgb}{0.3,0.6,0.7}	
\definecolor{mathematicaprintcolor}{rgb}{0.5,0.5,0.5}		
\definecolor{mathematicaerrorcolor}{rgb}{1,0,0}			
\definecolor{mathematicaechocolor}{rgb}{1,0.5,0}		
\definecolor{mathematicalinenumbercolor}{rgb}{0.5,0.5,0.5}	
\definecolor{mathematicainoutcolor}{rgb}{0.5,0.5,0.5}	

\colorlet{questionscolor}{greenery}					
\colorlet{linkcolor}{radiantorchid}					
	\colorlet{internallinkcolor}{linkcolor}				
	\colorlet{externallinkcolor}{linkcolor}				
\colorlet{embeddedcodecolor}{tangerinetango}			
\newcommand{\embeddedcodecolorname}{orange}		

\colorlet{chapternumbercolor}{ultraviolet}
\colorlet{chaptertitlecolor}{ultraviolet}

\hypersetup{
    colorlinks,
    linkcolor={internallinkcolor},
    urlcolor={externallinkcolor},
    pdfauthor={Roman Schmied},
    pdftitle={Using Mathematica for Quantum Mechanics - A Student's Manual}
    pdfdisplaydoctitle,
    pdfsubject={Quantum mechanical calculations with Wolfram Mathematica},
    pdfkeywords={quantum mechanics, Wolfram Mathematica}
}

\let\oldtheequation\theequation
    \makeatletter
    \def\tagform@#1{\maketag@@@{\ignorespaces#1\unskip\@@italiccorr}}
    \renewcommand{\theequation}{(\oldtheequation)}
    \makeatother

\newcommand{\emdash}{\nobreak---\nobreak\hskip0pt}
\hyphenchar\font=\string"7F

\newcounter{dummy}

\newboolean{smallfigures}
\setboolean{smallfigures}{true}

\newcommand{\listattachmentsname}{list of attached notebooks}
\newlistof{listofattachments}{att}{\listattachmentsname}
\newcounter{attachments}
\newcommand{\attachcode}[2]{\refstepcounter{attachments}{\color{embeddedcodecolor}[}\textattachfile{#1.nb}{code}{\color{embeddedcodecolor}]}\addcontentsline{att}{section}{\protect\numberline{\theattachments.}\textbf{#1.nb} --- #2}}
\attachfilesetup{description={Double-click to open in Mathematica.}, print=false, color=embeddedcodecolor, mimetype=application/vnd.wolfram.mathematica}

\makechapterstyle{graychapters}{
	\chapterstyle{default}

	}
\chapterstyle{graychapters}

\newcommand{\chapterpicture}[1]{
	\begin{center}
		\includegraphics[height=0.33\textheight]{#1}
	\end{center}}

\setsecnumdepth{subsection}
\maxtocdepth{subsection}

\newcommand{\op}[1]{\ensuremath{\hat{#1}}}
\newcommand{\vect}[1]{\ensuremath{\boldsymbol{\vec{#1}}}}
\newcommand{\opvect}[1]{\ensuremath{\boldsymbol{\hat{\vec{#1}}}}}
\newcommand{\dotvect}[1]{\ensuremath{\boldsymbol{\dot{\vec{#1}}}}}
\newcommand{\matr}[1]{\ensuremath{\boldsymbol{#1}}}

\newcommand{\bra}[1]{\ensuremath{\langle#1\rvert}}
\newcommand{\ket}[1]{\ensuremath{\lvert#1\rangle}}
\newcommand{\scp}[2]{\ensuremath{{\langle{#1}{\mid}{#2}\rangle}}}
\newcommand{\CG}[6]{\scp{#1,#2,#3,#4}{#1,#3,#5,#6}}
\newcommand{\me}[3]{\ensuremath{{\langle#1\lvert#2\rvert#3\rangle}}}
\newcommand{\avg}[1]{\ensuremath{{\langle#1\rangle}}}
\newcommand{\abs}[1]{\ensuremath{\lvert#1\rvert}}
\newcommand{\norm}[1]{\ensuremath{\lVert#1\rVert}}

\newcommand{\Ham}{\op{\mathcal{H}}}
\newcommand{\ii}{\ensuremath{\text{i}}}
\newcommand{\dagg}{\ensuremath{^{\dagger}}}
\newcommand{\ddd}[2]{\ensuremath{#1\kern-0.5pt#2}}		
\newcommand{\dd}[1][]{\ifthenelse{\isempty{#1}}{\ensuremath{\text{d}}}{\ddd{\text{d}}{#1}}}
\newcommand{\one}{\ensuremath{\mathds{1}}} 
\newcommand{\dsC}{\ensuremath{\mathds{C}}}
\newcommand{\dsN}{\ensuremath{\mathds{N}}}

\newcommand{\dsR}{\ensuremath{\mathds{R}}}

\newcommand{\makemathematicasymbol}[3]{\expandafter\newcommand\csname mm#1\endcsname{\protect\BeginAccSupp{method=pdfstringdef,ActualText=\textbackslash\relax[#3]}#2\protect\EndAccSupp{}}}
\makemathematicasymbol{alpha}{\textalpha}{Alpha}
\makemathematicasymbol{beta}{\textbeta}{Beta}
\makemathematicasymbol{gamma}{\textgamma}{Gamma}
\makemathematicasymbol{delta}{\textdelta}{Delta}
\makemathematicasymbol{epsilon}{\textepsilon}{Epsilon}
\makemathematicasymbol{zeta}{\textzeta}{Zeta}
\makemathematicasymbol{eta}{\texteta}{Eta}
\makemathematicasymbol{theta}{\texttheta}{Theta}
\makemathematicasymbol{iota}{\textiota}{Iota}
\makemathematicasymbol{kappa}{\textkappa}{Kappa}
\makemathematicasymbol{lambda}{\textlambda}{Lambda}
\makemathematicasymbol{mu}{\textmu}{Mu}
\makemathematicasymbol{nu}{\textnu}{Nu}
\makemathematicasymbol{xi}{\textxi}{Xi}
\makemathematicasymbol{omikron}{\textomikron}{Omicron}
\makemathematicasymbol{pi}{\textpi}{Pi}
\makemathematicasymbol{rho}{\textrho}{Rho}
\makemathematicasymbol{sigma}{\textsigma}{Sigma}
\makemathematicasymbol{tau}{\texttau}{Tau}
\makemathematicasymbol{upsilon}{\textupsilon}{Upsilon}
\makemathematicasymbol{phi}{\textphi}{Phi}
\makemathematicasymbol{chi}{\textchi}{Chi}
\makemathematicasymbol{psi}{\textpsi}{Psi}
\makemathematicasymbol{omega}{\textomega}{Omega}
\makemathematicasymbol{Alpha}{\textAlpha}{CapitalAlpha}
\makemathematicasymbol{Beta}{\textBeta}{CapitalBeta}
\makemathematicasymbol{Gamma}{\textGamma}{CapitalGamma}
\makemathematicasymbol{Delta}{\textDelta}{CapitalDelta}
\makemathematicasymbol{Epsilon}{\textEpsilon}{CapitalEpsilon}
\makemathematicasymbol{Zeta}{\textZeta}{CapitalZeta}
\makemathematicasymbol{Eta}{\textEta}{CapitalEta}
\makemathematicasymbol{Theta}{\textTheta}{CapitalTheta}
\makemathematicasymbol{Iota}{\textIota}{CapitalIota}
\makemathematicasymbol{Kappa}{\textKappa}{CapitalKappa}
\makemathematicasymbol{Lambda}{\textLambda}{CapitalLambda}
\makemathematicasymbol{Mu}{\textMu}{CapitalMu}
\makemathematicasymbol{Nu}{\textNu}{CapitalNu}
\makemathematicasymbol{Xi}{\textXi}{CapitalXi}
\makemathematicasymbol{Omikron}{\textOmikron}{CapitalOmicron}
\makemathematicasymbol{Pi}{\textPi}{CapitalPi}
\makemathematicasymbol{Rho}{\textRho}{CapitalRho}
\makemathematicasymbol{Sigma}{\textSigma}{CapitalSigma}
\makemathematicasymbol{Tau}{\textTau}{CapitalTau}
\makemathematicasymbol{Upsilon}{\textUpsilon}{CapitalUpsilon}
\makemathematicasymbol{Phi}{\textPhi}{CapitalPhi}
\makemathematicasymbol{Chi}{\textChi}{CapitalChi}		
\makemathematicasymbol{Psi}{\textPsi}{CapitalPsi}
\makemathematicasymbol{Omega}{\textOmega}{CapitalOmega}
\makemathematicasymbol{hbar}{\ensuremath{\hbar}}{HBar}
\makemathematicasymbol{infty}{\ensuremath{\infty}}{Infinity}
\makemathematicasymbol{times}{\ensuremath{\times}}{Times}

\newcommand{\ix}[1]{\ensuremath{_{\text{#1}}}}
\newcommand{\ex}[1]{\ensuremath{^{\text{#1}}}}

\newcommand{\ie}{\emph{i.e.}}

\newcommand{\etal}{\emph{et al.}}

\newcommand{\myurl}[1]{{\scriptsize{\url{#1}}}}

\newcommand{\Eunit}{\mathcal{E}}	
\newcommand{\tunit}{\mathcal{T}}	
\newcommand{\lunit}{\mathcal{L}}	
\newcommand{\munit}{\mathcal{M}}	

\DeclareMathOperator{\Tr}{Tr}

\DeclareMathOperator{\sinc}{sinc}

\DeclareMathOperator{\round}{round}
\DeclareMathOperator{\Ai}{Ai}

\newcommand{\R}{$^{\circledR}$}
\newcommand{\AdobeReader}{Adobe\R\ Acrobat\R\ Reader\R}
\newcommand{\AdobeReaderLink}{\href{https://acrobat.adobe.com/us/en/acrobat/pdf-reader.html}{\AdobeReader}}

\DefineVerbatimEnvironment{mathematica}{Verbatim}{gobble=1,numbers=left,numbersep=2mm,formatcom=\color{mathematicacolor},%
	fontfamily=tt,frame=single,commandchars=¤Ž}
\newcounter{mathincounter}
\newcounter{mathoutcounter} 

\newcommand{\mm}[1]{{\color{mathematicacolor}\texttt{#1}}}
\newlist{questions}{enumerate}{1}
\setlist[questions,1]{label={\color{questionscolor}\textbf{Q\arabic{chapter}.\arabic*}},resume}

\makeindex

\begin{document}

\frontmatter

\includepdf{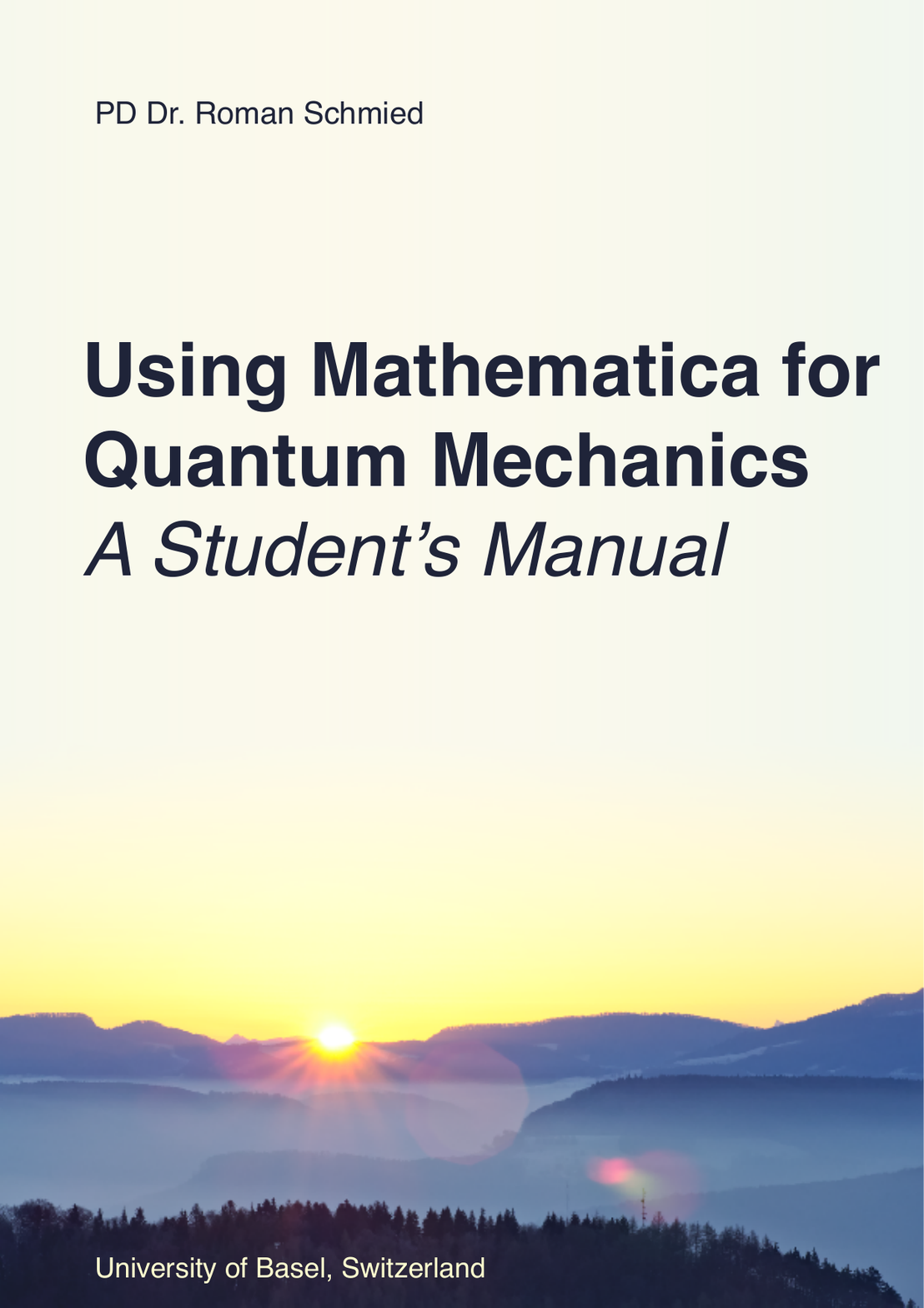}


\noindent
\textcopyright\ 2012--2019 Roman Schmied.\\
\\
compiled on \today\\
\\
published at \url{https://arxiv.org/abs/1403.7050} \qrcode[nolink,level=L,height=8mm]{https://arxiv.org/abs/1403.7050}\\
\\
This book accompanies a one-semester lecture given yearly at the Department of Physics, University of Basel, Switzerland.
The target audience are Master's and PhD students in physics and related fields.\\
\\
Wolfram Mathematica\R, the Wolfram language\R, and Wolfram Alpha\R\ are registered trademarks of Wolfram Research, Inc.\\
\AdobeReader\ is a registered trademark of Adobe Systems, Inc.\\
\vfill\noindent
Cover: First light of 2017, Aussichtsturm Liestal, Switzerland.\\
\textcopyright\ 2017 Roman Schmied.

\renewcommand\contentsname{contents}
\tableofcontents*



\chapter{preface}

\epigraph{The limits of my language mean the limits of my world.}{Ludwig Wittgenstein\index{Wittgenstein, Ludwig}}


\noindent
Learning quantum mechanics is difficult and counter-intuitive. The first lectures I heard were filled with strange concepts that had no relationship with the mechanics I knew, and it took me years of solving research problems until I acquired even a semblance of understanding and intuition.  This process is much like learning a new language, in which a solid mastery of the concepts and rules is required before new ideas and relationships can be expressed fluently.

The major difficulty in bridging the chasm between introductory quantum lectures, on the one hand, and advanced research topics, on the other, was for me the lack of such a language, or of a technical framework in which quantum ideas could be expressed and manipulated. On the one hand, I had the hand tools of algebraic notation, which are understandable but only serve to express very small systems and ideas; on the other hand I had diagrams, circuits, and quasi-phenomenological formulae that describe interesting research problems, but which are difficult to grasp with the mathematical detail I was looking for.

This book is an attempt to help students transform all of the concepts of quantum mechanics into concrete computer representations, which can be constructed, evaluated, analyzed, and hopefully understood at a deeper level than what is possible with more abstract representations.  It was written for a Master's and PhD lecture given yearly at the University of Basel, Switzerland. The goal is to give a language to the student in which to speak about quantum physics in more detail, and to start the student on a path of fluency in this language.
We will revisit most of the problems encountered in introductory quantum mechanics, focusing on computer implementations for finding analytical as well as numerical solutions and their visualization. On our journey we approach questions such as:
\begin{itemize}
	\item You already know how to calculate the energy eigenstates of a single particle in a simple one-dimensional potential. How can such calculations be generalized to non-trivial potentials, higher dimensions, and interacting particles?
	\item You have heard that quantum mechanics describes our everyday world just as well as classical mechanics does, but have you ever seen an example where such behavior is calculated in detail and where the transition from classical to quantum physics is evident?
	\item How can we describe the internal spin structure of particles? How does this internal structure couple to the particles' motion?
	\item What are qubits and quantum circuits, and how can they be assembled to simulate a future quantum computer?
\end{itemize}
Most of the calculations necessary to study and visualize such problems are too complicated to be done by hand. Even relatively simple problems, such as two interacting particles in a one-dimensional trap, do not have analytic solutions and require the use of computers for their solution and visualization. More complex problems scale exponentially with the number of degrees of freedom, and make the use of large computer simulations unavoidable.

The methods presented in this book do not pretend to solve large-scale quantum-mechanical problems in an efficient way; the focus here is more on developing a descriptive language. Once this language is established, it will provide the reader with the tools for understanding efficient large-scale calculations better.




\section{Why Mathematica?}\index{Mathematica!why?}

This book is written in the \emph{Wolfram language} of Mathematica (version 11); however, any other language such as Matlab or Python may be used with suitable translation, as the core ideas presented here are not specific to the Wolfram language. 

There are several reasons why Mathematica was chosen over other computer-algebra systems:
\begin{itemize}
	\item Mathematica is a very high-level programming environment, which allows the user to focus on \emph{what} s?he wants to do instead of \emph{how} it is done. The Wolfram language is extremely expressive and can perform deep calculations with very short and unencumbered programs.
	\item Mathematica supports a wide range of programming paradigms, which means that you can keep programming in your favorite style. See \autoref{sec:factorial15} for a concrete example.
	\item The \emph{Notebook} interface of Mathematica provides an interactive experience that holds programs, experimental code, results, and graphics in one place.
	\item Mathematica seamlessly mixes analytic and numerical facilities. For many calculations it allows you to push analytic evaluations as far as possible, and then continue with numerical evaluations by making only minimal changes.
	\item A very large number of algorithms for analytic and numerical calculations is included in the Mathematica kernel and its libraries.
\end{itemize}

\subsection{Mathematica source code}

Some sections of this book contain embedded Mathematica source code files, for direct evaluation by the reader (see \autopageref{sec:attached} for a list of embedded files). If your PDF reader supports embedded files, you will see a double-clickable \embeddedcodecolorname\ link here:
\attachcode{MathematicaExample}{a simple example}.
If all you see is a blank space between \embeddedcodecolorname\ square brackets, or a non-clickable \embeddedcodecolorname\ link, your PDF reader does not support embedded files; please switch to the \AdobeReaderLink.

\section{outline of discussed topics}

In five chapters, this book takes the student all the way to relatively complex numerical simulations of quantum circuits and interacting particles with spin:
\begin{description}
	\item[{\hyperref[chap:language]{Chapter~\ref*{chap:language}}}] gives an introduction to Mathematica and the Wolfram language, with a focus on techniques that will be useful for this book. This chapter can be safely skipped or replaced by an alternative introduction to Mathematica.
	\item[{\hyperref[chap:basis]{Chapter~\ref*{chap:basis}}}] makes the connection between quantum mechanics and vector/matrix algebra. In this chapter, the abstract concepts of quantum mechanics are converted into computer representations, which form the basis for the following chapters.
	\item[{\hyperref[chap:spin]{Chapter~\ref*{chap:spin}}}] discusses quantum systems with finite-dimensional Hilbert spaces, focusing on spin systems and qubits. These are the most basic quantum-mechanical elements and are ideal for making a first concrete use of the tools of \autoref{chap:basis}.
	\item[{\hyperref[chap:1D]{Chapter~\ref*{chap:1D}}}] discusses the quantum mechanics of particles moving in one- and several-dimensional space. We develop a real-space description of these particles' motion and interaction, and stay as close as possible to the classical understanding of particle motion in phase space.
	\item[{\hyperref[chap:spacespin]{Chapter~\ref*{chap:spacespin}}}] connects the topics of \autoref{chap:spin} and \autoref{chap:1D}, describing particles with spin that move through space.
\end{description}




\mainmatter


\chapter{Wolfram language overview}\index{Mathematica}\index{Wolfram language}
\label{chap:language}
\chapterpicture{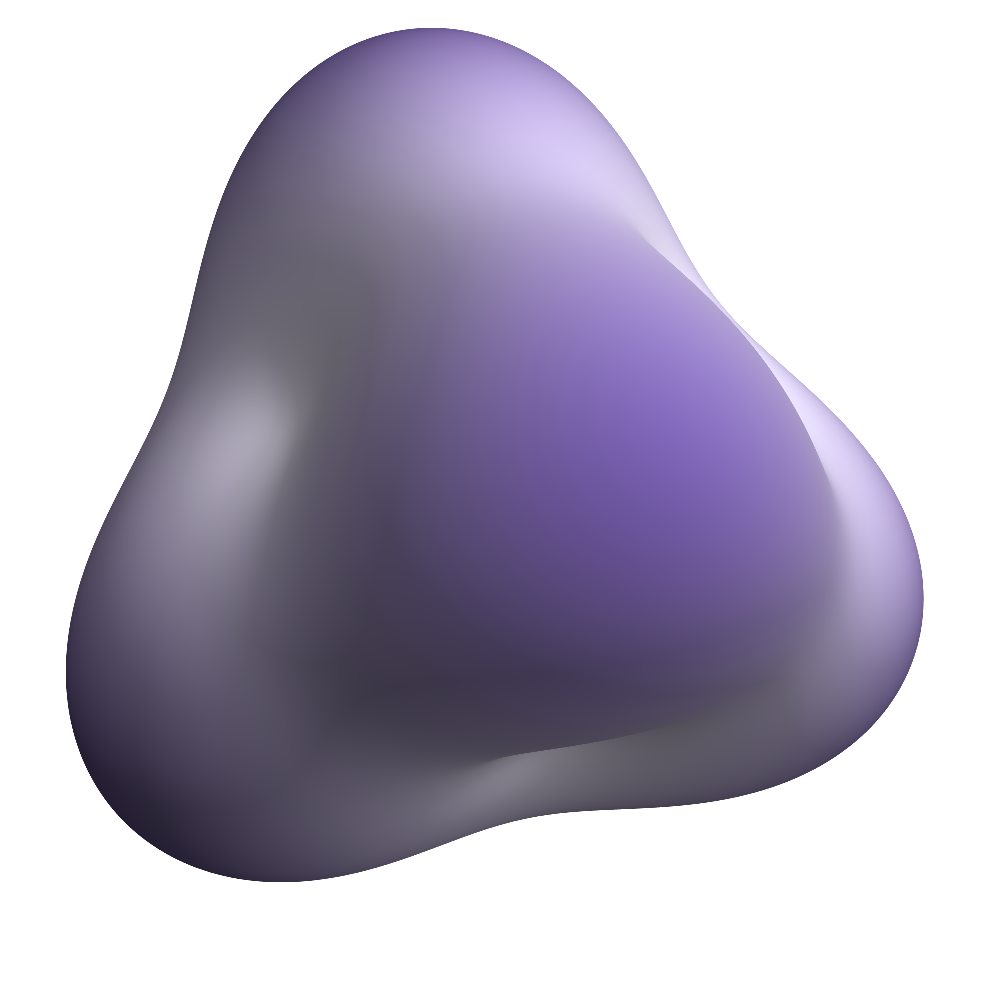}
The Wolfram language is a beautiful and handy tool for expressing a wide variety of technical thoughts. Wolfram Mathematica is the software that implements the Wolfram language.
In this chapter, we have a look at the most central parts of this language, without focusing on quantum mechanics yet. Students who are familiar with the Wolfram language may skip this chapter; others may prefer alternative introductions. Wolfram Research, the maker of Mathematica and the Wolfram language, provides many resources for learning:
\begin{itemize}
	\item \url{https://www.wolfram.com/mathematica/resources/} -- an overview of Mathematica resources to learn at your own pace
	\item \url{https://reference.wolfram.com/language/guide/LanguageOverview.html} -- an overview of the Wolfram language
	\item \url{https://www.wolfram.com/language/} -- the central resource for learning the Wolfram language
	\item \url{https://reference.wolfram.com/language/} -- the Mathematica documentation
\end{itemize}

\clearpage
\section{introduction}

Wolfram Mathematica is an interactive system for mathematical calculations. The Mathematica system is composed of two main components: the \emph{front end}\index{Mathematica!front end}, where you write the input in the Wolfram language, give execution commands, and see the output, and the \emph{kernel}\index{Mathematica!kernel}, which does the actual calculations.
\begin{center}
\includegraphics[width=0.5\textwidth]{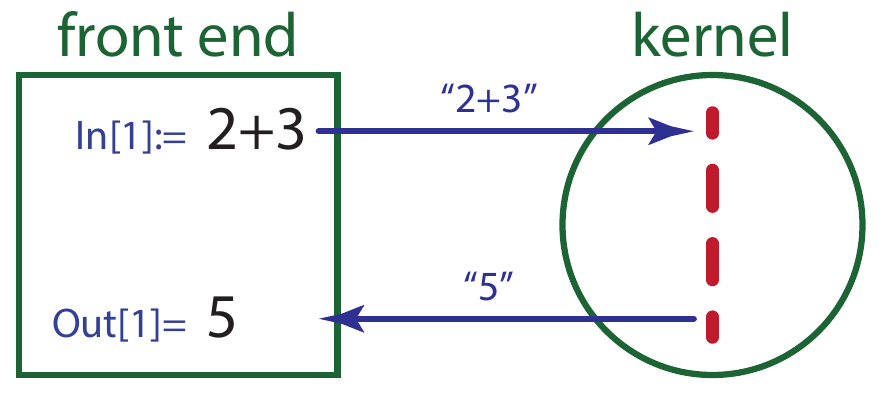}
\end{center}
This distinction is important to remember because the kernel remembers all the operations in the order they are sent to it, and this order may have nothing to do with the order in which these commands are displayed in the front end.

When you start Mathematica you see an empty ``notebook'' in which you can write commands. These commands are written in a mixture of text and mathematical symbols and structures, and it takes a bit of practice to master all the special input commands. In the beginning you can write all your input in pure text mode, if you prefer. Let's try an example: add the numbers $2+3$ by giving the input
\begin{mathematica}
	¤mathin 2+3
\end{mathematica}
and, with the cursor anywhere within the ``cell'' containing this text (look on the right edge of the notebook to see cell limits and groupings) you press ``shift-enter''. This sends the contents of this cell to the kernel, which executes it and returns a result that is displayed in the next cell:
\begin{mathematica}
	¤mathout 5
\end{mathematica}
If there are many input cells in a notebook, they only get executed in order if you select ``Evaluate Notebook'' from the ``Evaluation'' menu; otherwise you can execute the input cells in any order you wish by simply setting the cursor within one cell and pressing ``shift-enter''.

The definition of any function or symbol can be called up with the \mm{?} command:
\begin{mathematica}
	¤mathin ?Factorial
	¤mathnlp n! gives the factorial of n. >>
\end{mathematica}
The arrow $\gg$ that appears at the end of this informative text is a hyperlink into the documentation, where (usually) instructive examples are presented.

\subsection{exercises}
Do the following calculations in Mathematica, and try to understand their structure:
\begin{questions}
	\item\label{Q:zeta} Calculate the numerical value of the Riemann zeta function $\zeta(3)$ with
\begin{mathematica}
	¤mathin N[Zeta[3]]
\end{mathematica}
\pagenote[\ref{Q:zeta}]{
\begin{mathematica}
	¤protect¤mathin¤ N[Zeta[3]]
	¤protect¤mathout¤ 1.20206
\end{mathematica}}
	\item\label{Q:square} Square the previous result (\mm{\%}) with
\begin{mathematica}
	¤mathin %^2
\end{mathematica}
\pagenote[\ref{Q:square}]{
\begin{mathematica}
	¤protect¤mathin¤ %^2
	¤protect¤mathout¤ 1.44494
\end{mathematica}}
	\item\label{Q:integrate} Calculate $\int_0^{\infty} \sin(x)e^{-x}\dd[x]$ with
\begin{mathematica}
	¤mathin Integrate[Sin[x]*Exp[-x], {x, 0, Infinity}]
\end{mathematica}
\pagenote[\ref{Q:integrate}]{
\begin{mathematica}
	¤protect¤mathin¤ Integrate[Sin[x]*Exp[-x],¤ {x,¤ 0,¤ Infinity}]
	¤protect¤mathout¤ 1/2
\end{mathematica}}
	\item\label{Q:pi} Calculate the first 1000 digits of $\pi$ with
\begin{mathematica}
	¤mathin N[Pi, 1000]
\end{mathematica}
	or, equivalently, using the Greek symbol \mm{\mmpi=Pi},
\begin{mathematica}
	¤mathin N[¤mmpi, 1000]
\end{mathematica}
\pagenote[\ref{Q:pi}]{
\begin{mathematica}
	¤protect¤mathin¤ N[¤mmpi,¤ 1000]
	¤protect¤mathout¤ 3.141592653589793238462643383279502884197169399375105820974944592307816406286
	¤protect¤mathnl¤ 20899862803482534211706798214808651328230664709384460955058223172535940812848
	¤protect¤mathnl¤ 11174502841027019385211055596446229489549303819644288109756659334461284756482
	¤protect¤mathnl¤ 33786783165271201909145648566923460348610454326648213393607260249141273724587
	¤protect¤mathnl¤ 00660631558817488152092096282925409171536436789259036001133053054882046652138
	¤protect¤mathnl¤ 41469519415116094330572703657595919530921861173819326117931051185480744623799
	¤protect¤mathnl¤ 62749567351885752724891227938183011949129833673362440656643086021394946395224
	¤protect¤mathnl¤ 73719070217986094370277053921717629317675238467481846766940513200056812714526
	¤protect¤mathnl¤ 35608277857713427577896091736371787214684409012249534301465495853710507922796
	¤protect¤mathnl¤ 89258923542019956112129021960864034418159813629774771309960518707211349999998
	¤protect¤mathnl¤ 37297804995105973173281609631859502445945534690830264252230825334468503526193
	¤protect¤mathnl¤ 11881710100031378387528865875332083814206171776691473035982534904287554687311
	¤protect¤mathnl¤ 59562863882353787593751957781857780532171226806613001927876611195909216420199
\end{mathematica}}
	\item\label{Q:ClebschGordanTest} Calculate the analytic and numeric values of the Clebsch--Gordan coefficient $\scp{100,10;200,-12}{110,-2}$:\index{Clebsch-Gordan coefficient@Clebsch--Gordan coefficient}
\begin{mathematica}
	¤mathin ClebschGordan[{100, 10}, {200, -12}, {110, -2}]
\end{mathematica}
\pagenote[\ref{Q:ClebschGordanTest}]{
\begin{mathematica}
	¤protect¤mathin¤ ClebschGordan[{100,¤ 10},¤ {200,¤ -12},¤ {110,¤ -2}]
	¤protect¤mathout¤ 8261297798499109361013742279092521767681*
	¤protect¤mathnl¤ Sqrt[769248995636473/297224869222895274740285232180446271746289127347456291479
	¤protect¤mathnl¤ 57669733897130076853320942746928207329]/14
	¤protect¤mathin¤ ¤%¤ //N
	¤protect¤mathout¤ 0.0949317
\end{mathematica}}
	\item\label{Q:limit} Calculate the limit $\lim_{x\to0} \frac{\sin x}{x}$ with
\begin{mathematica}
	¤mathin Limit[Sin[x]/x, x -> 0]
\end{mathematica}
\pagenote[\ref{Q:limit}]{
\begin{mathematica}
	¤protect¤mathin¤ Limit[Sin[x]/x,¤ x¤ ->¤ 0]
	¤protect¤mathout¤ 1
\end{mathematica}}
	\item\label{Q:plot} Make a plot of the above function with
\begin{mathematica}
	¤mathin Plot[Sin[x]/x, {x, -20, 20}, PlotRange -> All]
\end{mathematica}
\pagenote[\ref{Q:plot}]{
\begin{mathematica}
	¤protect¤mathin¤ Plot[Sin[x]/x,¤ {x,¤ -20,¤ 20},¤ PlotRange¤ ->¤ All]
\end{mathematica}
	\protect\includegraphics[width=0.5\textwidth]{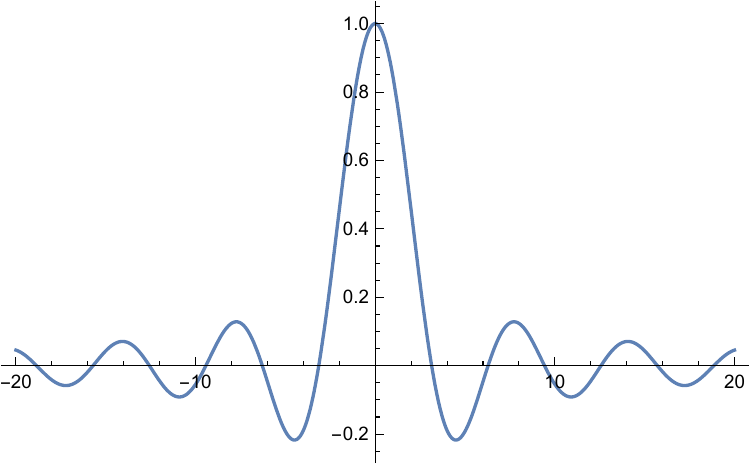}}
	\item\label{Q:Mandelbrot1} Draw a Mandelbrot set\index{Mandelbrot set} with
\begin{mathematica}
	¤mathin F[c_, imax_] := Abs[NestWhile[#^2+c&, 0., Abs[#]<=2&, 1, imax]] <= 2
	¤mathin With[{n = 100, imax = 1000},
	¤mathnl Graphics[Raster[Table[Boole[!F[x+I*y,imax]],{y,-2,2,1/n},{x,-2,2,1/n}]]]]
\end{mathematica}
\pagenote[\ref{Q:Mandelbrot1}]{
\begin{mathematica}
	¤protect¤mathin¤ F[c_,¤ imax_]¤ :=¤ Abs[NestWhile[#^2+c&,¤ 0.,¤ Abs[#]¤ <=¤ 2¤ &,¤ 1,¤ imax]]¤ <=¤ 2
	¤protect¤mathin¤ With[{n¤ =¤ 100,¤ imax¤ =¤ 1000},
	¤protect¤mathnl¤ ¤ ¤ Graphics[Raster[Table[Boole[!F[x+I*y,imax]],{y,-2,2,1/n},{x,-2,2,1/n}]]]]
\end{mathematica}
	\protect\includegraphics[width=0.5\textwidth]{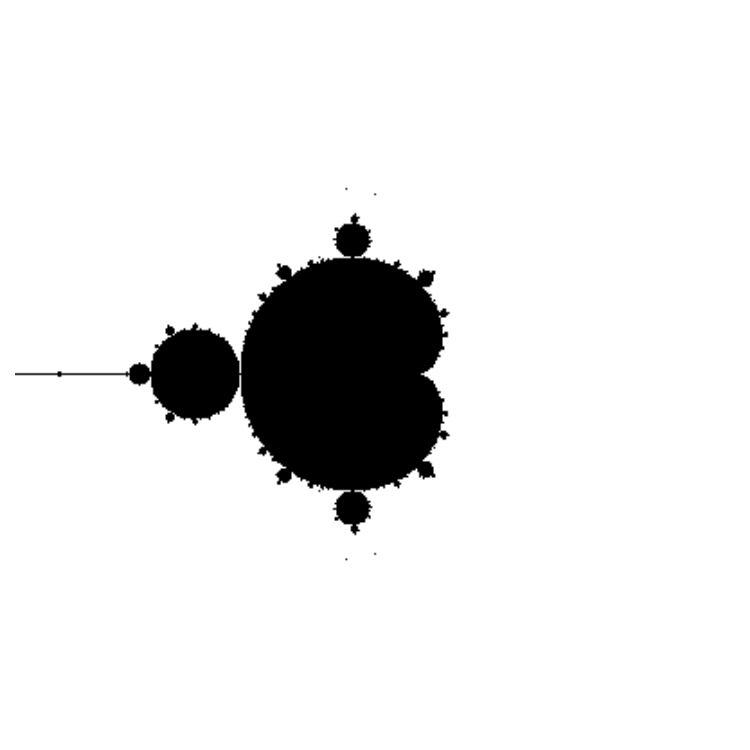}}
	\item\label{Q:Mandelbrot2} Do the same with a built-in function call:
\begin{mathematica}
	¤mathin MandelbrotSetPlot[]
\end{mathematica}
\pagenote[\ref{Q:Mandelbrot2}]{
\begin{mathematica}
	¤protect¤mathin¤ MandelbrotSetPlot[]
\end{mathematica}
	\protect\includegraphics[width=0.5\textwidth]{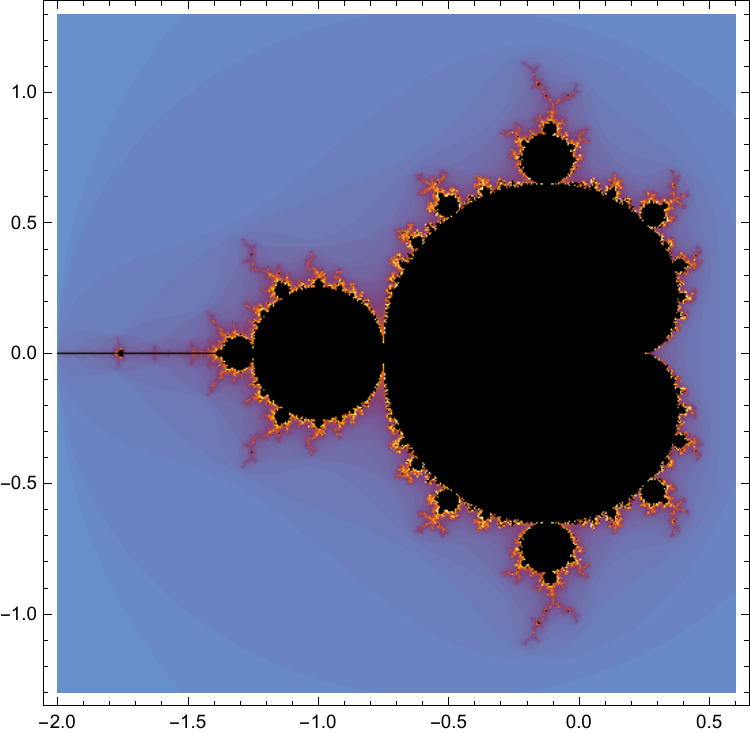}}
\end{questions}

\section{variables and assignments}\index{Mathematica!variable}
\label{sec:variables}
\myurl{https://reference.wolfram.com/language/howto/WorkWithVariablesAndFunctions.html}

\noindent Variables in the Wolfram language can be letters or words with uppercase or lowercase letters, including Greek symbols. Assigning a value to a variable is done with the $\mm{=}$ symbol,
\begin{mathematica}
	¤mathin a = 5
	¤mathout 5
\end{mathematica}
If you wish to suppress the output, then you must end the command with a semi-colon:
\begin{mathematica}
	¤mathin a = 5;
\end{mathematica}
The variable name can then be used anywhere in an expression:
\begin{mathematica}
	¤mathin a + 2
	¤mathout 7
\end{mathematica}

\subsection{immediate and delayed assignments}
\label{sec:assignments}
\myurl{https://reference.wolfram.com/language/tutorial/ImmediateAndDelayedDefinitions.html}

\noindent Consider the two commands\index{Mathematica!random number}
\begin{mathematica}
	¤mathin a = RandomReal[]
	¤mathout 0.38953
	¤mathin b := RandomReal[]
\end{mathematica}
(your random number will be different).

The first statement \mm{a=\dots} is an \emph{immediate assignment}\index{Mathematica!immediate assignment}, which means that its right-hand side is evaluated when you press shift-enter, produces a specific random value, and is assigned to the variable \mm{a} (and printed out). From now on, every time you use the variable \mm{a}, the \emph{exact same} number will be substituted. In this sense, the variable \mm{a} contains the number \num{0,38953} and has no memory of where it got this number from. You can check the definition of \mm{a} with \mm{?a}:
\begin{mathematica}
	¤mathin ?a
	¤mathnlp Global`a
	¤mathnlp a = 0.38953
\end{mathematica}
The definition \mm{b:=\dots} is a \emph{delayed assignment}\index{Mathematica!delayed assignment}, which means that when you press shift-enter the right-hand side is not evaluated but merely stored as a definition of \mm{b}.  From now on, every time you use the variable \mm{b}, its right-hand-side definition will be substituted and executed, resulting in a new random number each time.
You can check the definition of \mm{b} with
\begin{mathematica}
	¤mathin ?b
	¤mathnlp Global`b
	¤mathnlp b := RandomReal[]
\end{mathematica}
Let's compare the repeated performance of \mm{a} and \mm{b}:
\begin{mathematica}
	¤mathin {a, b}
	¤mathout {0.38953, 0.76226}
	¤mathin {a, b}
	¤mathout {0.38953, 0.982921}
	¤mathin {a, b}
	¤mathout {0.38953, 0.516703}
	¤mathin {a, b}
	¤mathout {0.38953, 0.0865169}
\end{mathematica}
If you are familiar with computer file systems, you can think of an immediate assignments as a \emph{hard link} (a direct link to a precomputed inode number) and a delayed assignment as a \emph{soft link} (symbolic link, textual instructions for how to find the linked target).

\subsection{exercises}

\begin{questions}
	\item\label{Q:delayedimmediate} Explain the difference between
\begin{mathematica}
	¤mathin x = u + v
\end{mathematica}
	and
\begin{mathematica}
	¤mathin y := u + v
\end{mathematica}
	In particular, distinguish the cases where \mm{u} and \mm{v} are already defined before \mm{x} and \mm{y} are defined, where they are defined only afterwards, and where they are defined before but change values after the definition of \mm{x} and \mm{y}.
	\pagenote[\ref{Q:delayedimmediate}]{In general, the definition of \mm{x} depends on the values of \mm{u} and \mm{v} at the time of the definition of \mm{x}, whereas \mm{y} depends on the values at the time of using the symbol \mm{y}. The second case below, however, needs special attention since the values of \mm{u} and \mm{v} are not defined at the time when \mm{x} is defined.
	\begin{itemize}
		\item When \mm{u} and \mm{v} are already defined before \mm{x} and \mm{y} are defined, then \mm{x} and \mm{y} return the same value:
\begin{mathematica}
	¤protect¤mathin¤ Clear[x,¤ y,¤ u,¤ v];
	¤protect¤mathin¤ u¤ =¤ 3;¤ v¤ =¤ 7;
	¤protect¤mathin¤ x¤ =¤ u+v;¤ y¤ :=¤ u+v;
	¤protect¤mathin¤ {x,¤ y}
	¤protect¤mathout¤ {10,¤ 10}
	¤protect¤mathin¤ ?x
	¤protect¤mathnlp¤ x=10
	¤protect¤mathin¤ ?y
	¤protect¤mathnlp¤ y:=u+v
\end{mathematica}
		\item When \mm{u} and \mm{v} are defined after \mm{x} and \mm{y} are defined, then \mm{x} and \mm{y} also return the same value. Notice, however, that the definition of \mm{x} is not static and thus depends on the values of \mm{u} and \mm{v} at the time of usage:
\begin{mathematica}
	¤protect¤mathin¤ Clear[x,¤ y,¤ u,¤ v];
	¤protect¤mathin¤ x¤ =¤ u+v;¤ y¤ :=¤ u+v;
	¤protect¤mathin¤ u¤ =¤ 3;¤ v¤ =¤ 7;
	¤protect¤mathin¤ {x,¤ y}
	¤protect¤mathout¤ {10,¤ 10}
	¤protect¤mathin¤ ?x
	¤protect¤mathnlp¤ x=u+v
	¤protect¤mathin¤ ?y
	¤protect¤mathnlp¤ y:=u+v
\end{mathematica}
		\item When \mm{u} and \mm{v} change values after \mm{x} and \mm{y} are defined, then \mm{x} and \mm{y} differ since only \mm{y} reflects the new values of \mm{u} and \mm{v}:
\begin{mathematica}
	¤protect¤mathin¤ Clear[x,¤ y,¤ u,¤ v];
	¤protect¤mathin¤ u¤ =¤ 3;¤ v¤ =¤ 7;
	¤protect¤mathin¤ x¤ =¤ u+v;¤ y¤ :=¤ u+v;
	¤protect¤mathin¤ u¤ =¤ 8;¤ v¤ =¤ 9;
	¤protect¤mathin¤ {x,¤ y}
	¤protect¤mathout¤ {10,¤ 17}
	¤protect¤mathin¤ ?x
	¤protect¤mathnlp¤ x=10
	¤protect¤mathin¤ ?y
	¤protect¤mathnlp¤ y:=u+v
\end{mathematica}}
\end{questions}

\section{four kinds of bracketing}\index{Mathematica!brackets}
\myurl{https://reference.wolfram.com/language/tutorial/TheFourKindsOfBracketingInTheWolframLanguage.html}

\noindent There are four types of brackets in the Wolfram language:
\begin{itemize}
	\item parentheses for grouping, for example in mathematical expressions:
\begin{mathematica}
	¤mathin 2*(3-7)
\end{mathematica}
	\item square brackets for function calls:\index{Mathematica!function}
\begin{mathematica}
	¤mathin Sin[0.2]
\end{mathematica}
	\item curly braces for lists:\index{Mathematica!list}
\begin{mathematica}
	¤mathin v = {a, b, c}
\end{mathematica}
	\item double square brackets for indexing within lists:\index{Mathematica!list} (see \autoref{sec:vecmat})
\begin{mathematica}
	¤mathin v[[2]]
\end{mathematica}
\end{itemize}

\section{prefix and postfix}

There are several ways of evaluating a function call in the Wolfram language, and we will see most of them in this lecture. As examples of function calls with a single argument, the main ways in which $\sin(0.2)$ and $\sqrt{2+3}$ can be calculated are
\begin{description}
	\item[standard notation] (infinite precedence):
\begin{mathematica}
	¤mathin Sin[0.2]
	¤mathout 0.198669
	¤mathin Sqrt[2+3]
	¤mathout Sqrt[5]
\end{mathematica}
	\item[prefix notation]\index{Mathematica!prefix notation} with \mm{@} (quite high precedence, higher than multiplication):
\begin{mathematica}
	¤mathin Sin @ 0.2
	¤mathout 0.198669
	¤mathin Sqrt @ 2+3
	¤mathout 3+Sqrt[2]
\end{mathematica}
	Notice how the high precedence of the \mm{@} operator effectively evaluates \mm{(Sqrt@2)+3}, not \mm{Sqrt@(2+3)}.
	\item[postfix notation]\index{Mathematica!postfix notation} with \mm{//} (quite low precedence, lower than addition):
\begin{mathematica}
	¤mathin 0.2 //Sin
	¤mathout 0.198669
	¤mathin 2+3 //Sqrt
	¤mathout Sqrt[5]
\end{mathematica}
	Notice how the low precedence of the \mm{//} operator effectively evaluates \mm{(2+3)//N}, not \mm{2+(3//N)}.

	Postfix notation is often used to transform the output of a calculation:
	\begin{itemize}
		\item Adding \mm{//N}\index{Mathematica!numerical evaluation} to the end of a command will convert the result to decimal representation, if possible.
		\item Adding \mm{//MatrixForm}\index{Mathematica!matrix!printing} to the end of a matrix calculation will display the matrix in a tabular form.
		\item Adding \mm{//Timing}\index{Mathematica!timing a calculation} to the end of a calculation will display the result together with the amount of time it took to execute.
	\end{itemize}
\end{description}
If you are not sure which form is appropriate, for example if you don't know the precedence of the involved operations, then you should use the standard notation or place parentheses where needed.

\subsection{exercises}

\begin{questions}
	\item\label{Q:Euler} Calculate the decimal value of Euler's constant $e$ (\mm{E}) using standard, prefix, and postfix notation.
	\pagenote[\ref{Q:Euler}]{
\begin{mathematica}
	¤protect¤mathin¤ N[E]
	¤protect¤mathout¤ 2.71828
	¤protect¤mathin¤ N@E
	¤protect¤mathout¤ 2.71828
	¤protect¤mathin¤ E¤ //N
	¤protect¤mathout¤ 2.71828
\end{mathematica}}
\end{questions}

\section{programming constructs}

When you program in the Wolfram language you can choose between a number of different programming paradigms, and you can mix these as you like. Depending on the chosen style, your program may run much faster or much slower.

\subsection{procedural programming}
\myurl{https://reference.wolfram.com/language/guide/ProceduralProgramming.html}

\noindent A subset of the Wolfram language behaves very similarly to C\index{C}, Python\index{Python}, Java\index{Java}, or other procedural programming languages. Be very careful to distinguish semi-colons, which separate commands within a single block of code, from commas, which separate different code blocks!
\begin{description}
	\item[Looping constructs]\index{Mathematica!loop} behave like in common programming languages:
\begin{mathematica}
	¤mathin For[i = 1, i <= 5, i++,
	¤mathnl   Print[i]]
	¤mathnlp 1
	¤mathnlp 2
	¤mathnlp 3
	¤mathnlp 4
	¤mathnlp 5
\end{mathematica}
Notice that \mm{i} is now a globally defined variable, which you can check with
\begin{mathematica}
	¤mathin ?i
	¤mathnlp Global`i
	¤mathnlp i=6
\end{mathematica}
The following, on the other hand, does not define the value of the variable \mm{j} in the global context:
\begin{mathematica}
	¤mathin Do[Print[j], {j, 1, 5}]
	¤mathnlp 1
	¤mathnlp 2
	¤mathnlp 3
	¤mathnlp 4
	¤mathnlp 5
	¤mathin ?j
	¤mathnlp Global`j
\end{mathematica}
In this sense, \mm{j} is a local variable in the \mm{Do} context.
The following, again, defines \mm{k} as a global variable:
\begin{mathematica}
	¤mathin k = 1;
	¤mathnl While[k <= 5,
	¤mathnl   Print[k];
	¤mathnl   k++]
	¤mathnlp 1
	¤mathnlp 2
	¤mathnlp 3
	¤mathnlp 4
	¤mathnlp 5
	¤mathin ?k
	¤mathnlp Global`k
	¤mathnlp k=6
\end{mathematica}
	\item[Conditional execution:]\index{Mathematica!conditional execution} The conditional statement \mm{If[condition, do-when-true, do-when-false]} follows the same logic as in every other programming language,
\begin{mathematica}
	¤mathin If[5! > 100,
	¤mathnl   Print["larger"],
	¤mathnl   Print["smaller or equal"]]
	¤mathnlp larger
\end{mathematica}
	Notice that the \mm{If} statement has a return value, similar to the ``?'' statement of C\index{C} and Java\index{Java}:
\begin{mathematica}
	¤mathin a = If[5! > 100, 1, -1]
	¤mathout 1
\end{mathematica}
	Apart from \emph{true} and \emph{false}, Mathematica statements can have a third state: \emph{unknown}. For example, the comparison \mm{x==0} evaluates to neither true nor false if \mm{x} is not defined. The fourth slot in the \mm{If} statement covers this case:
\begin{mathematica}
	¤mathin x == 0
	¤mathout x == 0
	¤mathin If[x == 0, "zero", "nonzero", "unknown"]
	¤mathout "unknown"
\end{mathematica}
	\item[Modularity:]\index{Mathematica!module} code can use local variables within a \emph{module}:
\begin{mathematica}
	¤mathin Module[{i},
	¤mathnl   i = 1;
	¤mathnl   While[i > 1/192, i = i/2];
	¤mathnl   i]
	¤mathout 1/256
\end{mathematica}
	After the execution of this code, the variable \mm{i} is still undefined in the global context.
\end{description}

\subsection{exercises}

\begin{questions}
	\item\label{Q:sum1} Write a program that sums all integers from \num{123} to \num{9968}. Use only local variables.
	\pagenote[\ref{Q:sum1}]{
\begin{mathematica}
	¤protect¤mathin¤ Total[Range[123,¤ 9968]]
	¤protect¤mathout¤ 49677993
\end{mathematica}}
	\item\label{Q:sum2} Write a program that sums consecutive integers, starting from \num{123}, until the sum is larger than \num{10000}. Return the largest integer in this sum. Use only local variables.
	\pagenote[\ref{Q:sum2}]{
\begin{mathematica}
	¤protect¤mathin¤ Module[{i},
	¤protect¤mathnl¤ ¤ ¤ i¤ =¤ 123;
	¤protect¤mathnl¤ ¤ ¤ s¤ =¤ 0;
	¤protect¤mathnl¤ ¤ ¤ While[s¤ <=¤ 10000,¤ s¤ +=¤ i;¤ i++];
	¤protect¤mathnl¤ ¤ ¤ i¤ -¤ 1]
	¤protect¤mathout¤ 187
\end{mathematica}}
\end{questions}

\subsection{functional programming}\index{Mathematica!functional programming}
\label{sec:functionalprogramming}
\myurl{https://reference.wolfram.com/language/guide/FunctionalProgramming.html}

\noindent Functional programming is a very powerful programming technique that can give large speedups in computation because it can often be parallelized over many computers or CPUs. In our context, we often use lists (vectors or matrices, see \autoref{sec:vecmat}) and want to apply functions to each one of their elements. 

The most common functional programming constructs are
\begin{description}
	\item[Anonymous functions:]\index{Mathematica!anonymous function}\footnote{See \url{https://en.wikipedia.org/wiki/Anonymous_functions}.} you can quickly define a function with parameters \mm{\#1}, \mm{\#2}, \mm{\#3}, etc., terminated with the \mm{\&} symbol: (the symbol \mm{\#} is an abbreviation for \mm{\#1})
\begin{mathematica}
	¤mathin¤labelŽmath:assignf f = #^2 &;
	¤mathin f[7]
	¤mathout 49
	¤mathin¤labelŽmath:assigng g = #1-#2 &;
	¤mathin g[88, 9]
	¤mathout 79
\end{mathematica}
		Functions and anonymous functions, for example \mm{\#\^{}2\&}, are first-class objects\footnote{See \url{https://en.wikipedia.org/wiki/First-class_citizen}.} just like numbers, matrices, etc. You can assign them to variables, as in \mm{\ref{math:assignf}} and \mm{\ref{math:assigng}} above; you can also use them directly as arguments to other functions, as for example in \mm{\ref{math:mapexample}} below; or you can use them as return values of other functions, as in \mm{\ref{math:Wigner}}.
		
		The symbol \mm{\#\#} stands for the sequence of all parameters of a function:
\begin{mathematica}
	¤mathin f = {1,2,3,##,4,5,6} &;
	¤mathin f[7,a,c]
	¤mathout {1,2,3,7,a,c,4,5,6}
\end{mathematica}
		The symbol \mm{\#0} stands for the function itself. This is useful for defining recursive anonymous functions (see \autoref{it:anonymousrecursion} of \autoref{sec:factorial15}).
	\item[Map \mm{/@}:] apply a function to each element of a list.
\begin{mathematica}
	¤mathin a = {1, 2, 3, 4, 5, 6, 7, 8};
	¤mathin¤labelŽmath:mapexample Map[#^2 &, a]
	¤mathout {1, 4, 9, 16, 25, 36, 49, 64}
	¤mathin #^2 & /@ a
	¤mathout {1, 4, 9, 16, 25, 36, 49, 64}
\end{mathematica}
		Notice how we have used the anonymous function \mm{\#\^{}2\&} here without ever giving it a name.
	\item[Apply \mm{@@}:] apply a function to an entire list and generate a single result. For example, applying \mm{Plus} to a list will calculate the sum of the list elements; applying \mm{Times} will calculate their product. This operation is also known as \emph{reduce}.\footnote{See \url{https://en.wikipedia.org/wiki/MapReduce}.}
\begin{mathematica}
	¤mathin a = {1, 2, 3, 4, 5, 6, 7, 8};
	¤mathin Apply[Plus, a]
	¤mathout 36
	¤mathin Plus @@ a
	¤mathout 36
	¤mathin Apply[Times, a]
	¤mathout 40320
	¤mathin Times @@ a
	¤mathout 40320
\end{mathematica}
\end{description}

\subsection{exercises}

\begin{questions}
	\item\label{Q:three} Write an anonymous function with three arguments that returns the product of these arguments.
	\pagenote[\ref{Q:three}]{
\begin{mathematica}
	¤protect¤mathin¤ f¤ =¤ #1*#2*#3¤ &;
\end{mathematica}}
	\item\label{Q:map} Given a list
\begin{mathematica}
	¤mathin a = {0.1, 0.9, 2.25, -1.9};
\end{mathematica}
	calculate $x\mapsto \sin(x^2)$ for each element of \mm{a} using the \mm{Map} operation.
	\pagenote[\ref{Q:map}]{
\begin{mathematica}
	¤protect¤mathin¤ a¤ =¤ {0.1,¤ 0.9,¤ 2.25,¤ -1.9};
	¤protect¤mathin¤ sa¤ =¤ Map[Sin[#]^2¤ &,¤ a]
	¤protect¤mathout¤ {0.00996671,¤ 0.613601,¤ 0.605398,¤ 0.895484}
\end{mathematica}}
	\item\label{Q:mapsum} Calculate the sum of all the results of \ref{Q:map}.
	\pagenote[\ref{Q:mapsum}]{The \mm{Total} function is the same as applying \mm{Plus} to a list:
\begin{mathematica}
	¤protect¤mathin¤ Apply[Plus,¤ sa]
	¤protect¤mathin¤ 2.12445
	¤protect¤mathin¤ Plus@@sa
	¤protect¤mathin¤ 2.12445
	¤protect¤mathin¤ Total[sa]
	¤protect¤mathin¤ 2.12445
\end{mathematica}}
\end{questions}

\section{function definitions}\index{Mathematica!function}
\label{sec:funcdef}
\myurl{https://reference.wolfram.com/language/tutorial/DefiningFunctions.html}

\noindent Functions are assignments (see \autoref{sec:variables}) with parameters. As for parameter-free assignments, we distinguish between \emph{immediate} and \emph{delayed} function definitions.

\subsection{immediate function definitions}\index{Mathematica!immediate assignment}
\label{sec:Set}
We start with \emph{immediate} definitions: a function $f(x)=\sin(x)/x$ is defined with
\begin{mathematica}
	¤mathin f[x_] = Sin[x]/x;
\end{mathematica}
Notice the underscore \mm{\_} symbol after the variable name \mm{x}: this underscore indicates a \emph{pattern}\index{Mathematica!pattern} (denoted by \mm{\_}) named \mm{x}, not the symbol \mm{x} itself. Whenever this function \mm{f} is called with any parameter value, this parameter value is inserted wherever \mm{x} appears on the right-hand side, as is expected for a function definition. You can find out how \mm{f} is defined with the \mm{?} operator:
\begin{mathematica}
	¤mathin ?f
	¤mathnlp Global`f
	¤mathnlp f[x_] = Sin[x]/x
\end{mathematica}
and you can ask for a function evaluation with
\begin{mathematica}
	¤mathin f[0.3]
	¤mathout 0.985067
	¤mathin f[0]
	¤matherrŽPowerInfinite expression 1/0 encountered.
	¤matherrŽInfinityIndeterminate expression 0 ComplexInfinity encountered.
	¤mathout Indeterminate
\end{mathematica}
Apparently the function cannot be evaluated for $x=0$. We can fix this by defining a special function value:
\begin{mathematica}
	¤mathin f[0] = 1;
\end{mathematica}
Notice that there is no underscore on the left-hand side, so there is no pattern definition. The full definition of \mm{f} is now
\begin{mathematica}
	¤mathin ?f
	¤mathnlp Global`f
	¤mathnlp f[0] = 1
	¤mathnlp f[x_] = Sin[x]/x
\end{mathematica}
If the function \mm{f} is called, then these definitions are checked in order of appearance in this list. For example, if we ask for \mm{f[0]}, then the first entry matches and the value 1 is returned. If we ask for \mm{f[0.3]}, then the first entry does not match (since 0 and 0.3 are not strictly equal), but the second entry matches since anything can be plugged into the pattern named \mm{x}. The result is $\sin(0.3)/0.3=\num{0,985067}$, which is what we expected.

\subsection{delayed function definitions}\index{Mathematica!delayed assignment}

Just like with delayed assignments (\autoref{sec:assignments}), we can define delayed function calls. For comparison, we define the two functions\index{Mathematica!random number}
\begin{mathematica}
	¤mathin g1[x_] = x + RandomReal[]
	¤mathout 0.949868 + x
	¤mathin g2[x_] := x + RandomReal[]
\end{mathematica}
Check their effective definitions with \mm{?g1} and \mm{?g2}, and notice that the definition of \mm{g1} was executed immediately when you pressed shift-enter and its result assigned to the function \mm{g1} (with a specific value for the random number, as printed out), whereas the definition of \mm{g2} was left unevaluated and is executed each time anew when you use the function \mm{g2}:
\begin{mathematica}
	¤mathin {g1[2], g2[2]}
	¤mathout {2.94987, 2.33811}
	¤mathin {g1[2], g2[2]}
	¤mathout {2.94987, 2.96273}
	¤mathin {g1[2], g2[2]}
	¤mathout {2.94987, 2.18215}
\end{mathematica}

\subsection{memoization: functions that remember their results}\index{Mathematica!remembering results}\index{memoization}
\label{sec:functionswhichremember}
\myurl{https://reference.wolfram.com/language/tutorial/FunctionsThatRememberValuesTheyHaveFound.html}

\noindent When we define a function that takes a long time to evaluate, we may wish to store its output values such that if the function is called with identical parameter values again, then we do not need to re-evaluate the function but can simply remember the already calculated result.\footnote{This is technically called \emph{memoization}: \url{https://en.wikipedia.org/wiki/Memoization}. A similar functionality can be achieved with Mathematica's \mm{Once} operator, which allows fine-grained control over the storage location, conditions, and duration of the persistent result.}
We can make use of the interplay between patterns and values, and between immediate and delayed assignments, to construct such a function that remembers its values from previous function calls.

See if you can understand the following definition.
\begin{mathematica}
	¤mathin F[x_] := F[x] = x^7
\end{mathematica}
If you ask for \mm{?F} then you will simply see this definition. Now call
\begin{mathematica}
	¤mathin F[2]
	¤mathout 128
\end{mathematica}
and ask for \mm{?F} again. You see that the specific immediate definition of \mm{F[2]=128} was added to the list of definitions, with the evaluated result 128 (which may have taken a long time to calculate in a more complicated function). The next time you call \mm{F[2]}, the specific definition of \mm{F[2]} will be found earlier in the definitions list than the general definition \mm{F[x\_]} and therefore the precomputed value of \mm{F[2]} will be returned.

When you re-define the function \mm{F} after making modifications to it, you must clear the associated remembered values in order for them to be re-computed at the next occasion. It is a good practice to prefix every definition of a memoizing function with a \mm{Clear} command:
\begin{mathematica}
	¤mathin Clear[F];
	¤mathin F[x_] := F[x] = x^9
\end{mathematica}
For function evaluations that take even longer, we may wish to save the accumulated results to a file in order to read them back at a later time. For the above example, we save all definitions associated with the symbol \mm{F} to the file \texttt{Fdef.mx} with\index{Mathematica!saving definitions}
\begin{mathematica}
	¤mathin SetDirectory[NotebookDirectory[]];
	¤mathin DumpSave["Fdef.mx", F];
\end{mathematica}
The next time we wish to continue the calculation, we define the function \mm{F} and load all of its already known values with
\begin{mathematica}
	¤mathin Clear[F];
	¤mathin F[x_] := F[x] = x^9
	¤mathin SetDirectory[NotebookDirectory[]];
	¤mathin Get["Fdef.mx"];
\end{mathematica}

\subsection{functions with conditions on their arguments}\index{Mathematica!pattern}
\label{sec:conditions}
\myurl{https://reference.wolfram.com/language/guide/Patterns.html}

\noindent The Wolfram language contains a powerful pattern language that we can use to define functions that only accept certain arguments. For function definitions we will use three main types of patterns:
\begin{description}
	\item[Anything-goes:] A function defined as
\begin{mathematica}
	¤mathin f[x_] := x^2
\end{mathematica}
	can be called with any sort of arguments, since the pattern \mm{x\_} can match \emph{anything}:
\begin{mathematica}
	¤mathin f[4]
	¤mathout 16
	¤mathin f[2.3-0.1I]
	¤mathout 5.28-0.46I
	¤mathin f[{1,2,3,4}]
	¤mathout {1,4,9,16}
	¤mathin f[y^2]
	¤mathout y^4
\end{mathematica}
	\item[Type-restricted:] A pattern like \mm{x\_Integer} will only match arguments of integer type. If the function is called with a non-matching argument, then the function is not executed:
\begin{mathematica}
	¤mathin g[x_Integer] := x-3
	¤mathin g[x_Rational] := x
	¤mathin g[x_Real] := x+3
	¤mathin g[x_Complex] := 0
	¤mathin g[7]
	¤mathout 4
	¤mathin g[7.1]
	¤mathout 10.1
	¤mathin g[2/3]
	¤mathout 2/3
	¤mathin g[2+3I]
	¤mathout 0
	¤mathin g[x]
	¤mathout g[x]
\end{mathematica}
	\item[Conditional:] Complicated conditions can be specified with the \mm{/;} operator:
\begin{mathematica}
	¤mathin h[x_/;x<=3] := x^2
	¤mathin h[x_/;x>3] := x-11
	¤mathin h[2]
	¤mathout 4
	¤mathin h[5]
	¤mathout -6
\end{mathematica}
	Conditions involving a single function call returning a Boolean value, for example \mm{x\_/;PrimeQ[x]}, can be abbreviated with \mm{x\_?PrimeQ}. Other useful ``question'' functions are \mm{IntegerQ}, \mm{NumericQ}, \mm{EvenQ}, \mm{OddQ}, etc. See \url{https://reference.wolfram.com/language/tutorial/PuttingConstraintsOnPatterns.html} for more information.
\end{description}

\subsection{functions with optional arguments}
\label{sec:optionalargument}
\myurl{https://reference.wolfram.com/language/tutorial/OptionalAndDefaultArguments.html}

\noindent Function arguments can be optional, indicated with the \mm{:} symbol. For each optional argument, a default value must be defined that is used whenever the function is called without the argument specified. The optional arguments must be the last ones in the arguments list. There can be arbitrarily many optional arguments.

As an example, the function
\begin{mathematica}
	¤mathin f[a_, b_:5] = {a,b}
\end{mathematica}
uses the default value $b=5$ whenever it is called with only one argument:
\begin{mathematica}
	¤mathin f[7]
	¤mathout {7,5}
\end{mathematica}
When called with two arguments, the second argument overrides the default value for $b$:
\begin{mathematica}
	¤mathin f[7,2]
	¤mathout {7,2}
\end{mathematica}


\section{rules and replacements}\index{Mathematica!rules}\index{Mathematica!replacements}
\label{sec:rulesandreplacements}
\myurl{https://reference.wolfram.com/language/tutorial/ApplyingTransformationRules.html}

\noindent We will often use replacement rules in the calculations of this course. A replacement rule is an instruction \mm{x -> y} that replaces any occurrence of the symbol (or pattern) \mm{x} with the symbol \mm{y}. We apply such a rule with the \mm{/.} or \mm{ReplaceAll} operator:
\begin{mathematica}
	¤mathin a + 2 /. a -> 7
	¤mathout 9
	¤mathin ReplaceAll[a + 2, a -> 7]
	¤mathout 9
	¤mathin c - d /. {c -> 2, d -> 8}
	¤mathout -6
	¤mathin ReplaceAll[c - d, {c -> 2, d -> 8}]
	¤mathout -6
\end{mathematica}
Rules can contain patterns, in the same way as we use them for defining the parameters of functions (\autoref{sec:funcdef}):
\begin{mathematica}
	¤mathin a + b /. x_ -> x^2
	¤mathout (a + b)^2
\end{mathematica}
Notice that here the pattern \mm{x\_} matched the entire expression \mm{a + b}, not the subexpressions \mm{a} and \mm{b}. To be more specific and do the replacement only at level 1 of this expression, we can write
\begin{mathematica}
	¤mathin Replace[a + b, x_ -> x^2, {1}]
	¤mathout a^2 + b^2
\end{mathematica}
Doing the replacement at level 0 gives again
\begin{mathematica}
	¤mathin Replace[a + b, x_ -> x^2, {0}]
	¤mathout (a + b)^2
\end{mathematica}
At other instances, restricted patterns can be used to achieve a desired result:
\begin{mathematica}
	¤mathin a + 2 /. x_Integer -> x^2
	¤mathout 4 + a
\end{mathematica}
Many Wolfram language functions return their results as replacement rules. For example, the result of solving an equation is a list of rules:
\begin{mathematica}
	¤mathin s = Solve[x^2 - 4 == 0, x]
	¤mathout {{x -> -2}, {x -> 2}}
\end{mathematica}
We can make use of these solutions with the replacement operator \mm{/.}, for example to check the solutions:
\begin{mathematica}
	¤mathin x^2 - 4 /. s
	¤mathout {0, 0}
\end{mathematica}

\subsection{immediate and delayed rules}
\label{sec:ruletypes}

Just as for assignments (\autoref{sec:assignments}) and functions (\autoref{sec:funcdef}), rules can be immediate or delayed. In an immediate rule of the form \mm{x -> y}, the value of \mm{y} is calculated once upon defining the rule. In a delayed rule of the form \mm{x :> y}, the value of \mm{y} is re-calculated every time the rule is applied. This can be important when the rule is supposed to perform an action. Here is an example: we replace \mm{c} by \mm{f} with
\begin{mathematica}
	¤mathin {a, b, c, d, c, a, c, b} /. c -> f
	¤mathout {a, b, f, d, f, a, f, b}
\end{mathematica}
We do the same while counting the number of replacements with
\begin{mathematica}
	¤mathin i = 0;
	¤mathin {a, b, c, d, c, a, c, b} /. c :> (i++; Echo[i, "replacement "]; f)
	¤mathechoŽreplacement 1
	¤mathechoŽreplacement 2
	¤mathechoŽreplacement 3
	¤mathout {a, b, f, d, f, a, f, b}
	¤mathin i
	¤mathout 3
\end{mathematica}
In this case, the delayed rule \mm{c :> (i++; Echo[i, "replacement "]; f)} is a list of commands enclosed in parentheses \mm{()} and separated by semicolons. The first command increments the replacement counter \mm{i}, the second prints a running commentary (see \autoref{sec:debug}), and the third gives the result of the replacement. The result of such a list of commands is always the last expression, in this case \mm{f}.

\subsection{repeated rule replacement}
\label{sec:repeatedrulereplacement}

The \mm{/.} operator uses the given list of replacement rules only once:
\begin{mathematica}
	¤mathin a /. {a -> b, b -> c}
	¤mathout b
\end{mathematica}
The \mm{//.} operator, on the other hand, uses the replacement rules repeatedly until the result no longer changes (in this case, after two applications):
\begin{mathematica}
	¤mathin a //. {a -> b, b -> c}
	¤mathout c
\end{mathematica}

\section{debugging and finding out how Mathematica expressions are evaluated}
\index{Mathematica!evaluation}\index{Mathematica!tracing}\index{Mathematica!debugging}
\label{sec:debug}
\myurl{https://reference.wolfram.com/language/guide/TuningAndDebugging.html}\\
\myurl{https://www.wolfram.com/language/elementary-introduction/2nd-ed/47-debugging-your-code.html}

\noindent The precise way Mathematica evaluates an expression depends on many details and can become very complicated.\footnote{See \url{https://reference.wolfram.com/language/tutorial/EvaluationOfExpressionsOverview.html}.} For finding out more about particular cases, especially when they aren't evaluated in the way that you were expecting, the \mm{Trace} command may be useful. This command gives a list of all intermediate results, which helps in understanding the way that Mathematica arrives at its output:
\begin{mathematica}
	¤mathin Trace[x - 3x + 1]
	¤mathout {{-(3x), -3x, -3x}, x-3x+1, 1-3x+x, 1-2x}
	¤mathin x = 5;
	¤mathin Trace[x - 3x + 1]
	¤mathout {{x, 5}, {{{x, 5}, 3¤mmtimes5, 15}, -15, -15}, 5-15+1, -9}
\end{mathematica}
A more verbose trace is achieved with \mm{TracePrint}:
\begin{mathematica}
	¤mathin TracePrint[y - 3y + 1]
	¤mathnlp y-3 y+1
	¤mathnlp  Plus
	¤mathnlp  y
	¤mathnlp  -(3 y)
	¤mathnlp   Times
	¤mathnlp   -1
	¤mathnlp   3 y
	¤mathnlp    Times
	¤mathnlp    3
	¤mathnlp    y
	¤mathnlp  -3 y
	¤mathnlp  -3 y
	¤mathnlp   Times
	¤mathnlp   -3
	¤mathnlp   y
	¤mathnlp  1
	¤mathnlp y-3 y+1
	¤mathnlp 1-3 y+y
	¤mathnlp 1-2 y
	¤mathnlp  Plus
	¤mathnlp  1
	¤mathnlp  -2 y
	¤mathnlp   Times
	¤mathnlp   -2
	¤mathnlp   y
	¤mathout 1 - 2 y
\end{mathematica}
It is very useful to print out intermediate results in a long calculation via the \mm{Echo} command, particularly during code development. Calling \mm{Echo[x,label]} prints \mm{x} with the given label, and returns \mm{x}; in this way, the \mm{Echo} command can be simply added to a calculation without perturbing it:
\begin{mathematica}
	¤mathin¤labelŽmath:echo Table[Echo[i!, "building table: "], {i, 3}]
	¤mathechoŽbuilding table: 1
	¤mathechoŽbuilding table: 2
	¤mathechoŽbuilding table: 6
	¤mathout {1, 2, 6}
\end{mathematica}
In order to run your code ``cleanly'' after debugging it with \mm{Echo}, you can either remove all instances of \mm{Echo}, or you can re-define \mm{Echo} to do nothing:
\begin{mathematica}
	¤mathin¤labelŽmath:unprotect Unprotect[Echo]; Echo = #1 &;
\end{mathematica}
Re-running the code of \mm{\ref{math:echo}} now gives just the result:
\begin{mathematica}
	¤mathin Table[Echo[i!, "building table: "], {i, 3}]
	¤mathout {1, 2, 6}
\end{mathematica}
Finally, it can be very insightful to study the ``full form''\index{Mathematica!full form} of expressions, especially when it does not match a pattern that you were expecting to match. For example, the internal full form of ratios depends strongly on the type of numerator or denominator:
\begin{mathematica}
	¤mathin¤labelŽmath:symbolicratio FullForm[a/b]
	¤mathout Times[a, Power[b, -1]]
	¤mathin¤labelŽmath:numericratio FullForm[1/2]
	¤mathout Rational[1, 2]
	¤mathin FullForm[a/2]
	¤mathout Times[Rational[1, 2], a]
	¤mathin FullForm[1/b]
	¤mathout Power[b, -1]
\end{mathematica}

\subsection{exercises}

\begin{questions}
	\item\label{Q:unprotect} Why do we need the \mm{Unprotect} command in \mm{\ref{math:unprotect}}?
	\pagenote[\ref{Q:unprotect}]{All built-in symbols, like \mm{Echo}, are protected in order to prevent accidental modification. Trying to modify \mm{Echo} without unprotecting it first gives an error:
\begin{mathematica}
	¤protect¤mathin¤ Echo¤ =¤ #1¤ &
	¤protect¤matherrŽSet¤ Symbol¤ Echo¤ is¤ Protected.
	¤protect¤mathout¤ #1¤ &
\end{mathematica}}
	\item\label{Q:ratiopattern} To replace a ratio $a/b$ by the function \mm{ratio[a,b]}, we could enter
\begin{mathematica}
	¤protect¤mathin a/b /. {x_/y_ -> ratio[x,y]}
	¤protect¤mathout ratio[a,b]
\end{mathematica}
	Why does this not work to replace the ratio $2/3$ by the function \mm{ratio[2,3]}?
\begin{mathematica}
	¤protect¤mathin 2/3 /. {x_/y_ -> ratio[x,y]}
	¤protect¤mathout 2/3
\end{mathematica}
	\pagenote[\ref{Q:ratiopattern}]{See \mm{\ref{math:symbolicratio}} and \mm{\ref{math:numericratio}}: the full forms of \mm{a/b} and \mm{x\_/y\_} are similar and match,
\begin{mathematica}
	¤protect¤mathin¤ FullForm[a/b]
	¤protect¤mathout¤ Times[a,¤ Power[b,¤ -1]]
	¤protect¤mathin¤ FullForm[x_/y_]
	¤protect¤mathout¤ Times[Pattern[x,¤ Blank[]],¤ Power[Pattern[y,¤ Blank[]],¤ -1]]
\end{mathematica}
while the full form of \mm{2/3} is different and does not match the pattern for replacements,
\begin{mathematica}
	¤protect¤mathin¤ FullForm[2/3]
	¤protect¤mathout¤ Rational[2,¤ 3]
\end{mathematica}}
\end{questions}

\section[many ways to define the factorial function]{\label{sec:factorial15}many ways to define the factorial function\hspace{\stretch{1}}\attachcode{FactorialDefinitions}{many ways to define the factorial function}}
\index{Mathematica!factorial}

The following list of definitions of the factorial function is based on the Wolfram demo \url{https://www.wolfram.com/training/videos/EDU002/}. Try to understand as many of these definitions as possible. What this means in practice is that for most problems you can pick the programming paradigm that suits your way of thinking best, instead of being forced into one way or another. The different paradigms have different advantages and disadvantages, which may become clearer to you as you become more familiar with them.

You must call \mm{Clear[f]} between different definitions!
\begin{enumerate}
	\item Define the function \mm{f} to be an alias of the built-in function \mm{Factorial}: calling \mm{f[5]} is now strictly the same thing as calling \mm{Factorial[5]}, which in turn is the same thing as calling  \mm{5!}.
\begin{mathematica}
	¤mathin f = Factorial;
\end{mathematica}
	\item A call to \mm{f} is forwarded to the function ``\mm{!}'': calling \mm{f[5]} triggers the evaluation of \mm{5!}.
\begin{mathematica}
	¤mathin f[n_] := n!
\end{mathematica}
	\item Use the mathematical definition $n!=\Gamma(n+1)$:
\begin{mathematica}
	¤mathin f[n_] := Gamma[n+1]
\end{mathematica}
	\item Use the mathematical definition $n!=\prod_{i=1}^n i$:
\begin{mathematica}
	¤mathin f[n_] := Product[i, {i,n}]
\end{mathematica}
	\item\label{it:recursion1} Rule-based recursion\index{Mathematica!recursion}, using the Wolfram language's built-in pattern-matching capabilities: calling \mm{f[5]} leads to a call of \mm{f[4]}, which leads to a call of \mm{f[3]}, and so on until \mm{f[1]} immediately returns the result 1, after which the program unrolls the recursion stack and does the necessary multiplications:
\begin{mathematica}
	¤mathin f[1] = 1;
	¤mathin f[n_] := n*f[n-1]
\end{mathematica}
	\item\label{it:recursion2} The same recursion\index{Mathematica!recursion} but without rules (no pattern-matching):
\begin{mathematica}
	¤mathin f[n_] := If[n == 1, 1, n*f[n-1]]
\end{mathematica}
	\item\label{it:anonymousrecursion} Define the same recursion\index{Mathematica!recursion} through functional programming\index{Mathematica!functional programming}: \mm{f} is a function whose name is \mm{\#0} and whose first (and only) argument is \mm{\#1}\index{Mathematica!anonymous function}. The end of the function definition is marked with \mm{\&}.
\begin{mathematica}
	¤mathin f = If[#1 == 1, 1, #1*#0[#1-1]]&;
\end{mathematica}
	\item procedural programming with a \mm{Do} loop\index{Mathematica!loop}:
\begin{mathematica}
	¤mathin f[n_] := Module[{t = 1},
	¤mathnl   Do[t = t*i, {i, n}];
	¤mathnl   t]
\end{mathematica}
	\item procedural programming with a \mm{For} loop\index{Mathematica!loop}: this is how you would compute factorials in procedural programming languages like C. It is a very precise step-by-step prescription of how exactly the computer is supposed to do the calculation.
\begin{mathematica}
	¤mathin f[n_] := Module[{t = 1, i},
	¤mathnl   For[i = 1, i <= n, i++,
	¤mathnl     t *= i];
	¤mathnl   t]
\end{mathematica}
	\item Make a list of the numbers $1\dots n$ (with \mm{Range[n]}) and then multiply them together at once, by applying the function \mm{Times} to this list. This is the most elegant way of multiplying all these numbers together, because both the generation of the list of integers and their multiplication are done with internally optimized methods. The programmer merely specifies \emph{what} he would like the computer to do, and not \emph{how} it is to be done.
\begin{mathematica}
	¤mathin f[n_] := Times @@ Range[n]
\end{mathematica}
	\item Make a list of the numbers $1\dots n$ and then multiply them together one after the other.
\begin{mathematica}
	¤mathin f[n_] := Fold[Times, 1, Range[n]]
\end{mathematica}
	\item Functional programming\index{Mathematica!functional programming}: make a list of functions $\{t\mapsto t, t\mapsto 2t, t\mapsto 3t, \dots, t\mapsto n t\}$, and then, starting with the number 1, apply each of these functions once.
\begin{mathematica}
	¤mathin f[n_] := Fold[#2[#1]&, 1, Array[Function[t, #1*t]&, n]]
\end{mathematica}
	\item Construct a list whose length we know to be $n!$:
\begin{mathematica}
	¤mathin f[n_] := Length[Permutations[Range[n]]]
\end{mathematica}
	\item\label{it:factreplace1} Use repeated pattern-based\index{Mathematica!pattern} replacement (\mm{//.}, see \autoref{sec:repeatedrulereplacement}) to find the factorial: start with the object $\{1,n\}$ and apply the given rule until the result no longer changes because the pattern no longer matches.
\begin{mathematica}
	¤mathin f[n_] := First[{1,n} //. {a_,b_/;b>0} :> {b*a,b-1}]
\end{mathematica}
	\item Build a string whose length is $n!$:
\begin{mathematica}
	¤mathin f[n_] := StringLength[Fold[StringJoin[Table[#1, {#2}]]&, "A", Range[n]]]
\end{mathematica}
	\item\label{it:factreplace2} Starting from the number $n$, repeatedly replace each number $m$ by a list containing $m$ times the number $m-1$. At the end, we have a list of lists of \dots\ of lists that overall contains $n!$ times the number 1. Flatten it out and count the number of elements.
\begin{mathematica}
	¤mathin f[n_] := Length[Flatten[n //. m_ /; m > 1 :> Table[m - 1, {m}]]]
\end{mathematica}
	\item Analytically calculate $\frac{\dd^n(x^n)}{\dd[x]^n}$, the $n\ex{th}$ derivative of $x^n$:
\begin{mathematica}
	¤mathin f[n_] := D[x^n, {x, n}]
\end{mathematica}
\end{enumerate}

\subsection{exercises}

\begin{questions}
	\item\label{Q:immdel} In which ones of the definitions of \autoref{sec:factorial15} can you replace a delayed assignment (\mm{:=}) with an immediate assignment (\mm{=}) or vice-versa? What changes if you do this replacement? (see \autoref{sec:assignments})
\pagenote[\ref{Q:immdel}]{Not all delayed assignments can be replaced by immediate ones. Whenever an immediate assignment can be used, it tends to be faster.
	\begin{enumerate}
		\item \mm{=} and \mm{:=} work equally well.
		\item \mm{=} and \mm{:=} work equally well.
		\item \mm{=} and \mm{:=} work equally well.
		\item \mm{=} and \mm{:=} work equally well. There is a significant difference though: while the delayed assignment executes as a product, the immediate assignment is simplified at the moment of definition to a factorial, which then executes much faster:
\begin{mathematica}
	¤protect¤mathin¤ f[n_]¤ =¤ Product[i,¤ {i,¤ n}]
	¤protect¤mathout¤ n!
\end{mathematica}
		\item Immediate assignment breaks the recursion, which cannot be executed at definition time.
		\item \mm{=} and \mm{:=} work equally well.
		\item \mm{=} and \mm{:=} work equally well.
		\item Immediate assignment breaks the \mm{Do} loop, which cannot be executed at definition time.
		\item Immediate assignment breaks the \mm{For} loop: since \mm{n} is not defined at definition time, the comparison \mm{i<=n} fails at the first iteration and the result is always \mm{f[n\_]=1}.
		\item Immediate assignment breaks the \mm{Range} command since \mm{n} is not defined at definition time.
		\item Immediate assignment breaks the \mm{Range} command since \mm{n} is not defined at definition time.
		\item Immediate assignment breaks the \mm{Array} command since \mm{n} is not defined at definition time.
		\item Immediate assignment breaks the \mm{Range} command since \mm{n} is not defined at definition time.
		\item Immediate assignment always gives \mm{f[n\_]=1} since the repeated replacement fails.
		\item Immediate assignment breaks the \mm{Range} command since \mm{n} is not defined at definition time.
		\item Immediate assignment always gives \mm{f[n\_]=1} since the repeated replacement fails.
		\item \mm{=} and \mm{:=} work equally well.
	\end{enumerate}}
	\item\label{Q:immdelrep} In which ones of the definitions of \autoref{sec:factorial15} can you replace a delayed rule (\mm{:>}) with an immediate rule (\mm{->}) or vice-versa? What changes if you do this replacement? (see \autoref{sec:ruletypes})
\pagenote[\ref{Q:immdelrep}]{Not all delayed rules can be replaced by immediate ones. Whenever an immediate rule can be used, it tends to be faster.
	\begin{enumerate}\setcounter{enumi}{\getrefnumber{it:factreplace1}-1}
		\item \mm{->} and \mm{:>} work equally well.
		\setcounter{enumi}{\getrefnumber{it:factreplace2}-1}
		\item Immediate rule (\mm{->}) breaks the \mm{Table} command since \mm{m} is not defined at definition time.
	\end{enumerate}}
	\item\label{Q:remember} Can you use the trick of \autoref{sec:functionswhichremember} for any of the definitions of \autoref{sec:factorial15}?
\pagenote[\ref{Q:remember}]{In the recursive definitions \ref{it:recursion1} and \ref{it:recursion2}, memoization gives a dramatic speedup, as it remembers intermediate results in the recursion. In the other examples, memoization only helps when the function is called repeatedly with the same argument.}
	\item\label{Q:fibonacci} Write two very different programs that calculate the first hundred Fibonacci numbers $\{1, 1, 2, 3, 5, 8, \dots\}$, where each number is the sum of the two preceding ones.
\pagenote[\ref{Q:fibonacci}]{Using a built-in function:
\begin{mathematica}
	¤protect¤mathin¤ Table[Fibonacci[n],¤ {n,¤ 100}]
\end{mathematica}
Even more directly, by using the \mm{Listable} attribute of the \mm{Fibonacci} function:
\begin{mathematica}
	¤protect¤mathin¤ Fibonacci[Range[100]]
\end{mathematica}
Recursive with memoization:
\begin{mathematica}
	¤protect¤mathin¤ g[1]¤ =¤ g[2]¤ =¤ 1;
	¤protect¤mathin¤ g[n_]¤ :=¤ g[n]¤ =¤ g[n-1]¤ +¤ g[n-2]
	¤protect¤mathin¤ Table[g[n],¤ {n,¤ 100}]
\end{mathematica}
Iterative construction of the list:
\begin{mathematica}
	¤protect¤mathin¤ L¤ =¤ {1,¤ 1};
	¤protect¤mathin¤ Do[AppendTo[L,¤ L[[-1]]¤ +¤ L[[-2]]],¤ {98}];
	¤protect¤mathin¤ L
\end{mathematica}}
\end{questions}

\section{vectors, matrices, tensors}
\label{sec:vecmat}

In this lecture we will use vectors and matrices to represent quantum states and operators, respectively. 

\subsection{vectors}\index{Mathematica!vector}
\label{sec:vectors}
\myurl{https://reference.wolfram.com/language/tutorial/VectorOperations.html}

\noindent In the Wolfram language, vectors are represented as lists\index{Mathematica!list} of objects, for example lists of real or complex numbers:
\begin{mathematica}
	¤mathin v = {1,2,3,2,1,7+I};
	¤mathin Length[v]
	¤mathout 6
\end{mathematica}
You can access any element by its index, using double brackets\index{Mathematica!brackets}, with the first element having index 1 (as in Fortran\index{Fortran} or Matlab\index{Matlab}), \emph{not} 0 (as in C\index{C}, Java\index{Java}, or Python\index{Python}):
\begin{mathematica}
	¤mathin v[[4]]
	¤mathout 2
\end{mathematica}
Negative indices count from the end of the list:
\begin{mathematica}
	¤mathin v[[-1]]
	¤mathout 7+I
\end{mathematica}
Lists can contain arbitrary elements (for example strings, graphics, expressions, lists, functions, etc.).

If two vectors $\vect{a}$ and $\vect{b}$ of equal length are defined, then their scalar product $\vect{a}^*\cdot\vect{b}$ is calculated with
\begin{mathematica}
	¤mathin a = {0.1, 0.2, 0.3 + 2I};
	¤mathin b = {-0.27I, 0, 2};
	¤mathin Conjugate[a].b
	¤mathout 0.6 - 4.027I
\end{mathematica}
Vectors of equal length can be element-wise added, subtracted, multiplied etc.\ with the usual operators:
\begin{mathematica}
	¤mathin a + b
	¤mathout {0.1 - 0.27I, 0.2, 2.3 + 2.I}
	¤mathin 2 a
	¤mathout {0.2, 0.4, 0.6 + 4.I}
\end{mathematica}

\subsection{matrices}\index{Mathematica!matrix}
\label{sec:matrices}
\myurl{https://reference.wolfram.com/language/tutorial/BasicMatrixOperations.html}

\noindent Matrices are lists of lists, where each sublist describes a row of the matrix:
\begin{mathematica}
	¤mathin M = {{3,2,7},{1,1,2},{0,-1,5},{2,2,1}};
	¤mathin Dimensions[M]
	¤mathout {4, 3}
\end{mathematica}
In this example, \mm{M} is a \num{4 x 3} matrix. Pretty-printing a matrix is done with the MatrixForm\index{Mathematica!matrix!printing} wrapper,
\begin{mathematica}
	¤mathin MatrixForm[M]
\end{mathematica}
Accessing matrix elements is analogous to accessing vector elements:\index{Mathematica!brackets}
\begin{mathematica}
	¤mathin M[[1,3]]
	¤mathout 7
	¤mathin M[[2]]
	¤mathout {1, 1, 2}
\end{mathematica}
Matrices can be transposed with \mm{Transpose[M]}.

Matrix--vector and matrix--matrix multiplications are done with the \mm{.} operator:
\begin{mathematica}
	¤mathin M.a
	¤mathout {2.8 + 14.I, 0.9 + 4.I, 1.3 + 10.I, 0.9 + 2.I}
\end{mathematica}

\subsection{sparse vectors and matrices}\index{Mathematica!matrix!sparse matrix}
\label{sec:sparsemat}
\myurl{https://reference.wolfram.com/language/guide/SparseArrays.html}

\noindent Large matrices can take up enormous amounts of computer memory. In practical situations we are often dealing with matrices that are ``sparse'', meaning that most of their entries are zero. A much more efficient way of storing them is therefore as a list of only their nonzero elements, using the \mm{SparseArray} function.

A given vector or matrix is converted to sparse representation with
\begin{mathematica}
	¤mathin M = {{0,3,0,0,0,0,0,0,0,0},
	¤mathnl      {0,0,0,-1,0,0,0,0,0,0},
	¤mathnl      {0,0,0,0,0,0,0,0,0,0}};
	¤mathin Ms = SparseArray[M]
	¤mathout SparseArray[<2>, {3, 10}]
\end{mathematica}
where the output shows that \mm{Ms} is a \num{3 x 10} sparse matrix with 2 non-zero entries. We could have entered this matrix more easily by giving the list of non-zero entries,
\begin{mathematica}
	¤mathin Ms = SparseArray[{{1, 2} -> 3, {2, 4} -> -1}, {3, 10}];
\end{mathematica}
which we can find out from
\begin{mathematica}
	¤mathin ArrayRules[Ms]
	¤mathout {{1, 2} -> 3, {2, 4} -> -1, {_, _} -> 0}
\end{mathematica}
which includes a specification of the default pattern\index{Mathematica!pattern} \mm{\{\_,\_\}}.
This sparse array is converted back into a normal array with
\begin{mathematica}
	¤mathin Normal[Ms]
	¤mathout {{0,3,0,0,0,0,0,0,0,0},
	¤mathnl  {0,0,0,-1,0,0,0,0,0,0},
	¤mathnl  {0,0,0,0,0,0,0,0,0,0}}
\end{mathematica}
Sparse arrays and vectors can be used just like full arrays and vectors (they are internally converted automatically whenever necessary). But for some linear algebra operations they can be much more efficient. A matrix multiplication of two sparse matrices, for example, scales only with the number of non-zero elements of the matrices, not with their size.

\subsection{matrix diagonalization}
\label{sec:diagonalization}

``Solving'' the time-independent Schr\"odinger equation, as we will be doing in \autoref{sec:TimeIndepSchr}, involves calculating the eigenvalues and eigenvectors of Hermitian\footnote{A complex matrix $\matr{H}$ is \emph{Hermitian} if $\matr{H}=\matr{H}\dagg$. See \url{https://en.wikipedia.org/wiki/Hermitian_matrix}.} matrices.

In what follows it is assumed that we have defined $\matr{H}$ as a Hermitian matrix. As an example we will use
\begin{mathematica}
	¤mathin H = {{0,   0.3, I,    0},
	¤mathnl      {0.3, 1,   0,    0},
	¤mathnl      {-I,  0,   1,    -0.2},
	¤mathnl      {0,   0,   -0.2, 3}};
\end{mathematica}

\subsubsection{eigenvalues}\index{Mathematica!matrix!eigenvalues}

The eigenvalues of a matrix \mm{H} are computed with
\begin{mathematica}
	¤mathin Eigenvalues[H]
	¤mathout {3.0237, 1.63842, 0.998322, -0.660442}
\end{mathematica}
Notice that these eigenvalues (energy values) are not necessarily sorted, even though in this example they appear in descending order. For a sorted list we use
\begin{mathematica}
	¤mathin Sort[Eigenvalues[H]]
	¤mathout {-0.660442, 0.998322, 1.63842, 3.0237}
\end{mathematica}
For very large matrices \mm{H}, and in particular for sparse matrices (see \autoref{sec:sparsemat}), it is computationally inefficient to calculate all eigenvalues. Further, we are often only interested in the lowest-energy eigenvalues and eigenvectors. There are very efficient algorithms for calculating extremal eigenvalues,\footnote{Arnoldi--Lanczos algorithm: \url{https://en.wikipedia.org/wiki/Lanczos_algorithm}.} which can be used by specifying options to the \mm{Eigenvalues} function: if we only need the largest two eigenvalue, for example, we call\index{Arnoldi algorithm}\index{Lanczos algorithm}
\begin{mathematica}
	¤mathin Eigenvalues[H, 2, Method -> {"Arnoldi",
	¤mathnl                              "Criteria" -> "RealPart",
	¤mathnl                              MaxIterations -> 10^6}]
	¤mathout {3.0237, 1.63842}
\end{mathematica}
There is no direct way to calculate the \emph{smallest} eigenvalues; but since the smallest eigenvalues of \mm{H} are the largest eigenvalues of \mm{-H} we can use
\begin{mathematica}
	¤mathin -Eigenvalues[-H, 2, Method -> {"Arnoldi",
	¤mathnl                                "Criteria" -> "RealPart",
	¤mathnl                                MaxIterations -> 10^6}]
	¤mathout {0.998322, -0.660442}
\end{mathematica}

\subsubsection{eigenvectors}\index{Mathematica!matrix!eigenvectors}
\label{sec:eigenvectors}
The eigenvectors of a matrix \mm{H} are computed with
\begin{mathematica}
	¤mathin Eigenvectors[H]
	¤mathout {{0.-0.0394613I, 0.-0.00584989I, -0.117564, 0.992264},
	¤mathnl  {0.+0.533642I, 0.+0.250762I, 0.799103, 0.117379},
	¤mathnl  {0.-0.0053472I, 0.+0.955923I, -0.292115, -0.029187},
	¤mathnl  {0.-0.844772I, 0.+0.152629I, 0.512134, 0.0279821}}
\end{mathematica}
In this case of a \num{4 x 4} matrix, this generates a list of four ortho-normal $4$-vectors.

Usually we are interested in calculating the eigenvalues and eigenvectors at the same time:
\begin{mathematica}
	¤mathin Eigensystem[H]
	¤mathout {{3.0237, 1.63842, 0.998322, -0.660442},
	¤mathnl  {{0.-0.0394613I, 0.-0.00584989I, -0.117564, 0.992264},
	¤mathnl   {0.+0.533642I, 0.+0.250762I, 0.799103, 0.117379},
	¤mathnl   {0.-0.0053472I, 0.+0.955923I, -0.292115, -0.029187},
	¤mathnl   {0.-0.844772I, 0.+0.152629I, 0.512134, 0.0279821}}}
\end{mathematica}
which generates a list containing the eigenvalues and the eigenvectors. The ordering of the elements in the eigenvalues list corresponds to the ordering in the eigenvectors list; but the sorting order is generally undefined. To generate a list of (eigenvalue, eigenvector) pairs in ascending order of eigenvalues, we calculate
\begin{mathematica}
	¤mathin Sort[Transpose[Eigensystem[H]]]
	¤mathout {{-0.660442, {0.-0.844772I, 0.+0.152629I, 0.512134, 0.0279821}},
	¤mathnl  {0.998322, {0.-0.0053472I, 0.+0.955923I, -0.292115, -0.029187}},
	¤mathnl  {1.63842, {0.+0.533642I, 0.+0.250762I, 0.799103, 0.117379}},
	¤mathnl  {3.0237, {0.-0.0394613I, 0.-0.00584989I, -0.117564, 0.992264}}}
\end{mathematica}
To generate a sorted list of eigenvalues \mm{eval} and a corresponding list of eigenvectors \mm{evec} we calculate
\begin{mathematica}
	¤mathin¤labelŽmath:sortev {eval,evec} = Transpose[Sort[Transpose[Eigensystem[H]]]];
	¤mathin eval
	¤mathout {-0.660442, 0.998322, 1.63842, 3.0237}
	¤mathin evec
	¤mathout {{0.-0.844772I, 0.+0.152629I, 0.512134, 0.0279821},
	¤mathnl  {0.-0.0053472I, 0.+0.955923I, -0.292115, -0.029187},
	¤mathnl  {0.+0.533642I, 0.+0.250762I, 0.799103, 0.117379},
	¤mathnl  {0.-0.0394613I, 0.-0.00584989I, -0.117564, 0.992264}}
\end{mathematica}
The trick with calculating only the lowest-energy eigenvalues can be applied to eigenvalue calculations as well, since the eigenvectors of \mm{-H} and \mm{H} are the same:
\begin{mathematica}
	¤mathin {eval,evec} = Transpose[Sort[Transpose[-Eigensystem[-H, 2,
	¤mathnl   Method -> {"Arnoldi", "Criteria" -> "RealPart", MaxIterations -> 10^6}]]]];
	¤mathin eval
	¤mathout {-0.660442, 0.998322}
	¤mathin evec
	¤mathout {{-0.733656+0.418794I, 0.132553-0.0756656I,
	¤mathnl   -0.253889-0.444771I, -0.0138721-0.0243015 I},
	¤mathnl  {-0.000575666-0.00531612I, 0.102912+0.950367I,
	¤mathnl   -0.290417+0.0314484I, -0.0290174+0.0031422I}}
\end{mathematica}
Notice that these eigenvectors are not the same as those calculated further above! This difference is due to arbitrary multiplications of the eigenvectors with phase factors $e^{\ii\varphi}$.

To check that the vectors in \mm{evec} are ortho-normalized, we calculate the matrix product
\begin{mathematica}
	¤mathin Conjugate[evec].Transpose[evec] //Chop //MatrixForm
\end{mathematica}
and verify that the matrix of scalar products is indeed equal to the unit matrix.

To check that the vectors in \mm{evec} are indeed eigenvectors of \mm{H}, we calculate all matrix elements of \mm{H} in this basis of eigenvectors:
\begin{mathematica}
	¤mathin Conjugate[evec].H.Transpose[evec] //Chop //MatrixForm
\end{mathematica}
and verify that the result is a diagonal matrix whose diagonal elements are exactly the eigenvalues \mm{eval}.

\subsection{tensor operations}\index{tensor}
\label{sec:tensors}
\myurl{https://reference.wolfram.com/language/guide/RearrangingAndRestructuringLists.html}

\noindent We have seen above that in the Wolfram language, a vector is a list of numbers (\autoref{sec:vectors}) and a matrix is a list of lists of numbers (\autoref{sec:matrices}). Higher-rank tensors are correspondingly represented as lists of lists of \dots of lists of numbers. In this section we describe general tools for working with tensors, which extend the methods used for vectors and matrices.
See \autoref{sec:rdm} for a concrete application of higher-rank tensors. We note that the sparse techniques of \autoref{sec:sparsemat} naturally extend to higher-rank tensors.

As an example, we start by defining a list (\ie, a vector) containing 24 elements:
\begin{mathematica}
	¤mathin v = Range[24]
	¤mathout {1,2,3,4,5,6,7,8,9,10,11,12,13,14,15,16,17,18,19,20,21,22,23,24}
	¤mathin Dimensions[v]
	¤mathout {24}
\end{mathematica}
We have chosen the elements in this vector to indicate their position in order to make the following transformations easier to understand.

\subsubsection{reshaping}

We reshape the list \mm{v} into a \num{2 x 3 x 4} tensor with
\begin{mathematica}
	¤mathin t = ArrayReshape[v, {2,3,4}]
	¤mathout {{{1,2,3,4},{5,6,7,8},{9,10,11,12}},
	¤mathnl  {{13,14,15,16},{17,18,19,20},{21,22,23,24}}}
	¤mathin Dimensions[t]
	¤mathout {2, 3, 4}
\end{mathematica}
Notice that the order of the elements has not changed; but they are now arranged as a list of lists of lists of numbers.
Alternatively, we could reshape \mm{v} into a \num{2 x 2 x 3 x 2} tensor with
\begin{mathematica}
	¤mathin u = ArrayReshape[v, {2,2,3,2}]
	¤mathout {{{{1,2},{3,4},{5,6}},{{7,8},{9,10},{11,12}}},
	¤mathnl  {{{13,14},{15,16},{17,18}},{{19,20},{21,22},{23,24}}}}
	¤mathin Dimensions[u]
	¤mathout {2, 2, 3, 2}
\end{mathematica}

\subsubsection{flattening}

The reverse operation is called flattening:
\begin{mathematica}
	¤mathin Flatten[t] == Flatten[u] == v
	¤mathout True
\end{mathematica}
Tensor flattening can be applied more specifically, without flattening the entire structure into a single list. As an example, in \mm{u} we flatten indices 1\&2 together and indices 3\&4 together, to find a \num{4 x 6} matrix that we could have calculated directly with \mm{ArrayReshape[v, \{4,6\}]}:
\begin{mathematica}
	¤mathin Flatten[u, {{1,2}, {3,4}}]
	¤mathout {{1,2,3,4,5,6},{7,8,9,10,11,12},{13,14,15,16,17,18},{19,20,21,22,23,24}}
	¤mathin % == ArrayReshape[v, {4,6}]
	¤mathout True
\end{mathematica}
We sometimes use the \mm{ArrayFlatten} command, which is just a special case of \mm{Flatten} with fixed arguments, flattening indices 1\&3 together and indices 2\&4 together:
\begin{mathematica}
	¤mathin ArrayFlatten[u] == Flatten[u, {{1,3}, {2,4}}]
	¤mathout True
\end{mathematica}

\subsubsection{transposing}

A tensor transposition is a re-ordering of a tensor's indices. For example,
\begin{mathematica}
	¤mathin tt = Transpose[t, {2,3,1}]
	¤mathout {{{1,5,9},{13,17,21}},{{2,6,10},{14,18,22}},
	¤mathnl  {{3,7,11},{15,19,23}},{{4,8,12},{16,20,24}}}
	¤mathin Dimensions[tt]
	¤mathout {4, 2, 3}
\end{mathematica}
generates a \num{4 x 2 x 3}-tensor \mm{tt}, where the first index of \mm{t} is the second index of \mm{tt}, the second index of \mm{t} is the third index of \mm{tt}, and the third index of \mm{t} is the first index of \mm{tt}; this order of index shuffling is given in the parameter list \mm{\{2,3,1\}} meaning $\{1\ex{st},2\ex{nd},3\ex{rd}\}\mapsto\{2\ex{nd},3\ex{rd},1\ex{st}\}$. More explicitly,
\begin{mathematica}
	¤mathin Table[t[[i,j,k]] == tt[[k,i,j]], {i,2}, {j,3}, {k,4}]
	¤mathout {{{True,True,True,True},{True,True,True,True},
	¤mathnl   {True,True,True,True}},{{True,True,True,True},
	¤mathnl   {True,True,True,True},{True,True,True,True}}}
\end{mathematica}

\subsubsection{contracting}

As a generalization of a scalar product, indices of equal length of a tensor can be contracted\index{tensor!contraction}. This is the operation of summing over an index that appears twice in the list of indices. For example, contracting indices 2 and 5 of the rank-6 tensor $X_{a,b,c,d,e,f}$ yields the rank-4 tensor with elements $Y_{a,c,d,f}=\sum_i X_{a,i,c,d,i,f}$.

For example, we can either contract indices 1\&2 in \mm{u}, or indices 1\&4, or indices 2\&4, since they are all of length 2:
\begin{mathematica}
	¤mathin TensorContract[u, {1, 2}]
	¤mathout {{20, 22}, {24, 26}, {28, 30}}
	¤mathin TensorContract[u, {1, 4}]
	¤mathout {{15, 19, 23}, {27, 31, 35}}
	¤mathin¤labelŽmath:contract TensorContract[u, {2, 4}]
	¤mathout {{9, 13, 17}, {33, 37, 41}}
\end{mathematica}

\subsection{exercises}

\begin{questions}
	\item\label{Q:Pauli} Calculate the eigenvalues and eigenvectors of the Pauli matrices:\\
		\url{https://en.wikipedia.org/wiki/Pauli_matrices}\\
		Are the eigenvectors ortho-normal? If not, find an ortho-normal set.
\pagenote[\ref{Q:Pauli}]{The eigenvectors are orthogonal, but not necessarily normalized.\par
\noindent Eigensystem of $\op{\sigma}_x$:
\begin{mathematica}
	¤protect¤mathin¤ {eval,¤ evec}¤ =¤ Eigensystem[PauliMatrix[1]]
	¤protect¤mathout¤ {{-1,¤ 1},¤ {{-1,¤ 1},¤ {1,¤ 1}}}
	¤protect¤mathin¤ Normalize¤ /@¤ evec
	¤protect¤mathout¤ {{-1/Sqrt[2],¤ 1/Sqrt[2]},¤ {1/Sqrt[2],¤ 1/Sqrt[2]}}
\end{mathematica}
Eigensystem of $\op{\sigma}_y$:
\begin{mathematica}
	¤protect¤mathin¤ {eval,¤ evec}¤ =¤ Eigensystem[PauliMatrix[2]]
	¤protect¤mathout¤ {{-1,¤ 1},¤ {{I,¤ 1},¤ {-I,¤ 1}}}
	¤protect¤mathin¤ Normalize¤ /@¤ evec
	¤protect¤mathout¤ {{I/Sqrt[2],¤ 1/Sqrt[2]},¤ {-I/Sqrt[2],¤ 1/Sqrt[2]}}
\end{mathematica}
Eigensystem of $\op{\sigma}_z$:
\begin{mathematica}
	¤protect¤mathin¤ {eval,¤ evec}¤ =¤ Eigensystem[PauliMatrix[3]]
	¤protect¤mathout¤ {{-1,¤ 1},¤ {{0,¤ 1},¤ {1,¤ 0}}}
\end{mathematica}}
	\item\label{Q:contract} After \mm{\ref{math:contract}}, try to contract indices 3\&4 in the tensor \mm{u}. What went wrong?
\pagenote[\ref{Q:contract}]{The tensor index dimensions do not match:
\begin{mathematica}
	¤protect¤mathin¤ TensorContract[u,¤ {3,¤ 4}]
	¤protect¤matherrŽTensorContract¤ Contraction¤ levels¤ {3,4}¤ have¤ different¤ dimensions¤ {3,2}.
\end{mathematica}}
\end{questions}

\section{complex numbers}\index{Mathematica!complex number}
\label{sec:complex}

By default all variables in the Wolfram language are assumed to be complex numbers, unless otherwise specified. All mathematical functions can take complex numbers as their input, often by analytic continuation.\footnote{See \url{https://en.wikipedia.org/wiki/Analytic_continuation}.}

The most commonly used functions on complex numbers are \mm{Conjugate}, \mm{Re}, \mm{Im}, \mm{Abs}, and \mm{Arg}. When applied to numerical arguments they do what we expect:
\begin{mathematica}
	¤mathin Conjugate[2 + 3I]
	¤mathout 2 - 3I
	¤mathin Im[0.7]
	¤mathout 0
\end{mathematica}
When applied to variable arguments, however, they fail and frustrate the inexperienced user:
\begin{mathematica}
	¤mathin Conjugate[x+I*y]
	¤mathout Conjugate[x] - I*Conjugate[y]
	¤mathin Im[a]
	¤mathout Im[a]
\end{mathematica}
This behavior is due to Mathematica not knowing that \mm{x}, \mm{y}, and \mm{a} in these examples are real-valued.
There are several ways around this, all involving \emph{assumptions}. The first is to use the \mm{ComplexExpand} function, which assumes that all variables are \emph{real}:
\begin{mathematica}
	¤mathin Conjugate[x+I*y] //ComplexExpand
	¤mathout x - I*y
	¤mathin Im[a] //ComplexExpand
	¤mathout 0
\end{mathematica}
The second is to use explicit \emph{local} assumptions,\index{Mathematica!assumptions} which may be more specific than assuming that all variables are real-valued:
\begin{mathematica}
	¤mathin Assuming[Element[x, Reals] && Element[y, Reals],
	¤mathnl   Conjugate[x + I*y] //FullSimplify]
	¤mathout x - I*y
	¤mathin Assuming[Element[a, Reals], Im[a] //FullSimplify]
	¤mathout 0
\end{mathematica}
The third is to use \emph{global} assumptions (in general, global system variables start with the \mm{\$} sign):
\begin{mathematica}
	¤mathin $Assumptions = Element[x, Reals] && Element[y, Reals] && Element[a, Reals];
	¤mathin Conjugate[x+I*y] //FullSimplify
	¤mathout x - I*y
	¤mathin Im[a] //FullSimplify
	¤mathout 0
\end{mathematica}

\section{units}\index{Mathematica!physical units}
\label{sec:units}
\myurl{https://reference.wolfram.com/language/tutorial/UnitsOverview.html}

\noindent The Wolfram language is capable of dealing with units of measure, as required for physical calculations. For example, we can make the assignment
\begin{mathematica}
	¤mathin s = Quantity[3, "m"];
\end{mathematica}
to specify that \mm{s} should be three meters. A large number of units can be used, as well as physical constants:
\begin{mathematica}
	¤mathin kB = Quantity["BoltzmannConstant"];
\end{mathematica}
will define the variable \mm{kB} to be Boltzmann's constant. Take note that complicated or slightly unusual quantities are evaluated through the online service \emph{Wolfram Alpha}\R, which means that you need an internet connection in order to evaluate them. For this and other reasons, unit calculations are very slow and to be avoided whenever possible.

If you are unsure whether your expression has been interpreted correctly, the full internal form\index{Mathematica!full form}
\begin{mathematica}
	¤mathin FullForm[kB]
	¤mathout Quantity[1, "BoltzmannConstant"]
\end{mathematica}
usually helps. Alternatively, converting to SI units can often clarify a definition:
\begin{mathematica}
	¤mathin UnitConvert[kB]
	¤mathout Quantity[1.38065*10^-23, "kg m^2/(s^2 K)"]
\end{mathematica}
In principle, we can use this mechanism to do all the calculations in this lecture with units; however, for the sake of generality (as many other computer programs cannot deal with units) when we do numerical calculations, we will convert every quantity into dimensionless form in what follows.

In order to eliminate units from a calculation, we must determine a set of units in which to express the relevant quantities. This means that every physical quantity $x$ is expressed as the product of a \emph{unit} and a \emph{dimensionless multiplier}. The actual calculations are performed only with the dimensionless multipliers. A smart choice of units can help in implementing a problem.

As an example we calculate the acceleration of an A380 airplane ($m=\SI{560}{t}$) due to its jet engines ($F=4\times\SI{311}{kN}$). The easiest way is to use the Wolfram language's built-in unit processing:
\begin{mathematica}
	¤mathin F = Quantity[4*311, "kN"];
	¤mathin m = Quantity[560, "t"];
	¤mathin a = UnitConvert[F/m, "m/s^2"] //N
	¤mathout¤labelŽmath:acceleration 2.22143 m/s^2
\end{mathematica}
This method is, however, much slower than using purely numerical calculations, and furthermore cannot be generalized to matrix and vector algebra.

Now we do the same calculation with dimensionless multipliers only. For this, we first set up a consistent set of units, for example the SI units:
\begin{mathematica}
	¤mathin ForceUnit = Quantity["Newtons"];
	¤mathin MassUnit = Quantity["Kilograms"];
	¤mathin AccelerationUnit = UnitConvert[ForceUnit/MassUnit]
	¤mathout 1 m/s^2
\end{mathematica}
It is important that these units are consistent with each other, \ie, that the product of the mass and acceleration units gives the force unit. The calculation is now effected with a simple numerical division \mm{a=F/m}:
\begin{mathematica}
	¤mathin F = Quantity[4*311, "kN"] / ForceUnit
	¤mathout 1244000
	¤mathin m = Quantity[560, "t"] / MassUnit
	¤mathout 560000
	¤mathin a = F/m //N
	¤mathout 2.22143
\end{mathematica}
This result of \num{2.22143} acceleration units, meaning \SI{2.22143}{m/s^2}, is the same as \mm{\ref{math:acceleration}}.

We can do this type of calculation in any consistent unit system: as a second example, we use the unit definitions
\begin{mathematica}
	¤mathin ForceUnit = Quantity["KiloNewtons"];
	¤mathin MassUnit = Quantity["AtomicMassUnit"];
	¤mathin AccelerationUnit = UnitConvert[ForceUnit/MassUnit]
	¤mathout 6.022141*10^29 m/s^2
\end{mathematica}
and calculate
\begin{mathematica}
	¤mathin F = Quantity[4*311, "kN"] / ForceUnit
	¤mathout 1244
	¤mathin m = Quantity[560, "t"] / MassUnit
	¤mathout 3.3723989*10^32
	¤mathin a = F/m //N
	¤mathout 3.68877*10^-30
\end{mathematica}
This result is again the same as \mm{\ref{math:acceleration}}, because \num{3.68877e-30} acceleration units are $\num{3.68877e-30}\times\SI{6.022141e29}{m/s^2}$.

It is not important which unit system we use. In practice, it is often convenient to use a system of units that yields dimensionless multipliers that are on the order of unity; but this is not a strict requirement.

\chapter{quantum mechanics: states and operators}
\label{chap:basis}
\restartlist{questions}
\chapterpicture{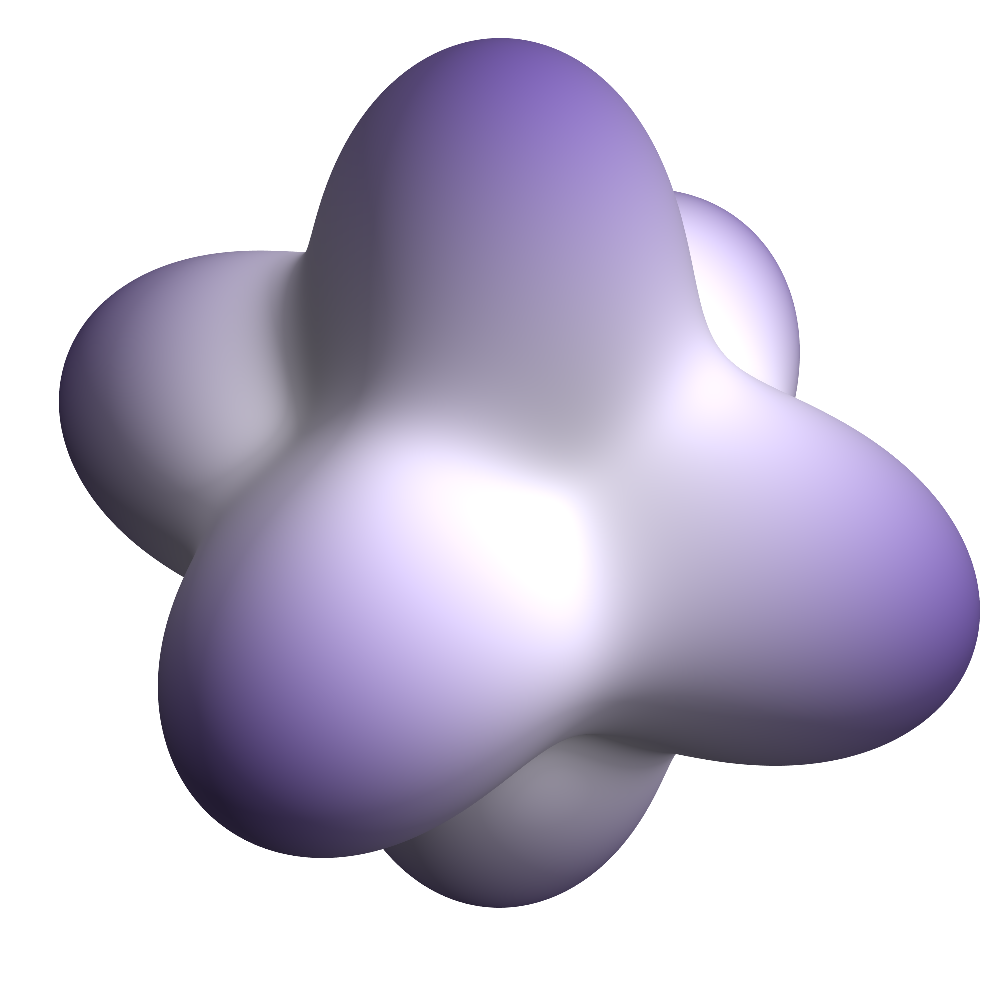}
If you are like most students of quantum mechanics, then you have begun your quantum studies by hearing stories about experiments such as
Young's double slit,\footnote{See \url{https://en.wikipedia.org/wiki/Double-slit_experiment}.}\index{double slit experiment}
the Stern--Gerlach spin quantization,\footnote{See \url{https://en.wikipedia.org/wiki/Stern-Gerlach_experiment}.}\index{Stern-Gerlach experiment@Stern--Gerlach experiment}
and
Heisenberg's uncertainty principle.\footnote{See \url{https://en.wikipedia.org/wiki/Uncertainty_principle}.}\index{uncertainty principle}
Many concepts and analogies are introduced to get an idea of what quantum mechanics is about and to begin to develop an intuition for it. Yet there is a large gap between this kind of qualitative understanding and being able to solve even the simplest quantum-mechanical problems on a computer, essentially because a computer only works with numbers, not with stories, analogies, or visualizations.

The goal of this chapter is to connect the fundamental quantum-mechanical concepts to representations that a computer can understand. We develop the tools that will be used in the remaining chapters to express and solve interesting quantum-mechanical problems.

\clearpage
\section{basis sets and representations}\index{basis set}
\label{sec:bases}

Quantum-mechanical problems are usually specified in terms of operators and quantum states. The quantum states are elements of a Hilbert space\index{Hilbert space}; the operators act on such vectors. How can these objects be represented on a computer, which only understands numbers but not Hilbert spaces?

In order to find a computer-representable form of these abstract objects, we assume that we know an ortho-normal\footnote{The following calculations can be extended to situations where the basis is not ortho-normal. For the scope of this lecture we are however not interested in this complication.} basis $\{\ket{i}\}_i$ of this Hilbert space, with scalar product $\scp{i}{j}=\delta_{i j}$. In \autoref{sec:basisconstruction} we will talk about how to construct such bases. For now we make the assumption that this basis is complete\index{completeness relation}, such that $\sum_i\ket{i}\bra{i}=\one$. We will see in \autoref{sec:incompletebasis} how to deal with incomplete basis sets.

Given any operator\index{operator} $\op{\mathcal{A}}$ acting on this Hilbert space, we use the completeness relation\index{completeness relation} twice to find
\begin{equation}
	\label{eq:1A1}
	\op{\mathcal{A}} = \one\cdot\op{\mathcal{A}}\cdot\one
	= \left[ \sum_i \ket{i}\bra{i} \right] \cdot\op{\mathcal{A}}\cdot \left[ \sum_j \ket{j}\bra{j} \right]
	= \sum_{i j}\underbrace{\me{i}{\op{\mathcal{A}}}{j}}_{A_{i j}} \,\, \ket{i}\bra{j}.
\end{equation}
We define a numerical matrix $\matr{A}$ with elements $A_{i j}=\me{i}{\op{\mathcal{A}}}{j}\in\dsC$ to rewrite this as
\begin{equation}
	\label{eq:opbasis}
	\op{\mathcal{A}}
	= \sum_{i j}A_{i j} \,\, \ket{i}\bra{j}.
\end{equation}
The same can be done with a state vector $\ket{\psi}$\index{quantum state}: using the completeness relation\index{completeness relation},
\begin{equation}
	\label{eq:1psi}
	\ket{\psi} = \one\cdot\ket{\psi} = \left[ \sum_i \ket{i}\bra{i} \right] \cdot\ket{\psi}
	= \sum_i \underbrace{\scp{i}{\psi}}_{\psi_i} \,\, \ket{i},
\end{equation}
and by defining a numerical vector $\vect{\psi}$ with elements $\psi_i=\scp{i}{\psi}\in\dsC$ the state vector is
\begin{equation}
	\label{eq:vectbasis}
	\ket{\psi}
	= \sum_i \psi_i \ket{i}.
\end{equation}
Both the matrix $\matr{A}$\index{Mathematica!matrix} and the vector $\vect{\psi}$\index{Mathematica!vector} are complex-valued objects which can be represented in any computer system. \autoref{eq:opbasis} and \autoref{eq:vectbasis} serve to convert between Hilbert-space representations and number-based (matrix/vector-based) representations. These equations are at the center of what it means to find a computer representation of a quantum-mechanical problem.

\subsection{incomplete basis sets}\index{basis set!incomplete}
\label{sec:incompletebasis}

For infinite-dimensional Hilbert spaces\index{Hilbert space} we must usually content ourselves with finite basis sets that approximate the low-energy physics (or, more generally, the physically relevant dynamics) of the problem. In practice this means that an orthonormal basis set may not be complete,\index{completeness relation}
\begin{equation}
	\sum_i \ket{i}\bra{i}=\op{P},
\end{equation}
which is the projector onto that subspace of the full Hilbert space which the basis is capable of describing. We denote $\op{Q}=\one-\op{P}$ as the complement of this projector: $\op{Q}$ is the projector onto the remainder of the Hilbert space that is left out of this truncated description. The equivalent of \autoref{eq:1A1} is then
\begin{multline}
	\label{eq:1A1inc}
	\op{\mathcal{A}} = \one\cdot\op{\mathcal{A}}\cdot\one
	= (\op{P}+\op{Q})\cdot\op{\mathcal{A}}\cdot(\op{P}+\op{Q})
	= \op{P}\cdot\op{\mathcal{A}}\cdot\op{P}
	+ \op{P}\cdot\op{\mathcal{A}}\cdot\op{Q}
	+ \op{Q}\cdot\op{\mathcal{A}}\cdot\op{P}
	+ \op{Q}\cdot\op{\mathcal{A}}\cdot\op{Q}\\
	= \underbrace{\sum_{i j}A_{i j} \,\, \ket{i}\bra{j}}\ix{within described subspace}
	+ \underbrace{\op{P}\cdot\op{\mathcal{A}}\cdot\op{Q}
	+ \op{Q}\cdot\op{\mathcal{A}}\cdot\op{P}}\ix{neglected coupling to (high-energy) part}
	+ \underbrace{\op{Q}\cdot\op{\mathcal{A}}\cdot\op{Q}}\ix{neglected (high-energy) part}
\end{multline}
In the same way, the equivalent of \autoref{eq:1psi} is
\begin{equation}
	\label{eq:1psiinc}
	\ket{\psi} = \one\cdot\ket{\psi} = (\op{P}+\op{Q})\cdot\ket{\psi}
	= \underbrace{\sum_i \psi_i \,\, \ket{i}}\ix{within described subspace} + \underbrace{\op{Q}\ket{\psi}}\ix{neglected (high-energy) part}
\end{equation}
Since $\op{Q}$ is the projector onto the neglected subspace, the component $\op{Q}\ket{\psi}$ of \autoref{eq:1psiinc} is the part of the quantum state $\ket{\psi}$ that is left out of the description in the truncated basis. In specific situations we will need to make sure that all terms involving $\op{Q}$ in \autoref{eq:1A1inc} and \autoref{eq:1psiinc} can be safely neglected. See \autoref{eq:poperator} for a problematic example of an operator expressed in a truncated basis.

\subsubsection{variational ground-state calculations}

Calculating the ground state of a Hamiltonian in an incomplete basis set is a special case of the variational method.\footnote{See \url{https://en.wikipedia.org/wiki/Variational_method_(quantum_mechanics)}.} As we will see for example in \autoref{sec:gravitywell}, the variational ground-state energy is always larger than the true ground-state energy. When we add more basis functions, the numerically calculated ground-state energy decreases monotonically. At the same time, the overlap (scalar product) of the numerically calculated ground state with the true ground state monotonically increases to unity. These convergence properties often allow us to judge whether or not a chosen computational basis set is sufficiently complete.

\subsection{exercises}

\begin{questions}
	\item\label{Q:spinhalf} We describe a spin-$1/2$ system in the basis $\mathcal{B}$ containing the two states
		\begin{align}
			\ket{\Uparrow_{\vartheta,\varphi}} &= \cos\left(\frac{\vartheta}{2}\right)\ket{\uparrow}
				+ e^{\ii\varphi}\sin\left(\frac{\vartheta}{2}\right)\ket{\downarrow}\nonumber\\
			\ket{\Downarrow_{\vartheta,\varphi}} &= -e^{-\ii\varphi}\sin\left(\frac{\vartheta}{2}\right)\ket{\uparrow}
				+ \cos\left(\frac{\vartheta}{2}\right)\ket{\downarrow}
		\end{align}
		\begin{enumerate}
			\item Show that the basis $\mathcal{B}=\{\ket{\Uparrow_{\vartheta,\varphi}},\ket{\Downarrow_{\vartheta,\varphi}}\}$ is orthonormal.
			\item Show that the basis $\mathcal{B}$ is complete: $\ket{\Uparrow_{\vartheta,\varphi}}\bra{\Uparrow_{\vartheta,\varphi}}+\ket{\Downarrow_{\vartheta,\varphi}}\bra{\Downarrow_{\vartheta,\varphi}}=\one$.
			\item Express the states $\ket{{\uparrow}}$ and $\ket{{\downarrow}}$ as vectors in the basis $\mathcal{B}$.
			\item Express the Pauli operators\index{Pauli matrices} $\op{\sigma}_x$, $\op{\sigma}_y$, $\op{\sigma}_z$ as matrices in the basis $\mathcal{B}$.
			\item Show that $\ket{\Uparrow_{\vartheta,\varphi}}$ and $\ket{\Downarrow_{\vartheta,\varphi}}$ are eigenvectors of $\op{\sigma}(\vartheta,\varphi) = \op{\sigma}_x\sin(\vartheta)\cos(\varphi)+\op{\sigma}_y\sin(\vartheta)\sin(\varphi)+\op{\sigma}_z\cos(\vartheta)$. What are the eigenvalues?
		\end{enumerate}
\pagenote[\ref{Q:spinhalf}]{We use the computational basis $\{\ket{{\uparrow}},\ket{{\downarrow}}\}$, in which the two given basis functions are
\begin{mathematica}
	¤protect¤mathin¤ up[¤mmtheta_,¤mmphi]¤ =¤ {Cos[¤mmtheta/2],¤ E^(I*¤mmphi)*Sin[¤mmtheta/2]};
	¤protect¤mathin¤ dn[¤mmtheta_,¤mmphi]¤ =¤ {-E^(-I*¤mmphi)*Sin[¤mmtheta/2],¤ Cos[¤mmtheta/2]};
\end{mathematica}
The corresponding $\bra{\Uparrow_{\vartheta,\varphi}}$ and $\bra{\Downarrow_{\vartheta,\varphi}}$ are calculated with \mm{Conjugate} (see \autoref{sec:complex}).
\begin{enumerate}
	\item Calculate $\scp{\Uparrow_{\vartheta,\varphi}}{\Uparrow_{\vartheta,\varphi}}=1$, $\scp{\Uparrow_{\vartheta,\varphi}}{\Downarrow_{\vartheta,\varphi}}=0$, $\scp{\Downarrow_{\vartheta,\varphi}}{\Uparrow_{\vartheta,\varphi}}=0$, $\scp{\Downarrow_{\vartheta,\varphi}}{\Downarrow_{\vartheta,\varphi}}=1$:
\begin{mathematica}
	¤protect¤mathin¤ Conjugate[up[¤mmtheta,¤mmphi]].up[¤mmtheta,¤mmphi]¤ //ComplexExpand¤ //FullSimplify
	¤protect¤mathout¤ 1
	¤protect¤mathin¤ Conjugate[up[¤mmtheta,¤mmphi]].dn[¤mmtheta,¤mmphi]¤ //ComplexExpand¤ //FullSimplify
	¤protect¤mathout¤ 0
	¤protect¤mathin¤ Conjugate[dn[¤mmtheta,¤mmphi]].up[¤mmtheta,¤mmphi]¤ //ComplexExpand¤ //FullSimplify
	¤protect¤mathout¤ 0
	¤protect¤mathin¤ Conjugate[dn[¤mmtheta,¤mmphi]].dn[¤mmtheta,¤mmphi]¤ //ComplexExpand¤ //FullSimplify
	¤protect¤mathout¤ 1
\end{mathematica}
	\item Construct the ket-bra products with \mm{KroneckerProduct}:
\begin{mathematica}
	¤protect¤mathin¤ KroneckerProduct[up[¤mmtheta,¤mmphi],¤ Conjugate[up[¤mmtheta,¤mmphi]]]¤ +
	¤protect¤mathnl¤ KroneckerProduct[dn[¤mmtheta,¤mmphi],¤ Conjugate[dn[¤mmtheta,¤mmphi]]]¤ //
	¤protect¤mathnl¤ ComplexExpand¤ //FullSimplify
	¤protect¤mathout¤ {{1,¤ 0},¤ {0,¤ 1}}
\end{mathematica}
	\item $\ket{\uparrow}=\ket{\Uparrow_{\vartheta,\varphi}}\scp{\Uparrow_{\vartheta,\varphi}}{\uparrow}+\ket{\Downarrow_{\vartheta,\varphi}}\scp{\Downarrow_{\vartheta,\varphi}}{\uparrow}=\cos(\vartheta/2)\ket{\Uparrow_{\vartheta,\varphi}}-e^{\ii\varphi}\sin(\vartheta/2)\ket{\Downarrow_{\vartheta,\varphi}}$:
\begin{mathematica}
	¤protect¤mathin¤ Cos[¤mmtheta/2]*up[¤mmtheta,¤mmphi]¤ -¤ E^(I*¤mmphi)*Sin[¤mmtheta/2]*dn[¤mmtheta,¤mmphi]¤ //FullSimplify
	¤protect¤mathout¤ {1,¤ 0}
\end{mathematica}
	$\ket{\downarrow}=\ket{\Uparrow_{\vartheta,\varphi}}\scp{\Uparrow_{\vartheta,\varphi}}{\downarrow}+\ket{\Downarrow(\vartheta,\varphi)}\scp{\Downarrow_{\vartheta,\varphi}}{\downarrow}=e^{-\ii\varphi}\sin(\vartheta/2)\ket{\Uparrow_{\vartheta,\varphi}}+\cos(\vartheta/2)\ket{\Downarrow_{\vartheta,\varphi}}$:
\begin{mathematica}
	¤protect¤mathin¤ E^(-I*¤mmphi)*Sin[¤mmtheta/2]*up[¤mmtheta,¤mmphi]¤ +¤ Cos[¤mmtheta/2]*dn[¤mmtheta,¤mmphi]¤ //FullSimplify
	¤protect¤mathout¤ {0,¤ 1}
\end{mathematica}
	\item The Pauli operators are defined in Mathematica in our computational basis with the \mm{PauliMatrix} command.\par
\noindent The matrix elements of the Pauli operator $\op{\sigma}_x$ are
\begin{mathematica}
	¤protect¤mathin¤ sx¤ =¤ PauliMatrix[1];
	¤protect¤mathin¤ Conjugate[up[¤mmtheta,¤mmphi]].sx.up[¤mmtheta,¤mmphi]¤ //ComplexExpand¤ //FullSimplify
	¤protect¤mathout¤ Sin[¤mmtheta]*Cos[¤mmphi]
	¤protect¤mathin¤ Conjugate[up[¤mmtheta,¤mmphi]].sx.dn[¤mmtheta,¤mmphi]¤ //ComplexExpand¤ //FullSimplify
	¤protect¤mathout¤ Exp[-I*¤mmphi]*(Cos[¤mmtheta]*Cos[¤mmphi]+I*Sin[¤mmphi])
	¤protect¤mathin¤ Conjugate[dn[¤mmtheta,¤mmphi]].sx.up[¤mmtheta,¤mmphi]¤ //ComplexExpand¤ //FullSimplify
	¤protect¤mathout¤ Exp[I*¤mmphi]*(Cos[¤mmtheta]*Cos[¤mmphi]-I*Sin[¤mmphi])
	¤protect¤mathin¤ Conjugate[dn[¤mmtheta,¤mmphi]].sx.dn[¤mmtheta,¤mmphi]¤ //ComplexExpand¤ //FullSimplify
	¤protect¤mathout¤ -Sin[¤mmtheta]*Cos[¤mmphi]
	¤protect¤mathin¤ sx¤ ==¤ Sin[¤mmtheta]*Cos[¤mmphi]¤ *¤ KroneckerProduct[up[¤mmtheta,¤mmphi],¤ Conjugate[up[¤mmtheta,¤mmphi]]]¤ +
	¤protect¤mathnl¤       E^(-I*¤mmphi)*(Cos[¤mmtheta]*Cos[¤mmphi]+I*Sin[¤mmphi])¤ *
	¤protect¤mathnl¤ KroneckerProduct[up[¤mmtheta,¤mmphi],¤ Conjugate[dn[¤mmtheta,¤mmphi]]]¤ +
	¤protect¤mathnl¤       E^(I*¤mmphi)*(Cos[¤mmtheta]*Cos[¤mmphi]-I*Sin[¤mmphi])¤ *
	¤protect¤mathnl¤ KroneckerProduct[dn[¤mmtheta,¤mmphi],¤ Conjugate[up[¤mmtheta,¤mmphi]]]¤ -
	¤protect¤mathnl¤        Sin[¤mmtheta]*Cos[¤mmphi]¤ *¤ KroneckerProduct[dn[¤mmtheta,¤mmphi],¤ Conjugate[dn[¤mmtheta,¤mmphi]]]¤ //
	¤protect¤mathnl¤ ComplexExpand¤ //FullSimplify
	¤protect¤mathout¤ True
\end{mathematica}
The matrix elements of the Pauli operator $\op{\sigma}_y$ are
\begin{mathematica}
	¤protect¤mathin¤ sy¤ =¤ PauliMatrix[2];
	¤protect¤mathin¤ Conjugate[up[¤mmtheta,¤mmphi]].sy.up[¤mmtheta,¤mmphi]¤ //ComplexExpand¤ //FullSimplify
	¤protect¤mathout¤ Sin[¤mmtheta]*Sin[¤mmphi]
	¤protect¤mathin¤ Conjugate[up[¤mmtheta,¤mmphi]].sy.dn[¤mmtheta,¤mmphi]¤ //ComplexExpand¤ //FullSimplify
	¤protect¤mathout¤ Exp[-I*¤mmphi]*(Cos[¤mmtheta]*Sin[¤mmphi]-I*Cos[¤mmphi])
	¤protect¤mathin¤ Conjugate[dn[¤mmtheta,¤mmphi]].sy.up[¤mmtheta,¤mmphi]¤ //ComplexExpand¤ //FullSimplify
	¤protect¤mathout¤ Exp[I*¤mmphi]*(Cos[¤mmtheta]*Sin[¤mmphi]+I*Cos[¤mmphi])
	¤protect¤mathin¤ Conjugate[dn[¤mmtheta,¤mmphi]].sy.dn[¤mmtheta,¤mmphi]¤ //ComplexExpand¤ //FullSimplify
	¤protect¤mathout¤ -Sin[¤mmtheta]*Sin[¤mmphi]
	¤protect¤mathin¤ sy¤ ==¤ Sin[¤mmtheta]*Sin[¤mmphi]¤ *¤ KroneckerProduct[up[¤mmtheta,¤mmphi],¤ Conjugate[up[¤mmtheta,¤mmphi]]]¤ +
	¤protect¤mathnl¤       E^(-I*¤mmphi)*(Cos[¤mmtheta]*Sin[¤mmphi]-I*Cos[¤mmphi])¤ *
	¤protect¤mathnl¤ KroneckerProduct[up[¤mmtheta,¤mmphi],¤ Conjugate[dn[¤mmtheta,¤mmphi]]]¤ +
	¤protect¤mathnl¤       E^(I*¤mmphi)*(Cos[¤mmtheta]*Sin[¤mmphi]+I*Cos[¤mmphi])¤ *
	¤protect¤mathnl¤ KroneckerProduct[dn[¤mmtheta,¤mmphi],¤ Conjugate[up[¤mmtheta,¤mmphi]]]¤ -
	¤protect¤mathnl¤       Sin[¤mmtheta]*Sin[¤mmphi]¤ *¤ KroneckerProduct[dn[¤mmtheta,¤mmphi],¤ Conjugate[dn[¤mmtheta,¤mmphi]]]¤ //
	¤protect¤mathnl¤ ComplexExpand¤ //FullSimplify
	¤protect¤mathout¤ True
\end{mathematica}
The matrix elements of the Pauli operator $\op{\sigma}_z$ are
\begin{mathematica}
	¤protect¤mathin¤ sz¤ =¤ PauliMatrix[3];
	¤protect¤mathin¤ Conjugate[up[¤mmtheta,¤mmphi]].sz.up[¤mmtheta,¤mmphi]¤ //ComplexExpand¤ //FullSimplify
	¤protect¤mathout¤ Cos[¤mmtheta]
	¤protect¤mathin¤ Conjugate[up[¤mmtheta,¤mmphi]].sz.dn[¤mmtheta,¤mmphi]¤ //ComplexExpand¤ //FullSimplify
	¤protect¤mathout¤ -Exp[-I*¤mmphi]*Sin[¤mmtheta]
	¤protect¤mathin¤ Conjugate[dn[¤mmtheta,¤mmphi]].sz.up[¤mmtheta,¤mmphi]¤ //ComplexExpand¤ //FullSimplify
	¤protect¤mathout¤ -Exp[I*¤mmphi]*Sin[¤mmtheta]
	¤protect¤mathin¤ Conjugate[dn[¤mmtheta,¤mmphi]].sz.dn[¤mmtheta,¤mmphi]¤ //ComplexExpand¤ //FullSimplify
	¤protect¤mathout¤ -Cos[¤mmtheta]
	¤protect¤mathin¤ sz¤ ==¤ Cos[¤mmtheta]¤ *¤ KroneckerProduct[up[¤mmtheta,¤mmphi],¤ Conjugate[up[¤mmtheta,¤mmphi]]]¤ -
	¤protect¤mathnl¤       E^(-I*¤mmphi)*Sin[¤mmtheta]¤ *¤ KroneckerProduct[up[¤mmtheta,¤mmphi],¤ Conjugate[dn[¤mmtheta,¤mmphi]]]¤ -
	¤protect¤mathnl¤       E^(I*¤mmphi)*Sin[¤mmtheta]¤ *¤ KroneckerProduct[dn[¤mmtheta,¤mmphi],¤ Conjugate[up[¤mmtheta,¤mmphi]]]¤ -
	¤protect¤mathnl¤        Cos[¤mmtheta]¤ *¤ KroneckerProduct[dn[¤mmtheta,¤mmphi],¤ Conjugate[dn[¤mmtheta,¤mmphi]]]¤ //
	¤protect¤mathnl¤ ComplexExpand¤ //FullSimplify
	¤protect¤mathout¤ True
\end{mathematica}
	\item We check the eigenvalue equations with eigenvalues $\pm1$:
\begin{mathematica}
	¤protect¤mathin¤ s¤ =¤ sx*Sin[¤mmtheta]*Cos[¤mmphi]¤ +¤ sy*Sin[¤mmtheta]*Sin[¤mmphi]¤ +¤ sz*Cos[¤mmtheta];
	¤protect¤mathin¤ Eigenvalues[s]
	¤protect¤mathout¤ {-1,¤ 1}
	¤protect¤mathin¤ s.up[¤mmtheta,¤mmphi]¤ ==¤ up[¤mmtheta,¤mmphi]¤ //FullSimplify
	¤protect¤mathout¤ True
	¤protect¤mathin¤ s.dn[¤mmtheta,¤mmphi]¤ ==¤ -dn[¤mmtheta,¤mmphi]¤ //FullSimplify
	¤protect¤mathout¤ True
\end{mathematica}
\end{enumerate}}
	\item\label{Q:squarewellbasissum} The eigenstate basis for the description of the infinite square well of unit width is made up of the ortho-normalized functions
		\begin{equation}
			\label{eq:momentumwavecompleteness}
			\scp{x}{n} = \phi_n(x) = \sqrt{2}\sin(n\pi x)
		\end{equation}
		defined on the interval $[0,1]$, with $n\in\{1,2,3,\dots\}$.
		\begin{enumerate}
			\item Calculate the function $P_{\infty}(x,y)=\bra{x}\left[\sum_{n=1}^{\infty}\ket{n}\bra{n}\right]\ket{y}$.
			\item In computer-based calculations we limit the basis set to $n\in\{1,2,3,\dots,n\ix{max}\}$ for some large value of $n\ix{max}$. Using Mathematica, calculate the function $P_{n\ix{max}}(x,y)=\bra{x}\left[\sum_{n=1}^{n\ix{max}}\ket{n}\bra{n}\right]\ket{y}$ (use the \mm{Sum} function). Make a plot for $n\ix{max}=10$ (use the \mm{DensityPlot} function).
			\item What does the function $P$ represent?
		\end{enumerate}
\pagenote[\ref{Q:squarewellbasissum}]{
\begin{enumerate}
	\item Since $\sum_{n=1}^{\infty}\ket{n}\bra{n}=\one$, we have $P_{\infty}(x,y)=\bra{x}\one\ket{y}=\scp{x}{y}=\delta(x-y)$.
	\item $P_{n\ix{max}}(x,y)=\bra{x}\left[\sum_{n=1}^{n\ix{max}}\ket{n}\bra{n}\right]\ket{y}=\sum_{n=1}^{n\ix{max}}\scp{x}{n}\scp{n}{y}=2\sum_{n=1}^{n\ix{max}}\sin(n\pi x)\sin(n\pi y)$:
\begin{mathematica}
	¤protect¤mathin¤ With[{nmax¤ =¤ 10},
	¤protect¤mathnl¤ P[x_,¤ y_]¤ =¤ 2*Sum[Sin[n*¤mmpi*x]*Sin[n*¤mmpi*y],¤ {n,¤ nmax}];
	¤protect¤mathnl¤ DensityPlot[P[x,¤ y],¤ {x,¤ 0,¤ 1},¤ {y,¤ 0,¤ 1},
	¤protect¤mathnl¤ PlotRange¤ ->¤ All,¤ PlotPoints¤ ->¤ 2*nmax]]
\end{mathematica}
\begin{center}
\ifthenelse{\boolean{smallfigures}}%
{\includegraphics[width=0.5\textwidth]{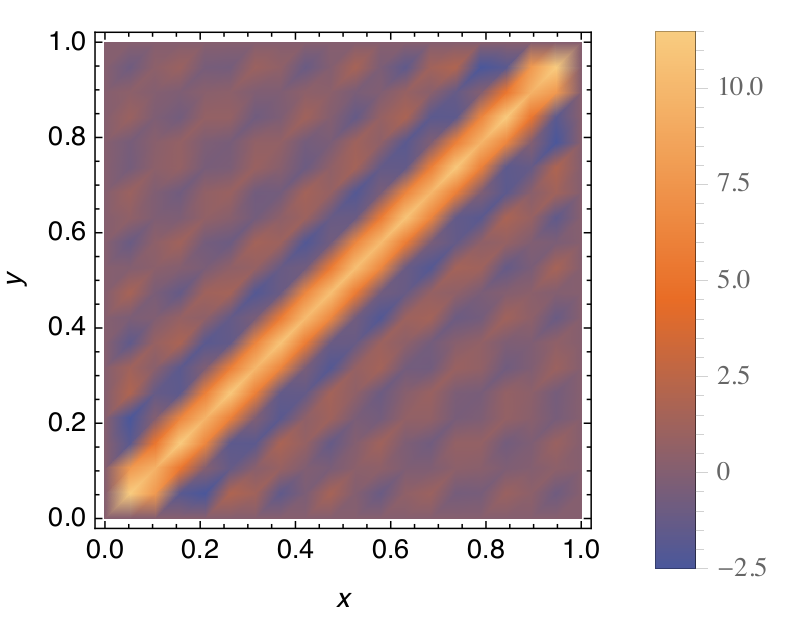}}%
{\includegraphics[width=0.5\textwidth]{sinesum}}
\end{center}
	\item The operator $\op{\Pi}_{n\ix{max}}=\sum_{n=1}^{n\ix{max}}\ket{n}\bra{n}$ is the projector onto the computational subspace (see \autoref{sec:incompletebasis}). The function $P_{n\ix{max}}(x,y)=\me{x}{\op{\Pi}_{n\ix{max}}}{y}$ is its real-space representation. Since the plot of $P_{n\ix{max}}(x,y)$ has a finite spatial resolution (\ie, no structure at length scales smaller than $1/n\ix{max}$), we see that this projection operator $\op{\Pi}_{n\ix{max}}$ is associated with a spatial smoothing operation.
\end{enumerate}}
\end{questions}

\section{time-independent Schr\"odinger equation}\index{Schroedinger equation@Schr\"odinger equation!time-independent}
\label{sec:TimeIndepSchr}

The time-independent Schr\"odinger equation is
\begin{equation}
	\label{eq:TimeIndepSchr}
	\Ham\ket{\psi}=E\ket{\psi}.
\end{equation}
As in \autoref{sec:bases} we use a computational basis to express the Hamiltonian operator $\Ham$ and the quantum state $\psi$ as
\begin{align}
	\Ham &= \sum_{i j} H_{i j} \,\, \ket{i}\bra{j}, &
	\ket{\psi} &= \sum_i \psi_i \,\,\ket{i}.
\end{align}
With these substitutions the Schr\"odinger equation becomes
\begin{align}
	\left[ \sum_{i j} H_{i j} \,\, \ket{i}\bra{j} \right] \left[ \sum_k \psi_k \,\, \ket{k} \right] &= E \left[ \sum_{\ell} \psi_{\ell} \,\, \ket{\ell} \right]\nonumber\\
	\sum_{i j k}H_{i j} \psi_k \,\, \underbrace{\scp{j}{k}}_{=\delta_{j k}} \,\, \ket{i} &= \sum_{\ell} E \psi_{\ell} \,\, \ket{\ell}\nonumber\\
	\sum_{i j}H_{i j}\psi_j \,\, \ket{i} &= \sum_{\ell} E \psi_{\ell} \,\, \ket{\ell}
\end{align}
Multiplying this equation by $\bra{m}$ from the left, and using the orthonormality of the basis set, gives
\begin{align}
	\bra{m}\sum_{i j}H_{i j}\psi_j \,\, \ket{i} &= \bra{m}\sum_{\ell} E \psi_{\ell} \,\, \ket{\ell}\nonumber\\
	\sum_{i j}H_{i j}\psi_j \underbrace{\scp{m}{i}}_{=\delta_{m i}} &= \sum_{\ell} E \psi_{\ell}\underbrace{\scp{m}{\ell}}_{=\delta_{m\ell}}\nonumber\\
	\sum_{j}H_{m j}\psi_j &= E \psi_m
\end{align}
In matrix notation this can be written as
\begin{equation}
	\label{eq:TimeIndepSchrMatrix}
	\boxed{\matr{H}\cdot\vect{\psi} = E\vect{\psi}.}
\end{equation}
This is the central equation of this lecture. It is the time-independent Schr\"odinger equation in a form that computers can understand, namely an eigenvalue equation in terms of numerical (complex) matrices and vectors.

If you think that there is no difference between \autoref{eq:TimeIndepSchr} and \autoref{eq:TimeIndepSchrMatrix}, then I invite you to re-read this section as I consider it extremely important for what follows in this course. You can think of \autoref{eq:TimeIndepSchr} as an abstract relationship between operators and vectors in Hilbert space,\index{Hilbert space} while \autoref{eq:TimeIndepSchrMatrix} is a \emph{numerical representation} of this relationship in a concrete basis set $\{\ket{i}\}_i$. They both contain the exact same information (since we converted one to the other in a few lines of mathematics) but they are conceptually very different, as one is understandable by a computer and the other is not.

%

\subsection{diagonalization}

The matrix form of \autoref{eq:TimeIndepSchrMatrix} of the Schr\"odinger equation is an eigenvalue equation as you know from linear algebra. Given a matrix of complex numbers $\matr{H}$ we can find the eigenvalues $E_i$ and eigenvectors $\vect{\psi}_i$ using Mathematica's built-in procedures, as described in \autoref{sec:diagonalization}.

\subsection{exercises}

\begin{questions}
	\item\label{Q:Hexpr} Express the spin-$1/2$ Hamiltonian
		\begin{equation}
			\label{eq:xyzham1}
			\Ham=\sin(\vartheta)\cos(\varphi) \op{\sigma}_x+\sin(\vartheta)\sin(\varphi) \op{\sigma}_y+\cos(\vartheta) \op{\sigma}_z
		\end{equation}
		in the basis $\{\ket{{\uparrow}},\ket{{\downarrow}}\}$, and calculate its eigenvalues and eigenvectors. NB: $\op{\sigma}_{x,y,z}$ are the Pauli operators.
\pagenote[\ref{Q:Hexpr}]{See \ref{Q:spinhalf}.}
\end{questions}

\section{time-dependent Schr\"odinger equation}\index{Schroedinger equation@Schr\"odinger equation!time-dependent}
\label{sec:TimeDepSchr}

The time-dependent Schr\"odinger equation is
\begin{equation}
	\label{eq:tdSchroedinger}
	\ii\hbar\frac{\dd}{\dd[t]}\ket{\psi(t)} = \Ham(t) \ket{\psi(t)},
\end{equation}
where the Hamiltonian $\Ham$ can have an explicit time dependence. This differential equation has the formal solution
\begin{equation}
	\label{eq:tdsesolgen}
	\ket{\psi(t)} = \op{\mathcal{U}}(t_0;t)\ket{\psi(t_0)}
\end{equation}
in terms of the \emph{propagator}\index{propagator}
\begin{multline}
	\label{eq:propagator}
	\op{\mathcal{U}}(t_0;t) = \one
		- \frac{\ii}{\hbar} \int_{t_0}^t \dd[t_1] \Ham(t_1)
		- \frac{1}{\hbar^2} \int_{t_0}^t \dd[t_1] \int_{t_0}^{t_1} \dd[t_2] \Ham(t_1) \Ham(t_2)
		+ \frac{\ii}{\hbar^3} \int_{t_0}^t \dd[t_1] \int_{t_0}^{t_1} \dd[t_2] \int_{t_0}^{t_2} \dd[t_3] \Ham(t_1) \Ham(t_2) \Ham(t_3)\\
		+ \frac{1}{\hbar^4} \int_{t_0}^t \dd[t_1] \int_{t_0}^{t_1} \dd[t_2] \int_{t_0}^{t_2} \dd[t_3] \int_{t_0}^{t_3} \dd[t_4] \Ham(t_1) \Ham(t_2) \Ham(t_3) \Ham(t_4)
		+ \dotsb
\end{multline}
that propagates any state from time $t_0$ to time $t$. An alternative form is given by the \emph{Magnus expansion}\index{Magnus expansion}\footnote{See \url{https://en.wikipedia.org/wiki/Magnus_expansion}.}
\begin{equation}
	\label{eq:propagatorME}
	\op{\mathcal{U}}(t_0;t) = \exp\left[\sum_{k=1}^{\infty} \op{\Omega}_k(t_0;t)\right]
\end{equation}
with the contributions
\begin{align}
	\op{\Omega}_1(t_0;t) &= -\frac{\ii}{\hbar} \int_{t_0}^t \dd[t_1] \op{\mathcal{H}}(t_1)\nonumber\\
	\op{\Omega}_2(t_0;t) &= -\frac{1}{2\hbar^2} \int_{t_0}^t \dd[t_1] \int_{t_0}^{t_1} \dd[t_2] [\op{\mathcal{H}}(t_1),\op{\mathcal{H}}(t_2)]\nonumber\\
	\op{\Omega}_3(t_0;t) &= \frac{\ii}{6\hbar^3} \int_{t_0}^t \dd[t_1] \int_{t_0}^{t_1} \dd[t_2] \int_{t_0}^{t_2} \dd[t_3] \left( [\op{\mathcal{H}}(t_1),[\op{\mathcal{H}}(t_2),\op{\mathcal{H}}(t_3)]]+[\op{\mathcal{H}}(t_3),[\op{\mathcal{H}}(t_2),\op{\mathcal{H}}(t_1)]]\right)\nonumber\\
	&\dotso
\end{align}
This expansion in terms of different-time commutators is often easier to evaluate than \autoref{eq:propagator}, especially when the contributions vanish for $k>k\ix{max}$ (see \autoref{sec:HHcommuting} for the case $k\ix{max}=1$). Even if higher-order contributions do not vanish entirely, they (usually) decrease in importance much more rapidly with increasing $k$ than those of \autoref{eq:propagator}. Also, even if the Magnus expansion is artificially truncated (neglecting higher-order terms), the quantum-mechanical evolution is still unitary; this is not the case for \autoref{eq:propagator}.

Notice that the exponential in \autoref{eq:propagatorME} haa an operator or a matrix as their argument: in Mathematica this matrix exponentiation is done with the \mm{MatrixExp} function.\index{Mathematica!matrix!exponential} It does not calculate the exponential element-by-element, but instead calculates
\begin{align}
	e^{\op{A}} &= \sum_{n=0}^{\infty} \frac{\op{A}^n}{n!}, & e^{\matr{A}} &= \sum_{n=0}^{\infty} \frac{\matr{A}^n}{n!}.
\end{align}

\subsection{time-independent basis}

We express the quantum state again in terms of the chosen basis, which is assumed to be time-independent. This leaves the time-dependence in the expansion coefficients,
\begin{align}
	\label{eq:psidecompt}
	\Ham(t) &= \sum_{i j} H_{i j}(t) \,\, \ket{i}\bra{j}, &
	\ket{\psi(t)} &= \sum_i \psi_i(t) \,\, \ket{i}.
\end{align}
Inserting these expressions into the time-dependent \hyperref[eq:tdSchroedinger]{Schr\"odinger equation~\ref*{eq:tdSchroedinger}} gives
\begin{equation}
	\ii\hbar \sum_i \dot{\psi}_i(t) \,\, \ket{i} = \left[ \sum_{j k} H_{j k}(t) \,\, \ket{j}\bra{k} \right] \sum_{\ell} \psi_{\ell}(t) \,\, \ket{\ell}
	= \sum_{j k} H_{j k}(t) \psi_k(t) \,\, \ket{j}.
\end{equation}
Multiplying with $\bra{m}$ from the left:
\begin{equation}
	\ii\hbar \dot{\psi}_{m}(t)
	= \sum_k H_{m k}(t) \psi_k(t)
\end{equation}
or, in matrix notation,
\begin{equation}
	\label{eq:tdSchrMat}
	\boxed{\ii\hbar \dotvect{\psi}(t) = \matr{H}(t)\cdot\vect{\psi}(t).}
\end{equation}
Since the matrix $\matr{H}(t)$ is supposedly known, this equation represents a system of coupled complex differential equations for the vector $\vect{\psi}(t)$, which can be solved on a computer.

\subsection{time-dependent basis: interaction picture}\index{interaction picture}

It can be advantageous to use a time-dependent basis. The most frequently used such basis is given by the interaction picture of quantum mechanics, where the Hamiltonian can be split into a time-independent principal part $\Ham_0$ and a small time-dependent part $\Ham_1$:
\begin{equation}
	\label{eq:H0H1}
	\Ham(t) = \Ham_0 + \Ham_1(t).
\end{equation}
Assuming that we can diagonalize $\Ham_0$, possibly numerically, such that the eigenfunctions satisfy $\Ham_0\ket{i}=E_i\ket{i}$, we propose the time-dependent basis
\begin{equation}
	\ket{i(t)} = e^{-\ii E_i t/\hbar}\ket{i}.
\end{equation}
If we express any quantum state in this basis as
\begin{equation}
	\label{eq:rotatingbasis}
	\ket{\psi(t)} = \sum_i \psi_i(t) \,\, \ket{i(t)} = \sum_i \psi_i(t) e^{-\ii E_i t/\hbar} \ket{i},
\end{equation}
the time-dependent Schr\"odinger equation\index{Schroedinger equation@Schr\"odinger equation!time-dependent} becomes
\begin{align}
	\sum_i \left[ \ii\hbar \dot{\psi}_i(t) + E_i \psi_i(t) \right] e^{-\ii E_i t/\hbar} \ket{i}
	&= \sum_j \psi_j(t) e^{-\ii E_j t/\hbar} E_j \,\, \ket{j} + \sum_j \psi_j(t) e^{-\ii E_j t/\hbar} \Ham_1(t) \,\, \ket{j}\nonumber\\
	\sum_i \ii\hbar \dot{\psi}_i(t) e^{-\ii E_i t/\hbar} \ket{i}
	&= \sum_j \psi_j(t) e^{-\ii E_j t/\hbar} \Ham_1(t) \,\, \ket{j}
\end{align}
Multiply by $\bra{k}$ from the left:
\begin{align}
	\label{eq:tdseint}
	\bra{k}\sum_i \ii\hbar \dot{\psi}_i(t) e^{-\ii E_i t/\hbar} \ket{i}
	&= \bra{k}\sum_j \psi_j(t) e^{-\ii E_j t/\hbar} \Ham_1(t) \,\, \ket{j}\nonumber\\
	\sum_i \ii\hbar \dot{\psi}_i(t) e^{-\ii E_i t/\hbar} \underbrace{\scp{k}{i}}_{=\delta_{k i}}
	&= \sum_j \psi_j(t) e^{-\ii E_j t/\hbar} \me{k}{\Ham_1(t)}{j}\nonumber\\
	\ii\hbar \dot{\psi}_k(t)
	&= \sum_j \psi_j(t) e^{-\ii (E_j-E_k) t/\hbar} \me{k}{\Ham_1(t)}{j}.
\end{align}
This is the same matrix/vector evolution expression as \autoref{eq:tdSchrMat}, except that here the Hamiltonian matrix elements must be defined as
\begin{equation}
	\label{eq:SchrHamIntPic}
	H_{i j}(t) = \me{i}{\Ham_1(t)}{j} e^{-\ii (E_j-E_i) t/\hbar}.
\end{equation}
We see immediately that if the interaction Hamiltonian vanishes [$\Ham_1(t)=0$], then the expansion coefficients $\psi_i(t)$ become time-independent, as expected since they are the coefficients of the eigenfunctions of the time-independent Schr\"odinger equation.

When a quantum-mechanical system is composed of different parts that have vastly different energy scales of their internal evolution $\Ham_0$, then the use of \autoref{eq:SchrHamIntPic} can have great numerical advantages. It turns out that the relevant interaction terms $H_{i j}(t)$ in the interaction picture will have relatively slowly evolving phases $\exp[-\ii(E_j-E_i)t/\hbar]$, on a time scale given by relative energy \emph{differences} and not by \emph{absolute} energies; this makes it possible to solve the coupled differential equations of \autoref{eq:tdSchrMat} numerically without using an absurdly small time step.

\subsection{special case: $\left[\Ham(t),\Ham(t')\right]=0$ $\forall (t,t')$}
\label{sec:HHcommuting}

If the Hamiltonian commutes with itself at different times, $\left[\Ham(t),\Ham(t')\right]=0$ $\forall (t,t')$, the \hyperref[eq:propagatorME]{propagator~\ref*{eq:propagatorME}} of \autoref{eq:tdSchroedinger} can be simplified to
\begin{equation}
	\label{eq:tdprop1}
	\op{\mathcal{U}}(t_0;t) = \exp\left[-\frac{\ii}{\hbar}\int_{t_0}^t\Ham(s)\dd[s]\right],
\end{equation}
and the corresponding solution of \autoref{eq:tdSchrMat} is
\begin{equation}
	\label{eq:tdSchrMatSol1}
	\vect{\psi}(t) = \exp\left[-\frac{\ii}{\hbar}\int_{t_0}^t\matr{H}(s)\dd[s]\right]\cdot\vect{\psi}(t_0).
\end{equation}
Again, these matrix exponentials are calculated with \mm{MatrixExp} in Mathematica.\index{Mathematica!matrix!exponential}

\subsection{special case: time-independent Hamiltonian}
\label{sec:timeindependentH}

In the special (but common) case where the Hamiltonian is time-independent, the integral in \autoref{eq:tdSchrMatSol1} can be evaluated immediately, and the solution is
\begin{equation}
	\label{eq:tdSchrMatSol2}
	\vect{\psi}(t) = \exp\left[-\frac{\ii(t-t_0)}{\hbar}\matr{H}\right]\cdot\vect{\psi}(t_0).
\end{equation}
If we have a specific Hamiltonian matrix \mm{H} defined, for example the matrix of \autoref{sec:diagonalization}, we can calculate the propagator $\matr{U}(\Delta t)=\exp[-\ii\matr{H}\Delta t/\hbar]$ for $\Delta t=t-t_0$ with\index{Mathematica!matrix!exponential}
\begin{mathematica}
	¤mathin¤labelŽmath:Ugeneral U[¤mmDeltaŽt_] = MatrixExp[-I*H*¤mmDeltaŽt/¤mmhbar]
\end{mathematica}
The resulting expression for \mm{U[\mmDelta{}t]} will in general be very long, and slow to compute. A more efficient definition is to matrix-exponentiate a numerical matrix for specific values of the propagation interval \mm{\mmDelta{}t}, using a delayed assignment:
\begin{mathematica}
	¤mathin¤labelŽmath:Unumerical U[¤mmDeltaŽt_?NumericQ] := MatrixExp[-I*H*N[¤mmDeltaŽt]/¤mmhbar]
\end{mathematica}

\subsection{exercises}

\begin{questions}
	\item\label{Q:propagator} Demonstrate that the \hyperref[eq:tdprop1]{propagator~\ref*{eq:tdprop1}} gives a \hyperref[eq:tdsesolgen]{quantum state~\ref*{eq:tdsesolgen}} that satisfies \autoref{eq:tdSchroedinger}.
\pagenote[\ref{Q:propagator}]{Inserting \autoref{eq:tdprop1} into \autoref{eq:tdsesolgen} gives the quantum state
\begin{equation}
	\ket{\psi(t)} = \exp\left[-\frac{\ii}{\hbar}\int_{t_0}^t\Ham(s)\dd[s]\right]\ket{\psi(t_0)}
\end{equation}
We calculate its time-derivative with the chain rule and $\frac{\dd}{\dd[x]}\left[\int_{f(x)}^{g(x)}h(x,y)\dd[y]\right]=h(x,g(x))g'(x)-h(x,f(x))f'(x)+\int_{f(x)}^{g(x)}\frac{\ddd{\partial}{h}(x,y)}{\ddd{\partial}{x}}\dd[y]$:
\begin{equation}
	\frac{\dd}{\dd[t]}\ket{\psi(t)} = -\frac{\ii}{\hbar}\Ham(t)\exp\left[-\frac{\ii}{\hbar}\int_{t_0}^t\Ham(s)\dd[s]\right]\ket{\psi(t_0)}
	= -\frac{\ii}{\hbar}\Ham(t)\ket{\psi(t)},
\end{equation}
which is the \hyperref[eq:tdSchroedinger]{Schr\"odinger equation~\ref*{eq:tdSchroedinger}}.}
	\item\label{Q:spinhalfprop} Calculate the propagator of the Hamiltonian of \ref{Q:Hexpr}.
\pagenote[\ref{Q:spinhalfprop}]{The Hamiltonian is
\begin{mathematica}
	¤protect¤mathin¤ {sx,sy,sz}¤ =¤ Table[PauliMatrix[i],¤ {i,¤ 3}];
	¤protect¤mathin¤ H¤ =¤ Sin[¤mmtheta]*Cos[¤mmphi]*sx¤ +¤ Sin[¤mmtheta]*Sin[¤mmphi]*sy¤ +¤ Cos[¤mmtheta]*sz¤ //FullSimplify
	¤protect¤mathout¤ {{Cos[¤mmtheta],¤ E^(-I*¤mmphi)*Sin[¤mmtheta]},¤ {E^(I*¤mmphi)*Sin[¤mmtheta],¤ -Cos[¤mmtheta]}}
\end{mathematica}
and the propagator is calculated from \autoref{eq:tdSchrMatSol2}
\begin{mathematica}
	¤protect¤mathin¤ U¤ =¤ MatrixExp[-I*(t-t0)/¤mmhbar*H]¤ //FullSimplify
	¤protect¤mathout¤ {{Cos[(t-t0)/¤mmhbar]-I*Cos[¤mmtheta]*Sin[(t-t0)/¤mmhbar],¤ -I*E^(-I*¤mmphi)*Sin[¤mmtheta]*Sin[(t-t0)/¤mmhbar]},
	¤protect¤mathnl¤ {-I*E^(I*¤mmphi)*Sin[¤mmtheta]*Sin[(t-t0)/¤mmhbar],¤ Cos[(t-t0)/¤mmhbar]+I*Cos[¤mmtheta]*Sin[(t-t0)/¤mmhbar]}}
\end{mathematica}}
	\item\label{Q:Uorder} After \mm{\ref{math:Ugeneral}} and \mm{\ref{math:Unumerical}}, check \mm{?U}. Which definition of \mm{U} comes first? Why?
\pagenote[\ref{Q:Uorder}]{The definitions are ordered with decreasing specificity:
\begin{mathematica}
	¤protect¤mathin¤ ?U
	¤protect¤mathnlp¤ Global`U
	¤protect¤mathnlp¤ U[¤mmtau_?NumericQ]¤ :=¤ MatrixExp[-I¤ H¤ N[¤mmtau]]
 	¤protect¤mathnlp¤ U[¤mmtau_]¤ =¤ MatrixExp[-I¤ H¤ ¤mmtau]
\end{mathematica}
	In this way, the more general definition \mm{\ref{math:Ugeneral}} does not override the more specific definition \mm{\ref{math:Unumerical}}.}
\end{questions}

\section{basis construction}\index{basis set!construction}
\label{sec:basisconstruction}

In principle, the choice of basis set $\{\ket{i}\}_i$ does not influence the way a computer program like Mathematica solves a quantum-mechanical problem. In practice, however, we always need a \emph{constructive} way to find some basis for a given quantum-mechanical problem. A basis that takes the system's Hamiltonian into account may give a computationally simpler description; but in complicated systems it is often more important to find \emph{any} way of constructing a usable basis set than finding the perfect one.

\subsection{description of a single degree of freedom}
\label{sec:singleDOF}

When we describe a single quantum-mechanical degree of freedom, it is often possible to deduce a useful basis set from knowledge of the Hilbert space\index{Hilbert space} itself. This is what we will be doing in \autoref{chap:spin} for spin systems, where the well-known Dicke basis\index{Dicke states} $\{\ket{S,M_S}\}_{M_S=-S}^S$ turns out to be very useful.

For more complicated degrees of freedom, we can find inspiration for a basis choice from an associated Hamiltonian. Such Hamiltonians describing a single degree of freedom are often so simple that they can be diagonalized by hand. If this is not the case, real-world Hamiltonians $\Ham$ can often be decomposed like \autoref{eq:H0H1} into a ``simple'' part $\Ham_0$ that is time-independent and can be diagonalized easily, and a ``difficult'' part $\Ham_1$ that usually contains complicated interactions and/or time-dependent terms but is of smaller magnitude.
A natural choice of basis set is the set of eigenstates of $\Ham_0$, or at least those eigenstates below a certain cutoff energy since they will be optimally suited to describe the complete low-energy behavior of the degree of freedom in question. This latter point is especially important for infinite-dimensional systems (\autoref{chap:1D}), where any computer representation will necessarily truncate the dimensionality, as discussed in \autoref{sec:incompletebasis}.

\subsubsection{examples of basis sets for single degrees of freedom:}

\begin{itemize}
	\item\emph{spin degree of freedom:} Dicke states $\ket{S,M_S}$\index{Dicke states} (see \autoref{chap:spin})
	\item\emph{translational degree of freedom:} square-well eigenstates\index{square well}, harmonic oscillator eigenstates\index{harmonic oscillator} (see \autoref{chap:1D})
	\item\emph{rotational degree of freedom:} spherical harmonics\index{spherical harmonics}
	\item\emph{atomic system:} hydrogen-like orbitals\index{hydrogen}
	\item\emph{translation-invariant system:} periodic plane waves\index{plane wave}
	\item\emph{periodic system (crystal):} periodic plane waves on the reciprocal lattice\index{reciprocal lattice}
\end{itemize}

\subsection{description of coupled degrees of freedom}
\label{sec:coupledDOF}

A broad range of quantum-mechanical systems of interest are governed by Hamiltonians of the form
\begin{equation}
	\label{eq:H0Hint}
	\Ham(t) = \left(\sum_{k=1}^N \Ham^{(k)}(t)\right) + \Ham\ix{int}(t),
\end{equation}
where $N$ individual degrees of freedom are governed by their individual Hamiltonians $\Ham^{(k)}(t)$, while their interactions\index{interaction} are described by $\Ham\ix{int}(t)$. This is a situation we will encounter repeatedly as we construct more complicated quantum-mechanical problems from simpler parts. A few simple examples are:
\begin{itemize}
	\item A set of $N$ interacting particles: the Hamiltonians $\Ham^{(k)}$ describe the individual particles, while $\Ham\ix{int}$ describes their interactions (see \autoref{sec:Ising}).
	\item A single particle moving in three spatial degrees of freedom: the three Hamiltonians $\Ham^{(x)}=-\frac{\hbar^2}{2m}\frac{\partial^2}{\ddd{\partial}{x}^2}$, $\Ham^{(y)}=-\frac{\hbar^2}{2m}\frac{\partial^2}{\ddd{\partial}{y}^2}$, $\Ham^{(z)}=-\frac{\hbar^2}{2m}\frac{\partial^2}{\ddd{\partial}{z}^2}$ describe the kinetic energy\index{operator!kinetic} in the three directions, while $\Ham\ix{int}$ contains the potential energy\index{operator!potential}, which usually couples these three degrees of freedom (see \autoref{sec:BEC}).
	\item A single particle with internal (spin) \index{spin} and external (motional) degrees of freedom, which are coupled through a state-dependent potential in $\Ham\ix{int}$ (see \autoref{chap:spacespin}).
\end{itemize}
The existence of individual Hamiltonians $\Ham^{(k)}$ assumes that the Hilbert space of the complete system has a tensor-product\index{tensor!product} structure
\begin{equation}
	\label{eq:HilbertTP}
	V = V^{(1)} \otimes V^{(2)} \otimes \dots \otimes V^{(N)},
\end{equation}
where each Hamiltonian $\Ham^{(k)}$ acts only in a single component space,
\begin{equation}
	\label{eq:Hamk}
	\Ham^{(k)} = \one^{(1)} \otimes \one^{(2)} \otimes \dots \otimes \one^{(k-1)} \otimes \op{h}^{(k)} \otimes \one^{(k+1)} \otimes \dots \otimes \one^{(N)}.
\end{equation}
Further, if we are able to construct bases $\{\ket{i_k}^{(k)}\}_{i_k=1}^{n_k}$ for all of the component Hilbert spaces $V^{(k)}$, as in \autoref{sec:singleDOF}, then we can construct a basis for the full Hilbert space $V$ by taking all possible tensor products of basis functions:\index{basis set}
\begin{equation}
	\label{eq:tpbf}
	\ket{i_1,i_2,\dots,i_N} = \ket{i_1}^{(1)}\otimes\ket{i_2}^{(2)}\otimes\dots\otimes\ket{i_N}^{(N)}.
\end{equation}
This basis will have $\prod_{k=1}^N n_k$ elements, which can easily become a very large number for composite systems.

\subsubsection{quantum states}\index{quantum state}

A product state of the complete system
\begin{equation}
	\label{eq:productwf}
	\ket{\psi} = \ket{\psi_1}^{(1)} \otimes \ket{\psi_2}^{(2)} \otimes \dots \otimes \ket{\psi_N}^{(N)}
\end{equation}
can be described in the following way. First, each single-particle state is decomposed in its own basis as in \autoref{eq:vectbasis},
\begin{equation}
	\ket{\psi_k}^{(k)} = \sum_{i_k=1}^{n_k} \psi_{i_k}^{(k)} \ket{i_k}^{(k)}.
\end{equation}
Inserting these expansions into \autoref{eq:productwf} gives the expansion into the \hyperref[eq:tpbf]{basis functions~\ref*{eq:tpbf}} of the full system,
\begin{multline}
	\label{eq:tensorproductstate}
	\ket{\psi} = \left[ \sum_{i_1=1}^{n_1} \psi_{i_1}^{(1)} \ket{i_1}^{(1)} \right] \otimes
		\left[ \sum_{i_2=1}^{n_2} \psi_{i_2}^{(2)} \ket{i_2}^{(2)} \right] \otimes
		\dots
		\otimes \left[ \sum_{i_N=1}^{n_N} \psi_{i_N}^{(N)} \ket{i_N}^{(N)} \right]\\
	= \sum_{i_1=1}^{n_1}\sum_{i_2=1}^{n_2}\dots\sum_{i_N=1}^{n_N}
		\left[ \psi_{i_1}^{(1)}\psi_{i_2}^{(2)}\dots \psi_{i_N}^{(N)} \right]
		\ket{i_1,i_2,\dots,i_N}
\end{multline}
In Mathematica, such a state tensor product can be calculated as follows. For example, assume that \mm{\mmpsi1} is a vector containing the expansion of $\ket{\psi_1}^{(1)}$ in its basis, and similarly for \mm{\mmpsi2} and \mm{\mmpsi3}. The vector \mm{\mmpsi} of expansion coefficients of the full state $\ket{\psi}=\ket{\psi_1}^{(1)}\otimes\ket{\psi_2}^{(2)}\otimes\ket{\psi_3}^{(3)}$ is calculated with\index{Mathematica!Kronecker product}
\begin{mathematica}
	¤mathin¤labelŽmath:psi ¤mmpsi = Flatten[KroneckerProduct[¤mmpsi1, ¤mmpsi2, ¤mmpsi3]]
\end{mathematica}
See \ref{it:jointstate} for a numerical example.

More generally, any state can be written as
\begin{equation}
	\ket{\psi}
	= \sum_{i_1=1}^{n_1}\sum_{i_2=1}^{n_2}\dots\sum_{i_N=1}^{n_N}
		\psi_{i_1,i_2,\dots,i_N}
		\ket{i_1,i_2,\dots,i_N},
\end{equation}
of which \autoref{eq:tensorproductstate} is a special case with $\psi_{i_1,i_2,\dots,i_N}=\psi_{i_1}^{(1)}\psi_{i_2}^{(2)}\dotsm\psi_{i_N}^{(N)}$.

\subsubsection{operators}\index{operator}

If the Hilbert space has the tensor-product structure of \autoref{eq:HilbertTP}, then the operators acting on this full space are often given as tensor products as well,
\begin{equation}
	\label{eq:productop}
	\op{A} = \op{a}_1^{(1)} \otimes \op{a}_2^{(2)} \otimes \dots \otimes \op{a}_N^{(N)},
\end{equation}
or as a sum over such products.
If every single-particle operator is decomposed in its own basis as in \autoref{eq:opbasis},
\begin{equation}
	\op{a}_k^{(k)} = \sum_{i_k=1}^{n_k} \sum_{j_k=1}^{n_k} a_{i_k,j_k}^{(k)}\ket{i_k}^{(k)}\bra{j_k}^{(k)},
\end{equation}
inserting these expressions into \autoref{eq:productop} gives the expansion into the \hyperref[eq:tpbf]{basis functions~\ref*{eq:tpbf}} of the full system,
\small
\begin{multline}
	\label{eq:tensorproductoperator}
	\op{A} =
		\left[ \sum_{i_1=1}^{n_1} \sum_{j_1=1}^{n_1} a_{i_1,j_1}^{(1)}\ket{i_1}^{(1)}\bra{j_1}^{(1)} \right]
		\otimes \left[ \sum_{i_2=1}^{n_2} \sum_{j_2=1}^{n_2} a_{i_2,j_2}^{(2)}\ket{i_2}^{(2)}\bra{j_2}^{(2)} \right]
		\otimes \dots \otimes \left[ \sum_{i_N=1}^{n_N} \sum_{j_N=1}^{n_N} a_{i_N,j_N}^{(N)}\ket{i_N}^{(N)}\bra{j_N}^{(N)} \right]\\
		= \sum_{i_1=1}^{n_1}\sum_{j_1=1}^{n_1}\sum_{i_2=1}^{n_2}\sum_{j_2=1}^{n_2}\dots\sum_{i_N=1}^{n_N}\sum_{j_N=1}^{n_N}
		\left[ a_{i_1,j_1}^{(1)}a_{i_2,j_2}^{(2)}\dots a_{i_N,j_N}^{(N)} \right]
		\ket{i_1,i_2,\dots,i_N}
		\bra{j_1,j_2,\dots,j_N}.
\end{multline}
\normalsize
In Mathematica, such an operator tensor product can be calculated similarly to \mm{\ref{math:psi}} above. For example, assume that \mm{a1} is a matrix containing the expansion of $\op{a}_1^{(1)}$ in its basis, and similarly for \mm{a2} and \mm{a3}. The matrix \mm{A} of expansion coefficients of the full operator $\op{A}=\op{a}_1^{(1)}\otimes\op{a}_2^{(2)}\otimes\op{a}_3^{(3)}$ is calculated with\index{Mathematica!Kronecker product}\index{product state}
\begin{mathematica}
	¤mathin¤labelŽmath:A A = KroneckerProduct[a1, a2, a3]
\end{mathematica}
Often we need to construct operators which act only on one of the component spaces, as in \autoref{eq:Hamk}. For example, in a 3-composite system the subsystem Hamiltonians $\op{h}^{(1)}$, $\op{h}^{(2)}$, and $\op{h}^{(3)}$ are first expanded to the full Hilbert space,
\begin{mathematica}
	¤mathin H1 = KroneckerProduct[h1,
	¤mathnl                       IdentityMatrix[Dimensions[h2]],
	¤mathnl                       IdentityMatrix[Dimensions[h3]]];
	¤mathin H2 = KroneckerProduct[IdentityMatrix[Dimensions[h1]],
	¤mathnl                       h2,
	¤mathnl                       IdentityMatrix[Dimensions[h3]]];
	¤mathin H3 = KroneckerProduct[IdentityMatrix[Dimensions[h1]],
	¤mathnl                       IdentityMatrix[Dimensions[h2]],
	¤mathnl                       h3];
\end{mathematica}
where \mm{IdentityMatrix[Dimensions[h1]]}\index{Mathematica!matrix!identity matrix} generates a unit matrix of size equal to that of \mm{h1}. In this way, the matrices \mm{H1}, \mm{H2}, \mm{H3} are of equal size and can be added together, even if \mm{h1}, \mm{h2}, \mm{h3} all have different sizes (expressed in Hilbert spaces of different dimensions):
\begin{mathematica}
	¤mathin H = H1 + H2 + H3;
\end{mathematica}
More generally, any operator can be written as
\begin{equation}
\label{eq:operatorgeneral}
	\op{A}
		= \sum_{i_1=1}^{n_1}\sum_{j_1=1}^{n_1}\sum_{i_2=1}^{n_2}\sum_{j_2=1}^{n_2}\dots\sum_{i_N=1}^{n_N}\sum_{j_N=1}^{n_N}
		a_{i_1,j_1,i_2,j_2,\dots,i_N,j_N}
		\ket{i_1,i_2,\dots,i_N}
		\bra{j_1,j_2,\dots,j_N},
\end{equation}
of which \autoref{eq:tensorproductoperator} is a special case with $a_{i_1,j_1,i_2,j_2,\dots,i_N,j_N}=a_{i_1,j_1}^{(1)}a_{i_2,j_2}^{(2)}\dotsm a_{i_N,j_N}^{(N)}$.

\subsection[reduced density matrices]{\label{sec:rdm}reduced density matrices\hspace{\stretch{1}}\attachcode{ReducedDensityMatrix}{reduced density matrices}}\index{density matrix!reduced|see {partial trace}}

In this section we calculate reduced density matrices by partial tracing. We start with the most general tripartite case, and then specialize to the more common bipartite case.

Assume that our quantum-mechanical system is composed of three parts A, B, C, and that its Hilbert space is a tensor product of the three associated Hilbert spaces with dimensions $d\ix{A}$, $d\ix{B}$, $d\ix{C}$: $V=V^{(A)}\otimes V^{(B)}\otimes V^{(C)}$. Similar to \autoref{eq:operatorgeneral}, any state of this system can be written as a density matrix
\begin{equation}
	\label{eq:dmABC}
	\op{\rho}\ix{ABC} = \sum_{i,i'=1}^{d\ix{A}}\sum_{j,j'=1}^{d\ix{B}}\sum_{k,k'=1}^{d\ix{C}}
		\rho_{i,j,k,i',j',k'} \ket{i\ix{A},j\ix{B},k\ix{C}}\bra{i'\ix{A},j'\ix{B},k'\ix{C}},
\end{equation}
where we use the basis states $\ket{i\ix{A},j\ix{B},k\ix{C}} = \ket{i}^{(A)}\otimes\ket{j}^{(B)}\otimes\ket{k}^{(C)}$ defined in terms of the three basis sets of the three component Hilbert spaces.

We calculate a reduced density matrix $\op{\rho}\ix{AC}=\Tr\ix{B}\op{\rho}\ix{ABC}$, which describes what happens to our knowledge of the subsystems A and C when we forget about subsystem B. For example, we could be studying a system of three particles, and take an interest in the state of particles A and C after we have lost particle B.
This reduced density matrix is defined as a partial trace\index{partial trace},
\begin{multline}
	\label{eq:dmAC}
	\op{\rho}\ix{AC} = \sum_{j''=1}^{d\ix{B}} \me{j''\ix{B}}{\op{\rho}\ix{ABC}}{j''\ix{B}}
	= \sum_{j''=1}^{d\ix{B}} \me{j''\ix{B}}{\left[\sum_{i,i'=1}^{d\ix{A}}\sum_{j,j'=1}^{d\ix{B}}\sum_{k,k'=1}^{d\ix{C}}
		\rho_{i,j,k,i',j',k'} \ket{i\ix{A},j\ix{B},k\ix{C}}\bra{i'\ix{A},j'\ix{B},k'\ix{C}}\right]}{j''\ix{B}}\\
	= \sum_{j''=1}^{d\ix{B}} \sum_{i,i'=1}^{d\ix{A}}\sum_{j,j'=1}^{d\ix{B}}\sum_{k,k'=1}^{d\ix{C}}
		\rho_{i,j,k,i',j',k'}\scp{j''\ix{B}}{i\ix{A},j\ix{B},k\ix{C}}\scp{i'\ix{A},j'\ix{B},k'\ix{C}}{j''\ix{B}}
	= \sum_{j''=1}^{d\ix{B}} \sum_{i,i'=1}^{d\ix{A}}\sum_{j,j'=1}^{d\ix{B}}\sum_{k,k'=1}^{d\ix{C}}
		\rho_{i,j,k,i',j',k'}\left[\delta_{j'',j}\ket{i\ix{A},k\ix{C}}\right]\left[\delta_{j'',j'}\bra{i'\ix{A},k'\ix{C}}\right]\\
	= \sum_{i,i'=1}^{d\ix{A}}\sum_{k,k'=1}^{d\ix{C}} \left[ \sum_{j=1}^{d\ix{B}}
		\rho_{i,j,k,i',j,k'} \right] \ket{i\ix{A},k\ix{C}}\bra{i'\ix{A},k'\ix{C}},
\end{multline}
which makes no reference to subsystem B. It only describes the joint system AC that is left after forgetting about subsystem B.

In Mathematica, we mostly use flattened basis sets, that is, our basis set for the joint Hilbert space of subsystems A, B, C is a flat list of length $d=d\ix{A}d\ix{B}d\ix{C}$:
\begin{equation}
	\label{eq:basislistABC}
	\{\ket{1\ix{A},1\ix{B},1\ix{C}},\ket{1\ix{A},1\ix{B},2\ix{C}},\dots,\ket{1\ix{A},1\ix{B},d\ix{C}},\ket{1\ix{A},2\ix{B},1\ix{C}},\ket{1\ix{A},2\ix{B},2\ix{C}},
	\dots,\ket{1\ix{A},2\ix{B},d\ix{C}},\dots,\ket{d\ix{A},d\ix{B},d\ix{C}}\}.
\end{equation}
In \autoref{sec:tensors} we have seen how lists and tensors can be re-shaped. As we will see below, these tools are used to switch between representations involving indices $(i,j,k)$ (\ie, lists with three indices, rank-three tensors) corresponding to \autoref{eq:dmABC}, and lists involving a single flattened-out index corresponding more to \autoref{eq:basislistABC}.

In practical calculations, any density matrix \mm{\mmrho ABC} of the joint system is given as a $d\times d$ matrix whose element $(u,v)$ is the prefactor of the contribution $\ket{u}\bra{v}$ with the indices $u$ and $v$ addressing elements in the flat list of \autoref{eq:basislistABC}. In order to calculate a reduced density matrix, we first reshape this $d\times d$ density matrix \mm{\mmrho ABC} into a rank-six tensor \mm{R} with dimensions $d\ix{A}\times d\ix{B}\times d\ix{C}\times d\ix{A}\times d\ix{B}\times d\ix{C}$, and with elements $r_{i,j,k,i',j',k'}$ of \autoref{eq:dmABC}:
\begin{mathematica}
	¤mathin¤labelŽmath:restructure2 R = ArrayReshape[¤mmrhoŽABC, {dA,dB,dC,dA,dB,dC}]
\end{mathematica}
Next, we contract\index{tensor!contraction} indices 2 and 5 of \mm{R} in order to do the partial trace over subsystem B, as is done in \autoref{eq:dmAC} (effectively setting $j=j'$ and summing over $j$). We find a rank-4 tensor \mm{S} with dimensions $d\ix{A}\times d\ix{C}\times d\ix{A}\times d\ix{C}$:
\begin{mathematica}
	¤mathin S = TensorContract[R, {2,5}]
\end{mathematica}
Finally, we flatten out this tensor again (simultaneously combining indices 1\&2 and 3\&4) to find the $d\ix{A}d\ix{C}\times d\ix{A}d\ix{C}$ reduced density matrix \mm{\mmrho AC}:
\begin{mathematica}
	¤mathin¤labelŽmath:flatten2 ¤mmrhoŽAC = Flatten[S, {{1,2}, {3,4}}]
\end{mathematica}
We assemble all of these steps into a generally usable function:
\begin{mathematica}
	¤mathin¤labelŽmath:rdmrho rdm[¤mmrhoŽABC_?MatrixQ, {dA_Integer /; dA >= 1,
	¤mathnl                     dB_Integer /; dB >= 1,
	¤mathnl                     dC_Integer /; dC >= 1}] /; 
	¤mathnl   Dimensions[¤mmrhoŽABC] == {dA*dB*dC, dA*dB*dC} := 
	¤mathnl     Flatten[TensorContract[ArrayReshape[¤mmrhoŽABC, {dA,dB,dC,dA,dB,dC}], {2,5}],
	¤mathnl       {{1,2}, {3,4}}]
\end{mathematica}
When our system is in a pure state, $\op{\rho}\ix{ABC}=\ket{\psi}\bra{\psi}$, this procedure can be simplified greatly. This is particularly important for large system dimensions, where calculating the full density matrix $\op{\rho}\ix{ABC}$ may be impossible due to memory constraints. For this, we assume that $\ket{\psi}=\sum_{i=1}^{d\ix{A}}\sum_{j=1}^{d\ix{B}}\sum_{k=1}^{d\ix{C}} \psi_{i,j,k}\ket{i\ix{A},j\ix{B},k\ix{C}}$, and therefore $\rho_{i,j,k,i',j',k'}=\psi_{i,j,k}\psi_{i',j',k'}^*$. Again, in Mathematica the coefficients of a state vector \mm{\mmpsi ABC} are a flat list referring to the elements of the flat basis of \autoref{eq:basislistABC}, and so we start by constructing a rank-3 tensor \mm{P} with dimensions $d\ix{A}\times d\ix{B}\times d\ix{C}$, whose elements are exactly the $\psi_{i,j,k}$, similar to \mm{\ref{math:restructure2}}:
\begin{mathematica}
	¤mathin P = ArrayReshape[¤mmpsiŽABC, {dA,dB,dC}]
\end{mathematica}
We transpose this rank-three tensor into a $d\ix{A}\times d\ix{C}\times d\ix{B}$ tensor \mm{P1} and a $d\ix{B}\times d\ix{A}\times d\ix{C}$ tensor \mm{P2} by changing the order of the indices:
\begin{mathematica}
	¤mathin P1 = Transpose[P, {1, 3, 2}]
	¤mathin¤labelŽmath:P2transpose P2 = Transpose[P, {2, 1, 3}]
\end{mathematica}
Now we can contract the index $j\ix{B}$ by a dot product, to find a rank-4 tensor \mm{Q} with dimensions $d\ix{A}\times d\ix{C}\times d\ix{A}\times d\ix{C}$:
\begin{mathematica}
	¤mathin¤labelŽmath:QP1P2 Q = P1 . Conjugate[P2]
\end{mathematica}
Finally we flatten \mm{Q} into the $d\ix{A}d\ix{C}\times d\ix{A}d\ix{C}$ reduced density matrix \mm{\mmrho AC} by combining indices 1\&2 and 3\&4:
\begin{mathematica}
	¤mathin ¤mmrhoŽAC = Flatten[Q, {{1,2}, {3,4}}]
\end{mathematica}
We assemble all of these steps into a generally usable function that extends the definition of \mm{\ref{math:rdmrho}}:
\begin{mathematica}
	¤mathin¤labelŽmath:rdmpsi rdm[¤mmpsiŽABC_?VectorQ, {dA_Integer /; dA >= 1,
	¤mathnl                     dB_Integer /; dB >= 1,
	¤mathnl                     dC_Integer /; dC >= 1}] /;
	¤mathnl   Length[¤mmpsiŽABC] == dA*dB*dC := 
	¤mathnl     With[{P = ArrayReshape[¤mmpsiŽABC, {dA,dB,dC}]},
	¤mathnl       Flatten[Transpose[P, {1,3,2}].ConjugateTranspose[P], {{1,2}, {3,4}}]]
\end{mathematica}
Notice that we have merged the transposition of \mm{\ref{math:P2transpose}} and the complex-conjugation of \mm{\ref{math:QP1P2}} into a single call of the \mm{ConjugateTranspose} function.

\subsubsection{bipartite systems}

Consider now the more common case of a bipartite system composed of only two subsystems A and B. We can still use the definitions developed above for tripartite (ABC) structures by introducing a trivial third subsystem with dimension $d\ix{C}=1$. This trivial subsystem will not change anything since it must always be in its one and only possible state. Therefore, given a density matrix \mm{\mmrho AB} of the joint system AB, we calculate the reduced density matrices of subsystems A and B with
\begin{mathematica}
	¤mathin ¤mmrhoŽA = rdm[¤mmrhoŽAB, {dA,dB,1}];
	¤mathin ¤mmrhoŽB = rdm[¤mmrhoŽAB, {1,dA,dB}];
\end{mathematica}
respectively, since it is always the middle subsystem of a given list of three subsystems that is eliminated through partial tracing.
In typical Mathematica fashion, we define a \mm{traceout} function that traces out the first $d$ dimensions if $\mm{d}>0$ and the last $d$ dimensions if $\mm{d}<0$:
\begin{mathematica}
	¤mathin¤labelŽmath:traceoutrho1 traceout[¤mmrho_?MatrixQ, d_Integer /; d >= 1] /;
	¤mathnl   Length[¤mmrho] == Length[Transpose[¤mmrho]] && Divisible[Length[¤mmrho], d] :=
	¤mathnl     rdm[¤mmrho, {1, d, Length[¤mmrho]/d}]
	¤mathin¤labelŽmath:traceoutrho2 traceout[¤mmrho_?MatrixQ, d_Integer /; d <= -1] /;
	¤mathnl   Length[¤mmrho] == Length[Transpose[¤mmrho]] && Divisible[Length[¤mmrho], -d] :=
	¤mathnl     rdm[¤mmrho, {Length[¤mmrho]/(-d), -d, 1}]
	¤mathin¤labelŽmath:traceoutpsi1 traceout[¤mmpsi_?VectorQ, d_Integer /; d >= 1] /; Divisible[Length[¤mmpsi], d] :=
	¤mathnl     rdm[¤mmpsi, {1, d, Length[¤mmpsi]/d}]
	¤mathin¤labelŽmath:traceoutpsi2 traceout[¤mmpsi_?VectorQ, d_Integer /; d <= -1] /; Divisible[Length[¤mmpsi], -d] :=
	¤mathnl     rdm[¤mmpsi, {Length[¤mmpsi]/(-d), -d, 1}]
\end{mathematica}

\subsection{exercises}

\begin{questions}
	\item\label{Q:tensorproduct} Two particles of mass $m$ are moving in a three-dimensional harmonic potential $V(r)=\frac12 m \omega^2 r^2$ with $r=\sqrt{x^2+y^2+z^2}$, and interacting via $s$-wave scattering $V\ix{int}=g \delta^3(\vect{r}_1-\vect{r}_2)$.
		\begin{enumerate}
			\item Write down the Hamiltonian of this system.
			\item Propose a basis set in which we can describe the quantum mechanics of this system.
			\item Calculate the matrix elements of the Hamiltonian in this basis set.
		\end{enumerate}
\pagenote[\ref{Q:tensorproduct}]{
\begin{enumerate}
	\item The Hamiltonian is
\begin{multline}
	\Ham = -\frac{\hbar^2}{2m}\left(\frac{\partial}{\ddd{\partial}{x}_1^2}+\frac{\partial}{\ddd{\partial}{y}_1^2}+\frac{\partial}{\ddd{\partial}{z}_1^2}+\frac{\partial}{\ddd{\partial}{x}_2^2}+\frac{\partial}{\ddd{\partial}{y}_2^2}+\frac{\partial}{\ddd{\partial}{z}_2^2}\right)\\
	+ \frac12 m \omega^2(x_1^2+y_1^2+z_1^2+x_2^2+y_2^2+z_2^2)
	+ g \delta(x_1-x_2)\delta(y_1-y_2)\delta(z_1-z_2)
\end{multline}
	\item For example, we could use the harmonic-oscillator basis functions that diagonalize the six degrees of freedom in the absence of coupling ($g=0$): the states $\ket{n}$ for which
\begin{equation}
	\left[-\frac{\hbar^2}{2m}\frac{\partial}{\ddd{\partial}{x}^2}+\frac12 m \omega^2 x^2\right]\ket{n}=\hbar\omega(n+\frac12)\ket{n},
\end{equation}
	with $n\in\dsN$.
	Explicitly, the position representations of these states are
\begin{equation}
	\scp{x}{n} = \phi_n(x) = x_0^{-1/2}\frac{H_n(x/x_0)}{\sqrt{2^nn!\sqrt{\pi}}}e^{-\frac{x^2}{2x_0^2}}.
\end{equation}
	For the six degrees of freedom we therefore propose the basis functions $\ket{n_{x_1},n_{y_1},n_{z_1},n_{x_2},n_{y_2},n_{z_2}}$.
	\item The matrix elements are
\begin{multline}
	\me{n_{x_1},n_{y_1},n_{z_1},n_{x_2},n_{y_2},n_{z_2}}{\Ham}{n'_{x_1},n'_{y_1},n'_{z_1},n'_{x_2},n'_{y_2},n'_{z_2}}\\
	=\hbar\omega\delta_{n_{x_1},n'_{x_1}}\delta_{n_{y_1},n'_{y_1}}\delta_{n_{z_1},n'_{z_1}}\delta_{n_{x_2},n'_{x_2}}\delta_{n_{y_2},n'_{y_2}}\delta_{n_{z_2},n'_{z_2}}
	(n_{x_1}+n_{y_1}+n_{z_1}+n_{x_2}+n_{y_2}+n_{z_2}+3)\\
	+ g\bra{n_{x_1},n_{y_1},n_{z_1},n_{x_2},n_{y_2},n_{z_2}}
	\int_{-\infty}^{\infty} \dd[x]_1\dd[y]_1\dd[z]_1\dd[x]_2\dd[y]_2\dd[z]_2\ket{x_1}\bra{x_1}\otimes\ket{y_1}\bra{y_1}\otimes\ket{z_1}\bra{z_1}\otimes\ket{x_2}\bra{x_2}\otimes\ket{y_2}\bra{y_2}\otimes\ket{z_2}\bra{z_2}\\
	\delta(x_1-x_2)\delta(y_1-y_2)\delta(z_1-z_2)
	\ket{n'_{x_1},n'_{y_1},n'_{z_1},n'_{x_2},n'_{y_2},n'_{z_2}}\\
	=\hbar\omega\delta_{n_{x_1},n'_{x_1}}\delta_{n_{y_1},n'_{y_1}}\delta_{n_{z_1},n'_{z_1}}\delta_{n_{x_2},n'_{x_2}}\delta_{n_{y_2},n'_{y_2}}\delta_{n_{z_2},n'_{z_2}}
	(n_{x_1}+n_{y_1}+n_{z_1}+n_{x_2}+n_{y_2}+n_{z_2}+3)\\
	+g\int_{-\infty}^{\infty} \dd[x]_1\dd[y]_1\dd[z]_1\dd[x]_2\dd[y]_2\dd[z]_2
	\delta(x_1-x_2)\delta(y_1-y_2)\delta(z_1-z_2)\\
	\times
	\phi_{n_{x_1}}(x_1)\phi_{n'_{x_1}}(x_1)
	\phi_{n_{y_1}}(y_1)\phi_{n'_{y_1}}(y_1)
	\phi_{n_{z_1}}(z_1)\phi_{n'_{z_1}}(z_1)
	\phi_{n_{x_2}}(x_2)\phi_{n'_{x_2}}(x_2)
	\phi_{n_{y_2}}(y_2)\phi_{n'_{y_2}}(y_2)
	\phi_{n_{z_2}}(z_2)\phi_{n'_{z_2}}(z_2)\\
	=\hbar\omega\delta_{n_{x_1},n'_{x_1}}\delta_{n_{y_1},n'_{y_1}}\delta_{n_{z_1},n'_{z_1}}\delta_{n_{x_2},n'_{x_2}}\delta_{n_{y_2},n'_{y_2}}\delta_{n_{z_2},n'_{z_2}}
	(n_{x_1}+n_{y_1}+n_{z_1}+n_{x_2}+n_{y_2}+n_{z_2}+3)\\
	+g \left[ \int_{-\infty}^{\infty} \dd[x] \phi_{n_{x_1}}(x)\phi_{n'_{x_1}}(x)\phi_{n_{x_2}}(x)\phi_{n'_{x_2}}(x) \right]
	\left[ \int_{-\infty}^{\infty} \dd[y] \phi_{n_{y_1}}(y)\phi_{n'_{y_1}}(y)\phi_{n_{y_2}}(y)\phi_{n'_{y_2}}(y) \right]
	\left[ \int_{-\infty}^{\infty} \dd[z] \phi_{n_{z_1}}(z)\phi_{n'_{z_1}}(z)\phi_{n_{z_2}}(z)\phi_{n'_{z_2}}(z) \right]\\
	=\hbar\omega\delta_{n_{x_1},n'_{x_1}}\delta_{n_{y_1},n'_{y_1}}\delta_{n_{z_1},n'_{z_1}}\delta_{n_{x_2},n'_{x_2}}\delta_{n_{y_2},n'_{y_2}}\delta_{n_{z_2},n'_{z_2}}
	(n_{x_1}+n_{y_1}+n_{z_1}+n_{x_2}+n_{y_2}+n_{z_2}+3)\\
	+\frac{g}{x_0^3} R_{n_{x_1},n'_{x_1},n_{x_2},n'_{x_2}} R_{n_{y_1},n'_{y_1},n_{y_2},n'_{y_2}} R_{n_{z_1},n'_{z_1},n_{z_2},n'_{z_2}}.
\end{multline}
	The required dimensionless integrals over products of four harmonic-oscillator eigenstates,
\begin{equation}
	R_{a,b,c,d} = x_0\int_{-\infty}^{\infty} \dd[x] \phi_a(x) \phi_b(x) \phi_c(x) \phi_d(x)
		= \int_{-\infty}^{\infty} \dd[\xi] \frac{H_a(\xi)H_b(\xi)H_c(\xi)H_d(\xi)}{\pi\sqrt{2^{a+b+c+d}a!b!c!d!}}e^{-2\xi^2},
\end{equation}
	can either be calculated by analytic integration,
\begin{mathematica}
	¤protect¤mathin¤ ¤mmphi[n_,¤ x_]¤ =¤ HermiteH[n,¤ x]/Sqrt[2^n*n!*Sqrt[¤mmpi]]*E^(-x^2/2);
	¤protect¤mathin¤ R[a_Integer/;a>=0,¤ b_Integer/;b>=0,¤ c_Integer/;c>=0,¤ d_Integer/;d>=0]¤ :=
	¤protect¤mathnl¤ ¤ ¤ Integrate[¤mmphi[a,x]*¤mmphi[b,x]*¤mmphi[c,x]*¤mmphi[d,x],¤ {x,¤ -¤mminfty,¤ ¤mminfty}]
\end{mathematica}
	or by an explicit but hypergeometric formula\footnote{See \url{http://www.ph.unimelb.edu.au/~jnnewn/cm-seminar-results/report/AnalyticIntegralOfFourHermites.pdf}.} (much faster),
\begin{mathematica}
	¤protect¤mathin¤ R[a_Integer/;a>=0,¤ b_Integer/;b>=0,¤ c_Integer/;c>=0,¤ d_Integer/;d>=0]¤ :=
	¤protect¤mathnl¤ ¤ ¤ If[OddQ[a+b+c+d],¤ 0,
	¤protect¤mathnl¤ ¤ ¤ ¤ ¤ 1/¤mmpi*(-1)^((a+b-c+d)/2)*Sqrt[c!/(2a!b!d!)]*
	¤protect¤mathnl¤ ¤ ¤ ¤ ¤ ¤ ¤ Gamma[(1+a-b+c-d)/2]*Gamma[(1-a+b+c-d)/2]*
	¤protect¤mathnl¤ ¤ ¤ ¤ ¤ ¤ ¤ HypergeometricPFQRegularized[{(1+a-b+c-d)/2,(1-a+b+c-d)/2,-d},
	¤protect¤mathnl¤ ¤ ¤ ¤ ¤ ¤ ¤ ¤ ¤ ¤ ¤ ¤ ¤ ¤ ¤ ¤ ¤ ¤ ¤ ¤ ¤ ¤ ¤ ¤ ¤ ¤ ¤ ¤ ¤ ¤ ¤ ¤ ¤ ¤ ¤ ¤ {1+c-d,(1-a-b+c-d)/2},1]]
\end{mathematica}
\end{enumerate}}
	\item\label{Q:kronecker1} Calculate \mm{\mmpsi} in \mm{\ref{math:psi}} without using \mm{KroneckerProduct}, but using the \mm{Table} command instead.
\pagenote[\ref{Q:kronecker1}]{
\begin{mathematica}
	¤protect¤mathin¤ ¤mmpsi¤ =¤ Flatten[Table[¤mmpsi1[[i1]]*¤mmpsi2[[i2]]*¤mmpsi3[[i3]],
	¤protect¤mathnl¤ {i1,¤ Length[¤mmpsi1]},¤ {i2,¤ Length[¤mmpsi2]},¤ {i3,¤ Length[¤mmpsi3]}]]
\end{mathematica}}
	\item\label{Q:kronecker2} Calculate \mm{A} in \mm{\ref{math:A}} without using \mm{KroneckerProduct}, but using the \mm{Table} command instead.
\pagenote[\ref{Q:kronecker2}]{
\begin{mathematica}
	¤protect¤mathin¤ A¤ =¤ Flatten[Table[a1[[i1,j1]]*a2[[i2,j2]]*a3[[i3,j3]],
	¤protect¤mathnl¤ {i1,¤ Length[a1]},¤ {i2,¤ Length[a2]},¤ {i3,¤ Length[a3]},
	¤protect¤mathnl¤ {j1,¤ Length[Transpose[a1]]},¤ {j2,¤ Length[Transpose[a2]]},
	¤protect¤mathnl¤ {j3,¤ Length[Transpose[a3]]}],¤ {{1,2,3},¤ {4,5,6}}]
\end{mathematica}}
	\item\label{it:jointstate} Given two spin-$1/2$ particles in states
		\begin{align}
			\label{eq:productstateexample}
			\ket{\psi}^{(1)} &= 0.8\ket{{\uparrow}}-0.6\ket{{\downarrow}}, &
			\ket{\psi}^{(2)} &= 0.6\ii\ket{{\uparrow}}+0.8\ket{{\downarrow}},
		\end{align}
		use the \mm{KroneckerProduct} function to calculate the joint state $\ket{\psi}=\ket{\psi}^{(1)}\otimes\ket{\psi}^{(2)}$, and compare the result to a manual calculation. In which order do the coefficients appear in the result of \mm{KroneckerProduct}?
\pagenote[\ref{it:jointstate}]{Manual calculation:
\begin{equation}
	\ket{\psi} = \left[0.8\ket{{\uparrow}}-0.6\ket{{\downarrow}}\right] \otimes \left[ 0.6\ii\ket{{\uparrow}}+0.8\ket{{\downarrow}} \right]
	= 0.48\ii\ket{{\uparrow\uparrow}}+0.64\ket{{\uparrow\downarrow}}-0.36\ii\ket{{\downarrow\uparrow}}-0.48\ket{{\downarrow\downarrow}},
\end{equation}
where $\ket{{\uparrow\downarrow}}=\ket{{\uparrow}}\otimes\ket{{\downarrow}}$ etc.
In Mathematica, using the computational basis $\{\ket{{\uparrow}},\ket{{\downarrow}}\}$, in this order:
\begin{mathematica}
	¤protect¤mathin¤ ¤mmpsi1¤ =¤ {0.8,¤ -0.6};
	¤protect¤mathin¤ ¤mmpsi2¤ =¤ {0.6*I,¤ 0.8};
	¤protect¤mathin¤ ¤mmpsi¤ =¤ Flatten[KroneckerProduct[¤mmpsi1,¤ ¤mmpsi2]]
	¤protect¤mathout¤ {0.+0.48*I,¤ 0.64+0.*I,¤ 0.-0.36*I,¤ -0.48+0.*I}
\end{mathematica}
	The ordering of the joint basis in the \mm{Kroneckerproduct} result is therefore $\{\ket{{\uparrow\uparrow}},\ket{{\uparrow\downarrow}},\ket{{\downarrow\uparrow}},\ket{{\downarrow\downarrow}}\}$.}
	\item\label{Q:productstate} For the state of \autoref{eq:productstateexample}, calculate the reduced density matrices $\rho^{(1)}$ and $\rho^{(2)}$ by tracing out the other subsystem. Compare them to the density matrices $\ket{\psi}^{(1)}\bra{\psi}^{(1)}$ and $\ket{\psi}^{(2)}\bra{\psi}^{(2)}$. What do you notice?
\pagenote[\ref{Q:productstate}]{We calculate the reduced density matrices with the \mm{traceout} command of \mm{\ref{math:traceoutpsi1}} and \mm{\ref{math:traceoutpsi2}}:
\begin{mathematica}
	¤protect¤mathin¤ ¤mmrho1¤ =¤ traceout[¤mmpsi,¤ -2]
	¤protect¤mathout¤ {{0.64+0.*I,¤ -0.48+0.*I},¤ {-0.48+0.*I,¤ 0.36+0.*I}}
	¤protect¤mathin¤ ¤mmrho2¤ =¤ traceout[¤mmpsi,¤ 2]
	¤protect¤mathout¤ {{0.36+0.*I,¤ 0.+0.48*I},¤ {0.-0.48*I,¤ 0.64+0.*I}}
\end{mathematica}
	Since $\ket{\psi}$ is a product state, these reduced density matrices are equal to the pure states of the subsystems:
\begin{mathematica}
	¤protect¤mathin¤ ¤mmrho1¤ ==¤ KroneckerProduct[¤mmpsi1,¤ Conjugate[¤mmpsi1]]
	¤protect¤mathout¤ True
	¤protect¤mathin¤ ¤mmrho2¤ ==¤ KroneckerProduct[¤mmpsi2,¤ Conjugate[¤mmpsi2]]
	¤protect¤mathout¤ True
\end{mathematica}}
\end{questions}
See also \ref{Q:Supup} and \ref{Q:Ssinglet}.


\chapter{spin and angular momentum}\index{spin}\index{angular momentum}
\label{chap:spin}
\restartlist{questions}
\chapterpicture{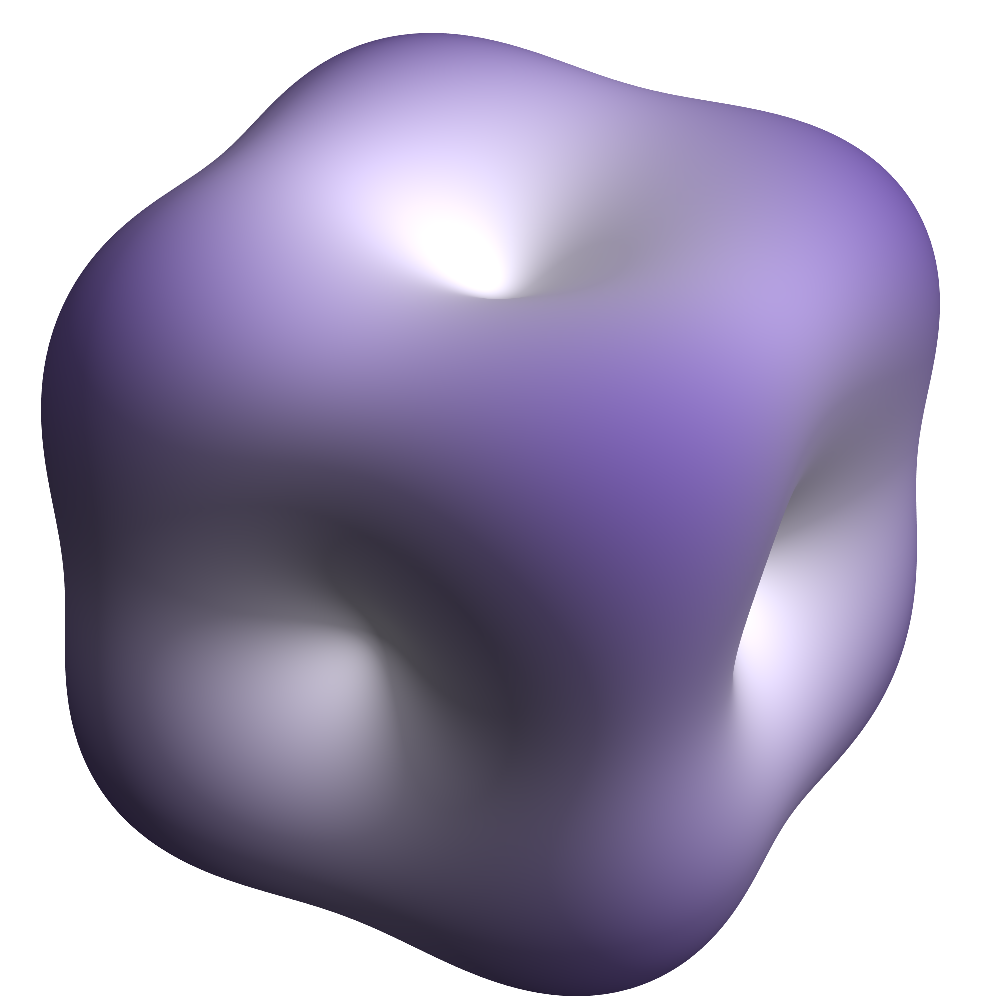}
In this chapter we put together everything we have studied so far\emdash Mathematica, quantum mechanics, computational bases, units\emdash to study simple quantum systems.
We start our explorations of quantum mechanics with the description of angular momentum. The reason for this choice is that, in contrast to the mechanically more intuitive linear motion (\autoref{chap:1D}), rotational motion is described with finite-dimensional Hilbert spaces and thus lends itself as a relatively simple starting point. As applications we look at the hyperfine structure of alkali atoms, lattice spin models, and quantum circuits.

\clearpage
\section[quantum-mechanical spin and angular momentum operators]{\label{sec:spinoperators}quantum-mechanical spin and angular momentum operators\hspace{\stretch{1}}\attachcode{SpinOperators}{spin and angular momentum operators}}

A classical rotational motion is described by its angular momentum, which is a three-dimensional pseudovector\index{pseudovector}\footnote{See \url{https://en.wikipedia.org/wiki/Pseudovector}.} whose direction indicates the rotation axis and whose length gives the rotational momentum. For an isolated system, the angular momentum is conserved and is thus very useful in the description of the system's state.

In quantum mechanics, angular momentum is equally described by a three-dimensional pseudovector operator $\opvect{S}$, with operator elements (in Cartesian coordinates) $\opvect{S}=(\op{S}_x,\op{S}_y,\op{S}_z)$. The joint eigenstates of the squared angular momentum magnitude $\norm{\opvect{S}}^2=\op{S}^2=\op{S}_x^2+\op{S}_y^2+\op{S}_z^2$ and of the $z$-component $\op{S}_z$ are called the Dicke states\index{Dicke states} $\ket{S,M}$, and satisfy
\begin{subequations}
\label{eq:spinopme1}
\begin{align}
	\op{S}^2\ket{S,M} &= S(S+1) \,\, \ket{S,M}\\
	\op{S}_z\ket{S,M} &= M \,\, \ket{S,M}
\end{align}
\end{subequations}
For every integer or half-integer value of the angular momentum $S\in\{0,\frac12,1,\frac32,2,\ldots\}$, there is a set of $2S+1$ Dicke states $\ket{S,M}$ with $M\in\{-S,-S+1,\ldots,S-1,S\}$ that form a basis for the description of the rotation axis orientation. These states also satisfy the following relationships with respect to the $x$- and $y$-components of the angular momentum:
\begin{subequations}
\label{eq:spinopme2}
\begin{align}
	\op{S}_+\ket{S,M} &= \sqrt{S(S+1)-M(M+1)} \,\, \ket{S,M+1} && \text{raising operator}\\
	\op{S}_-\ket{S,M} &= \sqrt{S(S+1)-M(M-1)} \,\, \ket{S,M-1} && \text{lowering operator}\\
	\op{S}_{\pm} &= \op{S}_x \pm \ii \op{S}_y && \text{Cartesian components}
\end{align}
\end{subequations}
As you know, quantum mechanics is not limited to spins  or angular momenta of length $S=1/2$.

In Mathematica we represent these operators in the Dicke basis as follows, with the elements of the basis set ordered with decreasing projection quantum number $M$:\index{angular momentum}
\begin{mathematica}
	¤mathin SpinQ[S_] := IntegerQ[2S] && S>=0
	¤mathin splus[0] = {{0}} //SparseArray;
	¤mathin splus[S_?SpinQ] := splus[S] =
	¤mathnl   SparseArray[Band[{1,2}] -> Table[Sqrt[S(S+1)-M(M+1)],
	¤mathnl     {M,S-1,-S,-1}], {2S+1,2S+1}]
	¤mathin sminus[S_?SpinQ] := Transpose[splus[S]]
	¤mathin sx[S_?SpinQ] := sx[S] = (splus[S]+sminus[S])/2
	¤mathin sy[S_?SpinQ] := sy[S] = (splus[S]-sminus[S])/(2I)
	¤mathin sz[S_?SpinQ] := sz[S] = SparseArray[Band[{1,1}]->Range[S,-S,-1], {2S+1,2S+1}]
	¤mathin id[S_?SpinQ] := id[S] = IdentityMatrix[2S+1, SparseArray]
\end{mathematica}
\begin{itemize}
	\item Notice that we have defined all these matrix representations as \emph{sparse} matrices\index{Mathematica!matrix!sparse matrix} (see \autoref{sec:sparsemat}), which will make larger calculations much more efficient later on. Further, all definitions are memoizing (see \autoref{sec:functionswhichremember}) to reduce execution time when they are used repeatedly.
	\item The function \mm{SpinQ[S]} yields \mm{True} only if \mm{S} is a nonnegative half-integer value and can therefore represent a physically valid spin. In general, functions ending in \mm{...Q} are \emph{questions} on the character of an argument (see \autoref{sec:conditions}).
	\item The operator $\op{S}_+$, defined with \mm{splus[S]}, contains only one off-diagonal band of non-zero values. The \mm{SparseArray} matrix constructor\index{Mathematica!matrix!sparse matrix} allows building such banded matrices by simply specifying the starting point of the band and a vector with the elements of the nonzero band.
	\item The operator $\op{S}_z$, defined with \mm{sz[S]}, shows you the ordering of the basis elements since it has the projection quantum numbers on the diagonal.
	\item The last operator \mm{id[S]} is the unit operator operating on a spin of length \mm{S}, and will be used below for tensor-product definitions. Note that the \mm{IdentityMatrix}\index{Mathematica!matrix!identity matrix} function usually returns a full matrix, which is not suitable for large-scale calculations. By giving it a \mm{SparseArray} option, it returns a sparse identity matrix of desired size.
	\item All these matrices can be displayed with, for example,\index{Mathematica!matrix!printing}
\begin{mathematica}
	¤mathin sx[3/2] //Normal
	¤mathout {{0, Sqrt[3]/2, 0, 0},
	¤mathnl  {Sqrt[3]/2, 0, 1, 0},
	¤mathnl  {0, 1, 0, Sqrt[3]/2},
	¤mathnl  {0, 0, Sqrt[3]/2, 0}}
\end{mathematica}
	or, for a more traditional view,
\begin{mathematica}
	¤mathin sx[3/2] //MatrixForm
\end{mathematica}
\end{itemize}

\subsection{exercises}

\begin{questions}
	\item\label{Q:spinhalfPauli} Verify that for $S=1/2$ the above Mathematica definitions give the Pauli matrices\index{Pauli matrices}: $\op{S}_i=\frac12 \op{\sigma}_i$ for $i=x,y,z$.
\pagenote[\ref{Q:spinhalfPauli}]{
\begin{mathematica}
	¤protect¤mathin¤ sx[1/2]¤ ==¤ 1/2*PauliMatrix[1]
	¤protect¤mathout¤ True
	¤protect¤mathin¤ sy[1/2]¤ ==¤ 1/2*PauliMatrix[2]
	¤protect¤mathout¤ True
	¤protect¤mathin¤ sz[1/2]¤ ==¤ 1/2*PauliMatrix[3]
	¤protect¤mathout¤ True
\end{mathematica}}
	\item\label{Q:spinverify} Verify in Mathematica that for given integer or half-integer $S$, the three operators (matrices) $\opvect{S}=\{\op{S}_x,\op{S}_y,\op{S}_z\}$ behave like a quantum-mechanical pseudovector of length $\norm{\opvect{S}}=\sqrt{S(S+1)}$:
		\begin{enumerate}
			\item Show that $[\op{S}_x,\op{S}_y]=\ii\op{S}_z$, $[\op{S}_y,\op{S}_z]=\ii\op{S}_x$, and $[\op{S}_z,\op{S}_x]=\ii\op{S}_y$.
			\item Show that $\op{S}_x^2+\op{S}_y^2+\op{S}_z^2=S(S+1)\one$.
			\item What is the largest value of $S$ for which you can do these verifications within one minute (each) on your computer? \emph{Hint:} use the \mm{Timing} function.\index{Mathematica!timing a calculation}
		\end{enumerate}
\pagenote[\ref{Q:spinverify}]{We only check up to $S=10$:
\begin{enumerate}
	\item commutators:
\begin{mathematica}
	¤protect¤mathin¤ Table[sx[S].sy[S]-sy[S].sx[S]¤ ==¤ I*sz[S],¤ {S,¤ 0,¤ 10,¤ 1/2}]
	¤protect¤mathout¤ {True,True,True,True,True,True,True,True,True,True,True,True,True,
	¤protect¤mathnl¤ ¤ True,True,True,True,True,True,True,True}
	¤protect¤mathin¤ Table[sy[S].sz[S]-sz[S].sy[S]¤ ==¤ I*sx[S],¤ {S,¤ 0,¤ 10,¤ 1/2}]
	¤protect¤mathout¤ {True,True,True,True,True,True,True,True,True,True,True,True,True,
	¤protect¤mathnl¤ ¤ True,True,True,True,True,True,True,True}
	¤protect¤mathin¤ Table[sz[S].sx[S]-sx[S].sz[S]¤ ==¤ I*sy[S],¤ {S,¤ 0,¤ 10,¤ 1/2}]
	¤protect¤mathout¤ {True,True,True,True,True,True,True,True,True,True,True,True,True,
	¤protect¤mathnl¤ ¤ True,True,True,True,True,True,True,True}
\end{mathematica}
	\item spin length:
\begin{mathematica}
	¤protect¤mathin¤ Table[sx[S].sx[S]+sy[S].sy[S]+sz[S].sz[S]¤ ==¤ S*(S+1)*id[S],¤ {S,0,10,1/2}]
	¤protect¤mathout¤ {True,True,True,True,True,True,True,True,True,True,True,True,True,
	¤protect¤mathnl¤ True,True,True,True,True,True,True,True}
\end{mathematica}
	\item Make sure to quit the Mathematica kernel before loading the spin-operator definitions and executing the following commands. On a MacBook Pro (Retina, 13-inch, Early 2015) with a \SI{3.1}{GHz} Intel Core i7 CPU and \SI{16}{GB} \SI{1867}{MHz} DDR3 RAM, the limit is around $S=\num{e5}$ for all verifications:
\begin{mathematica}
	¤protect¤mathin¤ s=100000;
	¤protect¤mathin¤ sx[s].sy[s]-sy[s].sx[s]¤ ==¤ I*sz[s]¤ //Timing
	¤protect¤mathout¤ {54.3985,¤ True}
	¤protect¤mathin¤ sy[s].sz[s]-sz[s].sy[s]¤ ==¤ I*sx[s]¤ //Timing
	¤protect¤mathout¤ {58.4917,¤ True}
	¤protect¤mathin¤ sz[s].sx[s]-sx[s].sz[s]¤ ==¤ I*sy[s]¤ //Timing
	¤protect¤mathout¤ {57.8856,¤ True}
	¤protect¤mathin¤ sx[s].sx[s]+sy[s].sy[s]+sz[s].sz[s]¤ ==¤ s*(s+1)*id[s]¤ //Timing
	¤protect¤mathout¤ {33.5487,¤ True}
\end{mathematica}
\end{enumerate}}
	\item\label{Q:rotations} The operators $\op{S}_{x,y,z}$ are the generators of rotations:\index{rotation} a rotation by an angle $\alpha$ around the axis given by a normalized vector $\vect{n}$ is done with the operator $\op{R}_{\vect{n}}(\alpha) = \exp(-\ii\alpha\vect{n}\cdot\opvect{S})$. Set $\vect{n}=\{\sin(\vartheta)\cos(\varphi),\sin(\vartheta)\sin(\varphi),\cos(\vartheta)\}$ and calculate the operator $\op{R}_{\vect{n}}(\alpha)$ explicitly for $S=0$, $S=1/2$, and $S=1$. Check that for $\alpha=0$ you find the unit operator.
\pagenote[\ref{Q:rotations}]{The expressions rapidly increase in complexity with increasing $S$:
\begin{mathematica}
	¤protect¤mathin¤ n¤ =¤ {Sin[¤mmtheta]*Cos[¤mmphi],¤ Sin[¤mmtheta]*Sin[¤mmphi],¤ Cos[¤mmtheta]};
	¤protect¤mathin¤ With[{S=0},¤ MatrixExp[-I*¤mmalpha*n.{sx[S],sy[S],sz[S]}]¤ //FullSimplify]
	¤protect¤mathout¤ {{1}}
	¤protect¤mathin¤ With[{S=1/2},¤ MatrixExp[-I*¤mmalpha*n.{sx[S],sy[S],sz[S]}]¤ //FullSimplify]
	¤protect¤mathout¤ {{Cos[¤mmalpha/2]-I*Cos[¤mmtheta]*Sin[¤mmalpha/2],¤ Sin[¤mmalpha/2]*Sin[¤mmtheta]*(-I*Cos[¤mmphi]-Sin[¤mmphi])},
	¤protect¤mathnl¤ ¤ {Sin[¤mmalpha/2]*Sin[¤mmtheta]*(-I*Cos[¤mmphi]+Sin[¤mmphi]),¤ Cos[¤mmalpha/2]+I*Cos[¤mmtheta]*Sin[¤mmalpha/2]}}
	¤protect¤mathin¤ ¤%¤ /.¤ ¤mmalpha¤ ->¤ 0
	¤protect¤mathout¤ {{1,¤ 0},¤ {0,¤ 1}}
	¤protect¤mathin¤ With[{S=1},¤ MatrixExp[-I*¤mmalpha*n.{sx[S],sy[S],sz[S]}]¤ //FullSimplify]
	¤protect¤mathout¤ {{(Cos[¤mmalpha/2]-I*Cos[¤mmtheta]*Sin[¤mmalpha/2])^2,
	¤protect¤mathnl¤ ¤ ¤ E^(-I*¤mmphi)*((-1+Cos[¤mmalpha])*Cos[¤mmtheta]-I*Sin[¤mmalpha])*Sin[¤mmtheta]/Sqrt[2],
	¤protect¤mathnl¤ ¤ ¤ -E^(-2I*¤mmphi)*Sin[¤mmalpha/2]^2*Sin[¤mmtheta]^2},
	¤protect¤mathnl¤ ¤ {Sqrt[2]*E^(-I*¤mmalpha)*Sin[¤mmalpha/2]*(Cos[¤mmalpha/2]-I*Cos[¤mmtheta]*Sin[¤mmalpha/2])
	¤protect¤mathnl¤ ¤ ¤ ¤ ¤ *Sin[¤mmtheta]*(-I*Cos[¤mmalpha+¤mmphi]+Sin[¤mmalpha+¤mmphi]),
	¤protect¤mathnl¤ ¤ ¤ Cos[¤mmalpha/2]^2+Cos[2¤mmtheta]*Sin[¤mmalpha/2]^2,
	¤protect¤mathnl¤ ¤ ¤ E^(-I*¤mmphi)*(Cos[¤mmtheta]-Cos[¤mmalpha]*Cos[¤mmtheta]-I*Sin[¤mmalpha])*Sin[¤mmtheta]/Sqrt[2]},
	¤protect¤mathnl¤ ¤ {-E^(2I*¤mmphi)*Sin[¤mmalpha/2]^2*Sin[¤mmtheta]^2,
	¤protect¤mathnl¤ ¤ ¤ -E^(I*¤mmphi)*((-1+Cos[¤mmalpha])*Cos[¤mmtheta]+I*Sin[¤mmalpha])*Sin[¤mmtheta]/Sqrt[2],
	¤protect¤mathnl¤ ¤ ¤ (Cos[¤mmalpha/2]+I*Cos[¤mmtheta]*Sin[¤mmalpha/2])^2}}
	¤protect¤mathin¤ ¤%¤ /.¤ ¤mmalpha¤ ->¤ 0
	¤protect¤mathout¤ {{1,¤ 0,¤ 0},¤ {0,¤ 1,¤ 0},¤ {0,¤ 0,¤ 1}}
\end{mathematica}}
\end{questions}

\section[spin-1/2 electron in a dc magnetic field]{\label{sec:electronspinB}spin-1/2 electron in a dc magnetic field\hspace{\stretch{1}}\attachcode{Electron}{spin-1/2 electron in a dc magnetic field}}\index{magnetic field}\index{electron}

As a first example we look at a single spin $S=1/2$. We use the basis containing the two states $\ket{\uparrow}=\ket{\frac12,\frac12}$ and $\ket{\downarrow}=\ket{\frac12,-\frac12}$, which we know to be eigenstates of the operators $\op{S}^2$ and $\op{S}_z$. The matrix expressions of the operators relevant for this system are given by the Pauli matrices\index{Pauli matrices} divided by two,
\begin{align}
	\matr{S}_x &= \frac12\begin{pmatrix} 0&1\\1&0 \end{pmatrix} = \frac12\matr{\sigma}_x &
	\matr{S}_y &= \frac12\begin{pmatrix} 0&-\ii\\\ii&0 \end{pmatrix} = \frac12\matr{\sigma}_y &
	\matr{S}_z &= \frac12\begin{pmatrix} 1&0\\0&-1 \end{pmatrix} = \frac12\matr{\sigma}_z
\end{align}
In Mathematica we enter these as
\begin{mathematica}
	¤mathin Sx = sx[1/2]; Sy = sy[1/2]; Sz = sz[1/2];
\end{mathematica}
using the general definitions of angular momentum operators given in \autoref{sec:spinoperators}. Alternatively, we can write
\begin{mathematica}
	¤mathin {Sx,Sy,Sz} = (1/2) * Table[PauliMatrix[i], {i,1,3}];
\end{mathematica}
As a Hamiltonian we use the coupling of this electron spin to an external magnetic field, $\Ham=-\opvect{\mu}\cdot\vect{B}$. The magnetic moment of the electron is $\opvect{\mu}=\mu\ix{B}g\ix{e} \opvect{S}$ in terms of its spin $\opvect{S}$, the Bohr magneton\index{Bohr magneton} $\mu\ix{B}=\SI{9.27400968(20)e-24}{J/T}$, and the electron's $g$-factor\index{g-factor@$g$-factor} $g\ix{e}=\num{-2,0023193043622(15)}$.\footnote{Notice that the magnetic moment of the electron is anti-parallel to its spin ($g\ix{e}<0$). The reason for this is the electron's negative electric charge. When the electron spin is parallel to the magnetic field, the electron's energy is higher than when they are anti-parallel.} The Hamiltonian is therefore
\begin{equation}
	\label{eq:spinhalfham}
	\Ham=-\mu\ix{B} g\ix{e} (\op{S}_x B_x + \op{S}_y B_y + \op{S}_z B_z).
\end{equation}
In our chosen matrix representation this Hamiltonian is
\begin{equation}
	\matr{H}=-\mu\ix{B} g\ix{e} (\matr{S}_x B_x + \matr{S}_y B_y + \matr{S}_z B_z)
	= -\frac12 \mu\ix{B} g\ix{e}
		\begin{pmatrix}
			B_z & B_x-\ii B_y \\
			B_x+\ii B_y & -B_z
		\end{pmatrix}.
\end{equation}
In order to implement this Hamiltonian, we first define a system of units. Here we express magnetic field strengths in Gauss and energies in MHz times Planck's constant (it is common to express energies in units of frequency, where the conversion is sometimes implicitly done via Planck's constant):
\begin{mathematica}
	¤mathin¤labelŽmath:magneticfieldunit MagneticFieldUnit = Quantity["Gausses"];
	¤mathin EnergyUnit = Quantity["PlanckConstant"]*Quantity["MHz"] //UnitConvert;
\end{mathematica}
In this unit system, the Bohr magneton is approximately \SI{1.4}{MHz/G}:
\begin{mathematica}
	¤mathin ¤mmmuŽB = Quantity["BohrMagneton"]/(EnergyUnit/MagneticFieldUnit) //UnitConvert
	¤mathout 1.3996245
\end{mathematica}
We define the electron's $g$-factor with
\begin{mathematica}
	¤mathin ge = UnitConvert["ElectronGFactor"]
	¤mathout -2.00231930436
\end{mathematica}
The Hamiltonian of \autoref{eq:spinhalfham} is then
\begin{mathematica}
	¤mathin H[Bx_, By_, Bz_] = -¤mmmuŽB * ge * (Sx*Bx+Sy*By+Sz*Bz)
\end{mathematica}

\subsubsection{natural units}
An alternative choice of units, called \emph{natural units}, is designed to simplify a calculation by making the numerical value of the largest possible number of quantities equal to 1. In the present case, this would be achieved by relating the field and energy units to each other in such a way that the Bohr magneton becomes equal to 1:
\begin{mathematica}
	¤mathin¤labelŽmath:magneticfieldunitN MagneticFieldUnit = Quantity["Gausses"];
	¤mathin¤labelŽmath:naturalunits EnergyUnit = MagneticFieldUnit * Quantity["BohrMagneton"] //UnitConvert;
	¤mathin ¤mmmuŽB = Quantity["BohrMagneton"]/(EnergyUnit/MagneticFieldUnit) //UnitConvert
	¤mathout 1.0000000
\end{mathematica}
In this way, calculations can often be simplified substantially because the Hamiltonian effectively becomes much simpler than it looks in other unit systems. We will be coming back to this point in future calculations.

\subsection{time-independent Schr\"odinger equation}\index{Schroedinger equation@Schr\"odinger equation!time-independent}
\label{sec:electronspinBtidse}

The time-independent Schr\"odinger equation for our spin-$1/2$ problem is, from \autoref{eq:TimeIndepSchrMatrix},
\begin{equation}
	\label{eq:spin1schr1}
	-\frac12 \mu\ix{B} g\ix{e}
		\begin{pmatrix}
			B_z & B_x-\ii B_y \\
			B_x+\ii B_y & -B_z
		\end{pmatrix}
	\cdot \vect{\psi} = E \vect{\psi}
\end{equation}
The eigenvalues of the Hamiltonian (in our chosen energy units) and eigenvectors are calculated with:\index{Mathematica!matrix!eigenvalues}\index{Mathematica!matrix!eigenvectors}
\begin{mathematica}
	¤mathin Eigensystem[H[Bx,By,Bz]]
\end{mathematica}
As described in \autoref{sec:diagonalization} the output is a list with two entries, the first being a list of eigenvalues and the second a list of associated eigenvectors. As long as the Hamiltonian matrix is Hermitian, the eigenvalues will all be real-valued; but the eigenvectors can be complex. Since the Hilbert space\index{Hilbert space} of this spin problem has dimension 2, and the basis contains two vectors, there are necessarily two eigenvalues and two associated eigenvectors of length 2. The eigenvalues can be called $E_{\pm} = \pm \frac12 \mu\ix{B} g\ix{e} \norm{\vect{B}}$. The list of eigenvalues is given in the Mathematica output as $\{E'_-,E'_+\}$. 
Notice that these eigenvalues only depend on the magnitude of the magnetic field, and not on its direction. This is to be expected: since there is no preferred axis in this system, there cannot be any directional dependence. The choice of the basis as the eigenstates of the $\op{S}_z$ operator was entirely arbitrary, and therefore the energy eigenvalues cannot depend on the orientation of the magnetic field with respect to this quantization axis.

The associated eigenvectors are
\begin{equation}
	\vect{\psi}_{\pm} = \{\frac{B_z\pm\norm{\vect{B}}}{B_x+\ii B_y},1\},
\end{equation}
which Mathematica returns as a list of lists, $\{\vect{\psi}_-,\vect{\psi}_+\}$. Notice that these eigenvectors are not normalized.

\subsection{exercises}

\begin{questions}
	\item\label{Q:electronenergies} Calculate the eigenvalues (in units of J) and eigenvectors (ortho-normalized) of an electron spin in a magnetic field of \SI{1}{T} in the $x$-direction.
\pagenote[\ref{Q:electronenergies}]{In the unit system of \mm{\ref{math:magneticfieldunitN}} we have
\begin{mathematica}
	¤protect¤mathin¤ {eval,¤ evec}¤ =¤ Eigensystem[H[Quantity[1,"Teslas"]/MagneticFieldUnit,¤ 0,¤ 0]]
	¤protect¤mathout¤ {{-14012.476,¤ 14012.476},¤ {{-0.7071068,¤ 0.7071068},¤ {0.7071068,¤ 0.7071068}}}
\end{mathematica}
	To convert the energy eigenvalues to Joules (or Yoctojoules), we use
\begin{mathematica}
	¤protect¤mathin¤ UnitConvert[eval*EnergyUnit,¤ "Yoctojoules"]
	¤protect¤mathout¤ {-9.284765¤ Yoctojoules,¤ 9.284765¤ Yoctojoules}
\end{mathematica}
	The corresponding eigenvectors are in the $\pm x$ direction:
	\begin{itemize}
		\item ground state: $E_-=-\SI{9.28e-24}{J}=-\SI{9.28}{yJ}$; $\ket{\psi_-}=\ket{-x}=\frac{\ket{\uparrow}-\ket{\downarrow}}{\sqrt{2}}$
		\item excited state: $E_+=+\SI{9.28e-24}{J}=+\SI{9.28}{yJ}$; $\ket{\psi_+}=\ket{+x}=\frac{\ket{\uparrow}+\ket{\downarrow}}{\sqrt{2}}$
	\end{itemize}}
	\item\label{Q:Bdirection} Set $\vect{B}=B[\vect{e}_x\sin(\vartheta)\cos(\varphi)+\vect{e}_y\sin(\vartheta)\sin(\varphi)+\vect{e}_z\cos(\vartheta)]$ and calculate the eigenvalues and normalized eigenvectors of the electron spin Hamiltonian.
\pagenote[\ref{Q:Bdirection}]{See also \ref{Q:spinhalf} and \ref{Q:Hexpr}.
\begin{mathematica}
	¤protect¤mathin¤ Bvec¤ =¤ B*{Sin[¤mmtheta]*Cos[¤mmphi],¤ Sin[¤mmtheta]*Sin[¤mmphi],¤ Cos[¤mmtheta]};
	¤protect¤mathin¤ Svec¤ =¤ {sx[1/2],¤ sy[1/2],¤ sz[1/2]};
	¤protect¤mathin¤ H¤ =¤ -¤mmmuŽB*ge*Bvec.Svec¤ //FullSimplify;
	¤protect¤mathin¤ {eval,¤ evec}¤ =¤ Eigensystem[H];
	¤protect¤mathin¤ eval
	¤protect¤mathout¤ {-B*ge*¤mmmuŽB/2,¤ B*ge*¤mmmuŽB/2}
	¤protect¤mathin¤ Assuming[0<¤mmtheta<¤mmpi,¤ ComplexExpand[Normalize¤ /@¤ evec]¤ //FullSimplify]
	¤protect¤mathout¤ {{E^(-I*¤mmphi)*Cos[¤mmtheta/2],¤ Sin[¤mmtheta/2]},¤ {-E^(-I*¤mmphi)*Sin[¤mmtheta/2],¤ Cos[¤mmtheta/2]}}
\end{mathematica}}
\end{questions}

\section[coupled spin systems: $^{87}$Rb hyperfine structure]{\label{sec:Rb87}coupled spin systems: $^{87}$Rb hyperfine structure\hspace{\stretch{1}}\attachcode{Rubidium87}{$^{87}$Rb hyperfine structure}}\index{magnetic field}\index{spin}\index{Rubidium-87}

Ground-state Rubidium-87 atoms consist of a nucleus with spin $I=3/2$, a single valence electron (spin $S=1/2$, orbital angular momentum $L=0$, and therefore total spin $J=1/2$), and 36 core electrons that do not contribute any angular momentum. In a magnetic field along the $z$-axis, the effective Hamiltonian of this system is\footnote{See \url{http://steck.us/alkalidata/rubidium87numbers.pdf}.}
\begin{equation}
	\label{eq:Rb87Hamiltonian}
	\Ham = \Ham_0 + h A\ix{hfs}\, \opvect{I}\cdot\opvect{J} - \mu\ix{B} B_z(g_I\op{I}_z+g_S\op{S}_z+g_L\op{L}_z),
\end{equation}
where $h$ is Planck's constant,\index{Planck's constant} $\mu\ix{B}$ is the Bohr magneton,\index{Bohr magneton} $A\ix{hfs}=\SI{3,417341305452145(45)}{GHz}$ is the spin--spin coupling constant\index{hyperfine interaction} in the ground state of $^{87}$Rb, $g_I=+\num{0,0009951414(10)}$ is the nuclear $g$-factor,\index{g-factor@$g$-factor} $g_S=\num{-2,0023193043622(15)}$ is the electron spin $g$-factor, and $g_L=\num{-0,99999369}$ is the electron orbital $g$-factor.

The first part $\Ham_0$ of \autoref{eq:Rb87Hamiltonian} contains all electrostatic interactions, core electrons, nuclear interactions etc. We will assume that the system is in the ground state of $\Ham_0$, which means that the valence electron is in the $5^2\text{S}_{1/2}$ state and the nucleus is deexcited. This ground state is eight-fold degenerate and consists of the four magnetic sublevels of the $I=3/2$ nuclear spin,\index{nuclear spin} the two sublevels of the $S=1/2$ electronic spin, and the single level of the $L=0$ angular momentum. The basis for the description of this atom is therefore the tensor product basis\index{tensor!product} of a spin-$3/2$, a spin-$1/2$, and a spin-$0$.\footnote{The spin-0 subsystem is trivial and could be left out in principle. It is included here to show the method in a more general way.}

The spin operators acting on this composite system are defined as in \autoref{sec:coupledDOF}. For example, the nuclear-spin operator $\op{I}_x$ is extended to the composite system by acting trivially on the electron spin and orbital angular momenta, $\op{I}_x \mapsto \op{I}_x\otimes\one\otimes\one$. The electron-spin operators are defined accordingly, for example $\op{S}_x \mapsto \one \otimes \op{S}_x\otimes\one$. The electron orbital angular momentum operators are, for example, $\op{L}_x \mapsto \one\otimes\one\otimes\op{L}_x$. In Mathematica these operators are defined with\index{Mathematica!Kronecker product}
\begin{mathematica}
	¤mathin Ix = KroneckerProduct[sx[3/2], id[1/2], id[0]];
	¤mathin Iy = KroneckerProduct[sy[3/2], id[1/2], id[0]];
	¤mathin Iz = KroneckerProduct[sz[3/2], id[1/2], id[0]];
	¤mathin Sx = KroneckerProduct[id[3/2], sx[1/2], id[0]];
	¤mathin Sy = KroneckerProduct[id[3/2], sy[1/2], id[0]];
	¤mathin Sz = KroneckerProduct[id[3/2], sz[1/2], id[0]];
	¤mathin Lx = KroneckerProduct[id[3/2], id[1/2], sx[0]];
	¤mathin Ly = KroneckerProduct[id[3/2], id[1/2], sy[0]];
	¤mathin Lz = KroneckerProduct[id[3/2], id[1/2], sz[0]];
\end{mathematica}
The total electron angular momentum is $\opvect{J}=\opvect{S}+\opvect{L}$:
\begin{mathematica}
	¤mathin Jx = Sx + Lx;  Jy = Sy + Ly;  Jz = Sz + Lz;
\end{mathematica}
The total angular momentum of the $^{87}$Rb atom is $\opvect{F}=\opvect{I}+\opvect{J}$:
\begin{mathematica}
	¤mathin Fx = Ix + Jx;  Fy = Iy + Jy;  Fz = Iz + Jz;
\end{mathematica}
Before defining the system's Hamiltonian, we declare a system of units. Any system will work here, so we stay with units commonly used in atomic physics: magnetic fields are expressed in Gauss, while energies are expressed in MHz times Planck's constant. As time unit we choose the microsecond:
\begin{mathematica}
	¤mathin MagneticFieldUnit = Quantity["Gausses"];
	¤mathin EnergyUnit = Quantity["PlanckConstant"] * Quantity["Megahertz"];
	¤mathin TimeUnit = Quantity["Microseconds"];
\end{mathematica}
The numerical values of the Bohr Magneton and the reduced Planck constant in these units are
\begin{mathematica}
	¤mathin ¤mmmuŽBn = Quantity["BohrMagneton"]/(EnergyUnit/MagneticFieldUnit)
	¤mathout 1.3996245
	¤mathin ¤mmhbarŽn = Quantity["ReducedPlanckConstant"]/(EnergyUnit*TimeUnit)
	¤mathout 0.15915494
\end{mathematica}
Using these definitions we define the hyperfine Hamiltonian with magnetic field in the $z$-direction as
\begin{mathematica}
	¤mathin Hhf = A(Ix.Jx+Iy.Jy+Iz.Jz) - ¤mmmuŽB*Bz*(gI*Iz+gS*Sz+gL*Lz);
	¤mathin hfc = {¤mmmuŽB -> ¤mmmuŽBn, ¤mmhbar -> ¤mmhbarŽn,
	¤mathnl   A->Quantity["PlanckConstant"]*Quantity[3.417341305452145,"GHz"]/EnergyUnit,
	¤mathnl   gS -> -2.0023193043622,
	¤mathnl   gL -> -0.99999369,
	¤mathnl   gI -> +0.0009951414};
\end{mathematica}
This yields the Hamiltonian as an \num{8 x 8} matrix, and we can calculate its eigenvalues and eigenvectors with
\begin{mathematica}
	¤mathin {eval, evec} = Eigensystem[Hhf] //FullSimplify;
\end{mathematica}
We plot the energy eigenvalues with\index{Mathematica!plotting}
\begin{mathematica}
	¤mathin Plot[Evaluate[eval /. hfc], {Bz, 0, 3000},
	¤mathnl   Frame -> True, FrameLabel -> {"Bz / G", "E / MHz"}]
\end{mathematica}
\begin{center}
\includegraphics[width=0.5\textwidth]{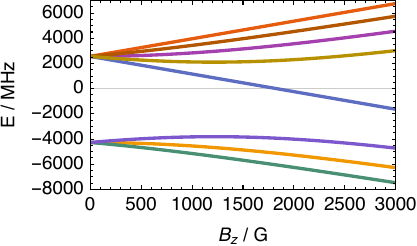}
\end{center}

\subsection{eigenstate analysis}\index{Mathematica!matrix!eigenvalues}\index{Mathematica!matrix!eigenvectors}

In this section we analyze the results \mm{eval} and \mm{evec} from the Hamiltonian diagonalization above. For this we first need to define \emph{ortho-normalized} eigenvectors since in general we cannot assume \mm{evec} to be ortho-normalized.

In general we can always define an ortho-normalized eigenvector set with\index{Mathematica!vector!orthogonal vectors}
\begin{mathematica}
	¤mathin nevec = Orthogonalize[evec]
\end{mathematica}
The problem with this definition is, however, immediately apparent if you look at the output given by Mathematica: since no assumptions on the reality of the variables were made, the orthogonalization is done in too much generality and quickly becomes unwieldy. Even using \mm{Assuming} and \mm{ComplexExpand}, as in \autoref{sec:complex}, does not give satisfactory results. But if we notice that the eigenvectors in \mm{evec} are all purely real-values, and are already orthogonal, then a simple vector-by-vector normalization is sufficient for calculating an ortho-normalized eigenvector set:
\begin{mathematica}
	¤mathin nevec = #/Sqrt[#.#] & /@ evec;
	¤mathin¤labelŽmath:orthonorm nevec . Transpose[nevec] //FullSimplify
\end{mathematica}
The fact that \mm{\ref{math:orthonorm}} finds a unit matrix implies that the vectors in \mm{nevec} are ortho-normal.

\subsubsection{field-free limit}

In the field-free limit $B_z=0$ the energy levels are
\begin{mathematica}
	¤mathin Assuming[A > 0, Limit[eval, Bz -> 0]]
	¤mathout {3A/4, 3A/4, -5A/4, 3A/4, -5A/4, 3A/4, -5A/4, 3A/4}
\end{mathematica}
We see that the level with energy $-\frac54A$ is three-fold degenerate while the level with energy $\frac34A$ is five-fold degenerate. This is also visible in the eigenvalue plot above. Considering that we have coupled two spins of lengths $I=\frac32$ and $J=\frac12$, we expect the composite system to have either total spin $F=1$ (three sublevels) or $F=2$ (five sublevels); we can make the tentative assignment that the $F=1$ level is at energy $E_1=-\frac54A$ and the $F=2$ level at $E_2=\frac34A$.

In order to demonstrate this assignment we express the matrix elements of the operators $\op{F}^2$ and $\op{F}_z$ in the field-free eigenstates, making sure to normalize these eigenstates before taking the limit $B_z\to0$:
\begin{mathematica}
	¤mathin nevec0 = Assuming[A > 0, Limit[nevec, Bz -> 0]];
	¤mathin nevec0 . (Fx.Fx+Fy.Fy+Fz.Fz) . Transpose[nevec0]
	¤mathin nevec0 . Fz . Transpose[nevec0]
\end{mathematica}
Notice that in this calculations we have used the fact that all eigenvectors are real, which may not always be the case for other Hamiltonians.
We see that the field-free normalized eigenvectors \mm{nevec0} are eigenvectors of both $\op{F}^2$ and $\op{F}_z$, and from looking at the eigenvalues we can identify them as
\begin{equation}
	\label{eq:Rb87identify}
	\{\ket{2,2},\ket{2,-2},\ket{1,0},\ket{2,0},\ket{1,1},\ket{2,1},\ket{1,-1},\ket{2,-1}\}
\end{equation}
in the notation $\ket{F,M_F}$.
These labels are often used to identify the energy eigenstates even for small $B_z\neq 0$.

\subsubsection{low-field limit}

For small magnetic fields, we series-expand the energy eigenvalues to first order in $B_z$:
\begin{mathematica}
	¤mathin Assuming[A > 0, Series[eval, {Bz, 0, 1}] //FullSimplify]
\end{mathematica}
From these low-field terms, in combination with the field-free level assignment, we see that the $F=1$ and $F=2$ levels have effective $g$-factors of $g_1=(-g_S+5g_I)/4\approx\num{0,501824}$ and $g_2=-(-g_S-3g_I)/4\approx\num{-0,499833}$, respectively, so that their energy eigenvalues follow the form
\begin{equation}
	\label{eq:Rb87lowfield}
	E_{F,M_F}(B_z) = E_F(0) - \mu\ix{B}M_Fg_FB_z+\mathcal{O}(B_z^2).
\end{equation}
These energy shifts due to the magnetic field are called \emph{Zeeman shifts}.\index{Zeeman shift!dc}

\subsubsection{high-field limit}

The energy eigenvalues in the high-field limit are infinite; but we can calculate their lowest-order series expansions with
\begin{mathematica}
	¤mathin Assuming[¤mmmuŽB > 0 && gS < -gI < 0,
	¤mathnl   Series[eval, {Bz, Infinity, 0}] //FullSimplify]
\end{mathematica}
From these expansions we can already identify the states in the eigenvalue plot above.

In order to calculate the eigenstates in the high-field limit we must again make sure to normalize the states \emph{before} taking the limit $B_z\to\infty$:\footnote{Note that in \mm{\ref{math:nevecinf}} we use two stages of assumptions, using the assumption $A>0$ only in \mm{FullSimplify} but not in \mm{Limit}. This is done in order to work around an inconsistency in Mathematica 11.3.0.0, and may be simplified in a future edition.}
\begin{mathematica}
	¤mathin¤labelŽmath:nevecinf nevecinf = Assuming[¤mmmuŽB > 0 && gS < -gI < 0,
	¤mathnl   FullSimplify[Limit[nevec, Bz -> Infinity], A > 0]]
	¤mathout {{1,  0, 0,  0, 0,  0, 0, 0},
	¤mathnl  {0,  0, 0,  0, 0,  0, 0, 1},
	¤mathnl  {0,  0, 0, -1, 0,  0, 0, 0},
	¤mathnl  {0,  0, 0,  0, 1,  0, 0, 0},
	¤mathnl  {0,  -1, 0,  0, 0, 0, 0, 0},
	¤mathnl  {0,  0, 1,  0, 0,  0, 0, 0},
	¤mathnl  {0, 0, 0,  0, 0,  -1, 0, 0},
	¤mathnl  {0,  0, 0,  0, 0,  0, 1, 0}}
\end{mathematica}
From this we immediately identify the high-field eigenstates as our basis states in a different order,
\begin{equation}
	\{\ket{\tfrac32,\tfrac12},
	\ket{-\tfrac32,-\tfrac12},
	\ket{\tfrac12,-\tfrac12},
	\ket{-\tfrac12,\tfrac12},
	\ket{\tfrac32,-\tfrac12},
	\ket{\tfrac12,\tfrac12},
	\ket{-\tfrac12,-\tfrac12},
	\ket{-\tfrac32,\tfrac12}\}
\end{equation}
where we have used the abbreviation $\ket{M_I,M_J} = \ket{\tfrac32,M_I}\otimes\ket{\tfrac12,M_J}$. You can verify this assignment by looking at the matrix elements of the $\op{I}_z$ and $\op{J}_z$ operators with
\begin{mathematica}
	¤mathin nevecinf . Iz . Transpose[nevecinf]
	¤mathin nevecinf . Jz . Transpose[nevecinf]
\end{mathematica}

\subsection{``magic'' magnetic field}\index{Rubidium-87!magic field}

The energy eigenvalues of the low-field states $\ket{1,-1}$ and $\ket{2,1}$ have almost the same first-order magnetic field dependence since $g_1\approx-g_2$ (see low-field limit above). If we plot their energy difference as a function of magnetic field we find an extremal point:
\begin{mathematica}
	¤mathin Plot[eval[[6]]-eval[[7]]-2A /. hfc, {Bz, 0, 6}]
\end{mathematica}
\begin{center}
\includegraphics[width=0.5\textwidth]{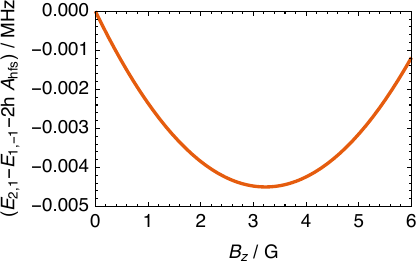}
\end{center}
At the ``magic'' field strength $B_0=\SI{3,22896}{G}$ the energy difference is independent of the magnetic field (to first order):\index{Mathematica!minimization}
\begin{mathematica}
	¤mathin NMinimize[eval[[6]] - eval[[7]] - 2 A /. hfc, Bz]
	¤mathout {-0.00449737, {Bz -> 3.22896}}
\end{mathematica}
This is an important discovery for quantum information\index{quantum information} science with $^{87}$Rb atoms. If we store a qubit\index{qubit} in the state $\ket{\vartheta,\varphi}=\cos(\vartheta/2)\ket{1,-1}+e^{\ii\varphi}\sin(\vartheta/2)\ket{2,1}$ and tune the magnetic field exactly to the magic value, then the experimentally unavoidable magnetic-field fluctuations will not lead to fluctuations of the energy difference between the two atomic levels and thus will not lead to qubit decoherence\index{decoherence}. Very long qubit coherence times can be achieved in this way.

For the present case where $\abs{g_I}\ll\abs{g_S}$, the magic field is approximately $B_z\approx\frac{16A g_I}{3\mu\ix{B}g_S^2}$.

\subsection{coupling to an oscillating magnetic field}\index{oscillating field}

In this section we study the coupling of a $^{87}$Rb atom to a weak oscillating magnetic field. Such a field could be the magnetic part of an electromagnetic wave, whose electric field does not couple to our atom in the electronic ground state. This calculation is a template for more general situations where a quantum-mechanical system is driven by an oscillating field.

The $^{87}$Rb hyperfine Hamiltonian in the presence of an oscillating magnetic field is
\begin{equation}
	\label{eq:Rb87HamiltonianOsc}
	\Ham(t) = \underbrace{h A\ix{hfs}\, \opvect{I}\cdot\opvect{J}
	- \mu\ix{B} B_z(g_I\op{I}_z+g_S\op{S}_z+g_L\op{L}_z)}_{\Ham_0}
	- \cos(\omega t) \times \underbrace{\mu\ix{B} \vect{B}\ex{ac}\cdot(g_I\opvect{I}+g_S\opvect{S}+g_L\opvect{L})}_{-\Ham_1}
\end{equation}
where the static magnetic field is assumed to be in the $z$ direction, as before. Unfortunately, $[\Ham(t),\Ham(t')]=[\Ham_1,\Ham_0]\left(\cos(\omega t)-\cos(\omega t')\right)\neq 0$ in general, so we cannot use the exact solution of \autoref{eq:tdSchrMatSol1} of the time-dependent Schr\"odinger equation. In fact, the time-dependent Schr\"odinger equation of this system has no analytic solution at all. In what follows we will calculate approximate solutions.

%
%
%
%

Since we have diagonalized the time-independent Hamiltonian $\Ham_0$ already, we use its eigenstates as a basis for calculating the effect of the oscillating perturbation $\Ham_1(t)$. In general, calling $\{\ket{i}\}_{i=1}^8$ the set of eigenstates of $\Ham_0$, with $\Ham_0\ket{i}=E_i\ket{i}$ for $i\in\{1\dots8\}$, we expand the general hyperfine state as in \autoref{eq:rotatingbasis},
\begin{equation}
	\label{eq:generalhyperfine}
	\ket{\psi(t)} = \sum_{i=1}^8 \psi_i(t)e^{-\ii E_i t/\hbar}\ket{i}.
\end{equation}
The time-dependent Schr\"odinger equation for the expansion coefficients $\psi_i(t)$ in this interaction picture is given in \autoref{eq:tdseint}:\index{Schroedinger equation@Schr\"odinger equation!time-dependent} for $i=1\dots8$ we have
\begin{equation}
	\label{eq:tdseintRb}
	\ii\hbar \dot{\psi}_i(t)
	= \sum_{j=1}^8 \psi_j(t) e^{-\ii (E_j-E_i) t/\hbar} \cos(\omega t) \me{i}{\Ham_1}{j}
	= \frac12 \sum_{j=1}^8 \psi_j(t) \left[ e^{-\ii \left(\frac{E_j-E_i}{\hbar}-\omega\right) t}+ e^{\ii \left(\frac{E_i-E_j}{\hbar}-\omega\right) t}\right] T_{i j},
\end{equation}
where we have replaced $\cos(\omega t) = \frac12 e^{\ii\omega t}+\frac12 e^{-\ii\omega t}$ and defined
\begin{equation}
	T_{i j}=\me{i}{\Ham_1}{j}=-\me{i}{\left[ \mu\ix{B} \vect{B}\ex{ac}\cdot(g_I\opvect{I}+g_S\opvect{S}+g_L\opvect{L})\right]}{j}.
\end{equation}
From \autoref{eq:tdseintRb} we can proceed in various ways:
\begin{description}
	\item[Transition matrix elements:]\index{transition matrix elements} The time-independent matrix elements $T_{i j}$ of the perturbation Hamiltonian are called the \emph{transition matrix elements} and describe how the populations of the different eigenstates of $\Ham_0$ are coupled through the oscillating field. We calculate them in Mathematica as follows:
\begin{mathematica}
	¤mathin H0 = A*(Ix.Jx + Iy.Jy + Iz.Jz) - ¤mmmuŽB*Bz*(gS*Sz + gL*Lz + gI*Iz);
	¤mathin H1 = -¤mmmuŽB*(gS*(Bacx*Sx + Bacy*Sy + Bacz*Sz)
	¤mathnl         + gI*(Bacx*Ix + Bacy*Iy + Bacz*Iz)
	¤mathnl         + gL*(Bacx*Lx + Bacy*Ly + Bacz*Lz));
	¤mathin H[t_] = H0 + H1*Cos[¤mmomega*t];
	¤mathin {eval, evec} = Eigensystem[H0] //FullSimplify;
	¤mathin nevec = Map[#/Sqrt[#.#] &, evec];
	¤mathin¤labelŽmath:T T = Assuming[A > 0, nevec.H1.Transpose[nevec] //FullSimplify];
\end{mathematica}
	Looking at this matrix $\matr{T}$ we see that not all energy levels are directly coupled by an oscillating magnetic field. For example, $T_{1,2}=0$ indicates that the populations of the states $\ket{1}$ and $\ket{2}$ can only be coupled indirectly through other states, but not directly (hint: check \mm{T[[1,2]]}).
	\item[Numerical solution:]\index{Mathematica!differential equation} 
	\autoref{eq:tdseintRb} is a series of linear coupled differential equations, which we write down explicitly in Mathematica with
\begin{mathematica}
	¤mathin deqs = Table[I*¤mmhbar*Subscript[¤mmpsi,i]¤textquotesingleŽ[t] ==
	¤mathnl        Sum[Subscript[¤mmpsi,j][t]*Exp[-I*(eval[[j]]-eval[[i]])*t/¤mmhbar]
	¤mathnl          *Cos[¤mmomega*t]*T[[i,j]], {j, 8}], {i,8}];
\end{mathematica}
	Assuming concrete conditions, for example the initial state $\ket{\psi(t=0)}=\ket{F=2,M_F=-2}$ which is the second eigenstate \mm{nevec[[2]]} [see \autoref{eq:Rb87identify}], and magnetic fields $B_z=\SI{3,22896}{G}$, $B_x\ex{ac}=\SI{100}{mG}$, $B_y\ex{ac}=B_z\ex{ac}=0$, and an ac field angular frequency of $\omega=2\pi\times\SI{6827,9}{MHz}$, we can find the time-dependent state $\ket{\psi(t)}$ with
\begin{mathematica}
	¤mathin¤labelŽmath:Rb87timesol S = NDSolve[Join[deqs /. hfc /.{Bz->3.22896, Bacx->0.1, Bacy->0, Bacz->0,
	¤mathnl        ¤mmomega->2*¤mmpi*6827.9},
	¤mathnl      {Subscript[¤mmpsi,1][0]==0,Subscript[¤mmpsi,2][0]==1,
	¤mathnl       Subscript[¤mmpsi,3][0]==0,Subscript[¤mmpsi,4][0]==0,
	¤mathnl       Subscript[¤mmpsi,5][0]==0,Subscript[¤mmpsi,6][0]==0,
	¤mathnl       Subscript[¤mmpsi,7][0]==0,Subscript[¤mmpsi,8][0]==0}],
	¤mathnl      Table[Subscript[¤mmpsi,i][t],{i,8}], {t, 0, 30},
	¤mathnl      MaxStepSize->10^(-5), MaxSteps->10^7]
\end{mathematica}
	Notice that the maximum step size in this numerical solution is very small (\num{e-5} time units or \SI{10}{ps}), since it needs to capture the fast oscillations of more than \SI{6.8}{GHz}. As a result, a large number of numerical steps is required, which makes this way of studying the evolution very difficult in practice.
	
	We plot the resulting populations with\index{Mathematica!plotting}
\begin{mathematica}
	¤mathin¤labelŽmath:Rb87pop1 Plot[Evaluate[Abs[Subscript[¤mmpsi,2][t] /. S[[1]]]^2], {t, 0, 30}]
\end{mathematica}
\begin{center}
\includegraphics[width=0.5\textwidth]{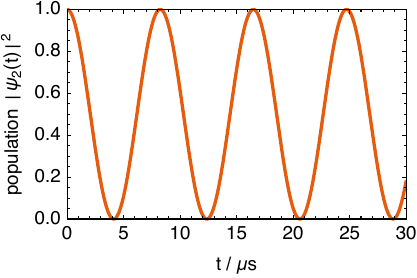}
\end{center}
\begin{mathematica}
	¤mathin¤labelŽmath:Rb87pop5 Plot[Evaluate[Abs[Subscript[¤mmpsi,7][t] /. S[[1]]]^2], {t, 0, 30}]
\end{mathematica}
\begin{center}
\includegraphics[width=0.5\textwidth]{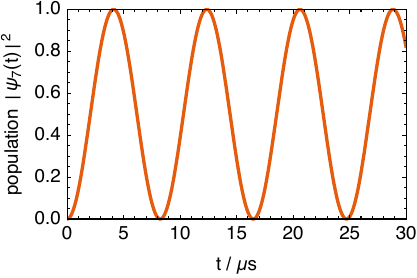}
\end{center}
	We see that the population is mostly sloshing between $\op{\mathcal{H}}_0$-eigenstates $\ket{2}\approx\ket{F=2,M_F=-2}$ and $\ket{7}\approx\ket{F=1,M_F=-1}$ [see \autoref{eq:Rb87identify}]. Each population oscillation takes about \SI{8,2}{\micro s} (the Rabi period), and we say that the Rabi frequency is about \SI{120}{kHz}.\label{lbl:Rabifreq}
	\item[Rotating-wave approximation:]\index{rotating-wave approximation} The time-dependent prefactor $\exp\left[-\ii \left(\frac{E_j-E_i}{\hbar}-\omega\right) t\right]+\exp\left[\ii \left(\frac{E_i-E_j}{\hbar}-\omega\right) t\right]$ of \autoref{eq:tdseintRb} oscillates very rapidly unless either $\frac{E_j-E_i}{\hbar}-\omega\approx0$ or $\frac{E_i-E_j}{\hbar}-\omega\approx0$, where one of its terms changes slowly in time. The \emph{rotating-wave approximation} (RWA) consists of neglecting all rapidly rotating terms in \autoref{eq:tdseintRb}. Assume that there is a single\footnote{The following derivation is readily extended to situations where several pairs of states have an energy difference approximately equal to $\hbar\omega$. In such a case we need to solve a larger system of coupled differential equations.} pair of states $\ket{i}$ and $\ket{j}$ such that $E_i-E_j\approx\hbar\omega$, with $E_i>E_j$, while all other states have an energy difference far from $\hbar\omega$. The RWA thus consists of simplifying \autoref{eq:tdseintRb} to
\begin{align}
	\label{eq:RWESchroedinger}
	\ii\hbar \dot{\psi}_i(t) &\approx \frac12 \psi_j(t) e^{\ii \left(\frac{E_i-E_j}{\hbar}-\omega\right) t} T_{i j} = \frac12 \psi_j(t) T_{i j} e^{-\ii \Delta t}\nonumber\\
	\ii\hbar \dot{\psi}_j(t) &\approx \frac12 \psi_i(t) e^{-\ii \left(\frac{E_i-E_j}{\hbar}-\omega\right) t} T_{j i} = \frac12 \psi_i(t) T_{j i} e^{\ii \Delta t}\nonumber\\
	\ii\hbar \dot{\psi}_k(t) &\approx 0 \text{ for $k\notin\{i,j\}$}
\end{align}
	with $T_{j i}=T_{i j}^*$ and the detuning\index{detuning} $\Delta=\omega-(E_i-E_j)/\hbar$. All other terms in \autoref{eq:tdseintRb} have been neglected because they rotate so fast in time that they ``average out'' to zero. This approximate system of differential equations has the exact solution
\begin{align}
	\label{eq:twolevelRWAsol}
	\psi_i(t) &= e^{-\frac{\ii}{2}\Delta t} \left[
		\psi_i(0)\cos\left(\frac{\Omega t}{2}\right)+\ii\left(\frac{\Delta}{\Omega}\psi_i(0)-\frac{T_{i j}}{\hbar\Omega}\psi_j(0)\right)\sin\left(\frac{\Omega t}{2}\right)\right]\nonumber\\
	\psi_j(t) &= e^{\frac{\ii}{2}\Delta t} \left[
		\psi_j(0)\cos\left(\frac{\Omega t}{2}\right)-\ii\left(\frac{\Delta}{\Omega}\psi_j(0)+\frac{T_{i j}^*}{\hbar\Omega}\psi_i(0)\right)\sin\left(\frac{\Omega t}{2}\right)\right]\nonumber\\
	\psi_k(t) &= \psi_k(0) \text{ for $k\notin\{i,j\}$}
\end{align}
in terms of the generalized Rabi frequency\index{Rabi frequency} $\Omega=\sqrt{\abs{T_{i j}}^2/\hbar^2+\Delta^2}$. We can see that the population sloshes back and forth (``Rabi oscillation'') between the two levels $\ket{i}$ and $\ket{j}$ with angular frequency $\Omega$, as we had seen numerically above.

We can verify this solution im Mathematica as follows. First we define
\begin{mathematica}
	¤mathin¤labelŽmath:Delta ¤mmDelta = ¤mmomega - (Ei-Ej)/¤mmhbar;
	¤mathin¤labelŽmath:Omega ¤mmOmega = Sqrt[Tij*Tji/¤mmhbar^2 + ¤mmDelta^2];
\end{mathematica}
and the solutions
\begin{mathematica}
	¤mathin ¤mmpsiŽi[t_] = E^(-I*¤mmDelta*t/2)*(¤mmpsiŽi0*Cos[¤mmOmega*t/2]+I*(¤mmDelta/¤mmOmega*¤mmpsiŽi0-Tij/(¤mmhbar*¤mmOmega)*¤mmpsiŽj0)
	¤mathnl            *Sin[¤mmOmega*t/2]);
	¤mathin ¤mmpsiŽj[t_] = E^(I*¤mmDelta*t/2)*(¤mmpsiŽj0*Cos[¤mmOmega*t/2]-I*(¤mmDelta/¤mmOmega*¤mmpsiŽj0+Tji/(¤mmhbar*¤mmOmega)*¤mmpsiŽi0)
	¤mathnl            *Sin[¤mmOmega*t/2]);
\end{mathematica}
With these definitions, we can check the \hyperref[{eq:RWESchroedinger}]{Schr\"odinger equations~\ref*{eq:RWESchroedinger}}:
\begin{mathematica}
	¤mathin FullSimplify[I*¤mmhbar*¤mmpsiŽi¤textquotesingleŽ[t] == (1/2) * ¤mmpsiŽj[t] * Exp[-I*¤mmDelta*t]*Tij]
	¤mathout True
	¤mathin FullSimplify[I*¤mmhbar*¤mmpsiŽj¤textquotesingleŽ[t] == (1/2) * ¤mmpsiŽi[t] * Exp[I*¤mmDelta*t]*Tji]
	¤mathout True
\end{mathematica}
as well as the initial conditions
\begin{mathematica}
	¤mathin ¤mmpsiŽi[0]
	¤mathout ¤mmpsiŽi0
	¤mathin ¤mmpsiŽj[0]
	¤mathout ¤mmpsiŽj0
\end{mathematica}

\item[dressed states:] If we insert the RWA solutions, \autoref{eq:twolevelRWAsol}, into the definition of the general hyperfine state, \autoref{eq:generalhyperfine}, and set all coefficients $\psi_k=0$ for $k\notin\{i,j\}$, and then write $\sin(z)=(e^{\ii z}-e^{-\ii z})/(2\ii)$ and $\cos(z)=(e^{\ii z}+e^{-\ii z})/2$, we find the state
\scriptsize
\begin{multline}
	\label{eq:dressedstates1}
	\ket{\psi(t)} \approx \psi_i(t)e^{-\ii E_i t/\hbar}\ket{i}+\psi_j(t)e^{-\ii E_j t/\hbar}\ket{j}\\
	= \frac12 e^{-\ii\left(E_i-\frac{\hbar(\Omega-\Delta)}{2}\right)t/\hbar} \left\{
		\left[\psi_i(0)\left(1+\frac{\Delta}{\Omega}\right)-\psi_j(0)\frac{T_{i j}}{\hbar\Omega}\right]\ket{i}
		+\left[\psi_j(0)\left(1-\frac{\Delta}{\Omega}\right)-\psi_i(0)\frac{T_{i j}^*}{\hbar\Omega}\right]e^{\ii\omega t}\ket{j}
	\right\}\\
	+ \frac12 e^{-\ii\left(E_i+\frac{\hbar(\Omega+\Delta)}{2}\right)t/\hbar} \left\{
		\left[\psi_i(0)\left(1-\frac{\Delta}{\Omega}\right)+\psi_j(0)\frac{T_{i j}}{\hbar\Omega}\right]\ket{i}
		+\left[\psi_j(0)\left(1+\frac{\Delta}{\Omega}\right)+\psi_i(0)\frac{T_{i j}^*}{\hbar\Omega}\right]e^{\ii\omega t}\ket{j}
	\right\}.
\end{multline}
\normalsize
In order to interpret this state more clearly, we need to expand our view of the problem to include the quantized driving field. For this we assume that the driving mode of the field (for example, the used mode of the electromagnetic field) in state $\ket{n}$ contains $n$ quanta of vibration (for example, photons), and has an energy of $E_n=n\hbar\omega$. The two states $\ket{i}$ and $\ket{j}$ describing our system, with $E_i-E_j\approx\hbar\omega$, actually correspond to states in the larger system containing the driving field. In this sense, we can say that the state $\ket{i,n}$, with the system in state $\ket{i}$ and the driving field containing $n$ quanta, is approximately resonant with the state $\ket{j,n+1}$, with the system in state $\ket{j}$ and the driving field containing $n+1$ quanta. A transition from $\ket{i}$ to $\ket{j}$ is actually a transition from $\ket{i,n}$ to $\ket{j,n+1}$, where one quantum is added simultaneously to the driving field in order to conserve energy (approximately). A transition from $\ket{j}$ to $\ket{i}$ corresponds to the system absorbing one quantum from the driving field.

The energy of the quantized driving field contributes an additional time dependence
\begin{align}
	\ket{i} &\mapsto \ket{i,n} e^{-\ii n\omega t}, &
	\ket{j} &\mapsto \ket{j,n+1} e^{-\ii (n+1)\omega t},
\end{align}
and \autoref{eq:dressedstates1} thus becomes
\begin{multline}
	\label{eq:dressedstates}
	\ket{\psi(t)} \approx\\
	\frac12 e^{-\ii\left(E_i+n\hbar\omega+\frac{\hbar(\Delta-\Omega)}{2}\right)t/\hbar} \left\{
		\left[\psi_i(0)\left(1+\frac{\Delta}{\Omega}\right)-\psi_j(0)\frac{T_{i j}}{\hbar\Omega}\right]\ket{i,n}
		+\left[\psi_j(0)\left(1-\frac{\Delta}{\Omega}\right)-\psi_i(0)\frac{T_{i j}^*}{\hbar\Omega}\right]\ket{j,n+1}
	\right\}\\
	+ \frac12 e^{-\ii\left(E_i+n\hbar\omega+\frac{\hbar(\Delta+\Omega)}{2}\right)t/\hbar} \left\{
		\left[\psi_i(0)\left(1-\frac{\Delta}{\Omega}\right)+\psi_j(0)\frac{T_{i j}}{\hbar\Omega}\right]\ket{i,n}
		+\left[\psi_j(0)\left(1+\frac{\Delta}{\Omega}\right)+\psi_i(0)\frac{T_{i j}^*}{\hbar\Omega}\right]\ket{j,n+1}
	\right\}
	\Bigg)\\
	= \frac12 e^{-\ii E_- t/\hbar}\ket{-}+\frac12 e^{-\ii E_+ t/\hbar}\ket{+}
\end{multline}
With this substitution, the state consists of two components, called \emph{dressed states},
\begin{equation}
	\label{eq:dressedstatespm}
	\ket{\pm} = 
		\left[\psi_i(0)\left(1\mp\frac{\Delta}{\Omega}\right)\pm\psi_j(0)\frac{T_{i j}}{\hbar\Omega}\right]\ket{i,n}
		+\left[\psi_j(0)\left(1\pm\frac{\Delta}{\Omega}\right)\pm\psi_i(0)\frac{T_{i j}^*}{\hbar\Omega}\right]\ket{j,n+1}.
\end{equation}
that are time-invariant apart from their energy (phase) prefactors.
These energy prefactors correspond to the effective energy of the dressed states in the presence of the oscillating field,\footnote{The instantaneous energy of a state is defined as $E=\avg{\op{H}}=\ii\hbar\avg{\frac{\partial}{\partial t}}$. For a state $\ket{\psi(t)}=e^{-\ii\omega t}\ket{\phi}$ the energy is $E=\ii\hbar\me{\psi(t)}{\frac{\partial}{\partial t}}{\psi(t)}=\ii\hbar\me{\phi}{e^{\ii\omega t}\frac{\partial}{\partial t}e^{-\ii\omega t}}{\phi}=\hbar\omega$.}
\begin{equation}
	E_{\pm} = E_i+n\hbar\omega+\frac{\hbar(\Delta\pm\Omega)}{2}=E_j+(n+1)\hbar\omega+\frac{\hbar(-\Delta\pm\Omega)}{2}.
\end{equation}

We look at these dressed states in two limits:
\begin{itemize}
	\item On resonance ($\Delta=0$), we have $\hbar\Omega=\abs{T_{i j}}$, and the dressed states of \autoref{eq:dressedstatespm} become
\begin{multline}
	\ket{\pm} = 
		\left[\psi_i(0)\pm\psi_j(0)\frac{T_{i j}}{\abs{T_{i j}}}\right]\ket{i,n}
		+\left[\psi_j(0)\pm\psi_i(0)\frac{T_{i j}^*}{\abs{T_{i j}}}\right]\ket{j,n+1}\\
	= \left[\psi_i(0)\pm\psi_j(0)\frac{T_{i j}}{\abs{T_{i j}}}\right]
		\left(\ket{i,n}\pm\frac{T_{i j}^*}{\abs{T_{i j}}}\ket{j,n+1}\right),
\end{multline}
which are equal mixtures of the original states $\ket{i,n}$ and $\ket{j,n+1}$.
	They have energies
\begin{equation}
	E_{\pm}=E_i+n\hbar\omega\pm\frac12\abs{T_{ij}}=E_j+(n+1)\hbar\omega\pm\frac12\abs{T_{ij}}
\end{equation}
in the presence of a resonant ac coupling field: the degeneracy of the levels $\ket{i,n}$ and $\ket{j,n+1}$ is lifted, and the dressed states are split by $E_+-E_-=\abs{T_{i j}}$.
	\item Far off-resonance ($\Delta\to\pm\infty$) we have $\Omega\approx\abs{\Delta}+\frac{\abs{T_{i j}}^2}{2\hbar^2\abs{\Delta}}$, and \autoref{eq:dressedstates} becomes
\begin{equation}
	\ket{\psi(t)} \approx
	e^{-\ii\left(E_i+n\hbar\omega-\frac{\abs{T_{i j}}^2}{4\hbar\Delta}\right)t/\hbar}\psi_i(0)\ket{i,n}
	+ e^{-\ii\left(E_j+(n+1)\hbar\omega+\frac{\abs{T_{i j}}^2}{4\Delta}\right)t/\hbar}\psi_j(0)\ket{j,n+1}.
\end{equation}
	(Hint: to verify this, look at the cases $\Delta\to+\infty$ and $\Delta\to-\infty$ separately).
	The energy levels $\ket{i,n}$ and $\ket{j,n+1}$ are thus shifted by $\mp\frac{\abs{T_{i j}}^2}{4\hbar\Delta}$, respectively, and there is no population transfer between the levels. That is, the dressed states become equal to the original states. Remember that we had assumed $E_i>E_j$:
	\begin{itemize}
		\item For a blue-detuned drive ($\Delta\to+\infty$), the upper level $\ket{i}$ is \emph{lowered} in energy by $\Delta E=\frac{\abs{T_{i j}}^2}{4\hbar\Delta}$ while the lower level $\ket{j}$ is \emph{raised} in energy by $\Delta E$.
		\item For a red-detuned drive ($\Delta\to-\infty$), the upper level $\ket{i}$ is \emph{raised} in energy by $\Delta E=\frac{\abs{T_{i j}}^2}{4\hbar\abs{\Delta}}$ while the lower level $\ket{j}$ is \emph{lowered} in energy by $\Delta E$.
	\end{itemize}
	These shifts are called \emph{ac Zeeman shifts}\index{Zeeman shift!ac} in this case, or \emph{level shifts}\index{level shift} more generally. When the oscillating field is a light field, level shifts are often called \emph{light shifts}\index{light shift} or \emph{ac Stark shifts}\index{Stark shift!ac}.

\end{itemize}
\end{description}

\subsection{exercises}

\begin{questions}
	\item\label{Q:IJF} Take two angular momenta, for example $I=3$ and $J=5$, and calculate the eigenvalues of the operators $\op{I}^2$, $\op{I}_z$, $\op{J}^2$, $\op{J}_z$, $\op{F}^2$, and $\op{F}_z$, where $\opvect{F}=\opvect{I}+\opvect{J}$.
\pagenote[\ref{Q:IJF}]{We define all operators in the combined Hilbert space of both spins, so that the operators $\opvect{F}$ can be defined by addition:
\begin{mathematica}
	¤protect¤mathin¤ With[{i=3,¤ j=5},
	¤protect¤mathnl¤ ¤ ¤ Ix¤ =¤ KroneckerProduct[sx[i],¤ id[j]];
	¤protect¤mathnl¤ ¤ ¤ Iy¤ =¤ KroneckerProduct[sy[i],¤ id[j]];
	¤protect¤mathnl¤ ¤ ¤ Iz¤ =¤ KroneckerProduct[sz[i],¤ id[j]];
	¤protect¤mathnl¤ ¤ ¤ Jx¤ =¤ KroneckerProduct[id[i],¤ sx[j]];
	¤protect¤mathnl¤ ¤ ¤ Jy¤ =¤ KroneckerProduct[id[i],¤ sy[j]];
	¤protect¤mathnl¤ ¤ ¤ Jz¤ =¤ KroneckerProduct[id[i],¤ sz[j]];
	¤protect¤mathnl¤ ¤ ¤ Fx=Ix+Jx;¤ Fy=Iy+Jy;¤ Fz=Iz+Jz;]
\end{mathematica}
	We calculate the eigenvalues in ascending order with \mm{Sort}. Remember that $\abs{I-J}\le F\le I+J$, and therefore $F\in\{2,3,4,5,6,7,8\}$ and $\avg{\op{F}^2}=F(F+1)\in\{6,12,20,30,42,56,72\}$.
\begin{mathematica}
	¤protect¤mathin¤ Ix.Ix¤ +¤ Iy.Iy¤ +¤ Iz.Iz¤ //Eigenvalues¤ //Sort
	¤protect¤mathout¤ {12,12,12,12,12,12,12,12,12,12,12,12,12,12,12,12,12,12,12,12,12,12,12,12,12,
	¤protect¤mathnl¤ 12,12,12,12,12,12,12,12,12,12,12,12,12,12,12,12,12,12,12,12,12,12,12,12,12,12,
	¤protect¤mathnl¤ 12,12,12,12,12,12,12,12,12,12,12,12,12,12,12,12,12,12,12,12,12,12,12,12,12,12}
	¤protect¤mathin¤ Iz¤ //Eigenvalues¤ //Sort
	¤protect¤mathout¤ {-3,-3,-3,-3,-3,-3,-3,-3,-3,-3,-3,-2,-2,-2,-2,-2,-2,-2,-2,-2,-2,-2,-1,-1,-1,
	¤protect¤mathnl¤ -1,-1,-1,-1,-1,-1,-1,-1,0,0,0,0,0,0,0,0,0,0,0,1,1,1,1,1,1,1,1,1,1,1,2,2,2,2,2,
	¤protect¤mathnl¤ 2,2,2,2,2,2,3,3,3,3,3,3,3,3,3,3,3}
	¤protect¤mathin¤ Jx.Jx¤ +¤ Jy.Jy¤ +¤ Jz.Jz¤ //Eigenvalues¤ //Sort
	¤protect¤mathout¤ {30,30,30,30,30,30,30,30,30,30,30,30,30,30,30,30,30,30,30,30,30,30,30,30,30,
	¤protect¤mathnl¤ 30,30,30,30,30,30,30,30,30,30,30,30,30,30,30,30,30,30,30,30,30,30,30,30,30,30,
	¤protect¤mathnl¤ 30,30,30,30,30,30,30,30,30,30,30,30,30,30,30,30,30,30,30,30,30,30,30,30,30,30}
	¤protect¤mathin¤ Jz¤ //Eigenvalues¤ //Sort
	¤protect¤mathout¤ {-5,-5,-5,-5,-5,-5,-5,-4,-4,-4,-4,-4,-4,-4,-3,-3,-3,-3,-3,-3,-3,-2,-2,-2,-2,
	¤protect¤mathnl¤ -2,-2,-2,-1,-1,-1,-1,-1,-1,-1,0,0,0,0,0,0,0,1,1,1,1,1,1,1,2,2,2,2,2,2,2,3,3,3,
	¤protect¤mathnl¤ 3,3,3,3,4,4,4,4,4,4,4,5,5,5,5,5,5,5}
	¤protect¤mathin¤ Fx.Fx¤ +¤ Fy.Fy¤ +¤ Fz.Fz¤ //Eigenvalues¤ //Sort
	¤protect¤mathout¤ {6,6,6,6,6,12,12,12,12,12,12,12,20,20,20,20,20,20,20,20,20,30,30,30,30,30,30,
	¤protect¤mathnl¤ 30,30,30,30,30,42,42,42,42,42,42,42,42,42,42,42,42,42,56,56,56,56,56,56,56,56,
	¤protect¤mathnl¤ 56,56,56,56,56,56,56,72,72,72,72,72,72,72,72,72,72,72,72,72,72,72,72,72}
	¤protect¤mathin¤ Fz¤ //Eigenvalues¤ //Sort
	¤protect¤mathout¤ {-8,-7,-7,-6,-6,-6,-5,-5,-5,-5,-4,-4,-4,-4,-4,-3,-3,-3,-3,-3,-3,-2,-2,-2,-2,
	¤protect¤mathnl¤ -2,-2,-2,-1,-1,-1,-1,-1,-1,-1,0,0,0,0,0,0,0,1,1,1,1,1,1,1,2,2,2,2,2,2,2,3,3,3,
	¤protect¤mathnl¤ 3,3,3,4,4,4,4,4,5,5,5,5,6,6,6,7,7,8}
\end{mathematica}}
	\item\label{Q:ClebschGordan} In \ref{Q:IJF} you have coupled two angular momenta but you have not used any Clebsch--Gordan coefficients. Why not? Where do these coefficients appear?\index{Clebsch-Gordan coefficient@Clebsch--Gordan coefficient}
\pagenote[\ref{Q:ClebschGordan}]{The Clebsch--Gordan coefficients $\CG{I}{M_I}{J}{M_J}{F}{M_F}$ serve to construct the states that simultaneously diagonalize $\op{I}^2$, $\op{J}^2$, $\op{F}^2$, and $\op{F}_z$ from those that simultaneously diagonalize $\op{I}^2$, $\op{I}_z$, $\op{J}^2$, and $\op{J}_z$: for $\abs{I-J}\le F\le I+J$ and $\abs{M_F}\le F$,
\begin{equation}
	\ket{I,J,F,M_F} = \sum_{M_I=-I}^I \sum_{M_J=-J}^J \CG{I}{M_I}{J}{M_J}{F}{M_F} \ket{I,M_I}\otimes\ket{J,M_J}
\end{equation}
	In Mathematica, $\mm{S[i,j,F,MF]}=\ket{I,J,F,M_F}$:
\begin{mathematica}
	¤protect¤mathin¤ S[i_,j_,F_,MF_]¤ :=¤ Sum[ClebschGordan[{i,Mi},{j,Mj},{F,MF}]*
	¤protect¤mathnl¤ ¤ ¤ Flatten[KroneckerProduct[SparseArray[i-Mi+1->1,¤ 2i+1],
	¤protect¤mathnl¤ ¤ ¤ ¤ ¤ ¤ ¤ ¤ ¤ ¤ ¤ ¤ ¤ ¤ ¤ ¤ ¤ ¤ ¤ ¤ ¤ ¤ ¤ ¤ ¤ ¤ ¤ ¤ SparseArray[j-Mj+1->1,¤ 2j+1]]],
	¤protect¤mathnl¤ ¤ ¤ {Mi,-i,i},¤ {Mj,-j,j}]	
\end{mathematica}
	Check that these diagonalize $\op{I}^2$, $\op{J}^2$, $\op{F}^2$, and $\op{F}_z$ simultaneously:
\begin{mathematica}
	¤protect¤mathin¤ With[{i=3,¤ j=5},
	¤protect¤mathnl¤ ¤ ¤ Table[(Ix.Ix+Iy.Iy+Iz.Iz).S[i,j,F,MF]¤ ==¤ i(i+1)*S[i,j,F,MF]¤ &&
	¤protect¤mathnl¤ ¤ ¤ ¤ ¤ ¤ ¤ ¤ ¤ (Jx.Jx+Jy.Jy+Jz.Jz).S[i,j,F,MF]¤ ==¤ j(j+1)*S[i,j,F,MF]¤ &&
	¤protect¤mathnl¤ ¤ ¤ ¤ ¤ ¤ ¤ ¤ ¤ (Fx.Fx+Fy.Fy+Fz.Fz).S[i,j,F,MF]¤ ==¤ F(F+1)*S[i,j,F,MF]¤ &&
	¤protect¤mathnl¤ ¤ ¤ ¤ ¤ ¤ ¤ ¤ ¤ ¤ ¤ ¤ ¤ ¤ ¤ ¤ ¤ ¤ ¤ ¤ ¤ ¤ ¤ ¤ ¤ ¤ Fz.S[i,j,F,MF]¤ ==¤ MF*S[i,j,F,MF],
	¤protect¤mathnl¤ ¤ ¤ ¤ ¤ {F,Abs[i-j],i+j},¤ {MF,-F,F}]
\end{mathematica}
	(disregard the warnings about \mm{ClebschGordan::phy}).
	
	When we use the basis of product states $\ket{I,M_I}\otimes\ket{J,M_J}$ to calculate the eigenvectors of the matrices \mm{Fx.Fx+Fy.Fy+Fz.Fz} and \mm{Fz}, using either \mm{Eigenvectors} or \mm{Eigensystem}, these Clebsch--Gordan coefficients appear naturally as coefficients of the resulting eigenvectors.}
	\item\label{Q:spinexpectation} For a spin of a certain length, for example $S=100$, take the state $\ket{S,S}$ (a spin pointing in the $+z$ direction) and calculate the expectation values $\avg{\op{S}_x}$, $\avg{\op{S}_y}$, $\avg{\op{S}_z}$, $\avg{\op{S}_x^2}-\avg{\op{S}_x}^2$, $\avg{\op{S}_y^2}-\avg{\op{S}_y}^2$, $\avg{\op{S}_z^2}-\avg{\op{S}_z}^2$. \emph{Hint:} the expectation value of an operator $\op{A}$ is $\me{S,S}{\op{A}}{S,S}$.
\pagenote[\ref{Q:spinexpectation}]{
\begin{mathematica}
	¤protect¤mathin¤ With[{S¤ =¤ 100},
	¤protect¤mathnl¤ ¤ ¤ ¤mmpsi¤ =¤ SparseArray[1->1,¤ 2S+1];
	¤protect¤mathnl¤ ¤ ¤ {x,y,z,¤ xx,yy,zz}¤ =¤ Conjugate[¤mmpsi].(¤#.¤mmpsi)¤&¤ /@
	¤protect¤mathnl¤ ¤ ¤ ¤ ¤ {sx[S],sy[S],sz[S],¤ sx[S].sx[S],sy[S].sy[S],sz[S].sz[S]};
	¤protect¤mathnl¤ ¤ ¤ {{x,y,z},¤ {xx-x^2,yy-y^2,zz-z^2}}]
	¤protect¤mathout¤ {{0,¤ 0,¤ 100},¤ {50,¤ 50,¤ 0}}
\end{mathematica}
	In general,
	\begin{align}
		\avg{\op{S}_x} &=0 & \avg{\op{S}_y} &=0 & \avg{\op{S}_z} &=S\nonumber\\
		\avg{\op{S}_x^2} &=S/2 & \avg{\op{S}_y^2} &=S/2 & \avg{\op{S}_z^2} &=S^2\nonumber\\
		\avg{\op{S}_x^2}-\avg{\op{S}_x}^2 &=S/2 & \avg{\op{S}_y^2}-\avg{\op{S}_y}^2 &=S/2 & \avg{\op{S}_z^2}-\avg{\op{S}_z}^2 &=0
	\end{align}}
	\item\label{Q:Rabifreq} Use \mm{\ref{math:Delta}} and \mm{\ref{math:Omega}} to calculate the detuning $\Delta$ and the generalized Rabi frequency $\Omega$ for the $^{87}$Rb solution of \mm{\ref{math:Rb87timesol}}, where the population oscillates between the levels $i=2$ and $j=7$. What is the oscillation period corresponding to $\Omega$? Does it match the plots of \mm{\ref{math:Rb87pop1}} and \mm{\ref{math:Rb87pop5}}?
\pagenote[\ref{Q:Rabifreq}]{
\begin{mathematica}
	¤protect¤mathin¤ {¤mmDelta,¤ ¤mmOmega}¤ /.
	¤protect¤mathnl¤ ¤ ¤ {Ei¤ ->¤ eval[[2]],¤ Ej¤ ->¤ eval[[7]],¤ Tij¤ ->¤ T[[2,7]],¤ Tji¤ ->¤ T[[7,2]]}¤ /.
	¤protect¤mathnl¤ ¤ ¤ hfc¤ /.¤ {Bz->3.22895,Bacx->0.1,Bacy->0,Bacz->0,¤mmomega->2¤mmpi*6827.9}
	¤protect¤mathout¤ {0.00476766,¤ 0.762616}
\end{mathematica}
	The oscillation period is $2\pi/\Omega=\SI{8,23899}{\micro s}$, which matches the full oscillation periods of the plots of \mm{\ref{math:Rb87pop1}} and \mm{\ref{math:Rb87pop5}}.}
	\item\label{Q:sodium23} Do the presented alkali atom calculation for $^{23}$Na: are there any magic field values?\\ \url{http://steck.us/alkalidata/sodiumnumbers.pdf}
\pagenote[\ref{Q:sodium23}]{$^{23}$Na has the same nuclear spin $I=3/2$ as $^{87}$Rb; they differ only in the constants:
	\begin{itemize}
		\item $A\ix{hfs}=\SI{885,81306440}{MHz}$
		\item $g_I=\num{0,00080461080}$
		\item $g_L=\num{-0,99997613}$
	\end{itemize}
	As a result, there is a magic field between the same states as for $^{87}$Rb, but at a field strength $B_z=\SI{0,676851}{G}\approx\frac{16A g_I}{3\mu\ix{B}g_S^2}$.}
	\item\label{Q:rubidium85} Do the presented alkali atom calculation for $^{85}$Rb: are there any magic field values?\\ \url{http://steck.us/alkalidata/rubidium85numbers.pdf}
\pagenote[\ref{Q:rubidium85}]{$^{85}$Rb has a nuclear spin $I=5/2$, which means that we must re-define the spin operators; all operators are now \num{12x12} matrices. Further, the constants to be used are
	\begin{itemize}
		\item $A\ix{hfs}=\SI{1,0119108130}{GHz}$
		\item $g_I=\num{0,00029364000}$
		\item $g_L=\num{-0,99999354}$
	\end{itemize}
	There are two magic fields:
	\begin{itemize}
		\item $B_z=\SI{0,357312}{G}\approx\frac{27A g_I}{4\mu\ix{B}g_S^2}$: the energy difference between $\ket{F=2,M_F=-1}$ and $\ket{F=3,M_F=1}$ is stationary.
		\item $B_z=\SI{1,14342}{G}\approx\frac{108A g_I}{5\mu\ix{B}g_S^2}$: the energy difference between $\ket{F=2,M_F=-2}$ and $\ket{F=3,M_F=2}$ is stationary.
	\end{itemize}}
	\item\label{Q:cesium133} Do the presented alkali atom calculation for $^{133}$Cs: are there any magic field values?\\ \url{http://steck.us/alkalidata/cesiumnumbers.pdf}
\pagenote[\ref{Q:cesium133}]{$^{133}$Cs has a nuclear spin $I=7/2$, which means that we must re-define the spin operators; all operators are now \num{16x16} matrices. Further, the constants to be used are
	\begin{itemize}
		\item $A\ix{hfs}=\SI{2,2981579425}{GHz}$
		\item $g_I=\num{0,00039885395}$
		\item $g_L=\num{-0,99999587}$
	\end{itemize}
	There are three magic fields:
	\begin{itemize}
		\item $B_z=\SI{1,39334}{G}\approx\frac{128A g_I}{15\mu\ix{B}g_S^2}$: the energy difference between $\ket{F=3,M_F=-1}$ and $\ket{F=4,M_F=1}$ is stationary.
		\item $B_z=\SI{3,48338}{G}\approx\frac{64A g_I}{3\mu\ix{B}g_S^2}$: the energy difference between $\ket{F=3,M_F=-2}$ and $\ket{F=4,M_F=2}$ is stationary.
		\item $B_z=\SI{8,9572}{G}\approx\frac{384A g_I}{7\mu\ix{B}g_S^2}$: the energy difference between $\ket{F=3,M_F=-3}$ and $\ket{F=4,M_F=3}$ is stationary.
	\end{itemize}}
	\item\label{Q:circularpol} Set $\vect{B}=0$ and $\vect{B}\ex{ac}=B(\vect{e}_x+\ii\vect{e}_y)$ in the expression for \mm{T} in \mm{\ref{math:T}}. Which transitions are allowed for such circularly-polarized light around the quantization axis? \emph{Hint:} use \autoref{eq:Rb87identify} to identify the states.
\pagenote[\ref{Q:circularpol}]{In the result from
\begin{mathematica}
	¤protect¤mathin¤ Assuming[A>0,¤ FullSimplify[T/.{Bx->0,By->0,Bz->0,Bacx->B,Bacy->I*B,Bacz->0}]]
\end{mathematica}
	we can see that the transitions $1\leftrightarrow5$, $1\leftrightarrow6$, $3\leftrightarrow7$, $3\leftrightarrow8$, $4\leftrightarrow7$, $4\leftrightarrow8$ are allowed. Using \autoref{eq:Rb87identify} we identify these transitions as $\ket{2,2}\leftrightarrow\ket{1,1}$, $\ket{2,2}\leftrightarrow\ket{2,1}$, $\ket{1,0}\leftrightarrow\ket{1,-1}$, $\ket{1,0}\leftrightarrow\ket{2,-1}$, $\ket{2,0}\leftrightarrow\ket{1,-1}$, $\ket{2,0}\leftrightarrow\ket{2,-1}$. These transitions are all $\Delta M_F=\pm1$.}
	\item\label{Q:linearpol} Set $\vect{B}=0$ and $\vect{B}\ex{ac}=B\vect{e}_z$ in the expression for \mm{T} in \mm{\ref{math:T}}. Which transitions are allowed for such linearly-polarized light along the quantization axis? \emph{Hint:} use \autoref{eq:Rb87identify} to identify the states.
\pagenote[\ref{Q:linearpol}]{In the result from
\begin{mathematica}
	¤protect¤mathin¤ Assuming[A>0,¤ FullSimplify[T/.{Bx->0,By->0,Bz->0,Bacx->0,Bacy->0,Bacz->B}]]
\end{mathematica}
	we can see that the transitions $3\leftrightarrow4$, $5\leftrightarrow6$, $7\leftrightarrow8$ are allowed. Using \autoref{eq:Rb87identify} we identify these transitions as $\ket{1,0}\leftrightarrow\ket{2,0}$, $\ket{1,1}\leftrightarrow\ket{2,1}$, $\ket{1,-1}\leftrightarrow\ket{2,-1}$. These transitions are all $\Delta M_F=0$.
	Further, the energy of levels 1, 2, 5, 6, 7, 8 (\ie, all levels with $M_F\neq0$) will be shifted by the non-zero diagonal elements of \mm{T}.}
\end{questions}

\section[coupled spin systems: Ising model in a transverse field]{\label{sec:Ising}coupled spin systems: Ising model in a transverse field\hspace{\stretch{1}}\attachcode{IsingModel}{Ising model in a transverse field}}\index{Ising model}

We now turn to larger numbers of coupled quantum-mechanical spins. A large class of such coupled spin systems can be described with Hamiltonians of the form
\begin{equation}
	\Ham = \sum_{k=1}^N \Ham^{(k)} + \sum_{k=1}^{N-1}\sum_{k'=k+1}^N \Ham\ix{int}^{(k,k')},
\end{equation}
where the $\Ham^{(k)}$ are single-spin Hamiltonians (for example couplings to a magnetic field) and the $\Ham\ix{int}^{(k,k')}$ are coupling Hamiltonians between two spins. Direct couplings between three or more spins can usually be neglected.

As an example we study the dimensionless ``transverse Ising'' Hamiltonian
\begin{equation}
	\Ham = -\frac{b}{2} \sum_{k=1}^N \op{S}_x^{(k)} - \sum_{k=1}^N \op{S}_z^{(k)} \op{S}_z^{(k+1)}
\end{equation}
acting on a ring of $N$ spin-$S$ systems where the ($N+1$)st spin is identified with the first spin. We can read off three limits from this Hamiltonian:
\begin{itemize}
	\item For $b\to\pm\infty$ the spin--spin coupling Hamiltonian can be neglected, and the ground state will have all spins aligned with the $\pm x$ direction,
	\begin{align}
		\label{eq:allupalldown}
		\ket{\psi_{+\infty}} &= \ket{+x}^{\otimes N}, & \ket{\psi_{-\infty}} &= \ket{-x}^{\otimes N}.
	\end{align}
	The system is therefore in a product state for $b\to\pm\infty$, which means that there is no entanglement between spins. In the basis of $\ket{S,M}$ Dicke states, \autoref{eq:spinopme1} and \autoref{eq:spinopme2}, the single-spin states making up these product states are
	\begin{subequations}
	\label{eq:xupdown}
	\begin{align}
		\ket{+x} &= 2^{-S} \sum_{M=-S}^S \sqrt{\binom{2S}{M+S}}\,\,\ket{S,M},\\
		\ket{-x} &= 2^{-S} \sum_{M=-S}^S (-1)^{M+S}\sqrt{\binom{2S}{M+S}}\,\,\ket{S,M},
	\end{align}
	\end{subequations}
	which are aligned with the $x$-axis in the sense that $\op{S}_x\ket{+x}=S\,\ket{+x}$ and $\op{S}_x\ket{-x}=-S\,\ket{-x}$.
	\item For $b=0$ the Hamiltonian contains only nearest-neighbor ferromagnetic spin--spin couplings $-\op{S}_z^{(k)} \op{S}_z^{(k+1)}$. We know that this Hamiltonian has two degenerate ground states: all spins pointing up or all spins pointing down,
	\begin{align}
		\ket{\psi_{0\uparrow}} &= \ket{+z}^{\otimes N}, &
		\ket{\psi_{0\downarrow}} &= \ket{-z}^{\otimes N},
	\end{align}
	where in the Dicke-state representation of \autoref{eq:spinopme1} we have $\ket{+z}=\ket{S,+S}$ and $\ket{-z}=\ket{S,-S}$.
	While these two states are product states, for $\abs{b}\ll1$ the perturbing Hamiltonian $-\frac{b}{2}\sum_{k=1}^N \op{S}_x^{(k)}$ is diagonal in the states $\frac{\ket{\psi_{0\uparrow}}\pm\ket{\psi_{0\downarrow}}}{\sqrt{2}}$, which are not product states. The exact ground state for $0<b\ll1$ is close to $\frac{\ket{\psi_{0\uparrow}}+\ket{\psi_{0\downarrow}}}{\sqrt{2}}$, and for $-1\ll b<0$ it is close to $\frac{\ket{\psi_{0\uparrow}}-\ket{\psi_{0\downarrow}}}{\sqrt{2}}$. These are both maximally entangled states (``Schr\"odinger cat states'').
\end{itemize}
Now we calculate the ground state $\ket{\psi_b}$ as a function of the parameter $b$, and compare the results to the above asymptotic limits.

\subsection{basis set}
\label{sec:Isingbasis}

The natural basis set for describing a set of $N$ coupled spins is the tensor-product basis\index{tensor!product} (see \autoref{sec:coupledDOF}). In this basis, the spin operators $\op{S}_{x,y,z}^{(k)}$ acting only on spin $k$ are defined as having a trivial action on all other spins, for example
\begin{equation}
	\op{S}_x^{(k)} \mapsto \underbrace{\one\otimes\one\otimes\dots\otimes\one}_{(k-1)}\otimes\op{S}_x\otimes\underbrace{\one\otimes\dots\otimes\one}_{(N-k)}.
\end{equation}
In Mathematica such single-spin-$S$ operators acting on spin $k$ out of a set of $N$ spins are defined as follows. First we define the operator acting as $\op{a}=\mm{a}$ on the $\mm{k}\ex{th}$ spin out of a set of \mm{n} spins, and trivially on all others:
\begin{mathematica}
	¤mathin¤labelŽmath:spinop op[S_?SpinQ, n_Integer, k_Integer, a_?MatrixQ] /;
	¤mathnl   1<=k<=n && Dimensions[a] == {2S+1,2S+1} :=
	¤mathnl   KroneckerProduct[IdentityMatrix[(2S+1)^(k-1), SparseArray],
	¤mathnl                    a,
	¤mathnl                    IdentityMatrix[(2S+1)^(n-k), SparseArray]]
\end{mathematica}
Next, we specialize this to $\op{a}=\op{S}_x$, $\op{S}_y$, $\op{S}_z$:
\begin{mathematica}
	¤mathin sx[S_?SpinQ, n_Integer, k_Integer] /; 1<=k<=n := op[S, n, k, sx[S]]
	¤mathin sy[S_?SpinQ, n_Integer, k_Integer] /; 1<=k<=n := op[S, n, k, sy[S]]
	¤mathin sz[S_?SpinQ, n_Integer, k_Integer] /; 1<=k<=n := op[S, n, k, sz[S]]
\end{mathematica}
Notice that we have used $\text{\mm{n}}=N$ because the symbol \mm{N} is already used internally in Mathematica.

From these we assemble the Hamiltonian:
\begin{mathematica}
	¤mathin H[S_?SpinQ, n_Integer/;n>=3, b_] := -b/2*Sum[sx[S, n, k], {k, n}] -
	¤mathnl   Sum[sz[S, n, k].sz[S, n, Mod[k+1,n,1]], {k, n}]
\end{mathematica}
The modulus \mm{Mod[k+1,n,1]} represents the periodicity of the spin ring and ensures that the index remains within $1\dots N$ (\ie, a modulus with offset 1).

\subsection{asymptotic ground states}

The asymptotic ground states for $b=0$ and $b\to\pm\infty$ mentioned above are all product states of the form $\ket{\psi}=\ket{\theta}^{\otimes N}$ where $\ket{\theta}$ is the state of a single spin. We form an $N$-particle tensor product state of such single-spin states with\index{Mathematica!Kronecker product}\index{product state}
\begin{mathematica}
	¤mathin productstate[¤mmtheta_?VectorQ, 1] = ¤mmtheta;
	¤mathin productstate[¤mmtheta_?VectorQ, n_Integer/;n>=2] :=
	¤mathnl    Flatten[KroneckerProduct @@ Table[¤mmtheta, n]]
\end{mathematica}
in accordance with \mm{\ref{math:psi}}; notice that the case $N=1$ requires special attention.

The particular single-spin states $\ket{+x}$, $\ket{-x}$, $\ket{+z}$, $\ket{-z}$ we will be using are
\begin{mathematica}
	¤mathin¤labelŽmath:xup xup[S_?SpinQ] := 2^(-S)*Table[Sqrt[Binomial[2S,M+S]],{M,S,-S,-1}]
	¤mathin¤labelŽmath:xdn xdn[S_?SpinQ] := 2^(-S)*Table[(-1)^(M+S)*Sqrt[Binomial[2S,M+S]], {M,S,-S,-1}]
	¤mathin zup[S_?SpinQ] := SparseArray[1 -> 1, 2S+1]
	¤mathin zdn[S_?SpinQ] := SparseArray[-1 -> 1, 2S+1]
\end{mathematica}
We can check that these are correct with
\begin{mathematica}
	¤mathin Table[sx[S].xup[S] == S*xup[S], {S, 0, 4, 1/2}]
	¤mathout {True, True, True, True, True, True, True, True, True}
	¤mathin Table[sx[S].xdn[S] == -S*xdn[S], {S, 0, 4, 1/2}]
	¤mathout {True, True, True, True, True, True, True, True, True}
	¤mathin Table[sz[S].zup[S] == S*zup[S], {S, 0, 4, 1/2}]
	¤mathout {True, True, True, True, True, True, True, True, True}
	¤mathin Table[sz[S].zdn[S] == -S*zdn[S], {S, 0, 4, 1/2}]
	¤mathout {True, True, True, True, True, True, True, True, True}
\end{mathematica}
From these we construct the product states
\begin{mathematica}
	¤mathin allxup[S_?SpinQ,n_Integer/;n>=1] := productstate[xup[S],n]
	¤mathin allxdn[S_?SpinQ,n_Integer/;n>=1] := productstate[xdn[S],n]
	¤mathin allzup[S_?SpinQ,n_Integer/;n>=1] := productstate[zup[S],n]
	¤mathin allzdn[S_?SpinQ,n_Integer/;n>=1] := productstate[zdn[S],n]
\end{mathematica}

\subsection{Hamiltonian diagonalization}

\newcommand{\IsingSize}{20}	
\newcommand{\SmallIsingSize}{14}	
We find the $m$ lowest-energy eigenstates of this Hamiltonian with the procedures described in \autoref{sec:diagonalization}: for example, with $S=1/2$ and $N=\IsingSize$,\footnote{The attached Mathematica code uses $N=\SmallIsingSize$ instead, since calculations with $N=\IsingSize$ take a long time.}
\begin{mathematica}
	¤mathin¤labelŽmath:HXY With[{S = 1/2, n = ¤IsingSize},
	¤mathnlc   (* Hamiltonian *)
	¤mathnl   h[b_] = H[S, n, b];
	¤mathnlc   (* two degenerate ground states for b=0 *)
	¤mathnl   gs0up = allzup[S, n];
	¤mathnl   gs0dn = allzdn[S, n];
	¤mathnlc   (* ground state for b=+Infinity *)
	¤mathnl   gsplusinf = allxup[S, n];
	¤mathnlc   (* ground state for b=-Infinity *)
	¤mathnl   gsminusinf = allxdn[S, n];
	¤mathnlc   (* numerically calculate lowest m eigenstates *) 
	¤mathnl   Clear[gs];
	¤labelŽmathline:numericalH¤mathnl   gs[b_?NumericQ, m_Integer /; m>=1] := gs[b, m] = -Eigensystem[-h[N[b]], m, 
	¤mathnl     Method -> {"Arnoldi", "Criteria" -> "RealPart", MaxIterations -> 10^6}] //
	¤labelŽmathline:sorting¤mathnl     Transpose //Sort //Transpose;
	¤labelŽmathline:observables¤mathnl ]
\end{mathematica}
Comments:
\begin{itemize}
	\item $\text{\mm{gs0up}}=\ket{\psi_{0\uparrow}}$ and $\text{\mm{gs0dn}}=\ket{\psi_{0\downarrow}}$ are the exact degenerate ground states for $b=0$; $\text{\mm{gsplusinf}}=\ket{\psi_{+\infty}}$ and $\text{\mm{gsminusinf}}=\ket{\psi_{-\infty}}$ are the exact nondegenerate ground states for $b=\pm\infty$.
	\item The function \mm{gs}, which calculates the \mm{m} lowest-lying eigenstates of the Hamiltonian, remembers its calculated values (see \autoref{sec:functionswhichremember}): this is important here because such eigenstate calculations can take a long time when \mm{n} is large.
	\item The function \mm{gs} numerically calculates the eigenvalues using \mm{h[N[b]]} as a Hamiltonian, which ensures that the Hamiltonian contains floating-point machine-precision numbers instead of exact numbers in case \mm{b} is given as an exact number. Calculating the eigenvalues and eigenvectors of a matrix of exact numbers takes extremely long (please try: on line~\ref*{mathline:numericalH} of \mm{\ref{math:HXY}} replace \mm{-Eigensystem[-h[N[b]], ...} with \mm{-Eigensystem[-h[b], ...} and compare the run time of \mm{gs[1, 2]} with that of \mm{gs[1.0, 2]}.).
	\item The operations \mm{//Transpose //Sort //Transpose} on line~\ref*{mathline:sorting} of \mm{\ref{math:HXY}} ensure that the eigenvalues (and associated eigenvectors) are sorted in ascending energy order (see \mm{\ref{math:sortev}}).
	\item When the ground state is degenerate, which happens here for $b\approx0$, the Arnoldi algorithm has some difficulty finding the correct degeneracy. This means that \mm{gs[0,2]} may return two non-degenerate eigenstates instead of the (correct) two degenerate ground states. This is a well-known problem that can be circumvented by calculating more eigenstates.
	\item A problem involving $N$ spin-$S$ systems leads to matrices of size $(2S+1)^N\times(2S+1)^N$. This scaling quickly becomes very problematic (even if we use sparse matrices) and is at the center of why quantum mechanics is difficult. Imagine a system composed of $N=\num{1000}$ spins $S=1/2$: its state vector is a list of $2^{\num{1000}}=\num{1.07e301}$ complex numbers! Comparing this to the fact that there are only about \num{e80} particles in the universe, we conclude that such a state vector could never be written down and therefore the Hilbert space\index{Hilbert space} method of quantum mechanics we are using here is fundamentally flawed. But as this is an introductory course, we will stick to this classical matrix-mechanics formalism and let the computer bear the weight of its complexity. Keep in mind, though, that this is not a viable strategy for large systems, as each doubling of computer capacity only allows us to add a single spin to the system, which, using Moore's law, allows us to add one spin every two years.\index{Moore's law}\footnote{Moore's law is the observation that over the history of computing hardware, the number of transistors on integrated circuits doubles approximately every two years. From \url{https://en.wikipedia.org/wiki/Moore's_law}.}

There are alternative formulations of quantum mechanics, notably the path-integral formalism,
\index{path integral} which partly circumvent this problem; but the computational difficulty is not eliminated, it is merely shifted. Modern developments such as tensor networks\footnote{Matrix product states and tensor networks: \url{https://en.wikipedia.org/wiki/Matrix_product_state}.}\index{tensor networks} try to limit the accessible Hilbert space by restricting calculations to a subspace where the entanglement between particles is bounded. This makes sense since almost all states of the huge Hilbert space are so complex and carry such complicated quantum-mechanical entanglement that (i) they would be extremely difficult to generate with realistic Hamiltonians, and (ii) they would decohere within very short time.
\end{itemize}

\subsection{analysis of the ground state}

\subsubsection{energy gap}\index{energy gap}

Much of the behavior of our Ising spin chain can be seen in a plot of the \emph{energy gap}, which is the energy difference between the ground state and the first excited state. With $\text{\mm{m}}=2$ we calculate the two lowest-lying energy levels and plot their energy difference as a function of the parameter $b$:
\begin{mathematica}
	¤mathin With[{bmax = 3, db = 1/64, m = 2},
	¤mathnl   ListLinePlot[Table[{b, gs[b,m][[1,2]]-gs[b,m][[1,1]]},
	¤mathnl     {b, -bmax, bmax, db}]]]
\end{mathematica}
Notice how the fact that the \mm{gs} function remembers its own results speeds up this calculation by a factor of 2 (see \autoref{sec:functionswhichremember}).
\begin{center}
\includegraphics[width=0.5\textwidth]{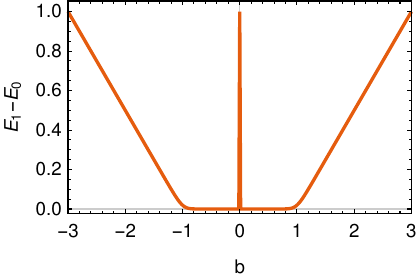}
\end{center}
Even in this small \IsingSize-spin simulation we can see that this gap is approximately
\begin{equation}
	\label{eq:IsingGap}
	E_1-E_0 \approx \begin{cases}
		0 & \text{if $\abs{b}<1$,}\\
		\frac{\abs{b}-1}{2} & \text{if $\abs{b}>1$.}
	\end{cases}
\end{equation}
This observation of a qualitative change in the excitation gap suggests that at $b=\pm 1$ the system undergoes a \emph{quantum phase transition}\index{quantum phase transition} (\ie, a phase transition induced by quantum fluctuations instead of thermal fluctuations). We note that the gap of \autoref{eq:IsingGap} is independent of the particle number $N$ and is therefore a \emph{global} property of the Ising spin ring, not a property of each individual spin (in which case it would scale with $N$).

\subsubsection{overlap with asymptotic states}

Once a ground state $\ket{\psi_b}$ has been calculated, we compute its overlap with the asymptotically known states using scalar products. Notice that for $b=0$ we calculate the scalar products with the states $\frac{\ket{\psi_{0\uparrow}}\pm\ket{\psi_{0\downarrow}}}{\sqrt{2}}$ as they are the approximate ground states for $\abs{b}\ll 1$.
\begin{mathematica}
	¤mathin With[{bmax = 3, db = 1/64, m = 2},
	¤mathnl   ListLinePlot[
	¤mathnl     Table[{{b, Abs[gsminusinf.gs[b,m][[2,1]]]^2},
	¤mathnl      {b, Abs[gsplusinf.gs[b, m][[2,1]]]^2},
	¤mathnl      {b, Abs[((gs0up-gs0dn)/Sqrt[2]).gs[b,m][[2,1]]]^2},
	¤mathnl      {b, Abs[((gs0up+gs0dn)/Sqrt[2]).gs[b,m][[2,1]]]^2},
	¤mathnl      {b, Abs[((gs0up-gs0dn)/Sqrt[2]).gs[b,m][[2,1]]]^2 +
	¤mathnl          Abs[((gs0up+gs0dn)/Sqrt[2]).gs[b,m][[2,1]]]^2}},
	¤mathnl        {b, -bmax, bmax, db}] //Transpose]]
\end{mathematica}
\begin{center}
\includegraphics[width=0.5\textwidth]{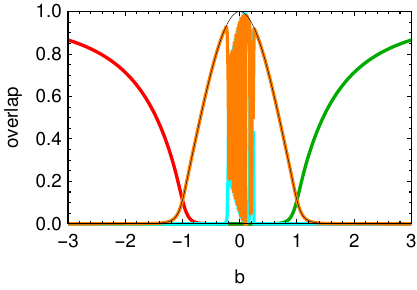}
\end{center}
Observations:
\begin{itemize}
	\item The overlap $\abs{\scp{\psi_b}{\psi_{-\infty}}}^2$ (red) approaches 1 as $b\to-\infty$.
	\item The overlap $\abs{\scp{\psi_b}{\psi_{+\infty}}}^2$ (green) approaches 1 as $b\to+\infty$.
	\item The overlap $\left|\langle\psi_b|\frac{\ket{\psi_{0\uparrow}}-\ket{\psi_{0\downarrow}}}{\sqrt{2}}\right|^2$ (cyan) is mostly negligible.
	\item The overlap $\left|\langle\psi_b|\frac{\ket{\psi_{0\uparrow}}+\ket{\psi_{0\downarrow}}}{\sqrt{2}}\right|^2$ (orange) approaches 1 as $b\to0$.
	\item The sum of these last two, $\left|\langle\psi_b|\frac{\ket{\psi_{0\uparrow}}-\ket{\psi_{0\downarrow}}}{\sqrt{2}}\right|^2+\left|\langle\psi_b|\frac{\ket{\psi_{0\uparrow}}+\ket{\psi_{0\downarrow}}}{\sqrt{2}}\right|^2 = \abs{\scp{\psi_b}{\psi_{0\uparrow}}}^2+\abs{\scp{\psi_b}{\psi_{0\downarrow}}}^2$ (thin black), approaches 1 as $b\to0$ and is less prone to numerical noise.
	\item If you redo this calculation with an \emph{odd} number of spins, you may find different overlaps with the $\frac{\ket{\psi_{0\uparrow}}\pm\ket{\psi_{0\downarrow}}}{\sqrt{2}}$ asymptotic states. Their sum, however, drawn in black, should be insensitive to the parity of $N$.
	\item For $\abs{b}\lesssim0.2$ the excitation gap (see above) is so small that the calculated ground-state eigenvector is no longer truly the ground state but becomes mixed with the first excited state due to numerical inaccuracies. This leads to the jumps in the orange and cyan curves (notice, however, that their sum, shown in black, is stable). If you redo this calculation with larger values for \mm{m}, you may get better results.
\end{itemize}

\subsubsection{magnetization}\index{magnetization}

Studying the ground state coefficients list directly is of limited use because of the large amount of information contained in its numerical representation. We gain more insight by studying specific observables, for example the magnetizations $\avg{\op{S}_x^{(k)}}$, $\avg{\op{S}_y^{(k)}}$, and $\avg{\op{S}_z^{(k)}}$.
We add the following definition to the \mm{With[]} clause in \mm{\ref{math:HXY}}:
\begin{mathematica}[firstnumber=\getrefnumber{mathline:observables}]
	¤mathnlc (* spin components expectation values *)
	¤mathnl Clear[mx,my,mz];
	¤mathnl mx[b_?NumericQ, m_Integer /; m >= 1, k_Integer] :=
	¤mathnl   mx[b, m, k] = With[{g = gs[b,m][[2,1]]},
	¤mathnl     Re[Conjugate[g].(sx[S, n, Mod[k, n, 1]].g)]];
	¤mathnl my[b_?NumericQ, m_Integer /; m >= 1, k_Integer] :=
	¤mathnl   my[b, m, k] = With[{g = gs[b,m][[2,1]]},
	¤mathnl     Re[Conjugate[g].(sy[S, n, Mod[k, n, 1]].g)]];
	¤mathnl mz[b_?NumericQ, m_Integer /; m >= 1, k_Integer] :=
	¤mathnl   mz[b, m, k] = With[{g = gs[b,m][[2,1]]},
	¤mathnl     Re[Conjugate[g].(sz[S, n, Mod[k, n, 1]].g)]];
	¤labelŽmathline:correlationsmore¤mathnl ]
\end{mathematica}
In our transverse Ising model only the $x$-component of the magnetization is nonzero. Due to the translational symmetry of the system we can look at the magnetization of any spin, for example the first one ($k=1$): $m_x(b)$ (blue) and $m_z(b)$ (orange, non-zero due to numerical inaccuracies)
\begin{center}
\label{fig:IsingMagnetization}
\includegraphics[width=0.5\textwidth]{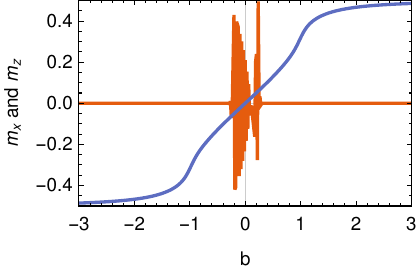}
\end{center}
We see that in the phases of large $\abs{b}$, the spins are almost entirely polarized, while in the phase $\abs{b}<1$ the $x$-magnetization is roughly proportional to $b$.

\subsubsection{spin--spin fluctuation correlations}\index{correlations}

Quantum-mechanical spins always fluctuate around their mean direction. In the example of \ref{Q:spinexpectation}, the state $\ket{S,S}$ points on average along the $+z$ direction in the sense that $\avg{\opvect{S}}=\me{S,S}{\opvect{S}}{S,S}=\{0,0,S\}$; but it fluctuates away from this axis as $\avg{\op{S}_x^2}=\avg{\op{S}_y^2}=S/2$.

By introducing the fluctuation operator $\opvect{\delta\!S}=\opvect{S}-\avg{\opvect{S}}$, we can interpret spin fluctuations through the expectation values $\avg{\opvect{\delta\!S}}=\{0,0,0\}$ (fluctuations always average to zero) and $\avg{(\opvect{\delta\!S})^2}=\avg{\opvect{\delta\!S}\cdot\opvect{\delta\!S}}=\avg{\opvect{S}\cdot\opvect{S}}-\avg{\opvect{S}}\cdot\avg{\opvect{S}}=S(S+1)-\norm{\avg{\opvect{S}}}^2$. Since the spin magnetization has length $0\le\norm{\avg{\opvect{S}}}\le S$, these fluctuations satisfy $S\le\avg{(\opvect{\delta\!S})^2}\le S(S+1)$: they are positive for every spin state.

When two (or more) spins are present, their quantum-mechanical fluctuations can become correlated. We quantify such spin--spin fluctuation correlations between two spins $k$ and $k'$ with the measure
\begin{equation}
	\label{eq:spinspincorr}
	C_{k,k'} = \avg{\opvect{\delta\!S}^{(k)}\cdot\opvect{\delta\!S}^{(k')}}
	= \avg{\opvect{S}^{(k)}\cdot\opvect{S}^{(k')}} - \avg{\opvect{S}^{(k)}}\cdot\avg{\opvect{S}^{(k')}},
\end{equation}
which has the form of a statistical covariance.\footnote{See \url{https://en.wikipedia.org/wiki/Covariance}.}
For any spin length $S$ (assuming $S^{(k)}=S^{(k')}$), the first term of \autoref{eq:spinspincorr} can be written as
\begin{equation}
	\avg{\opvect{S}^{(k)}\cdot\opvect{S}^{(k')}}
	= \frac{\avg{\big(\opvect{S}^{(k)}+\opvect{S}^{(k')}\big)^2}-\avg{\big(\opvect{S}^{(k)}\big)^2}-\avg{\big(\opvect{S}^{(k')}\big)^2}}{2}
	= \frac12\avg{\big(\opvect{S}^{(k)}+\opvect{S}^{(k')}\big)^2}-S(S+1),
\end{equation}
which allows us to predict its expectation value as a function of the total-spin quantum number describing the two spins-$S$. As this quantum number can be anywhere between $0$ and $2S$,  we have $0\le\avg{\big(\opvect{S}^{(k)}+\opvect{S}^{(k')}\big)^2}\le2S(2S+1)$. This expectation value is not restricted to integer values.
As a result we make the following observations:
\begin{itemize}
	\item $-S(S+1) \le C_{k,k'} \le S^2$: spin fluctuations can be correlated ($C_{k,k'} > 0$), anti-correlated ($C_{k,k'} < 0$), or uncorrelated ($C_{k,k'}=0$).
	\item The strongest correlations $C_{k,k'} = S^2$ are found when the two spins-$S$ form a joint spin-$2S$ and at the same time are unaligned ($\avg{\opvect{S}^{(k)}}\cdot\avg{\opvect{S}^{(k')}}=0$).
	\item The strongest anti-correlations $C_{k,k'}=-S(S+1)$ are found when the two spins-$S$ form a joint spin-0 (\ie, a spin-singlet). In this case, the magnetizations always vanish: $\avg{\opvect{S}^{(k)}}=\avg{\opvect{S}^{(k')}}=\{0,0,0\}$.
\end{itemize}
For the specific case $S=1/2$, which we use in the present calculations, two spins can form a joint singlet (total spin 0; $\avg{\big(\opvect{S}^{(k)}+\opvect{S}^{(k')}\big)^2}=0$), a joint triplet (total spin 1; $\avg{\big(\opvect{S}^{(k)}+\opvect{S}^{(k')}\big)^2}=2$), or a mixture of these ($0\le\avg{\big(\opvect{S}^{(k)}+\opvect{S}^{(k')}\big)^2}\le2$), and the correlation is restricted to the values $-\frac34 \le C_{k,k'} \le +\frac14$ for all states. Specific cases are:
\begin{itemize}
	\item In the pure joint singlet state $\frac{\ket{\uparrow\downarrow}-\ket{\downarrow\uparrow}}{\sqrt{2}}$ the correlation is precisely $C_{k,k'}=-\frac34$. A fluctuation of one spin implies a counter-fluctuation of the other in order to keep them anti-aligned and in a spin-0 joint state. Remember that the spin monogamy theorem states that if spins $k$ and $k'$ form a joint singlet, then both must be uncorrelated with all other spins in the system.
	\item In a pure joint triplet state, \ie, any mixture of the states $\ket{\uparrow\uparrow}$, $\ket{\downarrow\downarrow}$, and $\frac{\ket{\uparrow\downarrow}+\ket{\downarrow\uparrow}}{\sqrt{2}}$, the correlation is $0\le C_{k,k'}\le+\frac14$. A fluctuation of one spin implies a similar fluctuation of the other in order to keep them aligned and in a spin-1 joint state.
	\item The maximum correlation $C_{k,k'}=+\frac14$ is reached for unaligned triplet states, \ie, when $\avg{\opvect{S}^{(k)}}\cdot\avg{\opvect{S}^{(k')}}=0$. Examples include the states $\frac{\ket{\uparrow\uparrow}+\ket{\downarrow\downarrow}}{\sqrt{2}}$, $\frac{\ket{\uparrow\uparrow}-\ket{\downarrow\downarrow}}{\sqrt{2}}$, and $\frac{\ket{\uparrow\downarrow}+\ket{\downarrow\uparrow}}{\sqrt{2}}$.
	\item In the fully parallel triplet states $\ket{\uparrow\uparrow}$ and $\ket{\downarrow\downarrow}$, the magnetizations are aligned but their fluctuations are uncorrelated: $C_{k,k'}=0$, and hence $\avg{\opvect{S}^{(k)}\cdot\opvect{S}^{(k')}} = \avg{\opvect{S}^{(k)}}\cdot\avg{\opvect{S}^{(k')}}$.
\end{itemize}
In order to estimate these spin fluctuation correlations, we add the following definition to the \mm{With[]} clause in \mm{\ref{math:HXY}}:
\begin{mathematica}[firstnumber=\getrefnumber{mathline:correlationsmore}]
	¤mathnlc (* spin-spin correlation operator *)
	¤mathnl Clear[Cop];
	¤mathnl Cop[k1_Integer, k2_Integer] := Cop[k1, k2] =
	¤mathnl   With[{q1 = Mod[k1,n,1], q2 = Mod[k2,n,1]},
	¤mathnl     sx[S,n,q1].sx[S,n,q2] + sy[S,n,q1].sy[S,n,q2]
	¤mathnl       + sz[S,n,q1].sz[S,n,q2]];
	¤mathnlc (* spin-spin correlations *)
	¤mathnl Clear[c];
	¤mathnl c[b_?NumericQ,m_Integer/;m>=1,{k1_Integer,k2_Integer}] :=
	¤mathnl   c[b,m,{k1,k2}] = With[{g = gs[b,m][[2,1]]},
	¤mathnl     Re[Conjugate[g].(Cop[k1,k2].g)]-(mx[b,m,k1]*mx[b,m,k2]
	¤mathnl       +my[b,m,k1]*my[b,m,k2]+mz[b,m,k1]*mz[b,m,k2])];
	¤mathnl ]
\end{mathematica}
Since our spin ring is translationally invariant, we can simply plot $C_{\delta} = C_{1,1+\delta}$: for $N=\IsingSize$ and $\delta=1\dots10$ (top to bottom),
\begin{center}
\includegraphics[width=0.5\textwidth]{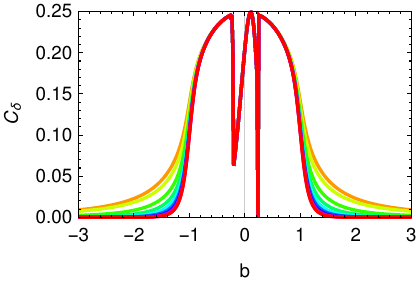}
\end{center}
Observations:
\begin{itemize}
	\item The spin fluctuations are maximally correlated ($C=+\frac14$) for $b=0$, in the ferromagnetic phase. They are all either pointing up or pointing down, so every spin is correlated with every other spin; keep in mind that the magnetization vanishes at the same time (\autopageref{fig:IsingMagnetization}). It is only the spin--spin interactions that correlate the spins' directions and therefore their fluctuations.
	\item The spin fluctuations are uncorrelated ($C\to0$) for $b\to\pm\infty$, in the paramagnetic phases. They are all pointing in the $+x$ direction for $b\gg1$ or in the $-x$ direction for $b\ll-1$, but they are doing so in an independent way and would keep pointing in that direction even if the spin--spin interactions were switched off. This means that the fluctuations of the spins' directions are uncorrelated.
\end{itemize}

\subsubsection{entropy of entanglement}\index{entropy of entanglement}\index{entanglement}

We know now that in the limits $b\to\pm\infty$ the spins are polarized (magnetized) but their fluctuations are uncorrelated, while close to $b=0$ they are unpolarized (unmagnetized) but their fluctuations are maximally correlated. Here we quantify these correlations with the \emph{entropy of entanglement}, which measures the entanglement of a single spin with the rest of the spin chain.

In a system composed of two subsystems A and B, the entropy of entanglement is  defined as the \emph{von Neumann entropy}\index{von Neumann entropy} of the reduced density matrix\index{partial trace} (see \autoref{sec:rdm}),
\begin{equation}
	S\ix{AB} = -\Tr\left(\op{\rho}\ix{A}\log_2\op{\rho}\ix{A}\right) = -\sum_i \lambda_i\log_2\lambda_i
\end{equation}
where the $\lambda_i$ are the eigenvalues of $\op{\rho}\ix{A}$ (or of $\op{\rho}\ix{B}$; the result is the same). Care must be taken with the case $\lambda_i=0$: we find $\lim_{\lambda\to0} \lambda\log_2\lambda=0$. For this we define the function
\begin{mathematica}
	¤mathin¤labelŽmath:entropy1 s[0|0.] = 0;
	¤mathin¤labelŽmath:entropy2 s[x_] = -x*Log[2, x];
\end{mathematica}
that uses Mathematica's pattern matching to separate out the special case $x=0$. Note that we use an alternative pattern\footnote{See \url{https://reference.wolfram.com/language/tutorial/PatternsInvolvingAlternatives.html}.}\index{Mathematica!pattern!alternative} \mm{0|0.}\ that matches both an analytic zero \mm{0} and a numeric zero \mm{0.}, which Mathematica distinguishes carefully.\footnote{Experiment: \mm{0==0.}\ yields \mm{True} (testing for semantic identity), whereas \mm{0===0.}\ yields \mm{False} (testing for symbolic identity).}

We define the entropy of entanglement of the first spin with the rest of the spin ring using the definition of \mm{\ref{math:traceoutpsi2}}, tracing out the last $(2S+1)^{N-1}$ degrees of freedom and leaving only the first $2S+1$ degrees of freedom of the first spin:
\begin{mathematica}
	¤mathin EE[S_?SpinQ, ¤mmpsi_] :=
	¤mathnl   Total[s /@ Re[Eigenvalues[traceout[¤mmpsi, -Length[¤mmpsi]/(2S+1)]]]]
\end{mathematica}
Observations:
\begin{itemize}
	\item Entanglement entropies of the known asymptotic ground states:
\begin{mathematica}
	¤mathin EE[1/2, (gs0up+gs0dn)/Sqrt[2]]
	¤mathout 1
	¤mathin EE[1/2, (gs0up-gs0dn)/Sqrt[2]]
	¤mathout 1
	¤mathin EE[1/2, gsplusinf]
	¤mathout 0
	¤mathin EE[1/2, gsminusinf]
	¤mathout 0
\end{mathematica}
	\item Entanglement entropy as a function of $b$: again the calculation is numerically difficult around $b\approx0$ because of the quasi-degeneracy.
\begin{mathematica}
	¤mathin With[{bmax = 3, db = 1/64, m = 2},
	¤mathnl   ListLinePlot[Table[{b, EE[1/2, gs[b,m][[2,1]]]},
	¤mathnl     {b, -bmax, bmax, db}], PlotRange -> {0, 1}]]
\end{mathematica}
\begin{center}
\includegraphics[width=0.5\textwidth]{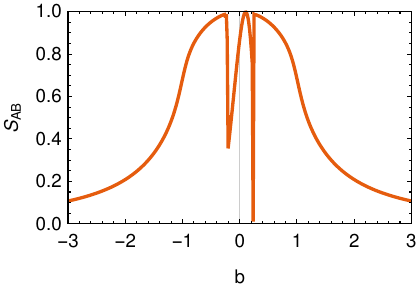}
\end{center}
	Notice that the quantum phase transitions at $b=\pm1$ are not visible in this plot.
\end{itemize}

\subsection{exercises}

\begin{questions}
	\item\label{Q:IsingSize} For $S=1/2$, what is the largest value of $N$ for which you can calculate the ground state of the transverse Ising model at the critical point $b=1$?
\pagenote[\ref{Q:IsingSize}]{On a MacBook Pro (Retina, 13-inch, Early 2015) with a \SI{3.1}{GHz} Intel Core i7 CPU and \SI{16}{GB} \SI{1867}{MHz} DDR3 RAM, it takes around 20 minutes (\mm{AbsoluteTiming[\mmgamma=gs[1,1];]}) to calculate the ground state \mm{gs[1,1]} with the definition of \mm{\ref{math:HXY}} and $N=22$. This calculation uses over \SI{24}{GB} of compressed RAM (\mm{MaxMemoryUsed[]}) and is the upper limit on $N$ for this computer. For $N=23$ the calculation runs out of memory.}
	\item\label{Q:Ising1} Study the transverse Ising model with $S=1$:\index{Ising model}
		\begin{enumerate}
			\item At which values of $b$ do you find quantum phase transitions?
			\item Characterize the ground state in terms of magnetization, spin--spin correlations, and entanglement entropy.
		\end{enumerate}
\pagenote[\ref{Q:Ising1}]{With the Mathematica code of \autoref{sec:Ising}, setting $S=1$ and $N=12$, we find a phase transition around $b=\pm2$. Similar to the $S=1/2$ case, the Ising model is gapless for $\abs{b}<2$ and gapped for $\abs{b}>2$. The correlations look qualitatively similar to the ones found for $S=1/2$.}
	\item\label{Q:XYmodel} Study the transverse XY model for $S=1/2$:\index{XY model}
\begin{equation}
	\label{eq:XYham}
	\Ham = -\frac{b}{2} \sum_{k=1}^N \op{S}_z^{(k)} - \sum_{k=1}^N \left(\op{S}_x^{(k)} \op{S}_x^{(k+1)}+\op{S}_y^{(k)} \op{S}_y^{(k+1)}\right)
\end{equation}
		\begin{enumerate}
			\item Guess the shape of the ground states for $b\pm\infty$ [notice that the first term in the Hamiltonian of \autoref{eq:XYham} is in the $z$-direction!] and compare to the numerical calculations.
			\item At which values of $b$ do you find quantum phase transitions?
			\item Characterize the ground state in terms of magnetization, spin--spin correlations, and entanglement entropy.
		\end{enumerate}
\pagenote[\ref{Q:XYmodel}]{
	\begin{enumerate}
		\item For $b\to\pm\infty$ the ground states are analogous to those of the transverse Ising model, \autoref{eq:allupalldown}, along the $\pm z$ axis:
		\begin{align}
			\label{eq:allupalldownz}
			\ket{\psi_{+\infty}} &= \ket{+z}^{\otimes N}, & \ket{\psi_{-\infty}} &= \ket{-z}^{\otimes N}.
		\end{align}
		Notice that, unlike the transverse Ising model, these asymptotic ground states are the exact ground states for $\abs{b}>2$, not just in the limits $b\to\pm\infty$.
		\item There are phase transitions at $b=\pm2$, recognizable in the ground-state gap.
		\item Since the states of \autoref{eq:allupalldownz} are product states, there are absolutely no correlations between the states of the spins for $\abs{b}>2$. For $\abs{b}<2$ the magnetization, spin--spin correlations, and entanglement entropy are qualitatively similar to those of the transverse Ising model. For $b=0$ the spin--spin correlations do not reach the full uniform 0.25 as for the Ising model, but rather they still decay with distance.
	\end{enumerate}}
	\item\label{Q:Heisenberg} Study the Heisenberg model for $S=1/2$:\index{Heisenberg model}
\begin{equation}
	\label{eq:HeisenbergHam}
	\Ham = -\frac{b}{2} \sum_{k=1}^N \op{S}_z^{(k)} - \sum_{k=1}^N \opvect{S}^{(k)}\cdot\opvect{S}^{(k+1)}
\end{equation}
		\begin{enumerate}
			\item Guess the shape of the ground states for $b\pm\infty$ [notice that the first term in the Hamiltonian of \autoref{eq:HeisenbergHam} is in the $z$-direction!] and compare to the numerical calculations.
			\item What is the ground-state degeneracy for $b=0$?
			\item At which values of $b$ do you find quantum phase transitions?
			\item Characterize the ground state in terms of magnetization, spin--spin correlations, and entanglement entropy.
		\end{enumerate}
\pagenote[\ref{Q:Heisenberg}]{
	\begin{enumerate}
		\item For $b\to\pm\infty$ the ground states are the same as \autoref{eq:allupalldownz}.
		\item At $b=0$ the ground-state degeneracy is $N+1$.
		\item For any $b>0$, $\ket{\psi_{+\infty}}$ is the exact ground state; for any $b<0$, $\ket{\psi_{-\infty}}$ is the exact ground state. There is a phase transition at $b=0$.
		\item Since the states of \autoref{eq:allupalldownz} are product states, there are absolutely no correlations between the states of the spins for any $b\neq0$.
	\end{enumerate}}
	\item\label{Q:Supup} Consider two spin-1/2 particles in the triplet state $\ket{\psi}=\ket{\uparrow\uparrow}$. Subsystem A is the first spin, and subsystem B is the second spin.
		\begin{enumerate}
			\item What is the density matrix $\op{\rho}_{AB}$ of this system?
			\item What is the reduced density matrix $\op{\rho}_{A}$ of subsystem A (the first spin)? Is this a pure state? If yes, what state?
			\item What is the reduced density matrix $\op{\rho}_{B}$ of subsystem B (the second spin)? Is this a pure state? If yes, what state?
			\item Calculate the von Neumann entropies of $\op{\rho}_{AB}$, $\op{\rho}_{A}$, and $\op{\rho}_{B}$.
		\end{enumerate}
\pagenote[\ref{Q:Supup}]{
	\begin{enumerate}
		\item $\op{\rho}_{AB} = \ket{\psi}\bra{\psi}=\ket{\uparrow\uparrow}\bra{\uparrow\uparrow}$:
\begin{mathematica}
	¤protect¤mathin¤ ¤mmpsi¤ =¤ Flatten[KroneckerProduct[{1,0},¤ {1,0}]]
	¤protect¤mathout¤ {1,¤ 0,¤ 0,¤ 0}
	¤protect¤mathin¤ ¤mmrhoŽAB¤ =¤ KroneckerProduct[¤mmpsi,¤ Conjugate[¤mmpsi]]
	¤protect¤mathout¤ {{1,0,0,0},¤ {0,0,0,0},¤ {0,0,0,0},¤ {0,0,0,0}}
\end{mathematica}
	\item $\op{\rho}_A=\Tr_B\op{\rho}_{AB}=\ket{\uparrow}\bra{\uparrow}$ is a pure state:
\begin{mathematica}
	¤protect¤mathin¤ ¤mmrhoŽA¤ =¤ traceout[¤mmrhoŽAB,¤ -2]
	¤protect¤mathout¤ {{1,0},¤ {0,0}}
	¤protect¤mathin¤ Tr[¤mmrhoŽA.¤mmrhoŽA]
	¤protect¤mathout¤ 1
\end{mathematica}
	\item $\op{\rho}_B=\Tr_A\op{\rho}_{AB}=\ket{\uparrow}\bra{\uparrow}$ is a pure state:
\begin{mathematica}
	¤protect¤mathin¤ ¤mmrhoŽB¤ =¤ traceout[¤mmrhoŽAB,¤ 2]
	¤protect¤mathout¤ {{1,0},¤ {0,0}}
	¤protect¤mathin¤ Tr[¤mmrhoŽB.¤mmrhoŽB]
	¤protect¤mathout¤ 1
\end{mathematica}
	\item Using \mm{\ref{math:entropy1}} and \mm{\ref{math:entropy2}}, we see that the entropy of entanglement is $S_A-S_{AB}=S_B-S_{AB}=0$ (no entanglement):
\begin{mathematica}
	¤protect¤mathin¤ SAB¤ =¤ Total[s¤ /@¤ Eigenvalues[¤mmrhoŽAB]]
	¤protect¤mathout¤ 0
	¤protect¤mathin¤ SA¤ =¤ Total[s¤ /@¤ Eigenvalues[¤mmrhoŽA]]
	¤protect¤mathout¤ 0
	¤protect¤mathin¤ SB¤ =¤ Total[s¤ /@¤ Eigenvalues[¤mmrhoŽB]]
	¤protect¤mathout¤ 0
\end{mathematica}
	\end{enumerate}}
	\item\label{Q:Ssinglet} Consider two spin-1/2 particles in the singlet state $\ket{\psi}=\frac{\ket{\uparrow\downarrow}-\ket{\downarrow\uparrow}}{\sqrt{2}}$. Subsystem A is the first spin, and subsystem B is the second spin.
		\begin{enumerate}
			\item What is the density matrix $\op{\rho}_{AB}$ of this system?
			\item What is the reduced density matrix $\op{\rho}_{A}$ of subsystem A (the first spin)? Is this a pure state? If yes, what state?
			\item What is the reduced density matrix $\op{\rho}_{B}$ of subsystem B (the second spin)? Is this a pure state? If yes, what state?
			\item Calculate the von Neumann entropies of $\op{\rho}_{AB}$, $\op{\rho}_{A}$, and $\op{\rho}_{B}$.
		\end{enumerate}
\pagenote[\ref{Q:Ssinglet}]{
	\begin{enumerate}
		\item $\op{\rho}_{AB} = \ket{\psi}\bra{\psi}=\frac{\ket{\uparrow\downarrow}-\ket{\downarrow\uparrow}}{\sqrt{2}}\frac{\bra{\uparrow\downarrow}-\bra{\downarrow\uparrow}}{\sqrt{2}}=\frac12(\ket{\uparrow\downarrow}\bra{\uparrow\downarrow}-\ket{\uparrow\downarrow}\bra{\downarrow\uparrow}-\ket{\downarrow\uparrow}\bra{\uparrow\downarrow}+\ket{\downarrow\uparrow}\bra{\downarrow\uparrow})$:
\begin{mathematica}
	¤protect¤mathin¤ ¤mmpsi¤ =¤ Flatten[KroneckerProduct[{1,0},¤ {0,1}]
	¤protect¤mathnl¤ ¤ ¤ -¤ KroneckerProduct[{0,1},¤ {1,0}]]/Sqrt[2]
	¤protect¤mathout¤ {0,¤ 1/Sqrt[2],¤ -1/Sqrt[2],¤ 0}
	¤protect¤mathin¤ ¤mmrhoŽAB¤ =¤ KroneckerProduct[¤mmpsi,¤ Conjugate[¤mmpsi]]
	¤protect¤mathout¤ {{0,0,0,0},¤ {0,1/2,-1/2,0},¤ {0,-1/2,1/2,0},¤ {0,0,0,0}}
\end{mathematica}
	\item $\op{\rho}_A=\Tr_B\op{\rho}_{AB}=\frac12(\ket{\uparrow}\bra{\uparrow}+\ket{\downarrow}\bra{\downarrow})$ is a mixed state:
\begin{mathematica}
	¤protect¤mathin¤ ¤mmrhoŽA¤ =¤ traceout[¤mmrhoŽAB,¤ -2]
	¤protect¤mathout¤ {{1/2,0},¤ {0,1/2}}
	¤protect¤mathin¤ Tr[¤mmrhoŽA.¤mmrhoŽA]
	¤protect¤mathout¤ 1/2
\end{mathematica}
	\item $\op{\rho}_B=\Tr_A\op{\rho}_{AB}=\frac12(\ket{\uparrow}\bra{\uparrow}+\ket{\downarrow}\bra{\downarrow})$ is a mixed state:
\begin{mathematica}
	¤protect¤mathin¤ ¤mmrhoŽB¤ =¤ traceout[¤mmrhoŽAB,¤ 2]
	¤protect¤mathout¤ {{1/2,0},¤ {0,1/2}}
	¤protect¤mathin¤ Tr[¤mmrhoŽB.¤mmrhoŽB]
	¤protect¤mathout¤ 1/2
\end{mathematica}
	\item Using \mm{\ref{math:entropy1}} and \mm{\ref{math:entropy2}}, we see that the entropy of entanglement is $S_A-S_{AB}=S_B-S_{AB}=1$ (maximal entanglement):
\begin{mathematica}
	¤protect¤mathin¤ SAB¤ =¤ Total[s¤ /@¤ Eigenvalues[¤mmrhoŽAB]]
	¤protect¤mathout¤ 0
	¤protect¤mathin¤ SA¤ =¤ Total[s¤ /@¤ Eigenvalues[¤mmrhoŽA]]
	¤protect¤mathout¤ 1
	¤protect¤mathin¤ SB¤ =¤ Total[s¤ /@¤ Eigenvalues[¤mmrhoŽB]]
	¤protect¤mathout¤ 1
\end{mathematica}
	\end{enumerate}}
\end{questions}

\section[coupled spin systems: quantum circuits]{\label{sec:quantumcircuits}coupled spin systems: quantum circuits\hspace{\stretch{1}}\attachcode{QuantumCircuits}{quantum gates and quantum circuits}}
\index{quantum circuit}

The computational structure developed so far in this chapter can be used to simulate quantum circuits, such as they are used to run quantum algorithms leading all the way to quantum computers.
In its simplest form, a quantum circuit contains a set of $N$ spin-1/2 quantum objects called \emph{qubits}\index{qubit}, on which a sequence of operations called \emph{quantum gates}\index{quantum gate} is executed.
In analogy to classical binary logic, the basis states of the qubits' Hilbert space are usually denoted as \ket{0} (replacing the spin-1/2 state \ket{\uparrow}) and \ket{1} (replacing \ket{\downarrow}).

In this section, we go through the steps of assembling quantum circuits and simulating their behavior on a classical computer.
Naturally, the matrix representation of quantum gates and circuits constructed here is neither efficient nor desirable for building an actual quantum computer. It is merely useful for acquiring a detailed understanding of the workings of quantum circuits and algorithms.

In what follows, we adhere strictly to Chapter 5 of Nielsen\&Chuang,\footnote{Michael A.\ Nielsen and Isaac L.\ Chuang: \emph{Quantum Computation and Quantum Information}, 10th Anniversary Edition, Cambridge University Press, Cambridge, UK (2010).} which provides many more details of the calculations, as well as further reading for the interested student.

\subsection{quantum gates}
\index{quantum gate}

Any quantum circuit can be constructed from a set of simple building blocks, similarly to a classical digital circuit. These building blocks are canonical quantum gates,\footnote{See \url{https://en.wikipedia.org/wiki/Quantum_logic_gate}.} of which we implement a useful subset here.

\subsubsection{single-qubit gates}
Single-qubit gates act on one specific qubit in a set:
\begin{itemize}
	\item The Pauli-$X$ gate {\Qcircuit @C=1em @R=.7em { & \gate{X} & \qw }} acts like $\op{\sigma}_x=\ket{1}\bra{0}+\ket{0}\bra{1}$ on the desired qubit, and has no effect on all other qubits. A single-qubit input state \ket{\psi\ix{in}} entering the gate from the left is transformed into the output state $\ket{\psi\ix{out}}=\op{\sigma}_x\ket{\psi\ix{in}}$ exiting the gate towards the right.
	\item The Pauli-$Y$ gate {\Qcircuit @C=1em @R=.7em { & \gate{Y} & \qw }} acts like $\op{\sigma}_y=\ii\ket{1}\bra{0}-\ii\ket{0}\bra{1}$ on the desired qubit, and has no effect on all other qubits.
	\item The Pauli-$Z$ gate {\Qcircuit @C=1em @R=.7em { & \gate{Z} & \qw }} acts like $\op{\sigma}_z=\ket{0}\bra{0}-\ket{1}\bra{1}$ on the desired qubit, and has no effect on all other qubits.
	\item The Hadamard gate {\Qcircuit @C=1em @R=.7em { & \gate{H} & \qw }} acts like $\frac{\op{\sigma}_x+\op{\sigma}_z}{\sqrt{2}}=\frac{\ket{0}\bra{0}+\ket{0}\bra{1}+\ket{1}\bra{0}-\ket{1}\bra{1}}{\sqrt{2}}$ on the desired qubit, and has no effect on all other qubits.
\end{itemize}
To implement these single-qubit gates in a general way, we proceed as in \mm{\ref{math:spinop}} by defining a matrix that represents the operator $\op{a}$ acting on the $k\ex{th}$ qubit in a set of $n$ qubits:
\begin{mathematica}
	¤mathin op[n_Integer, k_Integer, a_] /; 1<=k<=n && Dimensions[a]=={2,2} :=
	¤mathnl   KroneckerProduct[IdentityMatrix[2^(k-1), SparseArray],
	¤mathnl                    a,
	¤mathnl                    IdentityMatrix[2^(n-k), SparseArray]]
\end{mathematica}
This allows us to define the single-qubit Pauli and Hadamard gates with
\begin{mathematica}
	¤mathin {id, ¤mmsigmaŽx, ¤mmsigmaŽy, ¤mmsigmaŽz} = Table[SparseArray[PauliMatrix[i]], {i, 0, 3}];
	¤mathin X[n_Integer, k_Integer] /; 1<=k<=n := op[n, k, ¤mmsigmaŽx]
	¤mathin Y[n_Integer, k_Integer] /; 1<=k<=n := op[n, k, ¤mmsigmaŽy]
	¤mathin Z[n_Integer, k_Integer] /; 1<=k<=n := op[n, k, ¤mmsigmaŽz]
	¤mathin H[n_Integer, k_Integer] /; 1<=k<=n := op[n, k, (¤mmsigmaŽx+¤mmsigmaŽz)/Sqrt[2]]
\end{mathematica}
as well as the corresponding rotation operators $\op{R}_x(\phi)=\frac{1+e^{\ii\phi}}{2}\one+\frac{1-e^{\ii\phi}}{2}\sigma_x=e^{\ii\phi/2}e^{-\ii\phi\op{\sigma}_x/2}$ etc.\ that are also known as phase gates,
\begin{mathematica}
	¤mathin RX[n_Integer, k_Integer, ¤mmphi_] /; 1<=k<=n :=
	¤mathnl   op[n, k, (1+Exp[I*¤mmphi])/2*id + (1-Exp[I*¤mmphi])/2*¤mmsigmaŽx]
	¤mathin RY[n_Integer, k_Integer, ¤mmphi_] /; 1<=k<=n :=
	¤mathnl   op[n, k, (1+Exp[I*¤mmphi])/2*id + (1-Exp[I*¤mmphi])/2*¤mmsigmaŽy]
	¤mathin¤labelŽmath:gateRZ RZ[n_Integer, k_Integer, ¤mmphi_] /; 1<=k<=n :=
	¤mathnl   op[n, k, (1+Exp[I*¤mmphi])/2*id + (1-Exp[I*¤mmphi])/2*¤mmsigmaŽz]
\end{mathematica}

\subsubsection{two-qubit gates}

Interesting quantum circuits require operations that involve more than one qubit.

The SWAP gate exchanges the state of qubits $j$ and $k$ in a set of $n$ qubits:

\noindent
\hspace{\stretch{1}}
\Qcircuit @C=0.5em @R=1em {
	\lstick{j}	& \qw	& \qswap		& \qw	& \qw \\
	\lstick{k}	& \qw	& \qswap \qwx	& \qw	& \qw
}
\hspace{\stretch{1}}
\\

\noindent
Without going through complicated considerations over basis-set indices, we construct it through the definition $\text{SWAP}^{(jk)} = (\one^{(j)}\otimes\one^{(k)}+\op{\sigma}_x^{(j)}\otimes\op{\sigma}_x^{(k)}+\op{\sigma}_y^{(j)}\otimes\op{\sigma}_y^{(k)}+\op{\sigma}_z^{(j)}\otimes\op{\sigma}_z^{(k)})/2$ and building on the above Pauli gates:
\begin{mathematica}
	¤mathin SWAP[n_Integer, {j_Integer, k_Integer}] /; 1<=j<=n && 1<=k<=n && j!=k :=
	¤mathnl   (IdentityMatrix[2^n, SparseArray] +
	¤mathnl    X[n,j].X[n,k] + Y[n,j].Y[n,k] + Z[n,j].Z[n,k])/2
\end{mathematica}
The matrix representation of a two-qubit SWAP takes on the familiar form
\begin{mathematica}
	¤mathin SWAP[2, {1,2}] //Normal
	¤mathout {{1, 0, 0, 0},
	¤mathnl  {0, 0, 1, 0},
	¤mathnl  {0, 1, 0, 0},
	¤mathnl  {0, 0, 0, 1}}
\end{mathematica}
The square root of the SWAP gate is also sometimes used, and is defined similarly:
\begin{mathematica}
	¤mathin SQRTSWAP[n_Integer, {j_Integer, k_Integer}] /; 1<=j<=n && 1<=k<=n && j!=k :=
	¤mathnl   (3+I)/4 * IdentityMatrix[2^n, SparseArray] +
	¤mathnl   (1-I)/4 * (X[n,j].X[n,k] + Y[n,j].Y[n,k] + Z[n,j].Z[n,k])
\end{mathematica}
To define the controlled-NOT or CNOT gate, we first make a general definition for controlled gates. The $n$-qubit operator \mm{CTRL[n,\mmlambda,A]} acts like the operator $\op{A}$ if all qubits in the list $\lambda=\{i_1,i_2,\ldots,i_k\}$ are in the \ket{1} state, and has no action (acts like the identity operator on $n$ qubits) if any of the qubits in the list $\lambda$ are in the \ket{0} state:
\begin{equation}
	\label{eq:ctrlgate}
	\text{CTRL} = \left[\bigotimes_{j=1}^k\ket{1}\bra{1}^{(i_j)}\right] \cdot \op{A} + \left[\one-\bigotimes_{j=1}^k\ket{1}\bra{1}^{(i_j)}\right] \cdot \one
	= \one + \left[\bigotimes_{j=1}^k\ket{1}\bra{1}^{(i_j)}\right] \cdot (\op{A}-\one)
\end{equation}
Its circuit representation is

\noindent
\hspace{\stretch{1}}
\Qcircuit @C=0.5em @R=.7em {
	\lstick{i_1}		& \qw	&\ctrl{1}				& \qw	& \qw \\
	\lstick{i_2}		& \qw	& \ctrl{2}				& \qw	& \qw \\
				& \cdots	&					& \cdots	& \\
	\lstick{i_k}		& \qw	& \ctrl{1}				& \qw	& \qw \\
				& \qw	& \multigate{3}{\op{A}}	& \qw	& \qw \\
				& \qw	& \ghost{\op{A}}		& \qw	& \qw \\
				& \cdots	& \nghost{\op{A}}		& \cdots	& \\
				& \qw	& \ghost{\op{A}}		& \qw	& \qw
}
\hspace{\stretch{1}}
\\

\noindent
The bracket in the last expression of \autoref{eq:ctrlgate} is constructed with \mm{Apply[Dot, op[n,\#,P1]\&/@\mmlambda]} that first constructs a list of projection operators $\ket{1}\bra{1}^{(i_j)}$ for the control qubits, and then applies the \mm{Dot} operator to assemble them into the product $\bigotimes_{j=1}^k\ket{1}\bra{1}^{(i_j)}$.
\begin{mathematica}
	¤mathin P0 = (id + ¤mmsigmaŽz)/2 //SparseArray;  (* qubit projector ¤ketŽ0¤braŽ0 *)
	¤mathin P1 = (id - ¤mmsigmaŽz)/2 //SparseArray;  (* qubit projector ¤ketŽ1¤braŽ1 *)
	¤mathin CTRL[n_Integer, ¤mmlambda_ /; VectorQ[¤mmlambda,IntegerQ], A_] /;
	¤mathnl   (Unequal@@¤mmlambda) && Min[¤mmlambda]>=1 && Max[¤mmlambda]<=n && Dimensions[A]=={2^n,2^n} := 
	¤mathnl   IdentityMatrix[2^n, SparseArray] +
	¤mathnl   Apply[Dot, op[n,#,P1]&/@¤mmlambda].(A - IdentityMatrix[2^n, SparseArray])
\end{mathematica}
With this definition, the CNOT operator $\text{CNOT}^{(jk)}=\ket{0}\bra{0}^{(j)}\otimes\one^{(k)}+\ket{1}\bra{1}^{(j)}\otimes\sigma_x^{(k)}$

\noindent
\hspace{\stretch{1}}
\Qcircuit @C=0.5em @R=.7em {
	\lstick{j}	& \qw	& \ctrl{1}	& \qw	& \qw \\
	\lstick{k}	& \qw	& \targ	& \qw	& \qw
}
\hspace{\stretch{1}}
\\

\noindent
is simply the CTRL operator with a single element in the list $\lambda=\{j\}$ and a single-qubit $\op{A}=\op{\sigma}_x^{(k)}$ operator,
\begin{mathematica}
	¤mathin CNOT[n_Integer, j_Integer -> k_Integer] /; 1<=j<=n && 1<=k<=n && j!=k :=
	¤mathnl   CTRL[n, {j}, op[n, k, ¤mmsigmaŽx]]
\end{mathematica}
Notice that here we use the notation \mm{CNOT[n, j->k]} to indicate that qubit $j$ controls qubit $k$: this arrow notation \mm{->} is purely for syntactic beauty and has no further effects (it is a pattern like any other, with no unintended side effects).
The matrix representation of a two-qubit CNOT takes on the familiar form
\begin{mathematica}
	¤mathin CNOT[2, 1->2] //Normal
	¤mathout {{1, 0, 0, 0},
	¤mathnl  {0, 1, 0, 0},
	¤mathnl  {0, 0, 0, 1},
	¤mathnl  {0, 0, 1, 0}}
\end{mathematica}

\subsubsection{three-qubit gates}

For completeness, we define three-qubit gates that are sometimes useful in the construction of general quantum circuits.

The CCNOT gate or Toffoli gate is a controlled-NOT gate \mm{CCNOT[n, \{i,j\}->k]} with two controlling qubits, $i$ and $j$, and is defined in analogy to the CNOT gate:

\noindent
\hspace{\stretch{1}}
\Qcircuit @C=0.5em @R=.7em {
	\lstick{i}	& \qw	& \ctrl{1}	& \qw	& \qw \\
	\lstick{j}	& \qw	& \ctrl{1}	& \qw	& \qw \\
	\lstick{k}	& \qw	& \targ	& \qw	& \qw
}
\hspace{\stretch{1}}

\begin{mathematica}
	¤mathin CCNOT[n_Integer, {i_Integer, j_Integer} -> k_Integer] /;
	¤mathnl   1<=i<=n && 1<=j<=n && 1<=k<=n && Unequal[i,j,k] :=
	¤mathnl     CTRL[n, {i,j}, op[n, k, ¤mmsigmaŽx]]
\end{mathematica}
The controlled-SWAP gate or Fredkin gate \mm{CSWAP[n, i->\{j,k\}]} conditionally swaps two qubits, $j$ and $k$:

\noindent
\hspace{\stretch{1}}
\Qcircuit @C=0.5em @R=1em {
	\lstick{i}	& \qw	& \ctrl{1}		& \qw	& \qw \\
	\lstick{j}	& \qw	& \qswap		& \qw	& \qw \\
	\lstick{k}	& \qw	& \qswap \qwx	& \qw	& \qw
}
\hspace{\stretch{1}}

\begin{mathematica}
	¤mathin CSWAP[n_Integer, i_Integer -> {j_Integer, k_Integer}] /;
	¤mathnl   1<=i<=n && 1<=j<=n && 1<=k<=n && Unequal[i,j,k] :=
	¤mathnl     CTRL[n, {i}, SWAP[n, {j, k}]]
\end{mathematica}

\subsection{a simple quantum circuit}
\label{sec:simplecircuit}

As a simple example, we study the quantum circuit

\noindent
\hspace{\stretch{1}}
\Qcircuit @C=0.5em @R=1em {
	\lstick{\text{qubit 1: }\ket{0}}	& \qw	& \gate{H}		& \qw	&\ctrl{1}		& \qw	& \qw \\
	\lstick{\text{qubit 2: }\ket{0}}	& \qw	& \qw		& \qw	& \targ		& \qw	& \qw
}
\hspace{\stretch{1}}
\\

\noindent
The unitary operation corresponding to this circuit is a Hadamard gate on qubit 1, followed by a controlled-NOT gate where qubit 1 controls the inversion of qubit 2. In Mathematica this gate sequence needs to be written from right to left, because the gates are represented by matrices that will be applied to a state vector on their right:
\begin{mathematica}
	¤mathin S = CNOT[2, 1->2] . H[2, 1];
	¤mathin Normal[S]
	¤mathout¤labelŽmath:simplecircuit {{1/Sqrt[2], 0, 1/Sqrt[2], 0},
	¤mathnl  {0, 1/Sqrt[2], 0, 1/Sqrt[2]},
	¤mathnl  {0, 1/Sqrt[2], 0, -1/Sqrt[2]},
	¤mathnl  {1/Sqrt[2], 0, -1/Sqrt[2], 0}}
\end{mathematica}
The matrix representation of \mm{\ref{math:simplecircuit}} refers to the two-qubit basis set $\mathcal{B}_2=\{\ket{00},\ket{01},\ket{10},\ket{11}\}$, which we can inspect for any number of qubits with
\begin{mathematica}
	¤mathin¤labelŽmath:binarybasis B[n_Integer /; n>=1] := Tuples[{0, 1}, n]
	¤mathin B[2]
	¤mathout {{0,0}, {0,1}, {1,0}, {1,1}}
\end{mathematica}
The input state of our circuit is the product state $\ket{\psi\ix{in}}=\ket{0}\otimes\ket{0}=\ket{00}$, which is the first element of $\mathcal{B}_2$:
\begin{mathematica}
	¤mathin ¤mmpsiŽin = {1,0,0,0};
\end{mathematica}
The output state of our circuit follows from the application of \mm{S},
\begin{mathematica}
	¤mathin ¤mmpsiŽout = S . ¤mmpsiŽin
	¤mathout {1/Sqrt[2], 0, 0, 1/Sqrt[2]}
\end{mathematica}
Looking at the basis set $\mathcal{B}_2$ we identify this output state with the maximally entangled state $\ket{\psi\ix{out}}=\frac{\ket{00}+\ket{11}}{\sqrt{2}}$.
Projective measurements on the two qubits,

\noindent
\hspace{\stretch{1}}
\Qcircuit @C=0.5em @R=1em {
	\lstick{\text{qubit 1: }\ket{0}}	& \qw	& \gate{H}		& \qw	&\ctrl{1}		& \qw	& \meter & \rstick{\text{bit 1}}\\
	\lstick{\text{qubit 2: }\ket{0}}	& \qw	& \qw		& \qw	& \targ		& \qw	& \meter & \rstick{\text{bit 2}}
}
\hspace{\stretch{1}}
\\

\noindent
give 50\% probability of finding the classical result ``00'' and 50\% probability of finding ``11'', whereas the bit combinations ``01'' and ``10'' never occur:
\begin{mathematica}
	¤mathin Abs[¤mmpsiŽout]^2
	¤mathout¤labelŽmath:simplepsiout {1/2, 0, 0, 1/2}
\end{mathematica}
It is important to recognize that these four probabilities are insufficient to identify the state \ket{\psi\ix{out}}, even if many measurements are made, because any state whose diagonal density-matrix elements match \mm{\ref{math:simplepsiout}} gives these probabilities. Generally, in order to identify a two-qubit output state fully, a quantum-state tomography (QST)\footnote{See \url{https://en.wikipedia.org/wiki/Quantum_tomography}.}\index{quantum state tomography} must be performed, which involves applying further phase gates (qubit rotations) before the projective measurements and measuring all sixteen (fifteen non-trivial) expectation values $\me{\psi\ix{out}}{\op{\sigma}_1\otimes\op{\sigma}_2}{\psi\ix{out}}$ for $\op{\sigma}_1,\op{\sigma}_2\in\{\one,\op{\sigma}_x,\op{\sigma}_y,\op{\sigma}_z\}$, followed by an inversion procedure to estimate the density matrix:\footnote{\autoref{eq:twoqubittomo} is a direct inversion that may not result in a positive semi-definite density matrix if experimental noise is present. In such cases, more elaborate inversion procedures are available.}
\begin{align}
\label{eq:twoqubittomo}
	\op{\rho}
	&= \frac14 \sum_{\op{\sigma}_1\in\{\one,\op{\sigma}_x,\op{\sigma}_y,\op{\sigma}_z\}}\sum_{\op{\sigma}_2\in\{\one,\op{\sigma}_x,\op{\sigma}_y,\op{\sigma}_z\}}
	\me{\psi\ix{out}}{\op{\sigma}_1\otimes\op{\sigma}_2}{\psi\ix{out}}\cdot\op{\sigma}_1\otimes\op{\sigma}_2\nonumber\\
	&= \frac14 \begin{bmatrix}
	\one \one + \one z + z \one + z z &
	\one x - \ii \one y + z x - \ii z y &
	x \one + x z - \ii y \one - \ii y z &
	x x - \ii x y - \ii y x - y y \\
	\one x + \ii \one y + z x + \ii z y &
	\one \one - \one z + z \one - z z &
	x x + \ii x y - \ii y x + y y &
	x \one - x z - \ii y \one + \ii y z \\
	x \one + x z + \ii y \one + \ii y z &
	x x - \ii x y + \ii y x + y y &
	\one \one + \one z - z \one - z z &
	\one x - \ii \one y - z x + \ii z y \\
	x x + \ii x y + \ii y x - y y &
	x \one - x z + \ii y \one - \ii y z &
	\one x + \ii \one y - z x - \ii z y &
	\one \one - \one z - z \one + z z
	\end{bmatrix}
\end{align}
(abbreviating $x y=\me{\psi\ix{out}}{\op{\sigma}_x\otimes\op{\sigma}_y}{\psi\ix{out}}$ etc.)
A full QST on $n$ qubits requires measuring $4^n-1$ such expectation values, which makes the QST infeasible in general.

\subsection{application: the Quantum Fourier Transform}
\label{sec:QFT}
\index{quantum Fourier transform}

The discrete classical Fourier transform\index{discrete Fourier transform}\footnote{See \url{https://en.wikipedia.org/wiki/Discrete_Fourier_transform}.} (CFT) of a list of $N$ complex numbers $\vect{x}=\{x_0,x_1,\ldots,x_{N-1}\}$ is given by the list $\vect{y}=\{y_0,y_1,\ldots,y_{N-1}\}$ with elements
\begin{equation}
	y_j = \frac{1}{\sqrt{N}}\sum_{k=0}^{N-1} x_k e^{2\pi\ii j k/N}.
\end{equation}
It can be seen as a unitary matrix operation
\begin{align}
	\label{eq:CFT}
	\vect{y}&=\matr{F}\cdot\vect{x} & \text{with } F_{jk}&=e^{2\pi\ii j k/N}/\sqrt{N}.
\end{align}
With the Fast Fourier Transform\index{fast Fourier transform} (FFT) algorithm,\footnote{See \url{https://en.wikipedia.org/wiki/Fast_Fourier_transform}.} the computational effort of evaluating \autoref{eq:CFT} is of order $\mathcal{O}[N \log(N)]$.

The discrete Quantum Fourier Transform (QFT) is precisely the same transformation, except that the vectors $\vect{x}$ and $\vect{y}$ are encoded into quantum states. For this, a quantum system with Hilbert space dimension $N$ is described by a basis set $\{\ket{0},\ket{1},\ldots,\ket{N-1}\}$, and the states $\ket{x}=\sum_{j=0}^{N-1}x_j\ket{j}$ and $\ket{y}=\sum_{j=0}^{N-1}y_j\ket{j}$ are seen as related by the unitary QFT operator $\op{\mathcal{F}}$ such that
\begin{align}
	\label{eq:QFT}
	\ket{y}& =\op{\mathcal{F}}\ket{x} & \text{with } \me{j}{\op{\mathcal{F}}}{k} = F_{jk},
\end{align}
in analogy to \autoref{eq:CFT}. The idea of this section is that the QFT can be evaluated much faster than the CFT, even though both are mathematically equivalent.

We assume that $N=2^n$ is an integer power of two.\footnote{For all other cases, choose $n$ as the smallest integer $\ge \log_2(N)$ and set $x_N\ldots x_{2^n-1}$ to zero.} The Hilbert space of $n$ qubits has exactly $2^n=N$ dimensions, and therefore we use these $n$ qubits to encode the states $\ket{x}$ and $\ket{y}$ in the following way. The $2^n$ basis states $\mathcal{B}_n=\{\ket{00\ldots00},\ket{00\ldots01},\ket{00\ldots10},\ldots,\ket{11\ldots11}\}$ are, in our usual construction through tensor products (\autoref{sec:coupledDOF}), listed in increasing order when interpreted as binary numbers (see \mm{\ref{math:binarybasis}}). We give each basis state a new label equal to this binary number: $\ket{00\ldots00}=\ket{0}$, $\ket{00\ldots01}=\ket{1}$, $\ket{00\ldots10}=\ket{2}$, \ldots, $\ket{11\ldots11}=\ket{2^n-1}$, such that the state of the first qubit is the most significant bit (MSB) of the binary representation of the basis state's index, and the state of the $n\ex{th}$ qubit is the least significant bit (LSB) of the binary representation of the basis state's index. What follows below is a quantum circuit operating on these $n$ qubits that has the effect of the QFT operator $\op{\mathcal{F}}$, as expressed in this binary basis.

The Quantum Fourier Transform circuit is assembled from single-qubit Hadamard gates and two-qubit controlled $Z$-phase gates, where $\op{R}_k=\op{R}_z(2\pi/2^k)=\ket{0}\bra{0}+e^{2\pi\ii/2^k}\ket{1}\bra{1}$ using \mm{\ref{math:gateRZ}}:

\noindent
\hspace{\stretch{1}}
\Qcircuit @C=0.5em @R=.7em {
	\lstick{1}		& \gate{H}	& \gate{R_2}	& \gate{R_3}	& \push{\cdots} \qw	& \gate{R_{n-1}}	& \gate{R_{n}}	& \qw	& \qw	& \qw		& \push{\cdots} \qw	& \qw			& \qw			& \push{\cdots} \qw	& \qw	& \qw		& \qw	& \qw	& \qw	& \qswap \qwx[5]	& \qw		& \qw	&\rstick{1} \qw \\
	\lstick{2}		& \qw	& \ctrl{-1}		& \qw		& \push{\cdots} \qw	& \qw 			& \qw		& \qw	& \gate{H}	& \gate{R_2}	& \push{\cdots} \qw	& \gate{R_{n-2}}	& \gate{R_{n-1}}	& \push{\cdots} \qw	& \qw	& \qw		& \qw	& \qw	& \qw	& \qw		& \qswap \qwx[3]	& \qw	& \rstick{2} \qw \\
	\lstick{3}		& \qw	& \qw		& \ctrl{-2}		& \push{\cdots} \qw	& \qw 			& \qw		& \qw	& \qw	& \ctrl{-1}		& \push{\cdots} \qw	& \qw			& \qw			& \push{\cdots} \qw	& \qw	& \qw		& \qw	& \qw	& \qw	& \qw		& \qw		& \qswap \qwx[1]& \rstick{3} \qw \\
	\lstick{\cdots}	&		&			&			& \push{\cdots}		&				&			&		&		&			& \push{\cdots}		&				&				&				&		&			&		&		&		&			&			&		& \rstick{\cdots} \\
	\lstick{n-1}		& \qw	& \qw		& \qw		& \push{\cdots} \qw	& \ctrl{-4} 			& \qw		& \qw	& \qw	& \qw		& \push{\cdots} \qw	& \ctrl{-3}			& \qw			& \push{\cdots} \qw	& \gate{H}	& \gate{R_2}	& \qw	& \qw	& \qw	& \qw		& \qswap		& \qw	& \rstick{n-1} \qw \\
	\lstick{n}		& \qw	& \qw		& \qw		& \push{\cdots} \qw	& \qw 			& \ctrl{-5}		& \qw	& \qw	& \qw		& \push{\cdots} \qw	& \qw			& \ctrl{-4}			& \push{\cdots} \qw	& \qw	& \ctrl{-1}		& \qw	& \gate{H}	& \qw	& \qswap		& \qw		& \qw	& \rstick{n} \qw
	\gategroup{1}{2}{6}{7}{.7em}{--} \gategroup{2}{9}{6}{13}{.7em}{--} \gategroup{5}{15}{6}{16}{.7em}{--} \gategroup{6}{18}{6}{18}{.7em}{--}
}
\hspace{\stretch{1}}

\noindent
To construct the $i\ex{th}$ dashed block consisting of a Hadamard gate on qubit $i$ followed by $n-i$ controlled $Z$-phase gates, we remember that the application of matrix operators happens from right to left, in the reverse order from that shown in the circit diagram above. We first construct a list of the controlled \mm{RZ} operators and contract it by applying \mm{Dot}:
\begin{mathematica}
	¤mathin QFTblock[n_Integer, i_Integer] /; 1<=i<=n :=
	¤mathnl   Apply[Dot, Table[CTRL[n, {j}, RZ[n, i, 2¤mmpi/2^(j+1-i)]], {j, n, i+1, -1}]].
	¤mathnl   H[n,i]
\end{mathematica}
We assemble the $n$-qubit QFT operator from these dashed \mm{QFTblock} blocks and a set of SWAP operations that reverses the qubit order,
\begin{mathematica}
	¤mathin QFT[n_Integer] /; n>=1 := 
	¤mathnl   Apply[Dot, Table[SWAP[n, {i, n+1-i}], {i, 1, n/2}]].
	¤mathnl   Apply[Dot, Table[QFTblock[n, i], {i, n, 1, -1}]]
\end{mathematica}
The matrix representation of this \mm{QFT} operator is a $2^n\times2^n$ matix with element $(j,k)$ given by $2^{-n/2}e^{2\pi\ii j k/2^n}$, precisely as expected from \autoref{eq:QFT} with $N=2^n$. We check this relation for $n=1\dotsc6$ with
\begin{mathematica}
	¤mathin Table[QFT[n] == 2^(-n/2)*Table[Exp[2¤mmpi*I*j*k/2^n], {j,0,2^n-1}, {k,0,2^n-1}],
	¤mathnl   {n, 6}] //FullSimplify
	¤mathout {True, True, True, True, True, True}
\end{mathematica}
The resources used to construct this quantum circuit are
\begin{itemize}
	\item $n$ Hadamard gates,
	\item $\frac{n(n-1)}{2}$ controlled $Z$-phase gates, and
	\item $n/2$ swap gates.
\end{itemize}
In the present classical simulation of quantum circuits, each quantum gate is a sparse $2^n\times2^n$ matrix, usually containing $\mathcal{O}(2^n)$ nonzero matrix elements; applying such a simulated gate to a state therefore takes $\mathcal{O}(2^n)$ time, which makes the simulated QFT no faster than the classical FFT, which scales as $\mathcal{O}(2^nn)$. However, if we can construct a physical system in which these gates can be applied in a time that scales at most polynomially with $n$, then the QFT is a massive improvement over the scaling of the classical FFT. The development of such physical qubit/gate systems is the focus of much ongoing scientific research.

\subsection{application: quantum phase estimation}
\index{quantum phase estimation}

The QFT circuit of \autoref{sec:QFT} cannot be used by itself in practice, because it requires the preparation of an arbitrary quantum state containing an exponential number of parameters $x_j$, as well as a full quantum-state tomography to read out an exponential number of parameters $y_j$ describing the final state (see \autoref{sec:simplecircuit}). In this section we study a quantum circuit that uses the QFT as a component, and circumvents these exponential input/output bottlenecks.

Unitary matrices have eigenvalues that are of unit norm, and can be written as $e^{2\pi\ii\varphi}$ with $\varphi\in\dsR$. The question addressed here is: given a unitary operator $\op{U}_{\varphi}$ and an eigenstate $\ket{u}$ such that $\op{U}_{\varphi}\ket{u}=e^{2\pi\ii\varphi}\ket{u}$, can we estimate $\varphi$ efficiently, that is, to $t$ binary digits with an effort that scales polynomially with $t$?

The answer is yes, using the following quantum circuit that makes use of the Quantum Fourier Transform of \autoref{sec:QFT} but (i) starts with an initial product state that can be prepared with $\mathcal{O}(t)$ effort, and (ii) does not require a full quantum state tomography, but instead finishes with a simple projective measurement that takes $\mathcal{O}(t)$ effort.

\noindent
\hspace{\stretch{1}}
\Qcircuit @C=0.5em @R=.7em {
	\lstick{\ket{0}}	& \gate{H}	& \qw	& \qw		& \qw		& \push{\cdots} \qw	& \ctrl{5}			& \multigate{4}{\mathcal{F}^\dag}	& \meter	& \rstick{\text{MSB: weight $2^{t-1}$}} \\
	\lstick{\cdots}	& 		&		&			&			& \push{\cdots}		&				&							& \cdots	& \rstick{\ldots} \\
	\lstick{\ket{0}}	& \gate{H}	& \qw	& \qw		& \ctrl{3}		& \push{\cdots} \qw	& \qw			& \ghost{\mathcal{F}^\dag}		& \meter	& \rstick{\text{bit 3: weight $4=2^2$}}\\
	\lstick{\ket{0}}	& \gate{H}	& \qw	& \ctrl{2}		& \qw		& \push{\cdots} \qw	& \qw			& \ghost{\mathcal{F}^\dag}		& \meter	& \rstick{\text{bit 2: weight $2=2^1$}}\\
	\lstick{\ket{0}}	& \gate{H}	& \ctrl{1}	& \qw		& \qw		& \push{\cdots} \qw	& \qw			& \ghost{\mathcal{F}^\dag}		& \meter	& \rstick{\text{LSB: weight $1=2^0$}} \\
	\lstick{\ket{u}}	& {/} \qw	& \gate{U_{\varphi}}	& \gate{U_{\varphi}^2}	& \gate{U_{\varphi}^4}	& \push{\cdots} \qw	& \gate{U_{\varphi}^{2^{t-1}}}	& {/} \qw						& \qw	& \rstick{\ket{u}\text{ (not measured)}} \qw
}
\hspace{\stretch{1}}
\\

\noindent
To set up a quantum phase estimation in Mathematica, we begin by defining the unitary operator $\op{U}_{\varphi}$ and its eigenstate $\ket{u}$ with
\begin{mathematica}
	¤mathin¤labelŽmath:PhaseEstu u = {1};
	¤mathin¤labelŽmath:PhaseEstU U[¤mmphi_] = {{Exp[2¤mmpi*I*¤mmphi]}};
\end{mathematica}
and check that they satisfy $\op{U}_{\varphi}\ket{u}=e^{2\pi\ii\varphi}\ket{u}$ and $\scp{u}{u}=1$:
\begin{mathematica}
	¤mathin {U[¤mmphi].u === E^(2¤mmpi*I*¤mmphi)*u, Norm[u] == 1}
	¤mathout {True, True}
\end{mathematica}
Here we use a one-dimensional quantum system: the operator $\op{U}_{\varphi}$ is a $1\times1$ matrix, and the state $\ket{u}$ is a list of length 1. More generally, the Hilbert space of the system under test (SUT) can be arbitrarily large (see \ref{Q:PhaseEst2}), and more complex quantum circuits can be substituted for $\op{U}$ in more elaborate experiments.

In order to construct the phase estimation circuit, we will also need the unit operator acting on the SUT:
\begin{mathematica}
	¤mathin U0 = IdentityMatrix[Length[u], SparseArray];
\end{mathematica}
The controlled version of the $\op{U}_{\varphi}$ operator, where the $i\ex{th}$ qubit out of a set of $n$ qubits controls the application of $\op{U}_{\varphi}$ to the SUT, is $\op{U}_{\varphi}^{(i)}=\ket{0}\bra{0}^{(i)}\otimes\one+\ket{1}\bra{1}^{(i)}\otimes\op{U}_{\varphi}$. We use the tensor-product techniques of \autoref{sec:coupledDOF} to couple the qubits to the SUT:
\begin{mathematica}
	¤mathin CTRLU[n_Integer, i_Integer, ¤mmphi_] /; 1<=i<=n :=
	¤mathnl   KroneckerProduct[op[n,i,P0], U0] + KroneckerProduct[op[n,i,P1], U[¤mmphi]]
\end{mathematica}
The initial state of the phase estimation circuit is $\ket{\psi_0} = \ket{0}^{\otimes t}\otimes\ket{u}$. We know that the state $\ket{0}^{\otimes t}=\ket{00\ldots00}$ is the first basis state in the computational basis $\mathcal{B}_n$ (eigen-basis of $\op{\sigma}_z$), and construct it with \mm{SparseArray[1->1, 2\textasciicircum{}t]}. As an example, we work with $t=4$ qubits here:
\begin{mathematica}
	¤mathin t = 4;
	¤mathin ¤mmpsiŽ0 = Flatten[KroneckerProduct[SparseArray[1->1, 2^t], u]] //Normal;
\end{mathematica}
Applying a Hadamard gate to each qubit gives the state
\begin{mathematica}
	¤mathin ¤mmpsiŽ1 = KroneckerProduct[Apply[Dot, Table[H[t, i], {i, t}]], U0] . ¤mmpsiŽ0;
\end{mathematica}
Applying the controlled $\op{U}_{\varphi}^m=\op{U}_{m\varphi}$ operations sequentially then gives the state
\begin{mathematica}
	¤mathin ¤mmpsiŽ2[¤mmphi_] = Apply[Dot, Table[CTRLU[t, i, 2^(t-i)*¤mmphi], {i, t, 1, -1}]] . ¤mmpsiŽ1;
\end{mathematica}
Finally, an inverse QFT yields the phase-estimation state $\ket{\varepsilon_{\varphi}}$. Remember that the QFT is a unitary operation, and therefore its inverse is its Hermitian conjugate:
\begin{mathematica}
	¤mathin ¤mmepsilon[¤mmphi_] = KroneckerProduct[ConjugateTranspose[QFT[t]], U0] . ¤mmpsiŽ2[¤mmphi];
\end{mathematica}
We use the techniques of \autoref{sec:rdm}, in particular \mm{\ref{math:traceoutpsi2}}, to drop the component $\ket{u}$ at the end of the quantum circuit and find the reduced density matrix of the qubits. The diagonal elements of this reduced density matrix are the probabilities of finding the various basis states of $\mathcal{B}_n$ in a projective measurement as shown on the right of the above circuit:
\begin{mathematica}
	¤mathin prob[¤mmphi_?NumericQ] := Re[Diagonal[traceout[¤mmepsilon[N[¤mmphi]], -Length[u]]]]
\end{mathematica}
The first element of \mm{prob[\mmphi]} gives the probability of measurement outcomes $\{0,0,0,0\}$, that is, the probability that the qubits are in the joint state $\ket{0000}=\ket{0}\otimes\ket{0}\otimes\ket{0}\otimes\ket{0}$.
The second element of \mm{prob[\mmphi]} gives the probability of measurement outcomes $\{0,0,0,1\}$, that is, the probability that the qubits are in the joint state $\ket{0001}=\ket{0}\otimes\ket{0}\otimes\ket{0}\otimes\ket{1}$.
And so forth: the $j\ex{th}$ element of \mm{prob[\mmphi]} gives the probability of measurement outcomes corresponding to the binary representation of $j-1$.

The trick of this phase-estimation quantum circuit is that the information on $\varphi$ is contained in the state $\ket{\varepsilon_{\varphi}}$ in a way that can be extracted from these probabilities without doing a full quantum-state tomography. We get an idea of what this means by looking at the probabilities for the different measurement outcomes when $\varphi$ is an integer multiple of $2^{-t}$:
\begin{mathematica}
	¤mathin Table[prob[¤mmphi], {¤mmphi, 0, 1, 2^(-t)}] //Chop
	¤mathout¤labelŽmath:diagonalprob {{1., 0, 0, 0, 0, 0, 0, 0, 0, 0, 0, 0, 0, 0, 0, 0},
	¤mathnl  {0, 1., 0, 0, 0, 0, 0, 0, 0, 0, 0, 0, 0, 0, 0, 0},
	¤mathnl  {0, 0, 1., 0, 0, 0, 0, 0, 0, 0, 0, 0, 0, 0, 0, 0},
	¤mathnl  {0, 0, 0, 1., 0, 0, 0, 0, 0, 0, 0, 0, 0, 0, 0, 0},
	¤mathnl  {0, 0, 0, 0, 1., 0, 0, 0, 0, 0, 0, 0, 0, 0, 0, 0},
	¤mathnl  {0, 0, 0, 0, 0, 1., 0, 0, 0, 0, 0, 0, 0, 0, 0, 0},
	¤mathnl  {0, 0, 0, 0, 0, 0, 1., 0, 0, 0, 0, 0, 0, 0, 0, 0},
	¤mathnl  {0, 0, 0, 0, 0, 0, 0, 1., 0, 0, 0, 0, 0, 0, 0, 0},
	¤mathnl  {0, 0, 0, 0, 0, 0, 0, 0, 1., 0, 0, 0, 0, 0, 0, 0},
	¤mathnl  {0, 0, 0, 0, 0, 0, 0, 0, 0, 1., 0, 0, 0, 0, 0, 0},
	¤mathnl  {0, 0, 0, 0, 0, 0, 0, 0, 0, 0, 1., 0, 0, 0, 0, 0},
	¤mathnl  {0, 0, 0, 0, 0, 0, 0, 0, 0, 0, 0, 1., 0, 0, 0, 0},
	¤mathnl  {0, 0, 0, 0, 0, 0, 0, 0, 0, 0, 0, 0, 1., 0, 0, 0},
	¤mathnl  {0, 0, 0, 0, 0, 0, 0, 0, 0, 0, 0, 0, 0, 1., 0, 0},
	¤mathnl  {0, 0, 0, 0, 0, 0, 0, 0, 0, 0, 0, 0, 0, 0, 1., 0},
	¤mathnl  {0, 0, 0, 0, 0, 0, 0, 0, 0, 0, 0, 0, 0, 0, 0, 1.},
	¤mathnl  {1., 0, 0, 0, 0, 0, 0, 0, 0, 0, 0, 0, 0, 0, 0, 0}}
\end{mathematica}
Whenever $\varphi$ is an integer multiple of $2^{-t}=1/16$, we find that only one basis state is occupied, and therefore the outcomes of the projective measurements on the 4 qubits always give the same results, with no quantum fluctuation. A single projective measurement of all 4 qubits can be interpreted as a binary number $j\in\{0,1,2,\ldots,15\}$ that is related to the phase estimate as $\varphi=j/16$; no quantum-state tomography is required.


What happens when $\varphi$ is not an integer multiple of $2^{-t}$? It turns out that the basis state corresponding to the nearest integer multiple of $2^{-t}$ will be found most frequently in the projective measurements. For example, for $\varphi=0.2$ the probabilities for projecting $\ket{\varepsilon_{0.2}}$ into the 16 basis states are
\begin{mathematica}
	¤mathin prob[0.2]
	¤mathout {0., 0.01, 0.02, 0.88, 0.06, 0.01, 0., 0., 0., 0., 0., 0., 0., 0., 0., 0.}
\end{mathematica}
(rounded here to two decimals). The fourth basis state, which is $\ket{0011}$ corresponding to $\varphi=3/16$, will be found in about 88\% of all experiments, and so a plurality vote most likely yields $\varphi\approx3/16=\num{0.1875}$ as a fair estimate of the phase, with an upper bound on the error of $2^{-t-1}=1/32$; no quantum-state tomography required.
We extract the expected plurality-vote winner of a large number of experiments with
\begin{mathematica}
	¤mathin mostprobable[¤mmphi_?NumericQ] := (Ordering[prob[¤mmphi], -1][[1]] - 1)/2^t
\end{mathematica}
where the \mm{Ordering} function is used to give the position of the largest element:
\begin{mathematica}
	¤mathin mostprobable[0.2]
	¤mathout 3/16
\end{mathematica}
It can be shown that \mm{mostprobable[\mmphi]==Mod[Round[\mmphi, 2\textasciicircum(-t)], 1]}.
Increasing the number of qubits $t$ results in more precise estimates, while keeping the circuit complexity at $\mathcal{O}(t^2)$.

\subsection{exercises}

\begin{questions}
	\item\label{Q:SimpleCircuit} For the output state \ket{\psi\ix{out}} of \mm{\ref{math:simplepsiout}}, calculate all expectation values necessary to fill in \autoref{eq:twoqubittomo}.
\pagenote[\ref{Q:SimpleCircuit}]{Don't forget to complex-conjugate \mm{\mmpsi{}out} on the left, for generality:
\begin{mathematica}
	¤protect¤mathin¤ ¤mmpsiŽout¤ =¤ {1/Sqrt[2],¤ 0,¤ 0,¤ 1/Sqrt[2]};
	¤protect¤mathin¤ Table[Conjugate[¤mmpsiŽout].(KroneckerProduct[PauliMatrix[i],PauliMatrix[j]].¤mmpsiŽout),
	¤protect¤mathnl¤ ¤ ¤ {i,0,3},¤ {j,0,3}]
	¤protect¤mathout¤ {{1,¤ 0,¤ 0,¤ 0},¤ {0,¤ 1,¤ 0,¤ 0},¤ {0,¤ 0,¤ -1,¤ 0},¤ {0,¤ 0,¤ 0,¤ 1}}
\end{mathematica}
	We find $\me{\psi\ix{out}}{\one\otimes\one}{\psi\ix{out}}=1$ (normalization), $\me{\psi\ix{out}}{\op{\sigma}_x\otimes\op{\sigma}_x}{\psi\ix{out}}=1$, $\me{\psi\ix{out}}{\op{\sigma}_y\otimes\op{\sigma}_y}{\psi\ix{out}}=-1$, $\me{\psi\ix{out}}{\op{\sigma}_z\otimes\op{\sigma}_z}{\psi\ix{out}}=1$, and all others equal to zero. The density matrix is therefore
\begin{align}
	\op{\rho}
	&= \frac14 \left( \one\otimes\one + \op{\sigma}_x\otimes\op{\sigma}_x - \op{\sigma}_y\otimes\op{\sigma}_y + \op{\sigma}_z\otimes\op{\sigma}_z\right)\nonumber\\
	&=\frac12\left(\ket{00}\bra{00}+\ket{00}\bra{11}+\ket{11}\bra{00}+\ket{11}\bra{11}\right)
	=\frac{\ket{00}+\ket{11}}{\sqrt{2}}\frac{\bra{00}+\bra{11}}{\sqrt{2}}
\end{align}
	as expected.}
	\item\label{Q:PhaseEst2} Multi-dimensional phase estimation: set $u=\{1,1\}/\sqrt{2}$ and $\op{U}_{\varphi}=e^{2\pi\ii\varphi}\{\{1,0\},\{0,1\}\}$ (two-dimensional system under test) and show that the phase-estimation algorithm still works.
\pagenote[\ref{Q:PhaseEst2}]{Replace \mm{\ref{math:PhaseEstu}} and \mm{\ref{math:PhaseEstU}} with
\begin{mathematica}
	¤protect¤mathin¤ u¤ =¤ {1,1}/Sqrt[2];
	¤protect¤mathin¤ U[¤mmphi_]¤ =¤ Exp[2¤mmpi*I*¤mmphi]¤ *¤ {{1,0},{0,1}};
\end{mathematica}
	and re-evaluate the attached Mathematica notebook. All results remain unchanged.}
	\item\label{Q:PhaseEst2f} What happens if \ket{u} is not an eigenstate of $\op{U}_{\varphi}$? Set $u=\{1,1\}/\sqrt{2}$ and $\op{U}_{\varphi}=\{\{e^{2\pi\ii\varphi},0\},\{0,e^{4\pi\ii\varphi}\}\}$ (two-dimensional system with two different evolution frequencies) and re-evaluate the attached Mathematica script. Plot \mm{prob[\mmphi]} for a range of frequencies $\varphi$ using \mm{ListDensityPlot} and interpret the resulting figure.
\pagenote[\ref{Q:PhaseEst2f}]{Replace \mm{\ref{math:PhaseEstu}} and \mm{\ref{math:PhaseEstU}} with
\begin{mathematica}
	¤protect¤mathin¤ u¤ =¤ {1,1}/Sqrt[2];
	¤protect¤mathin¤ U[¤mmphi_]¤ =¤ {{Exp[2¤mmpi*I*¤mmphi],¤ 0},¤ {0,¤ Exp[4¤mmpi*I*¤mmphi]}};
\end{mathematica}
	and re-evaluate the attached Mathematica notebook. The probabilities for the different estimates of $\varphi$ show both frequencies simultaneously, and there is no cross-talk between them:
\begin{mathematica}
	¤protect¤mathin¤ ListDensityPlot[Transpose[Table[prob[¤mmphi],¤ {¤mmphi,0,1,1/256}]],
	¤protect¤mathnl¤ ¤ ¤ PlotRange->All,¤ DataRange->{{0,1},{0,1-2^-t}},
	¤protect¤mathnl¤ ¤ ¤ FrameLabel->{"setting¤ ¤mmphi","estimated¤ ¤mmphi"}]
\end{mathematica}
\begin{center}
\ifthenelse{\boolean{smallfigures}}%
{\includegraphics[width=0.5\textwidth]{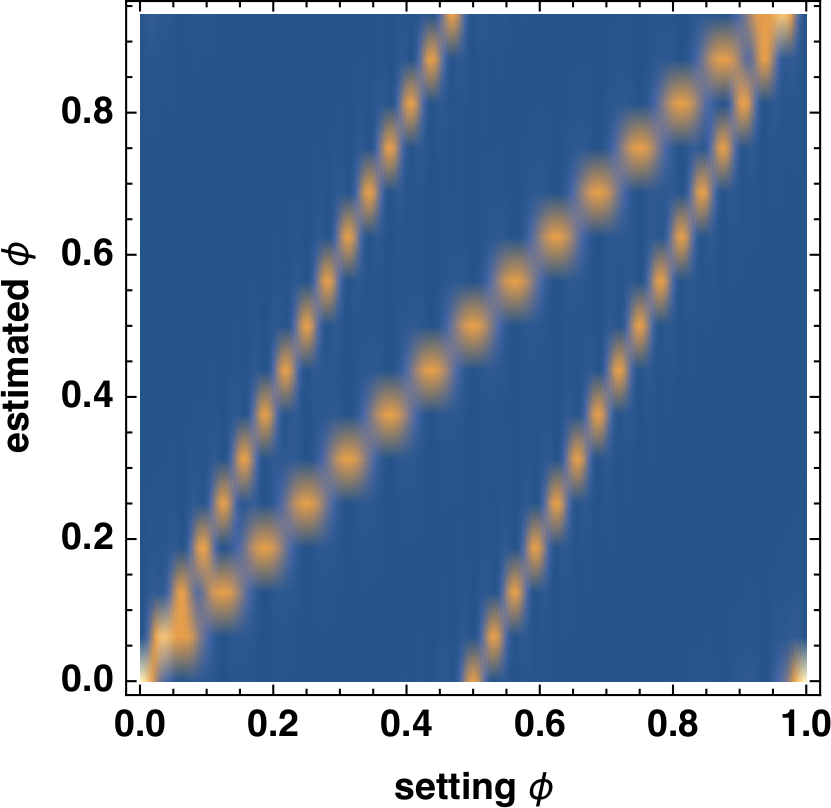}}%
{\includegraphics[width=0.5\textwidth]{phaseest_twofreq}}
\end{center}}
\end{questions}

%
%
%
%
%
%

\chapter{quantum motion in real space}
\label{chap:1D}
\restartlist{questions}
\chapterpicture{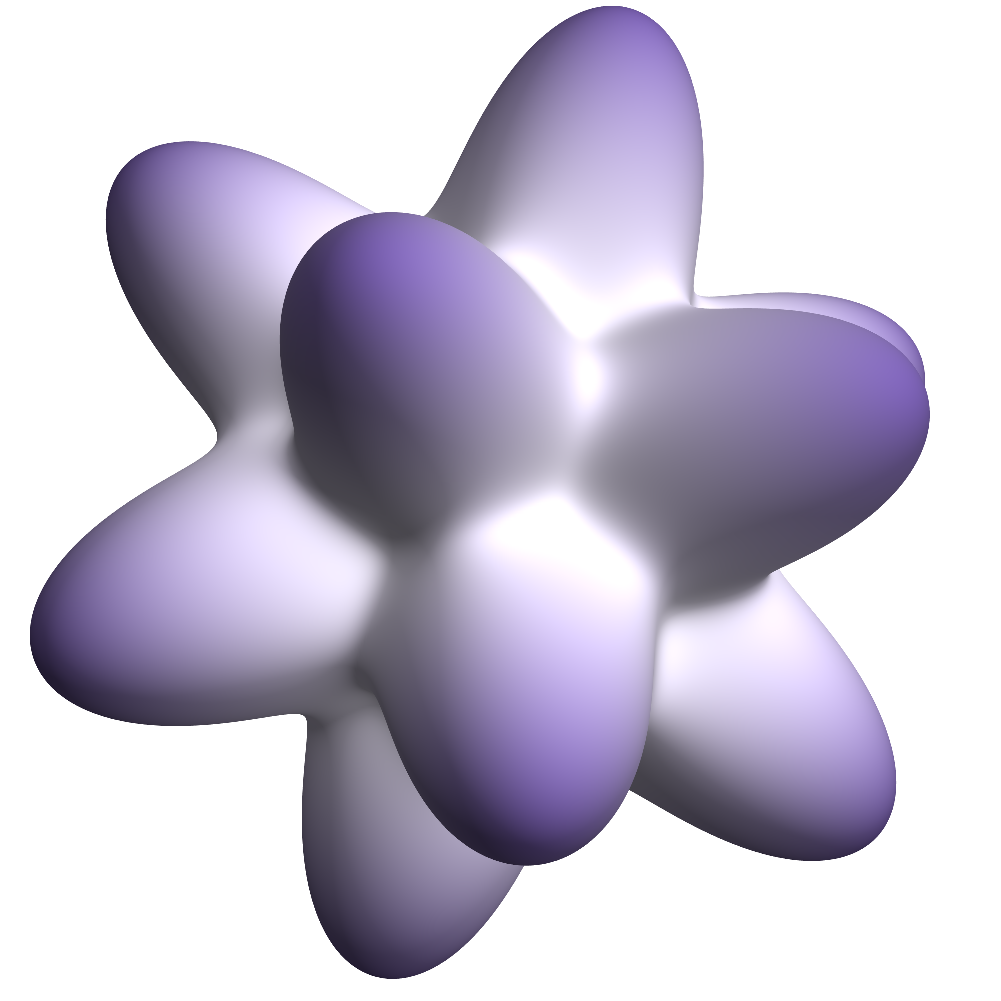}
So far we have studied the quantum formalism in the abstract (\autoref{chap:basis}) and in the context of rotational dynamics (\autoref{chap:spin}).
In this chapter we work with the spatial motion of point particles, which represents a kind of mechanics that is much closer to our everyday experience. Here, quantum states are called \emph{wavefunctions} and depend on the spatial coordinate(s). This apparent difference to the material covered in the previous chapters disappears when we express all wavefunctions in a basis set. We develop numerical methods for studying spatial dynamics that stay as close to a real-space description as quantum mechanics allows.

\clearpage
\section{one particle in one dimension}\index{real-space dynamics}
\label{sec:1part1dim}

A single particle moving in one dimension is governed by a Hamiltonian of the form
\begin{equation}
	\label{eq:1DHamiltonian}
	\Ham = \op{T} + \op{V}
\end{equation}
in terms of the kinetic operator $\op{T}$ and the potential operator $\op{V}$. These operators are usually expressed in the Dirac position basis set $\{\ket{x}\}_{x\in\dsR}$\index{basis set!position basis},\footnote{To be exact, the Dirac position basis set spans a space that is much larger than the Hilbert space of square-integrable smooth functions used in quantum mechanics. This can be seen by noting that this basis set has an uncountably infinite number of elements $\ket{x}$, while the dimension of the Hilbert space in question is only countably infinite [see \autoref{eq:momentumbasis1D} for a countably infinite basis set]. The underlying problem of the \emph{continuum}, which quantum mechanics attempts to resolve, is discussed with some of its philosophical origins and implications by Erwin Schr\"odinger in his essay ``Science and Humanism'' (Cambridge University Press, 1951, ISBN 978-0521575508).} which diagonalizes the position operator in the sense that $\op{x}\ket{x}=x\ket{x}$,\footnote{This eigenvalue equation is tricky: remember that $\op{x}$ is an operator, $\ket{x}$ is a state, and $x$ is a real number.} is ortho-normalized $\scp{x}{y}=\delta(x-y)$, and complete $\int_{-\infty}^{\infty}\ket{x}\bra{x}\dd[x]=\one$\index{completeness relation}. Using this Dirac basis, the explicit expressions for the operators in the Hamiltonian are
\begin{align}
	\label{eq:potop}
	\op{T} &= -\frac{\hbar^2}{2m} \int_{-\infty}^{\infty}\dd[x] \ket{x} \frac{\dd^2}{\dd[x]^2} \bra{x},&
	\op{V} &= \int_{-\infty}^{\infty} \dd[x] \ket{x} V(x) \bra{x},
\end{align}
where $m$ is the particle's mass and $V(x)$ is its potential. Single-particle states $\ket{\psi}$, on the other hand, are written in this basis as
\begin{equation}
	\label{eq:wavefunction}
	\ket{\psi} = \int_{-\infty}^{\infty} \dd[x] \psi(x) \ket{x},
\end{equation}
where $\psi(x)=\scp{x}{\psi}$ is the wavefunction.


In what follows we restrict the freedom of the particle to a domain $x\in\Omega=[0,a]$, where $a$ can be very large in order to approximately describe infinite systems (example: \autoref{sec:gravitywell}). This assumes the potential to be
\begin{equation}
	\label{eq:bottompotential}
	V(x) = \begin{cases}
		\infty & \text{for $x\le0$}\\
		W(x) & \text{for $0<x<a$}\\
		\infty & \text{for $x\ge a$}
	\end{cases}
\end{equation}
This restriction is necessary in order to achieve a finite representation of the system in a computer.

\subsubsection{exercises}

\begin{questions}
	\item\label{Q:Schroedinger} Insert \hyperref[eq:potop]{Equations~\ref*{eq:potop}} and \autoref{eq:wavefunction} into the time-independent Schr\"odinger equation $\Ham\ket{\psi}=E\ket{\psi}$. Use the ortho-normality of the Dirac basis to derive the usual form of the Schr\"odinger equation for a particle's wavefunction in 1D: $-\frac{\hbar^2}{2m}\psi''(x)+V(x)\psi(x)=E\psi(x)$.
\pagenote[\ref{Q:Schroedinger}]{Starting with the Schr\"odinger equation
\begin{equation}
	\left[
		-\frac{\hbar^2}{2m} \int_{-\infty}^{\infty}\dd[x] \ket{x} \frac{\dd^2}{\dd[x]^2} \bra{x}
		+ \int_{-\infty}^{\infty} \dd[x] \ket{x} V(x) \bra{x} \right] \ket{\psi}
		= E\ket{\psi},
\end{equation}
	we (i) leave away the bracket and (ii) multiply by $\bra{y}$ from the left ($y\in\dsR$):
\begin{equation}
		-\frac{\hbar^2}{2m} \int_{-\infty}^{\infty}\dd[x] \scp{y}{x} \frac{\dd^2}{\dd[x]^2} \scp{x}{\psi}
		+ \int_{-\infty}^{\infty} \dd[x] \scp{y}{x} V(x) \scp{x}{\psi}
		= E\scp{y}{\psi}
\end{equation}
	Remembering that $\scp{y}{x}=\delta(x-y)$ and $\scp{x}{\psi}=\psi(x)$:
\begin{equation}
		-\frac{\hbar^2}{2m} \int_{-\infty}^{\infty}\dd[x] \delta(x-y) \psi''(x)
		+ \int_{-\infty}^{\infty} \dd[x] \delta(x-y) V(x) \psi(x)
		= E\psi(y)
\end{equation}
	Simplify the integrals with the Dirac $\delta$-functions:
\begin{equation}
		-\frac{\hbar^2}{2m} \psi''(y)
		+ V(y) \psi(y)
		= E\psi(y)
\end{equation}
	Since this is valid for any $y\in\dsR$, it concludes the proof.}
	\item\label{Q:scalarproduct} Use \autoref{eq:wavefunction} to show that the scalar product between two states is given by the usual formula $\scp{\psi}{\chi}=\int_{-\infty}^{\infty}\psi^*(x)\chi(x)\dd[x]$.
\pagenote[\ref{Q:scalarproduct}]{
\begin{align}
	\scp{\psi}{\chi}
	&= \left[ \int_{-\infty}^{\infty} \dd[x] \psi^*(x) \bra{x} \right] \left[ \int_{-\infty}^{\infty} \dd[y] \chi(y) \ket{y} \right]\nonumber\\
	&= \int_{-\infty}^{\infty} \dd[x] \dd[y] \psi^*(x) \chi(y) \scp{x}{y}\nonumber\\
	&= \int_{-\infty}^{\infty} \dd[x] \dd[y] \psi^*(x) \chi(y) \delta(x-y)\nonumber\\
	&= \int_{-\infty}^{\infty} \dd[x] \psi^*(x) \chi(x)
\end{align}}
\end{questions}

%
%

\subsection{units}
\label{sec:1Dunits}

In order to proceed with implementing the \hyperref[eq:1DHamiltonian]{Hamiltonian~\ref*{eq:1DHamiltonian}}, we first need a consistent set of units (see \autoref{sec:units}) in which to express length, time, mass, and energy. Of these four units, only three are independent: expressions like the classical kinetic energy $E=\frac12mv^2$ indicate a fixed relationship between these four units.

A popular system of units is the International System of Units (SI),\footnote{See \url{https://en.wikipedia.org/wiki/International_System_of_Units}.} in which this consistency is built in:
\begin{mathematica}
	¤mathin LengthUnit = Quantity["Meters"];    (* choose freely *)
	¤mathin TimeUnit = Quantity["Seconds"];     (* choose freely *)
	¤mathin MassUnit = Quantity["Kilograms"];   (* choose freely *)
	¤mathin¤labelŽmath:Eunit EnergyUnit = MassUnit*LengthUnit^2/TimeUnit^2 //UnitConvert;
\end{mathematica}
The consistency of this set of definitions is seen in \mm{\ref{math:Eunit}}, making the energy unit depend on the other units, which in turn can be chosen freely.
Many other combinations are possible, as long as this consistency remains.

Another popular choice is to additionally couple the time and energy units through Planck's constant $\hbar$, and make both dependent on the length and mass units (thus reducing the system of units to only two degrees of freedom):
\begin{mathematica}
	¤mathin¤labelŽmath:coupledunits LengthUnit = Quantity["Meters"];    (* choose freely *)
	¤mathin MassUnit = Quantity["Kilograms"];   (* choose freely *)
	¤mathin TimeUnit =
	¤mathnl   MassUnit*LengthUnit^2/Quantity["ReducedPlanckConstant"] //UnitConvert;
	¤mathin¤labelŽmath:Etcoupling EnergyUnit = Quantity["ReducedPlanckConstant"]/TimeUnit //UnitConvert;
\end{mathematica}
This latter set of units is what we will be using in what follows, without restriction of generality.
We express the reduced Planck constant in these units with
\begin{mathematica}
	¤mathin¤labelŽmath:defhbar ¤mmhbar = Quantity["ReducedPlanckConstant"]/(EnergyUnit*TimeUnit) //UnitConvert //N
	¤mathout 1.
\end{mathematica}
which is equal to unity because of our chosen coupling between energy and time units; in other unit systems the value will be different. Note the use of \mm{//N} at the end of \mm{\ref{math:defhbar}} to force the result to be a pure machine-precision number instead of a variable-precision number that tracks the accuracy of the involved physical quantities.

To set the physical size $a$ of the computational box, for example to $a=\SI{5}{\micro m}$, we execute
\begin{mathematica}
	¤mathin¤labelŽmath:defa a = Quantity[5, "Micrometers"]/LengthUnit //UnitConvert //N;
\end{mathematica}
and to set the particle's mass $m$, for example to the neutron's mass,
\begin{mathematica}
	¤mathin¤labelŽmath:defm m = Quantity["NeutronMass"]/MassUnit //UnitConvert //N;
\end{mathematica}
In the calculations that follow, we will not be explicit about the system of units and the physical quantities. Instead, we will use direct dimensionless definitions such as
\begin{mathematica}
	¤mathin a = 30;   (* calculation box size in units of length *)
	¤mathin m = 1;    (* particle mass in units of mass *)
	¤mathin ¤mmhbar = 1;    (* value of ¤mmhbar assuming ¤refŽmath:Etcoupling *)
\end{mathematica}
These are to be replaced by \mm{\ref{math:defa}}, \mm{\ref{math:defm}}, and \mm{\ref{math:defhbar}} in a more concrete physical situation.

\subsection{computational basis functions}


In order to perform quantum-mechanical calculations of a particle moving in one dimension, we need a basis set that is more practical than the Dirac basis used to define the relevant operators and states above. Indeed, Dirac states $\ket{x}$ are difficult to represent in a computer because they are uncountable, densely spaced, and highly singular.


The most generally useful basis sets for computations are the \emph{momentum basis} and the \emph{finite-resolution position basis}, which we will look at in turn, and which will be shown to be related to each other by a type-I discrete sine transform.

\subsubsection{momentum basis}\index{basis set!momentum basis}
\label{sec:momentumbasis1D}

The simplest one-dimensional quantum-mechanical system of the type of \autoref{eq:1DHamiltonian} is the infinite square well with $W(x)=0$. Its energy eigenstates $\phi_n(x)$ for $n=1,2,3,\dots$ satisfy the Schr\"odinger equation $-\frac{\hbar^2}{2m}\phi_n''(x)=E_n\phi_n(x)$ (see \ref{Q:Schroedinger}) and the boundary conditions $\phi_n(0)=\phi_n(a)=0$ necessitated by \autoref{eq:bottompotential}. Their explicit normalized forms are
\begin{equation}
	\label{eq:momentumbasis1D}
	\scp{x}{n} = \phi_n(x) = \sqrt{\frac{2}{a}}\sin\left(\frac{n\pi x}{a}\right)
\end{equation}
with eigen-energies
\begin{equation}
	\label{eq:squarewellenergies}
	E_n=\frac{n^2\pi^2\hbar^2}{2m a^2}. 
\end{equation}
We know from the Sturm--Liouville theorem\index{Sturm-Liouville theorem@Sturm--Liouville theorem}\footnote{See \url{https://en.wikipedia.org/wiki/Sturm-Liouville_theory}.} that these functions form a complete set (see \ref{Q:squarewellbasissum}); further, we can use Mathematica to show that they are ortho-normalized:
\begin{mathematica}
	¤mathin ¤mmphi[a_, n_, x_] = Sqrt[2/a]*Sin[n*¤mmpi*x/a];
	¤mathin Table[Integrate[¤mmphi[a,n1,x]*¤mmphi[a,n2,x], {x, 0, a}],
	¤mathnl   {n1, 10}, {n2, 10}] //MatrixForm
\end{mathematica}
They are eigenstates of the squared momentum operator $\op{p}^2=\left(-\ii\hbar\frac{\dd}{\dd[x]}\right)^2= -\hbar^2\frac{\dd^2}{\dd[x]^2}$:
\begin{equation}
	\label{eq:momentumop}
	\op{p}^2\ket{n} = \frac{n^2\pi^2\hbar^2}{a^2}\ket{n},
\end{equation}
which we verify with
\begin{mathematica}
	¤mathin -¤mmhbar^2*D[¤mmphi[a,n,x], {x,2}] == (n^2*¤mmpi^2*¤mmhbar^2)/a^2*¤mmphi[a,n,x]
	¤mathout True
\end{mathematica}
This makes the kinetic\index{operator!kinetic} operator $\op{T}=\op{p}^2/(2m)$ diagonal in this basis:
\begin{align}
	\label{eq:kineticelements}
	\me{n}{\op{T}}{n'}&=E_n\delta_{n n'}, &
	\op{T} &= \sum_{n=1}^{\infty} \ket{n}E_n\bra{n}.
\end{align}
However, in general the potential energy, and most other operators that will appear later, are difficult to express in this momentum basis.

The momentum basis of \autoref{eq:momentumbasis1D} contains a countably infinite number of basis functions, which is a great advantage over the uncountably infinite cardinality of the Dirac basis set. In practical calculations, we restrict the computational basis to $n\in\{1\dots n\ix{max}\}$, which means that we only consider physical phenomena with excitation energies below $E_{n\ix{max}}= \frac{\pi^2\hbar^2}{2m a^2}n\ix{max}^2$ (see \autoref{sec:incompletebasis}). 
Here is an example of what these position-basis functions look like for $n\ix{max}=10$:
\begin{center}
	\includegraphics[width=\textwidth]{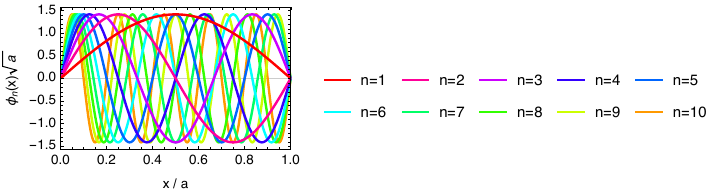}
\end{center}
Using the approximate completeness of the momentum basis, $\sum_{n=1}^{n\ix{max}}\ket{n}\bra{n}\approx\one$ (see \autoref{sec:incompletebasis}), the kinetic Hamiltonian thus becomes
\begin{equation}
	\label{eq:kineticoperator}
	\op{T} \approx \left[\sum_{n=1}^{n\ix{max}}\ket{n}\bra{n}\right] \op{T} \left[\sum_{n'=1}^{n\ix{max}}\ket{n'}\bra{n'}\right]
	= \sum_{n,n'=1}^{n\ix{max}}\ket{n}\me{n}{\op{T}}{n'}\bra{n'}
	= \sum_{n=1}^{n\ix{max}}\ket{n}E_n\bra{n}.
\end{equation}
We set up the kinetic Hamiltonian operator as a sparse diagonal matrix in the momentum basis with
\begin{mathematica}
	¤mathin nmax = 100;
	¤mathin¤labelŽmath:HkinM TM = SparseArray[Band[{1,1}] -> Range[nmax]^2*¤mmpi^2*¤mmhbar^2/(2*m*a^2)];
\end{mathematica}
where $n\ix{max}=100$ was chosen as an example.


\subsubsection{finite-resolution position basis}\index{basis set!finite-resolution position basis}
\label{sec:positionbasis1D}

Given an energy-limited momentum basis set $\{\ket{n}\}_{n=1}^{n\ix{max}}$ from above, we define a set of $n\ix{max}$ equally-spaced points
\begin{equation}
	x_j = j\cdot\Delta
\end{equation}
for $j\in\{1\dots n\ix{max}\}$, with spacing $\Delta=a/(n\ix{max}+1)$. These grid points fill the calculation range $x\in[0,a]$ uniformly without covering the end points.
We then define a new basis set as the closest possible representations of delta-functions at these points: for $j\in\{1\dots n\ix{max}\}$,
\begin{equation}
	\label{eq:positionkets}
	\ket{j} = \sqrt{\Delta} \sum_{n=1}^{n\ix{max}} \phi_n(x_j) \ket{n}.
\end{equation}
The spatial wavefunctions of these basis states are
\begin{equation}
	\label{eq:positionbasis1D}
	\scp{x}{j} = \vartheta_j(x) = \sqrt{\Delta} \sum_{n=1}^{n\ix{max}} \phi_n(x_j)\phi_n(x).
\end{equation}
Here is an example of what these position-basis functions look like for $n\ix{max}=10$:
\begin{center}
	\includegraphics[width=\textwidth]{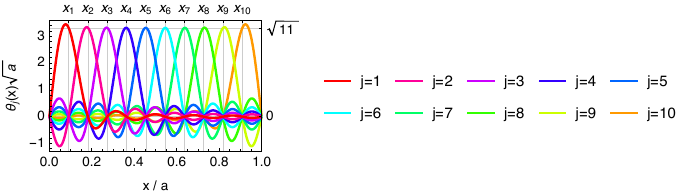}
\end{center}
This new basis set is also ortho-normal, $\scp{j}{j'} = \delta_{j j'}$, and it is strongly local in the sense that only the basis function $\vartheta_j(x)$ is nonzero at $x_j$, while all others vanish:
\begin{equation}
	\label{eq:positionbasislocality}
	\scp{x_{j'}}{j} = \vartheta_j(x_{j'}) = \delta_{j j'}/\sqrt{\Delta}.
\end{equation}
We define these basis functions in Mathematica with
\begin{mathematica}
	¤mathin nmax = 10;
	¤mathin ¤mmDelta = a/(nmax+1);
	¤mathin xx[j_] = j*¤mmDelta;
	¤mathin ¤mmtheta[j_, x_] = Sqrt[¤mmDelta]*Sum[¤mmphi[n,xx[j]]*¤mmphi[n,x], {n, nmax}];
\end{mathematica}
Since the basis function $\vartheta_j(x)=\mm{\mmtheta[j,x]}$ is the only one which is nonzero at $x_j=\mm{xx[j]}$, and it is close to zero everywhere else (exactly zero at the $x_{j'\neq j}$), we can usually make several approximations:
\begin{itemize}
	\item If a wavefunction is given as a vector $\vect{v}=\mm{v}$ in the position basis, $\ket{\psi} = \sum_{j=1}^{n\ix{max}} v_j\ket{j}$, then by \autoref{eq:positionbasislocality} the wavefunction is known at the grid points:
	\begin{equation}
		\psi(x_j) = \scp{x_j}{\psi} = \bra{x_j}\sum_{j'=1}^{n\ix{max}} v_{j'}\ket{j'} = \sum_{j'=1}^{n\ix{max}} v_{j'}\scp{x_j}{j'}
		= \sum_{j'=1}^{n\ix{max}} v_{j'}\delta_{j j'}/\sqrt{\Delta}=\frac{v_j}{\sqrt{\Delta}}.
	\end{equation}
	The density profile is thus given by the values of $\rho(x_j)=\abs{\psi(x_j)}^2=\abs{v_j}^2/\Delta$. This allows for very easy plotting of wavefunctions and  densities by linearly interpolating between these grid points (\ie, an interpolation of order 1):
\begin{mathematica}
	¤mathin¤labelŽmath:plotwf ListLinePlot[Transpose[{Table[xx[j], {j, nmax}], Abs[v]^2/¤mmDelta}]]
\end{mathematica}
	By the truncation of the basis at $n\ix{max}$, the wavefunction has no frequency components faster than one half-wave per grid-point spacing, and therefore we can be sure that this linear interpolation is a reasonably accurate representation of the wavefunction $\psi(x)$ and the density $\rho(x)=\abs{\scp{x}{\psi}}^2$, in particular as $n\ix{max}\to\infty$.
	\item An even simpler interpolation of order zero assumes that the wavefunction is constant over intervals $[-\Delta/2,\Delta/2]$ centered at each grid point. This primitive interpolation is used, for example, to calculate the Wigner quasi-probability distribution in \autoref{sec:Wigner}.
	\item Similarly, if a density operator is given by $\op{\rho}=\sum_{j,j'=1}^{n\ix{max}}R_{j,j'}\ket{j}\bra{j'}$, then the value of the density operator at a grid point $(x_j,x_{j'})$ is given by
		\begin{multline}
			\label{eq:densitymatrix}
			\rho(x_j,x_{j'}) = \me{x_j}{\op{\rho}}{x_{j'}} = \bra{x_j}\left[\sum_{j'',j''''=1}^{n\ix{max}}R_{j'',j'''}\ket{j''}\bra{j'''}\right]\ket{x_{j'}}
				= \sum_{j'',j''''=1}^{n\ix{max}}R_{j'',j'''}\scp{x_j}{j''}\scp{j'''}{x_{j'}}\\
				= \sum_{j'',j''''=1}^{n\ix{max}}R_{j'',j'''}\frac{\delta_{j,j''}}{\sqrt{\Delta}}\frac{\delta_{j',j'''}}{\sqrt{\Delta}}
				= \frac{R_{j,j'}}{\Delta}.
		\end{multline}
		That is, the coefficients $R_{j,j'}$ and the density values $\rho(x_j,x_{j'})$ are very closely related.
		The diagonal elements of this expression ($j=j'$) give the spatial density profile.
	\item For any function $f(x)$ that varies slowly (smoothly) over length scales of the grid spacing $\Delta$, we can make the approximation
		\begin{equation}
			\label{eq:fthetaapprox}
			f(x)\vartheta_j(x)\approx f(x_j)\vartheta_j(x).
		\end{equation}
		This approximation becomes exact on every grid point according to \autoref{eq:positionbasislocality}, and the assumed smoothness of $f(x)$ makes it a good estimate for any $x$.
\end{itemize}

\subsubsection{conversion between basis sets}

Within the approximation of a truncation at maximum energy $E_{n\ix{max}}$, we can express any wavefunction $\ket{\psi}$ in both basis sets of \autoref{eq:momentumbasis1D} and \autoref{eq:positionbasis1D}:
\begin{equation}
	\label{eq:posmombasis}
	\ket{\psi} = \sum_{n=1}^{n\ix{max}} u_n\ket{n} = \sum_{j=1}^{n\ix{max}} v_j\ket{j}
\end{equation}
Inserting the definition of \autoref{eq:positionkets} into \autoref{eq:posmombasis} we find
\begin{equation}
	\sum_{n=1}^{n\ix{max}} u_n\ket{n} = \sum_{j=1}^{n\ix{max}} v_j \left[ \sqrt{\Delta} \sum_{n'=1}^{n\ix{max}} \phi_{n'}(x_j) \ket{n'} \right]
	= \sum_{n'=1}^{n\ix{max}} \left[ \sqrt{\Delta} \sum_{j=1}^{n\ix{max}} v_j \phi_{n'}(x_j) \right] \ket{n'}
\end{equation}
and therefore, since the basis set $\{\ket{n}\}$ is ortho-normalized,
\begin{equation}
	\label{eq:uvconv}
	u_n
	= \sqrt{\Delta} \sum_{j=1}^{n\ix{max}} v_j \phi_n(x_j)
	= \sum_{j=1}^{n\ix{max}} X_{n j} v_j
\end{equation}
with the basis conversion coefficients
\begin{equation}
	\label{eq:Xnj}
	X_{n j} = \scp{n}{j} = \sqrt{\Delta} \phi_n(x_j)
	= \sqrt{\frac{a}{n\ix{max}+1}} \sqrt{\frac{2}{a}} \sin\left(\frac{n\pi x_j}{a}\right)
	= \sqrt{\frac{2}{n\ix{max}+1}} \sin\left(\frac{\pi n j}{n\ix{max}+1}\right).
\end{equation}
The inverse transformation is found from $\ket{n}=\sum_{j=1}^{n\ix{max}}\scp{j}{n}\ket{j}$ inserted into \autoref{eq:posmombasis}, giving
\begin{equation}
	\label{eq:vuconv}
	v_j = \sum_{n=1}^{n\ix{max}} X_{n j} u_n
\end{equation}
in terms of the same coefficients of \autoref{eq:Xnj}. Thus the transformations relating the vectors $\vect{u}$ (with components $u_n$) and $\vect{v}$ (with components $v_j$) are $\vect{v}=\matr{X}\cdot\vect{u}$ and $\vect{u}=\matr{X}\cdot\vect{v}$ in terms of the same symmetric orthogonal matrix $\matr{X}$ with coefficients $X_{n j}$.

We could calculate these coefficients with
\begin{mathematica}
	¤mathin X = Table[Sqrt[2/(nmax+1)]*Sin[¤mmpi*n*j/(nmax+1)], {n, nmax}, {j, nmax}] //N;
\end{mathematica}
but this is not very efficient, especially for large $n\ix{max}$.

It turns out that \autoref{eq:uvconv} and \autoref{eq:vuconv} relate the vectors $\vect{u}$ and $\vect{v}$ by a \emph{type-I discrete sine transform} (DST-I),\index{discrete sine transform} which Mathematica can evaluate very efficiently via a fast Fourier transform.\index{fast Fourier transform}\footnote{See \url{https://en.wikipedia.org/wiki/Discrete_sine_transform} and \url{https://en.wikipedia.org/wiki/Fast_Fourier_transform}. The precise meaning of the DST-I can be seen from its equivalent definition through a standard discrete Fourier transform of doubled length: for a complex vector \mm{v}, we can substitute \mm{FourierDST[v,1]} by \mm{DST1[v\_?VectorQ]:=-I*Fourier[Join[\{0\},v,\{0\},Reverse[-v]]][[2;;Length[v]+1]]}. In this sense it is the discrete Fourier transform of a list \mm{v} augmented with (i) zero boundary conditions and (ii) reflection anti-symmetry at the boundaries. Remember that the \mm{Fourier[]} transform assumes periodic boundary conditions, which are incorrect in the present setup, and need to be modified into a DST-I.}
Since the DST-I is its own inverse, we can use
\begin{mathematica}
	¤mathin¤labelŽmath:DSTuv v = FourierDST[u, 1];
	¤mathin¤labelŽmath:DSTvu u = FourierDST[v, 1];
\end{mathematica}
to effect such conversions. We will see a very useful application of this transformation when we study the time-dependent behavior of a particle in a potential (``split-step method'', \autoref{sec:dynamics1D}).\index{split-step method}

The matrix $\matr{X}$
is also useful for converting \emph{operator} representations between the basis sets: the momentum representation $\matr{U}$ and the position representation $\matr{V}$ of the same operator satisfy $\matr{V}=\matr{X}\cdot\matr{U}\cdot\matr{X}$ and $\matr{U}=\matr{X}\cdot\matr{V}\cdot\matr{X}$.
In practice, as above we can convert operators between the position and momentum representation with a two-dimensional type-I discrete sine transform:
\begin{mathematica}
	¤mathin V = FourierDST[U, 1];
	¤mathin U = FourierDST[V, 1];
\end{mathematica}
This easy conversion is very useful for the construction of the matrix representations of Hamiltonian operators, since the kinetic energy is diagonal in the momentum basis, \autoref{eq:momentumop}, while the potential energy operator is approximately diagonal in the position basis, \autoref{eq:Vdiagonal}.

\subsection{the position operator}\index{operator!position}
\label{sec:positionoperator}

The position operator $\op{x}=\int_{-\infty}^{\infty}\dd[x]\ket{x}x\bra{x}$ is one of the basic operators that is used frequently to construct Hamiltonians of moving particles. The exact expressions for the matrix elements of this operator in the momentum basis are
\begin{equation}
	\label{eq:xopexact}
	\me{n}{\op{x}}{n'} = \int_0^a\dd[x]\sqrt{\frac{2}{a}}\sin\left(\frac{n\pi x}{a}\right)x\sqrt{\frac{2}{a}}\sin\left(\frac{n'\pi x}{a}\right)
	= \begin{cases}
		\frac{a}{2} & \text{if $n=n'$}\\
		-\frac{8a n n'}{\pi^2(n^2-n'^2)^2} & \text{if $n-n'$ is odd}\\
		0 & \text{otherwise}
	\end{cases}
\end{equation}
This allows us to construct the exact matrix representations of the operator $\op{x}$ in both the momentum (\mm{xM}) and the position (\mm{xP}) bases:
\begin{mathematica}
	¤mathin¤labelŽmath:xMop xM = SparseArray[{
	¤mathnl        Band[{1,1}] -> a/2,
	¤mathnl        {n1_,n2_} /; OddQ[n1-n2] -> -8*a*n1*n2/(¤mmpi^2*(n1^2-n2^2)^2)},
	¤mathnl      {nmax,nmax}];
	¤mathin¤labelŽmath:xPop xP = FourierDST[xM, 1];
\end{mathematica}
A simple approximation of the position operator, which will be extremely useful in what follows, is found by observing that \mm{xP} is almost a diagonal matrix, with approximately the grid coordinates $x_1, x_2, \dotsc, x_{n\ix{max}}$ on the diagonal. This approximate form can be proved by using the locality of the position basis functions, \autoref{eq:fthetaapprox}:
	\begin{multline}
		x_{j j'} = \me{j}{\op{x}}{j'} = \me{j}{\left[\int_{-\infty}^{\infty}\dd[x] \ket{x}x\bra{x}\right]}{j'}
		= \int_0^a\dd[x] \scp{j}{x}x\scp{x}{j'}
		= \int_0^a\dd[x] \vartheta_j^*(x) x\vartheta_{j'}(x)\\
		\approx x_{j'} \int_0^a\dd[x] \vartheta_j^*(x) \vartheta_{j'}(x)
		= \delta_{j j'} x_j.
	\end{multline}
The resulting approximate diagonal form of the position operator in the position basis, found from the approximate completeness relation $\sum_{j=1}^{n\ix{max}}\ket{j}\bra{j}\approx\one$, is
	\begin{equation}
		\label{eq:positionoperator}
		\op{x}
		\approx \left[\sum_{j=1}^{n\ix{max}}\ket{j}\bra{j}\right]\op{x}\left[\sum_{j'=1}^{n\ix{max}}\ket{j'}\bra{j'}\right]
		= \sum_{j,j'=1}^{n\ix{max}} \ket{j} \me{j}{\op{x}}{j'}\bra{j'}
		\approx \sum_{j,j'=1}^{n\ix{max}} \ket{j} \delta_{j j'}x_j \bra{j'}
		= \sum_{j=1}^{n\ix{max}} \ket{j}x_j\bra{j}.
	\end{equation}
\begin{mathematica}
	¤mathin ¤mmDelta = a/(nmax+1);                          (* the grid spacing *)
	¤mathin¤labelŽmath:xopA xgrid = Range[nmax]*¤mmDelta;                   (* the computational grid *)
	¤mathin¤labelŽmath:xopB xP = SparseArray[Band[{1,1}] -> xgrid];  (* x operator, position basis *)
	¤mathin¤labelŽmath:xopC xM = FourierDST[xP, 1];                  (* x operator, momentum basis *)
\end{mathematica}

\subsection{the potential-energy operator}\index{operator!potential}
\label{sec:potentialenergy}

If a potential energy function $W(x)$ varies smoothly over length scales of the grid spacing $\Delta$, then the trick of \autoref{sec:positionoperator} allows us to approximate the matrix elements of this potential energy in the position basis,
	\begin{multline}
		\label{eq:Vdiagonal}
		V_{j j'} = \me{j}{\op{V}}{j'} = \me{j}{\left[\int_0^a\dd[x] \ket{x}W(x)\bra{x}\right]}{j'}
		= \int_0^a\dd[x] \scp{j}{x}W(x)\scp{x}{j'}
		= \int_0^a\dd[x] \vartheta_j^*(x) W(x) \vartheta_{j'}(x)\\
		\approx W(x_{j'}) \int_0^a\dd[x] \vartheta_j^*(x) \vartheta_{j'}(x)
		= \delta_{j j'} W(x_j),
	\end{multline}
	where we have used the definitions of \autoref{eq:potop} and \autoref{eq:bottompotential}.
	This is a massive simplification compared to the explicit evaluation of potential integrals for each specific potential energy function. The potential-energy operator thus becomes approximately
	\begin{equation}
		\label{eq:potentialoperator}
		\op{V}
		\approx \left[\sum_{j=1}^{n\ix{max}}\ket{j}\bra{j}\right]\op{V}\left[\sum_{j'=1}^{n\ix{max}}\ket{j'}\bra{j'}\right]
		= \sum_{j,j'=1}^{n\ix{max}} \ket{j} \me{j}{\op{V}}{j'} \bra{j'}
		\approx \sum_{j,j'=1}^{n\ix{max}} \ket{j} \delta_{j j'}W(x_j) \bra{j'}
		= \sum_{j=1}^{n\ix{max}} \ket{j}W(x_j)\bra{j}.
	\end{equation}
\begin{mathematica}
	¤mathin¤labelŽmath:VP Wgrid = Map[W, xgrid];           (* the potential on the computational grid *)
	¤mathin VP = SparseArray[Band[{1,1}]->Wgrid]; (* potential operator, position basis *)
	¤mathin VM = FourierDST[VP, 1];               (* potential operator, momentum basis *)
\end{mathematica}

\subsection{the kinetic-energy operator}\index{operator!kinetic}
\label{sec:kineticop}

The representation of the kinetic energy operator can be calculated very accurately with the description given above. We transform the definition of \mm{\ref{math:HkinM}} to the finite-resolution position basis with
\begin{mathematica}
	¤mathin¤labelŽmath:HkinP TP = FourierDST[TM, 1];   (* kinetic operator, position basis *)
\end{mathematica}
For large $n\ix{max}$ and small excitation energies the exact kinetic-energy operator can be replaced by the position-basis form
\begin{equation}
	\label{eq:Tapprox}
	\me{j}{\op{T}}{j'} \approx \frac{\hbar^2}{2m \Delta^2} \times
	\begin{cases}
		2 & \text{if $j=j'$,}\\
		-1 & \text{if $\abs{j-j'}=1$,}\\
		0 & \text{if $\abs{j-j'}\ge2$,}
	\end{cases}
\end{equation}
which corresponds to replacing the second derivative in the kinetic operator by the finite-differences expression $\psi''(x)\approx-\left[2\psi(x)-\psi(x-\Delta)-\psi(x+\Delta)\right]/\Delta^2$. While \autoref{eq:Tapprox} looks simple, it is ill suited for the calculations that will follow because (i) any matrix exponentials involving $\op{T}$ will be difficult to calculate, and (ii) it is not very accurate (higher-order finite-differences expressions\footnote{See \url{https://en.wikipedia.org/wiki/Finite_difference_coefficient} for explicit forms of higher-order finite-differences expressions that can be used to approximate derivatives.} are not much better). Thus we will not be using such approximations in what follows, and prefer the more useful and more accurate definition through \mm{\ref{math:HkinM}} and \mm{\ref{math:HkinP}}.

\subsection{the momentum operator}\index{operator!momentum}
\label{sec:momentumoperator}

The discussion has so far been conducted in terms of the kinetic-energy operator $\op{T}=\op{p}^2/(2m)$ without explicitly talking about the momentum operator $\op{p}=-\ii\hbar\frac{\dd}{\dd[x]}$. This was done because the matrix representation of the momentum operator is problematic.
A direct calculation of the matrix elements in the momentum basis yields
\begin{equation}
	\label{eq:poperator}
	\me{n}{\op{p}}{n'}
	= -\ii\hbar\int_0^a \dd[x] \phi_n(x) \frac{\dd\phi_{n'}(x)}{\dd[x]}
	= \frac{\hbar}{a}\times\begin{cases}
		\frac{4\ii n n'}{n'^2-n^2} & \text{if $n-n'$ is odd,}\\
		0 & \text{if $n-n'$ is even.}
	\end{cases}
\end{equation}
In Mathematica, this is implemented with the definition
\begin{mathematica}
	¤mathin¤labelŽmath:poperator pM = SparseArray[{n1_,n2_}/;OddQ[n1-n2]->(4*I*¤mmhbar*n1*n2)/(a*(n2^2-n1^2)),
	¤mathnl                  {nmax,nmax}];   (* momentum operator, momentum basis *)
	¤mathin pP = FourierDST[pM, 1];          (* momentum operator, position basis *)
\end{mathematica}
This result is, however, unsatisfactory, since (i) it generates a matrix that is not sparse, and (ii) for a finite basis size $n\le n\ix{max}<\infty$ it does not exactly generate the kinetic-energy operator $\op{T}=\op{p}^2/(2m)$ (see \ref{Q:momentumoperator}).
We will avoid using the momentum operator whenever possible, and use the kinetic-energy operator $\op{T}$ instead (see above). An example of the direct use of $\op{p}$ is given in \autoref{sec:Rashba}.

For large $n\ix{max}$ and small excitation energies the exact momentum operator can be replaced by the position-basis form
\begin{equation}
	\label{eq:papprox}
	\me{j}{\op{p}}{j'} \approx \frac{\ii\hbar}{2\Delta} \times
	\begin{cases}
		-1 & \text{if $j-j'=-1$,}\\
		+1 & \text{if $j-j'=+1$,}\\
		0 & \text{if $\abs{j-j'}\neq1$,}
	\end{cases}
\end{equation}
which corresponds to replacing the first derivative in the momentum operator by the finite-differences expression $\psi'(x)\approx\left[\psi(x+\Delta)-\psi(x-\Delta)\right]/(2\Delta)$. While \autoref{eq:papprox} looks simple, it is ill suited for the calculations that will follow because any matrix exponentials involving $\op{p}$ will still be difficult to calculate; further, the same finite-differences caveats as in \autoref{sec:kineticop} apply. Thus we will not be using such approximations in what follows, and prefer the more accurate definition through \mm{\ref{math:poperator}}.

\subsubsection{exercises}

\begin{questions}
	\item\label{Q:momentumoperator} Using $n\ix{max}=100$, calculate the matrix representations of the kinetic-energy operator $\op{T}$ and the momentum operator $\op{p}$ in the momentum basis. Compare the spectra of $\op{T}$ and $\op{p}^2/(2m)$ and notice the glaring differences, even at low energies. \emph{Hint:} use natural units such that \mm{a=m=\mmhbar=1} for simplicity.
\pagenote[\ref{Q:momentumoperator}]{
\begin{mathematica}
	¤protect¤mathin¤ a¤ =¤ m¤ =¤ ¤mmhbar¤ =¤ 1;¤ ¤ ¤ (*¤ natural¤ units¤ *)
	¤protect¤mathin¤ nmax¤ =¤ 100;
	¤protect¤mathin¤ TM¤ =¤ SparseArray[Band[{1,1}]¤ ->¤ Range[nmax]^2*¤mmpi^2*¤mmhbar^2/(2*m*a^2)];
	¤protect¤mathin¤ pM¤ =¤ SparseArray[{n1_,n2_}/;OddQ[n1-n2]->(4*I*¤mmhbar*n1*n2)/(a*(n2^2-n1^2)),
	¤protect¤mathnl¤ ¤ ¤ ¤ ¤ ¤ ¤ ¤ ¤ ¤ ¤ ¤ ¤ ¤ ¤ ¤ ¤ ¤ {nmax,nmax}];
	¤protect¤mathin¤ TM¤ //N¤ //Eigenvalues¤ //Sort
	¤protect¤mathout¤ {4.9348,¤ 19.7392,¤ 44.4132,¤ 78.9568,¤ 123.37,¤ 177.653,¤ ...,¤ 48366.,¤ 49348.}
	¤protect¤mathin¤ pM.pM/(2m)¤ //N¤ //Eigenvalues¤ //Sort
	¤protect¤mathout¤ {4.8183,¤ 4.8183,¤ 43.3646,¤ 43.3646,¤ 120.457,¤ 120.457,¤ ...,¤ 47257.4,¤ 47257.4}
\end{mathematica}
	The eigenvalues of $\op{T}$ are quadratically spaced, whereas those of $\op{p}^2/(2m)$ come in degenerate pairs (one involving only states of even $n$ and one only states of odd $n$) and thus never converge to the eigenvalues of $\op{T}$, even in the limit $n\ix{max}\to\infty$.}
	\item\label{Q:xpcommutator} Using $n\ix{max}=20$, calculate the matrix representations of the position operator $\op{x}$ and the momentum operator $\op{p}$ in the momentum basis. To what extent is the commutation relation $[\op{x},\op{p}]=\ii\hbar$ satisfied? \emph{Hint:} use natural units such that \mm{a=m=\mmhbar=1} for simplicity.
\pagenote[\ref{Q:xpcommutator}]{We use the more accurate form of the position operator from \mm{\ref{math:xMop}}:
\begin{mathematica}
	¤protect¤mathin¤ a¤ =¤ m¤ =¤ ¤mmhbar¤ =¤ 1;¤ ¤ ¤ (*¤ natural¤ units¤ *)
	¤protect¤mathin¤ nmax¤ =¤ 20;
	¤protect¤mathin¤ xM¤ =¤ SparseArray[{
	¤protect¤mathnl¤ ¤ ¤ ¤ ¤ ¤ ¤ ¤ Band[{1,1}]¤ ->¤ a/2,
	¤protect¤mathnl¤ ¤ ¤ ¤ ¤ ¤ ¤ ¤ {n1_,n2_}¤ /;¤ OddQ[n1-n2]¤ ->¤ -8*a*n1*n2/(¤mmpi^2*(n1^2-n2^2)^2)},
	¤protect¤mathnl¤ ¤ ¤ ¤ ¤ ¤ {nmax,nmax}];
	¤protect¤mathin¤ pM¤ =¤ SparseArray[{n1_,n2_}/;OddQ[n1-n2]->4*I*¤mmhbar*n1*n2/(a*(n2^2-n1^2)),
	¤protect¤mathnl¤ ¤ ¤ ¤ ¤ ¤ {nmax,nmax}];
	¤protect¤mathin¤ coM¤ =¤ xM.pM¤ -¤ pM.xM;¤ ¤ ¤ (*¤ commutator¤ [x,p]¤ in¤ the¤ momentum¤ basis¤ *)
	¤protect¤mathin¤ coM/¤mmhbar¤ //N¤ //MatrixForm
\end{mathematica}
	In the upper-left corner (low values of $n$) the result looks like the unit matrix multiplied by the imaginary unit $\ii$; but towards the lower-right corner (large values of $n$) it deviates dramatically from the correct expression. This is to be expected from the problematic nature of the momentum operator; see \autoref{sec:momentumoperator}.}
\end{questions}

\subsection[example: gravity well]{\label{sec:gravitywell}example: gravity well\hspace{\stretch{1}}\attachcode{GravityWell}{gravity well}}\index{gravity well}

As an example of a single particle moving in one spatial dimension, we study the gravity well. This problem can be solved analytically, which helps us to determine the accuracy of our numerical methods.

We assume that a particle of mass $m$ is free to move in the vertical direction $x$, where $x=0$ is the earth's surface and $x>0$ is up; the particle is forbidden from travelling below the earth's surface (\ie, it is restricted to $x>0$ at all times). There is no dissipation or friction. The Hamiltonian of the particle's motion is
\begin{equation}
	\label{eq:bounceHamFull}
	\Ham = -\frac{\hbar^2}{2m}\frac{\dd^2}{\dd[x]^2} + m g \op{x}.
\end{equation}
The wavefunction $\psi(x)$ of this particle must satisfy the boundary condition $\psi(x)=0$ $\forall x\le0$.

In what follows we use the length unit $\lunit=\big(\frac{\hbar^2}{m^2 g}\big)^{1/3}$, which is proportional to the size of the ground state of \autoref{eq:bounceHamFull}, as well as the mass unit $\munit=m$. Following \autoref{sec:1Dunits} we then define the time unit $\tunit=\munit\lunit^2/\hbar=\big(\frac{\hbar}{m g^2}\big)^{1/3}$ and the energy unit $\Eunit=\hbar/\tunit=(m g^2 \hbar^2)^{1/3}$. These natural units lead to simple expressions for the mass: $\mm{m}=\frac{m}{\munit}=1$, $\mm{\mmhbar}=\frac{\hbar}{\Eunit\tunit}=1$, and $\mm{g}=\frac{g}{\lunit/\tunit^2}=1$. As a result, we can set up the \hyperref[eq:bounceHamFull]{Hamiltonian~\ref*{eq:bounceHamFull}} without fixing the particle's mass and gravitational acceleration explicitly:
\begin{mathematica}
	¤mathin m = ¤mmhbar = g = 1;
\end{mathematica}
Other systems of units can be used in the same way by using the tools of \autoref{sec:1Dunits}: first define a consistent set of units, and then express the physical quantities in terms of these units. The gravitational acceleration\index{gravitational acceleration} in particular would be set with
\begin{mathematica}
	¤mathin g = Quantity["StandardAccelerationOfGravity"]/(LengthUnit/TimeUnit^2);
\end{mathematica}

\subsubsection{analytic quantum energy eigenstates}

The exact normalized eigenstates and associated energy eigenvalues of \autoref{eq:bounceHamFull} are
\begin{align}
	\label{eq:GravityWellExact}
	\psi_k(x) &= \begin{cases}
		\left(\frac{2m^2g}{\hbar^2}\right)^{1/6} \cdot
		\frac{\Ai\left[\alpha_k+x\cdot\left(\frac{2m^2g}{\hbar^2}\right)^{1/3}\right]}{\Ai'(\alpha_k)} & \text{if $x>0$}\\
		0 & \text{if $x\le0$}
	\end{cases} &
	E_k &= -\alpha_k\cdot\left(\frac{m g^2\hbar^2}{2}\right)^{1/3}
\end{align}
for $k=1,2,3,\dotsc$,
where $\Ai(z)=\mm{AiryAi[z]}$ is the Airy function,\index{Airy function} $\Ai'(z)$ its first derivative, and $\alpha_k=\mm{AiryAiZero[k]}$ its zeros: $\alpha_1\approx-2.33811$, $\alpha_2\approx-4.08795$, $\alpha_3\approx-5.52056$, etc.

For comparison to numerical calculations below, we define the exact eigenstates and eigen-energies with
\begin{mathematica}
	¤mathin¤labelŽmath:AiryStates ¤mmpsi[k_,x_] = (2*m^2*g/¤mmhbar^2)^(1/6)*AiryAi[AiryAiZero[k]+x*(2*m^2*g/¤mmhbar^2)^(1/3)]/
	¤mathnl               AiryAi¤textquotesingle[AiryAiZero[k]];
	¤mathin ¤mmepsilon[k_] = -AiryAiZero[k]*(m*g^2*¤mmhbar^2/2)^(1/3);
\end{mathematica}
The ground-state energy is, in our chosen energy unit $\Eunit$,
\begin{mathematica}
	¤mathin N[¤mmepsilon[1]]
	¤mathout¤labelŽmath:gravitywellgse 1.85576
\end{mathematica}
The lowest three energy eigenstates look thus:
\begin{center}
\includegraphics[width=0.5\textwidth]{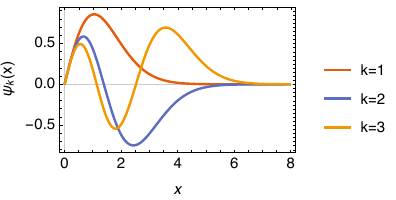}
\label{fig:AiryStates}
\end{center}

\subsubsection{numerical solution (I): momentum basis}

Our first numerical attempt to find the ground state of the gravity well relies on the momentum basis of states $\ket{n}$. For this approach we treat the Hamiltonian of the problem in the same way as discussed in \autoref{chap:basis} and \autoref{chap:spin}: we express each term of the Hamiltonian as a matrix in a fixed basis set.

Since our calculation will take place in a finite box $x\in[0,a]$, we must choose the box size $a$ large enough to contain most of the ground-state probability if we want to calculate it accurately. For the present calculation we choose $a=10\lunit$, which is sufficient (see figure above) since we picked the length unit $\lunit$ similar to the ground-state size:
\begin{mathematica}
	¤mathin a = 10;
\end{mathematica}
We only use a small number of basis functions here, to illustrate the method:
\begin{mathematica}[firstnumber=last]
	¤mathin nmax = 12;
\end{mathematica}
The matrix elements of the kinetic energy are set up following \mm{\ref{math:HkinM}}:
\begin{mathematica}[firstnumber=last]
	¤mathin TM = SparseArray[Band[{1,1}] -> Range[nmax]^2*¤mmpi^2*¤mmhbar^2/(2*m*a^2)];
\end{mathematica}
The matrix elements of the potential energy of \autoref{eq:bounceHamFull} are, from \mm{\ref{math:xMop}},
\begin{mathematica}[firstnumber=last]
	¤labelŽmathline:xM1L¤mathin¤labelŽmath:xM1 xM = SparseArray[{
	¤mathnl        Band[{1,1}] -> a/2,
	¤mathnl        {n1_,n2_} /; OddQ[n1-n2] -> -8*a*n1*n2/(¤mmpi^2*(n1^2-n2^2)^2)},
	¤mathnl      {nmax,nmax}];
	¤mathin VM = m*g*xM;
\end{mathematica}
The full Hamiltonian in the momentum representation is therefore
\begin{mathematica}[firstnumber=last]
	¤mathin HM = TM + VM;
\end{mathematica}
and the ground-state energy and wavefunction coefficients in the momentum representation
\begin{mathematica}[firstnumber=last]
	¤mathin gsM = -Eigensystem[-N[HM], 1,
	¤mathnl   Method -> {"Arnoldi", "Criteria" -> "RealPart", MaxIterations -> 10^6}];
\end{mathematica}
The ground state energy is
\begin{mathematica}[firstnumber=last]
	¤mathin gsM[[1, 1]]
	¤mathout 1.85608
\end{mathematica}
very close to the exact result of \mm{\ref{math:gravitywellgse}}.

The ground state wavefunction is defined as a sum over basis functions,
\begin{mathematica}[firstnumber=last]
	¤mathin ¤mmphi[n_, x_] = Sqrt[2/a]*Sin[n*¤mmpi*x/a];
	¤mathin¤labelŽmath:psi0 ¤mmpsi0[x_] = gsM[[2,1]] . Table[¤mmphi[n, x], {n, nmax}];
\end{mathematica}
We can calculate the overlap of this numerical ground state with the exact one given in \mm{\ref{math:AiryStates}}, $\abs{\scp{\psi_0}{\psi_1}}^2$:
\begin{mathematica}[firstnumber=last]
	¤mathin¤labelŽmath:overlapI Abs[NIntegrate[¤mmpsi0[x]*¤mmpsi[1,x], {x, 0, a}]]^2
	¤mathout 0.999965
\end{mathematica}
Even for $n\ix{max}=12$ this overlap is already very close to unity in magnitude.
It quickly approaches unity as $n\ix{max}$ increases, with the mismatch decreasing as $n\ix{max}^{-9}$ for this specific system.
The numerically calculated ground-state energy approaches the exact result from above, with the mismatch decreasing as $n\ix{max}^{-7}$ for this specific system.
These convergence properties, discussed in \autoref{sec:incompletebasis}, are very general and allow us to extrapolate many quantities to $n\ix{max}\to\infty$ by polynomial fits of numerically calculated quantities as functions of $n\ix{max}$.

\subsubsection{numerical solution (II): mixed basis}
\label{sec:mixedbasis}

The numerical method outlined above only works because we have an analytic expression for the matrix elements of the potential operator $\op{V}=m g\op{x}$, given in \autoref{eq:xopexact}. For a more general potential, the method of \autoref{eq:potentialoperator} is more useful, albeit less accurate. Here we re-do the numerical ground-state calculation in the position basis. The computation is set up in the same way as above,
\begin{mathematica}
	¤mathin a = 10;
	¤mathin nmax = 12;
	¤mathin ¤mmDelta = a/(nmax+1);        (* grid spacing *)
	¤mathin xgrid = Range[nmax]*¤mmDelta; (* the computational grid *)
\end{mathematica}
The matrix elements of the kinetic-energy operator in the position basis are calculated with a discrete sine transform,
\begin{mathematica}[firstnumber=last]
	¤mathin TM = SparseArray[Band[{1,1}] -> Range[nmax]^2*¤mmpi^2*¤mmhbar^2/(2*m*a^2)];
	¤mathin TP = FourierDST[TM, 1];
\end{mathematica}
The matrix elements of the potential energy of in \autoref{eq:bounceHamFull} are, from \mm{\ref{math:VP}},
\begin{mathematica}[firstnumber=last]
	¤mathin¤labelŽmath:potential W[x_] = m*g*x;        (* the potential function *)
	¤mathin Wgrid = Map[W, xgrid];(* the potential on the computational grid *)
	¤mathin VP = SparseArray[Band[{1,1}] -> Wgrid];
\end{mathematica}
The full Hamiltonian in the position representation is therefore
\begin{mathematica}[firstnumber=last]
	¤mathin HP = TP + VP;
\end{mathematica}
and the ground-state energy and wavefunction coefficients in the position representation
\begin{mathematica}[firstnumber=last]
	¤mathin gsP = -Eigensystem[-N[HP], 1,
	¤mathnl   Method -> {"Arnoldi", "Criteria" -> "RealPart", MaxIterations -> 10^6}];
\end{mathematica}
The ground state energy is now less close to the exact value than before, due to the additional approximation of \autoref{eq:potentialoperator}:
\begin{mathematica}[firstnumber=last]
	¤mathin gsP[[1, 1]]
	¤mathout 1.86372
\end{mathematica}
We therefore need a larger $n\ix{max}$ to achieve the same accuracy as in the first numerical calculation. The great advantage of the present calculation is, however, that it is easily generalized to arbitrary potential-energy functions in \mm{\ref{math:potential}}.

As shown in \mm{\ref{math:plotwf}}, the wavefunction can be plotted approximately with
\begin{mathematica}[firstnumber=last]
	¤mathin¤labelŽmath:gwgsc ¤mmgamma = Join[{{0,0}}, Transpose[{xgrid, gsP[[2,1]]/Sqrt[¤mmDelta]}], {{a,0}}];
	¤mathin ListLinePlot[¤mmgamma]
\end{mathematica}
where we have ``manually'' added the known boundary values $\gamma(0)=\gamma(a)=0$ to the list of numerically calculated wave-function values.
\begin{center}
\label{fig:GravityWellGsP}
\includegraphics[width=0.5\textwidth]{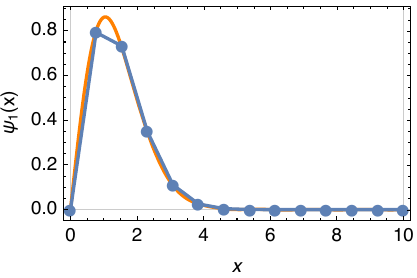}
\end{center}
You can see that even with $n\ix{max}=12$ grid points this ground-state wavefunction (blue lines interpolating between blue calculated points) looks remarkably close to the exact one (orange line, see plot on \autopageref{fig:AiryStates}).

If we need to go beyond linear interpolation, the precise wavefunction is calculated by converting to the momentum representation as in \mm{\ref{math:DSTvu}} and multiplying with the basis functions as in \mm{\ref{math:psi0}}:
\begin{mathematica}[firstnumber=last]
	¤mathin ¤mmphi[n_, x_] = Sqrt[2/a]*Sin[n*¤mmpi*x/a];
	¤mathin¤labelŽmath:gravitywellN2 ¤mmpsi0[x_] = FourierDST[gs[[2,1]],1] . Table[¤mmphi[n, x], {n, nmax}];
\end{mathematica}

\subsubsection{exercises}

\begin{questions}
	\item\label{Q:steppotN} What is the probability to find the particle below $\mm{x}=1$ (\ie, below $x=\lunit$) when it is in the ground state of the gravity well, \autoref{eq:bounceHamFull}? Calculate analytically, with numerical method I, and with numerical method II.
\pagenote[\ref{Q:steppotN}]{The exact probability is about 37.1\%:
\begin{mathematica}
	¤protect¤mathin¤ Integrate[¤mmpsi[1,x]^2,¤ {x,¤ 0,¤ 1}]¤ //N
	¤protect¤mathout¤ 0.37087
\end{mathematica}
	Using \mm{\ref{math:psi0}}, the first numerical method gives a good approximation of 37.0\%:
\begin{mathematica}
	¤protect¤mathin¤ Integrate[¤mmpsi0[x]^2,¤ {x,¤ 0,¤ 1}]
	¤protect¤mathout¤ 0.369801
\end{mathematica}
	Using \mm{\ref{math:gravitywellN2}}, the second numerical method gives an approximation of 36.2\%:
\begin{mathematica}
	¤protect¤mathin¤ Integrate[¤mmpsi0[x]^2,¤ {x,¤ 0,¤ 1}]
	¤protect¤mathout¤ 0.362126
\end{mathematica}
	Alternatively, we set up an interpolating function from the data of \mm{\ref{math:gwgsc}}, and integrate it numerically. The result depends on the interpolation order: 	higher-order interpolations tend to yield more accurate results.
\begin{mathematica}
	¤protect¤mathin¤ ¤mmpsi0i1¤ =¤ Interpolation[¤mmgamma,¤ InterpolationOrder¤ ->¤ 1];
	¤protect¤mathin¤ NIntegrate[¤mmpsi0i1[x]^2,¤ {x,¤ 0,¤ 1}]
	¤protect¤mathout¤ 0.302899
	¤protect¤mathin¤ ¤mmpsi0i2¤ =¤ Interpolation[¤mmgamma,¤ InterpolationOrder¤ ->¤ 2];
	¤protect¤mathin¤ NIntegrate[¤mmpsi0i2[x]^2,¤ {x,¤ 0,¤ 1}]
	¤protect¤mathout¤ 0.358003
	¤protect¤mathin¤ ¤mmpsi0i3¤ =¤ Interpolation[¤mmgamma,¤ InterpolationOrder¤ ->¤ 3];
	¤protect¤mathin¤ NIntegrate[¤mmpsi0i3[x]^2,¤ {x,¤ 0,¤ 1}]
	¤protect¤mathout¤ 0.3812
\end{mathematica}}
	\item\label{Q:neutronbounce} Calculate the mean height $\avg{\op{x}}$ in the ground state of the gravity well. How large is this quantity for a neutron in earth's gravitational field? \emph{Hint:} see \emph{Quantum states of neutrons in the Earth's gravitational field} by Valery V.\ Nesvizhevsky \etal, \href{https://doi.org/10.1038/415297a}{Nature 415, pages 297--299} (2002).
\pagenote[\ref{Q:neutronbounce}]{From \autoref{eq:GravityWellExact} the average height in state $k$ is
\begin{equation}
	\me{k}{\op{x}}{k} = \int_0^{\infty}\dd[x]\abs{\psi_k(x)}^2x
	= -\alpha_k\cdot\left(\frac{4\hbar^2}{27m^2g}\right)^{1/3},
\end{equation}
	which you can verify with \mm{\ref{math:AiryStates}} and
\begin{mathematica}
	¤protect¤mathin¤ Assuming[m>0&&g>0&&¤mmhbar>0,¤ Table[Integrate[¤mmpsi[k,x]^2*x,¤ {x,¤ 0,¤ ¤mminfty}],¤ {k,¤ 1,¤ 5}]]
\end{mathematica}
	For a neutron in earth's gravitational field this gives an average height of about \SI{9}{\micro m}:
\begin{mathematica}
	¤protect¤mathin¤ With[{k¤ =¤ 1,
	¤protect¤mathnl¤ ¤ ¤ ¤ ¤ ¤ ¤ m¤ =¤ Quantity["NeutronMass"],
	¤protect¤mathnl¤ ¤ ¤ ¤ ¤ ¤ ¤ g¤ =¤ Quantity["StandardAccelerationOfGravity"],
	¤protect¤mathnl¤ ¤ ¤ ¤ ¤ ¤ ¤ ¤mmhbar¤ =¤ Quantity["ReducedPlanckConstant"]},
	¤protect¤mathnl¤ ¤ ¤ UnitConvert[-AiryAiZero[k]*(4*¤mmhbar^2/(27*m^2*g))^(1/3),¤ "Micrometers"]]
	¤protect¤mathout¤ 9.147654¤ ¤mmmuŽm
\end{mathematica}}
	\item\label{Q:harmonicwellN} Calculate the energy levels and energy eigenstates of a particle in a harmonic potential, described by the Hamiltonian
\begin{equation}
	\Ham = -\frac{\hbar^2}{2m}\frac{\dd^2}{\dd[x]^2}+\frac12m\omega^2\op{x}^2.
\end{equation}
	Do the calculated energy levels match the analytically known values?
	\emph{Hint:} use the system of units given in \mm{\ref{math:coupledunits}}ff with a length unit of $\lunit=\sqrt{\hbar/(m\omega)}$, a mass unit $\munit=m$, and an energy unit $\Eunit=\hbar\omega$ (\ie, the natural units). Choose the calculation box with size $a=10\lunit$ and shift the minimum of the harmonic potential to the center of the calculation box.
\pagenote[\ref{Q:harmonicwellN}]{The exact energy levels are $E_n=\hbar\omega(n+1/2)$ with $n\in\dsN_0$.

In the given unit system, the mass is \mm{m=1}, Planck's constant is \mm{\mmhbar=1}, and the angular frequency is \mm{\mmomega=1/\mmhbar=1}.

We set up a calculation in the position basis with the mixed-basis numerical method:
\begin{mathematica}
	¤protected¤mathin¤ a¤ =¤ 10;¤ ¤ ¤ ¤ ¤ ¤ ¤ ¤ ¤ ¤ ¤ ¤ ¤ ¤ ¤ ¤ ¤ (*¤ calculation¤ box¤ size¤ *)
	¤protetced¤mathin¤ m¤ =¤ ¤mmhbar¤ =¤ ¤mmomega¤ =¤ 1;¤ ¤ ¤ ¤ ¤ ¤ ¤ ¤ ¤ (*¤ natural¤ units¤ *)
	¤protected¤mathin¤ nmax¤ =¤ 100;
	¤protected¤mathin¤ ¤mmDelta¤ =¤ a/(nmax+1);¤ ¤ ¤ ¤ ¤ ¤ ¤ ¤ (*¤ grid¤ spacing¤ *)
	¤protected¤mathin¤ xgrid¤ =¤ Range[nmax]*¤mmDelta;¤ (*¤ the¤ computational¤ grid¤ *)
	¤protected¤mathin¤ TP¤ =¤ FourierDST[SparseArray[Band[{1,1}]->Range[nmax]^2*¤mmpi^2*¤mmhbar^2/(2*m*a^2)],¤ 1];
	¤protected¤mathin¤ W[x_]¤ =¤ m*¤mmomega^2*(x-a/2)^2/2;¤ ¤ ¤ (*¤ the¤ potential¤ function,¤ centered¤ *)
	¤protected¤mathin¤ Wgrid¤ =¤ Map[W,¤ xgrid];¤ (*¤ the¤ potential¤ on¤ the¤ computational¤ grid¤ *)
	¤protected¤mathin¤ VP¤ =¤ SparseArray[Band[{1,1}]¤ ->¤ Wgrid];
	¤protected¤mathin¤ HP¤ =¤ TP¤ +¤ VP;
\end{mathematica}
We find the energy eigenvalues (in units of $\Eunit=\hbar\omega$) with
\begin{mathematica}
	¤protect¤mathin¤ Eigenvalues[HP]¤ //Sort
	¤protect¤mathout¤ {0.5,¤ 1.5,¤ 2.5,¤ 3.5,¤ 4.50001,¤ 5.5001,¤ 6.5006,¤ 7.50293,¤ 8.51147,¤ 9.53657,¤ ...}
\end{mathematica}
and see that at least the lowest eigenvalues match the analytic expression. Using a larger value of $n\ix{max}$ will give more accurate eigenstates and eigenvalues.}
\end{questions}

\subsection[the Wigner quasi-probability distribution]{\label{sec:Wigner}the Wigner quasi-probability distribution\hspace{\stretch{1}}\attachcode{WignerDistribution}{the Wigner quasi-probability distribution}}\index{Wigner distribution}

The Wigner quasi-probability distribution\footnote{See \url{https://en.wikipedia.org/wiki/Wigner_quasiprobability_distribution}.} of a wavefunction $\psi(x)$ is a real-valued distribution in phase space defined as
\begin{equation}
	\label{eq:Wigner1}
	W(x,k) = \frac{1}{\pi} \int_{-\infty}^{\infty}\dd[y] \psi(x-y)\psi^*(x+y)e^{2\ii k y},
\end{equation}
where $k=p/\hbar$ is the wavenumber, closely related to the momentum but in units of inverse length.
$W$ often makes it easier to interpret wavefunctions than simply plotting $\psi(x)$, especially when $\psi(x)$ is complex-valued. Time-dependent wavefunctions are often plotted as Wigner distribution movies, which makes it easier to track a particle as it moves through phase space. In the classical limit, the time-dependent Wigner distribution becomes the classical phase-space density that satisfies the Liouville equation.

For a quick and easy evaluation of the Wigner distribution, we approximate the wavefunction as piecewise constant, using \autoref{eq:positionbasislocality}: $\psi(x)\approx\psi(x_{[x/\Delta]})=v_{[x/\Delta]}/\sqrt{\Delta}$, where we have used the calculation grid spacing $\Delta=a/(n\ix{max}+1)$ and the nearest-integer rounding function $[z]=\round(z)$. This approximation will be valid as long as $\abs{k}\ll\pi/\Delta$. Inserting it into \autoref{eq:Wigner1}, and assuming that $x=x_j=j\Delta$ is a grid point (\ie, we will only sample the Wigner function on the spatial grid of the calculation), we can split the integral over $y$ into integrals over segments of length $\Delta$,
\begin{multline}
	\label{eq:Wigner2}
	W(x_j,k) = \frac{1}{\pi} \sum_{m=-\infty}^{\infty} \int_{-\Delta/2}^{\Delta/2}\dd[y] \psi[x_j-(m\Delta+y)]\psi^*[x_j+(m\Delta+y)]e^{2\ii k (m\Delta+y)}\\
	\approx \frac{1}{\pi} \sum_{m=-\infty}^{\infty} \int_{-\Delta/2}^{\Delta/2}\dd[y] \psi(x_{j-m})\psi^*(x_{j+m})e^{2\ii k (m\Delta+y)}
	= \frac{1}{\pi} \sum_{m=-\infty}^{\infty} \underbrace{\psi(x_{j-m})}_{\frac{v_{j-m}}{\sqrt{\Delta}}}\underbrace{\psi^*(x_{j+m})}_{\frac{v_{j+m}^*}{\sqrt{\Delta}}} e^{2\ii k m\Delta} \underbrace{\int_{-\Delta/2}^{\Delta/2}\dd[y] e^{2\ii k y}}_{\sin(k\Delta)/k}\\
	= \frac{\sinc(k\Delta)}{\pi} \sum_{m=-\min(j-1,n\ix{max}-j)}^{\min(j-1,n\ix{max}-j)} v_{j-m}v^*_{j+m} e^{2\ii k m\Delta},
\end{multline}
where $\sinc(z)=\sin(z)/z$.
The following Mathematica code converts a coefficient vector $\vect{v}$ of length $n\ix{max}$ into a function of the dimensionless momentum $\kappa=a\cdot k$ that calculates $W(x_j,k)$ for every grid point $j=0,1,\dots,n\ix{max}+1$ (including the boundary grid points that are usually left out of our calculations):
\begin{mathematica}
	¤mathin¤labelŽmath:Wigner WignerDistribution[v_?VectorQ] := With[{nmax = Length[v]},
	¤mathnl   Function[¤mmkappa, Evaluate[Sinc[¤mmkappa/(nmax+1)]/¤mmpi*Table[
	¤mathnl     Sum[v[[j-m]]*Conjugate[v[[j+m]]]*Exp[2*I*¤mmkappa*m/(nmax+1)],
	¤mathnl     {m,-Min[j-1,nmax-j],Min[j-1,nmax-j]}]//Re//ComplexExpand, {j,0,nmax+1}]]]]
\end{mathematica}
Notice that this function \mm{WignerDistribution} returns an anonymous function\index{Mathematica!anonymous function} (see \autoref{sec:functionalprogramming}) of one parameter, which in turn returns a list of values. As an example of its use, we make a 2D plot of the Wigner distribution on the interval $x\in[x\ix{min},x\ix{max}]$:\footnote{This procedure works for situations other than the usual $x\ix{min}=0$ and $x\ix{max}=a$.}
\begin{mathematica}
	¤mathin¤labelŽmath:WignerPlot WignerDistributionPlot[Y_,
	¤mathnl       {xmin_?NumericQ, xmax_?NumericQ} /; xmax > xmin] :=
	¤mathnl Module[{nmax, qmax, w, W},
	¤mathnlc   (* number of grid points *)
	¤mathnl   nmax = Length[Y];
	¤mathnlc   (* calculate the Wigner distribution *)
	¤mathnl   w = WignerDistribution[Y];
	¤mathnlc   (* evaluate it on the natural dimensionless momentum grid *)
	¤mathnl   qmax = Floor[nmax/2];
	¤mathnl   W = Table[w[q*¤mmpi], {q, -qmax, qmax}];
	¤mathnlc   (* make a plot *)
	¤mathnl   ArrayPlot[W, FrameTicks->Automatic, AspectRatio -> 1/GoldenRatio,
	¤mathnl     DataRange->{{xmin,xmax},qmax*¤mmpi/(xmax-xmin)*{-1,1}},
	¤mathnl     ColorFunctionScaling -> False,
	¤mathnl     ColorFunction -> (Blend[{Blue, White, Red}, (¤mmpi*#+1)/2]&)]]
\end{mathematica}
Notice that we evaluate the Wigner distribution only up to momenta $\pm n\ix{max}\pi/(2a)$, which is the Nyquist limit\index{Nyquist frequency} in this finite-resolution system.\footnote{See \url{https://en.wikipedia.org/wiki/Nyquist_frequency}.} The color scheme is chosen such that the Wigner distribution values range $[-\frac{1}{\pi},+\frac{1}{\pi}]$ is mapped onto the colors blended from blue, white, and red, such that negative Wigner values are shown in shades of blue while positive values are shown in shades of red.

As an example, we plot the Wigner distribution of the numerical ground-state wavefunction shown on \autopageref{fig:GravityWellGsP}: on the left, the exact distribution from \autoref{eq:Wigner1}; on the right, the grid evaluation of \mm{\ref{math:Wigner}} and \mm{\ref{math:WignerPlot}} (calculated with $n\ix{max}=40$) with
\begin{mathematica}
	¤mathin WignerDistributionPlot[gsP[[2, 1]], {0, a}]
\end{mathematica}
\begin{center}
\includegraphics[height=0.15\textheight]{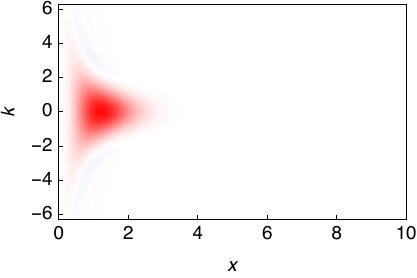}
\includegraphics[height=0.15\textheight]{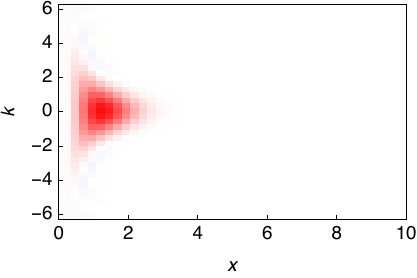}
\includegraphics[height=0.15\textheight]{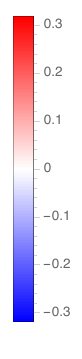}
\end{center}

\subsubsection{extension to density operators}\index{density matrix}

If the state of the system is not pure, but given as a density matrix $\rho(x,x')=\me{x}{\op{\rho}}{x'}$ instead of as a wavefunction $\psi(x)=\scp{x}{\psi}$, then we do not have the option of plotting the wavefunction and we can only resort to the Wigner distribution for a graphical representation.

Noticing that \autoref{eq:Wigner1} contains the term
\begin{equation}
	\psi(x-y)\psi^*(x+y)=\scp{x-y}{\psi}\scp{\psi}{x+y}=\me{x-y}{\op{\rho}}{x+y}=\rho(x-y,x+y)
\end{equation}
in terms of the pure-state density operator $\op{\rho}=\ket{\psi}\bra{\psi}$, the definition of the Wigner distribution is generalized to
\begin{equation}
	\label{eq:Wignerrho}
	W(x,k) = \frac{1}{\pi} \int_{-\infty}^{\infty}\dd[y] \rho(x-y,x+y)e^{2\ii k y}.
\end{equation}
We can make the same approximations as in \autoref{eq:Wigner2} to calculate the Wigner function on a spatial grid point $x=x_j$ [see \autoref{eq:densitymatrix}]:
\begin{multline}
	W(x_j,k) = \frac{1}{\pi} \sum_{m=-\infty}^{\infty} \int_{-\Delta/2}^{\Delta/2}\dd[y] \rho[x_j-(m\Delta+y),x_j+(m\Delta+y)]e^{2\ii k (m\Delta+y)}\\
	\approx \frac{1}{\pi} \sum_{m=-\infty}^{\infty} \int_{-\Delta/2}^{\Delta/2}\dd[y] \rho(x_{j-m},x_{j+m})e^{2\ii k (m\Delta+y)}\\
	= \frac{\sinc(k\Delta)}{\pi} \sum_{m=-\min(j-1,n\ix{max}-j)}^{\min(j-1,n\ix{max}-j)} R_{j-m,j+m} e^{2\ii k m\Delta}.
\end{multline}
In analogy to \mm{\ref{math:Wigner}} we define
\begin{mathematica}
	¤mathin¤labelŽmath:WignerR WignerDistribution[R_ /; MatrixQ[R, NumericQ] &&
	¤mathnl   Length[R] == Length[Transpose[R]]] :=
	¤mathnl With[{n = Length[R]},
	¤mathnl   Function[k, Evaluate[Sinc[k/(n+1)]/¤mmpi*Table[
	¤mathnl     Sum[R[[j-m,j+m]]*Exp[2*I*k*m/(n+1)],
	¤mathnl     {m,-Min[j-1,n-j],Min[j-1,n-j]}]//Re//ComplexExpand, {j,0,n+1}]]]]
\end{mathematica}
For a pure state, the density matrix has the coefficients $R_{j,j'} = v_j v_{j'}^*$, and the definitions of \mm{\ref{math:Wigner}} and \mm{\ref{math:WignerR}} thus give exactly the same result if we use
\begin{mathematica}
	¤mathin R = KroneckerProduct[v, Conjugate[v]]
\end{mathematica}
In addition, the 2D plotting function of \mm{\ref{math:WignerPlot}} also works when called with a density matrix as first parameter.

\subsubsection{exercises}

\begin{questions}
	\item\label{Q:WignerExcited} Plot the Wigner distribution of the first excited state of the gravity well. What do you notice?
\pagenote[\ref{Q:WignerExcited}]{The excited-state Wigner distribution has a significant negative region around its center:
\begin{mathematica}
	¤protect¤mathin¤ gsP¤ =¤ Transpose[Sort[Transpose[-Eigensystem[-N[HP],¤ 2,
	¤protect¤mathnl¤ ¤ ¤ Method->{"Arnoldi",¤ "Criteria"->"RealPart",¤ MaxIterations->10^6}]]]];
	¤protect¤mathin¤ WignerDistributionPlot[gsP[[2,¤ 2]],¤ {0,¤ a}]
\end{mathematica}
\begin{center}
\includegraphics[height=0.15\textheight]{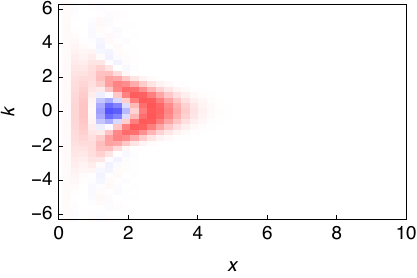}
\includegraphics[height=0.15\textheight]{WignerDistribution_legend}
\end{center}}
\end{questions}

\subsection[1D dynamics in the square well]{\label{sec:dynamics1D}1D dynamics in the square well\hspace{\stretch{1}}\attachcode{ParticleDynamics}{single-particle dynamics in 1D}}\index{Schroedinger equation@Schr\"odinger equation!time-dependent}

Assume again a single particle of mass $m$ moving in a one-dimensional potential, with the time-independent Hamiltonian given in \autoref{eq:1DHamiltonian}.
The motion is again restricted to $x\in[0,a]$. We want to study the time-dependent wavefunction $\psi(x,t)=\scp{x}{\psi(t)}$ given in \autoref{eq:tdSchrMatSol2},\index{propagator}
\begin{equation}
	\label{eq:tdsesol}
	\ket{\psi(t)} = \exp\left[-\frac{\ii(t-t_0)}{\hbar}\Ham\right]\ket{\psi(t_0)}.
\end{equation}
The simplest way of computing this propagation is to express the wavefunction and the Hamiltonian in a particular basis and use a matrix exponentiation to find the time dependence of the expansion coefficients of the wavefunction. For example, if we use the finite-resolution position basis, we have seen on \autopageref{sec:mixedbasis} how to find the matrix representation of the Hamiltonian, \mm{HP}. For a given initial wavefunction represented by a position-basis coefficient vector \mm{v0} we can then define\index{Mathematica!matrix!matrix exponential}
\begin{mathematica}
	¤mathin¤labelŽmath:directprop v[¤mmDeltaŽt_?NumericQ] := MatrixExp[-I*HP*¤mmDeltaŽt/¤mmhbar].v0
\end{mathematica}
as the propagation over a time interval $\ddd{\Delta}{t}=t-t_0$.
If you try this, you will see that calculating $\ket{\psi(t)}$ in this way is not very efficient, because the matrix exponentiation is a numerically difficult operation.

A much more efficient method can be found by
using the Trotter expansion\index{Trotter expansion}
\begin{multline}
	\label{eq:Trotter}
	e^{\lambda(X+Y)} = e^{\frac{\lambda}{2}X} e^{\lambda Y} e^{\frac{\lambda}{2}X}
	\times e^{\frac{\lambda^3}{24}[X,[X,Y]]+\frac{\lambda^3}{12}[Y,[X,Y]]}
	\times e^{-\frac{\lambda^4}{48}[X,[X,[X,Y]]]-\frac{\lambda^4}{16}[X,[Y,[X,Y]]]-\frac{\lambda^4}{24}[Y,[Y,[X,Y]]]}\dotsm\\
	\approx e^{\frac{\lambda}{2}X} e^{\lambda Y} e^{\frac{\lambda}{2}X},
\end{multline}
where the approximation is valid for small $\lambda$ since the neglected terms are of third and higher orders in $\lambda$ (notice that there is no second-order term in $\lambda$!). Setting $\lambda=-\frac{\ii(t-t_0)}{M\hbar}$ for some large integer $M$, as well as $X=\op{V}$ and $Y=\op{T}$, we find
\begin{equation}
	\label{eq:splitstep}
	\ket{\psi(t)}
	= e^{M\lambda\Ham}\ket{\psi(t_0)}
	= \left[e^{\lambda\Ham}\right]^M\ket{\psi(t_0)}
	= \left[e^{\lambda(\op{T}+\op{V})}\right]^M\ket{\psi(t_0)}
	\stackrel{\stackrel{\text{Trotter \autoref{eq:Trotter}}}{\downarrow}}{=}
	\lim_{M\to\infty} \left[e^{\frac{\lambda}{2}\op{V}}e^{\lambda\op{T}}e^{\frac{\lambda}{2}\op{V}}\right]^M\ket{\psi(t_0)}.
\end{equation}
This can be evaluated very efficiently. We express the potential Hamiltonian in the finite-resolution position basis, \autoref{eq:potentialoperator}, the kinetic Hamiltonian in the momentum basis, \autoref{eq:kineticoperator}, and the time-dependent wavefunction in both bases of \autoref{eq:posmombasis}:
\begin{subequations}
\begin{align}
	\ket{\psi(t)} &= \sum_{n=1}^{n\ix{max}} u_n(t)\ket{n} = \sum_{j=1}^{n\ix{max}} v_j(t)\ket{j}\\
	\label{eq:opVwell}\op{V} &\approx \sum_{j=1}^{n\ix{max}} W(x_j)\ket{j}\bra{j}\\
	\label{eq:opTwell}\op{T} &\approx \sum_{n=1}^{n\ix{max}} \frac{n^2\pi^2\hbar^2}{2ma^2} \ket{n}\bra{n}
\end{align}
\end{subequations}
The expansion coefficients of the wavefunction are related by a type-I discrete sine transform, see \autoref{eq:uvconv}, \autoref{eq:vuconv}, \mm{\ref{math:DSTuv}}, and \mm{\ref{math:DSTvu}}.

The great advantage of the diagonal matrices of \autoref{eq:opVwell} and \autoref{eq:opTwell} is that algebra with diagonal matrices is as simple as algebra with scalars, but applied to the diagonal elements one-by-one. In particular, for any diagonal matrix $D=\sum_j d_j\ket{j}\bra{j}$ the integer matrix powers are $D^k=\sum_j d_j^k\ket{j}\bra{j}$, and matrix exponentionals are calculated by exponentiating each diagonal element separately: $\exp(D)=\sum_{k=0}^{\infty}D^k/k!=\sum_{k=0}^{\infty}(\sum_j d_j^k\ket{j}\bra{j})/k!=\sum_j(\sum_{k=0}^{\infty}d_j^k/k!)\ket{j}\bra{j}=\sum_j\exp(d_j)\ket{j}\bra{j}$. As a result, 
\begin{equation}
	e^{\frac{\lambda}{2}\op{V}}
	= \sum_{j=1}^{n\ix{max}} e^{\frac{\lambda}{2} W(x_j)}\ket{j}\bra{j},
\end{equation}
and the action of the potential Hamiltonian thus becomes straightforward:
\begin{equation}
	e^{\frac{\lambda}{2}\op{V}}\ket{\psi(t)}
	= \left[ \sum_{j=1}^{n\ix{max}} e^{\frac{\lambda}{2} W(x_j)}\ket{j}\bra{j} \right] \left[ \sum_{j'=1}^{n\ix{max}} v_{j'}(t)\ket{j'} \right]
	= \sum_{j,j'=1}^{n\ix{max}} e^{\frac{\lambda}{2} W(x_j)}v_{j'}(t)\ket{j}\underbrace{\scp{j}{j'}}_{\delta_{j j'}}
	= \sum_{j=1}^{n\ix{max}} \underbrace{\left[ e^{\frac{\lambda}{2} W(x_j)}v_j(t) \right]}_{v_j'} \ket{j},
\end{equation}
which is an element-by-element multiplication of the coefficients of the wavefunction with the exponentials of the potential\emdash no matrix operations are required. The expansion coefficients (position basis) after propagation with the potential Hamiltonian for a ``time'' step $\lambda/2$ are therefore
\begin{equation}
	v_j' = e^{\frac{\lambda}{2} W(x_j)} v_j.
\end{equation}
The action of the kinetic Hamiltonian in the momentum representation is found in exactly the same way:
\begin{equation}
	e^{\lambda\op{T}}\ket{\psi(t)}
	= \left[ \sum_{n=1}^{n\ix{max}} e^{\lambda \frac{n^2\pi^2\hbar^2}{2ma^2}}\ket{n}\bra{n} \right] \left[ \sum_{n'=1}^{n\ix{max}} u_{n'}(t)\ket{n'} \right]
	= \sum_{n=1}^{n\ix{max}} \underbrace{\left[ e^{\lambda \frac{n^2\pi^2\hbar^2}{2ma^2}}u_n(t) \right]}_{u_n'} \ket{n}.
\end{equation}
The expansion coefficients (momentum basis) after propagation with the kinetic Hamiltonian for a ``time'' step $\lambda$ are therefore
\begin{equation}
	\label{eq:kinevoln}
	u_n' = e^{\lambda\frac{n^2\pi^2\hbar^2}{2ma^2}} u_n.
\end{equation}
We know that a type-I discrete sine transform brings the wavefunction from the finite-resolution position basis to the momentum basis and vice-versa. The propagation under the kinetic Hamiltonian thus consists of
\begin{enumerate}
	\item a type-I discrete sine transform to calculate the coefficients $v_j \mapsto u_n$,
	\item an element-by-element multiplication, \autoref{eq:kinevoln}, to find the coefficients $u_n\mapsto u_n'$,
	\item and a second type-I discrete sine transform to calculate the coefficients $u_n'\mapsto v_j'$.
\end{enumerate}
Here we assemble all these pieces into a program that propagates a state $\ket{\psi(t_0)}$, which is given as a coefficient vector $\vect{v}$ in the finite-resolution position basis, forward in time to $t=t_0+\ddd{\Delta}{t}$.
First, for reference, a procedure for the exact propagation by matrix exponentiation and matrix--vector multiplication, as in \mm{\ref{math:directprop}}:
\begin{mathematica}
	¤mathin VP = SparseArray[Band[{1,1}]->Wgrid];
	¤mathin TM = SparseArray[Band[{1,1}]->Range[nmax]^2*¤mmpi^2*¤mmhbar^2/(2*m*a^2)];
	¤mathin TP = FourierDST[TM, 1];
	¤mathin HP = TP + VP;
	¤mathin propExact[¤mmDeltaŽt_?NumericQ, v0_ /; VectorQ[v0, NumericQ]] :=
	¤mathnl   MatrixExp[-I*HP*N[¤mmDeltaŽt/¤mmhbar]].v0
\end{mathematica}
Next, an iterative procedure that propagates by $M$ small steps via the Trotter approximation, \autoref{eq:Trotter}:
\begin{mathematica}
	¤mathin¤labelŽmath:splitstep propApprox[¤mmDeltaŽt_?NumericQ, M_Integer /; M >= 1,
	¤mathnl            v0_ /; VectorQ[v0, NumericQ]] :=
	¤mathnl   Module[{¤mmlambda, Ke, Pe2, propKin, propPot2, prop},
	¤mathnlc     (* compute the ¤mmlambda constant *)
	¤mathnl     ¤mmlambda = -I*N[¤mmDeltaŽt/(M*¤mmhbar)];
	¤mathnlc     (* compute the diagonal elements of exp[¤mmlambda*T] *)
	¤mathnl     Ke = Exp[¤mmlambda*Range[nmax]^2*¤mmpi^2*¤mmhbar^2/(2*m*a^2)];
	¤mathnlc     (* propagate by a full time-step with T *)
	¤mathnl     propKin[v_] := FourierDST[Ke*FourierDST[v, 1], 1];
	¤mathnlc     (* compute the diagonal elements of exp[¤mmlambda*V/2] *)
	¤labelŽmathline:potexp¤mathnl     Pe2 = Exp[¤mmlambda/2*Wgrid];
	¤mathnlc     (* propagate by a half time-step with V *)
	¤mathnl     propPot2[v_] := Pe2*v;
	¤mathnlc     (* propagate by a full time-step by H=T+V *)
	¤mathnlc     (* using the Trotter approximation *)
	¤mathnl     prop[v_] := propPot2[propKin[propPot2[v]]];
	¤mathnlc     (* step-by-step propagation *)
	¤labelŽmathline:1Ddyn¤mathnl     Nest[prop, v0, M]]
\end{mathematica}
Notice that there are no basis functions, integrals, etc.\ involved in this calculation; everything is done in terms of the values of the wavefunction on the grid $x_1\dots x_{n\ix{max}}$. This efficient method is called \emph{split-step propagation}.\index{split-step method}

The \mm{Nest} command\index{Mathematica!nesting function calls} ``nests'' a function call: for example, \mm{Nest[f,x,3]} calculates $f(f(f(x)))))$. We use this on line~\ref*{mathline:1Ddyn} of \mm{\ref{math:splitstep}} to repeatedly propagate by small time steps via the Trotter approximation. Since this algorithm internally calculates the wavefunction at all the intermediate times $t=t_0+\frac{m}{M}(t-t_0)$ for $m=1,2,3,\dots,M$, we can modify our program in order to follow this time evolution. To achieve this we simply replace the \mm{Nest} command with \mm{NestList},\label{pp:NestList} which is similar to \mm{Nest} but returns all intermediate results: for example, \mm{NestList[f,x,3]} returns the list $\{x, f(x), f(f(x)), f(f(f(x)))\}$. We replace last line of the code above with
\begin{mathematica}[firstnumber=\getrefnumber{mathline:1Ddyn}]
	¤mathnl     Transpose[{Range[0, M]/M*¤mmDeltaŽt, NestList[prop, v0, M]}]]
\end{mathematica}
which now returns a list of pairs containing (i) the time and (ii) the wavefunction at the corresponding time.

\subsubsection{example: bouncing in the gravity well}

As an example of particle dynamics, we return to the gravity well of \autoref{sec:gravitywell}.
Classically, if we drop a particle from height $x_0$ at $t=0$ under the influence of gravity, then its trajectory is $x(t)=x_0-\frac12 g t^2$, until it reaches the earth's surface ($x=0$) at time $t_1=\sqrt{2x_0/g}$.
We plot this classical bouncing trajectory for a scaled starting height $x_0=15$ with
\begin{mathematica}
	¤mathin¤labelŽmath:classicalbounce With[{x0 = 15, ¤mmDeltaŽt = 50}, {t1 = Sqrt[2*x0/g]},
	¤mathnl   Plot[x0 - Mod[t, 2*t1, -t1]^2/2, {t, 0, ¤mmDeltaŽt}]]
\end{mathematica}
In order to simulate a quantum particle bouncing along this trajectory, we start at the same height $x_0=15$ but assume that the particle initially has a wavefunction of root-mean-square width $\sigma=1$: the initial state in the position basis is
\begin{mathematica}
	¤mathin x0 = 15;             (* starting height *)
	¤mathin ¤mmsigma = 1;               (* starting width *)
	¤mathin t1 = Sqrt[2*x0/g];   (* classical bounce time *)
	¤mathin vv = Normalize[N[Exp[-((xgrid-x0)/(2*¤mmsigma))^2]]];  (* starting state *)
	¤mathin ListLinePlot[Join[{{0,0}},Transpose[{xgrid,vv/Sqrt[¤mmDelta]}],{{a,0}}],
	¤mathnl   PlotRange->All]
\end{mathematica}
\begin{center}
\includegraphics[width=0.5\textwidth]{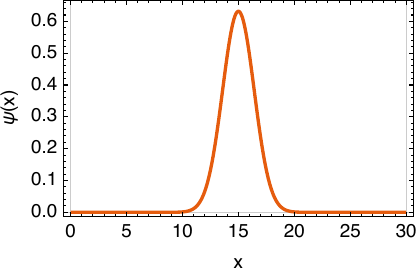}
\end{center}
We propagate this particle in time for $\ddd{\Delta}{t}=50$ time units, using $M=1000$ time steps, and plot the time-dependent density $\rho(x,t)=\abs{\psi(x,t)}^2=\abs{\scp{x}{\psi(t)}}^2$ using the trick of \autoref{eq:positionbasislocality}:
\begin{mathematica}
	¤mathin With[{¤mmDeltaŽt = 50, M = 1000},
	¤mathnl   ¤mmrho = ArrayPad[Abs[propApprox[¤mmDeltaŽt, M, vv][[All,2]]]^2/¤mmDelta, {{0, 0}, {1, 1}}];
	¤mathnl   ArrayPlot[Reverse[Transpose[¤mmrho]], DataRange -> {{0, ¤mmDeltaŽt}, {0, a}}]
\end{mathematica}
\begin{center}
\includegraphics[width=0.6\textwidth]{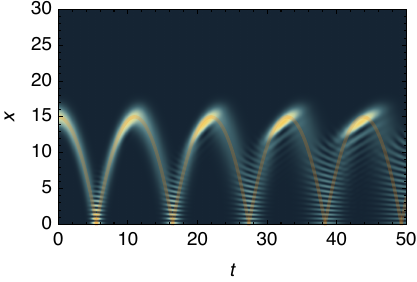}
\end{center}
The orange overlaid curve shows the classical particle trajectory from \mm{\ref{math:classicalbounce}}, which the quantum particle follows approximately while self-interfering during the reflections.

To study the correspondence between classical and quantum-mechanical motion more quantitatively, we can calculate and plot time-dependent quantities such as the time-dependent mean position: using \autoref{eq:positionoperator},
\begin{equation}
	\avg{\op{x}}(t)=\me{\psi(t)}{\op{x}}{\psi(t)}
	\approx \me{\psi(t)}{\left[ \sum_{j=1}^{n\ix{max}}\ket{j}x_j\bra{j}\right]}{\psi(t)}
	= \sum_{j=1}^{n\ix{max}} x_j \abs{v_j(t)}^2.
\end{equation}
\begin{mathematica}
	¤mathin With[{¤mmDeltaŽt = 50, M = 1000},
	¤mathnl   ListLinePlot[{#[[1]], Abs[#[[2]]]^2.xgrid} & /@ propApprox[¤mmDeltaŽt, M, vv]]]
\end{mathematica}
\begin{center}
\includegraphics[width=0.5\textwidth]{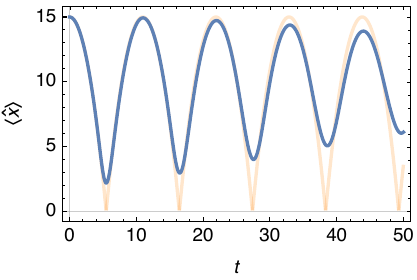}
\end{center}
Here the quantum deviations from the classical trajectory (orange) become apparent.

\subsection{1D dynamics in a time-dependent potential}\index{split-step method}\index{Schroedinger equation@Schr\"odinger equation!time-dependent}
\label{sec:dynamics1Dtd}

While the direct propagation of \autoref{eq:tdSchrMatSol2} only works for time-independent Hamiltonians, the split-step method of \mm{\ref{math:splitstep}} can be extended to time-dependent Hamiltonians, in particular to time-dependent potentials $W(x,t)$. For this, we assume that the potential varies slowly enough in time that it is almost constant during a Trotter step $\ddd{\Delta}{t}/M$; this assumption usually becomes exact as $M\to\infty$.
\begin{mathematica}
	¤mathin¤labelŽmath:splitstepT propApprox[Wt_,
	¤mathnl            ¤mmDeltaŽt_?NumericQ, M_Integer /; M >= 1,
	¤mathnl            v0_ /; VectorQ[v0, NumericQ]] :=
	¤mathnl   Module[{¤mmlambda, Ke, propKin, propPot2, prop},
	¤mathnlc     (* compute the ¤mmlambda constant *)
	¤mathnl     ¤mmlambda = -I*N[¤mmDeltaŽt/(M*¤mmhbar)];
	¤mathnlc     (* compute the diagonal elements of exp[¤mmlambda*T] *)
	¤mathnl     Ke = Exp[¤mmlambda*Range[nmax]^2*¤mmpi^2/2];
	¤mathnlc     (* propagate by a full time-step with T *)
	¤mathnl     propKin[v_] := FourierDST[Ke*FourierDST[v, 1], 1];
	¤mathnlc     (* propagate by a half time-step with V *)
	¤mathnlc     (* evaluating the potential at time t *)
	¤labelŽmathline:1Dproppot2¤mathnl     propPot2[t_, v_] := Exp[¤mmlambda/2*(Wt[#,t]&/@xgrid)]*v;
	¤mathnlc     (* propagate by a full time-step by H=T+V *)
	¤mathnlc     (* using the Trotter approximation *)
	¤mathnlc     (* starting at time t *)
	¤mathnl     prop[v_, t_] := propPot2[t+3¤mmDeltaŽt/(4M), propKin[propPot2[t+¤mmDeltaŽt/(4M), v]]];
	¤mathnlc     (* step-by-step propagation *)
	¤labelŽmathline:nestfold¤mathnl     Transpose[{Range[0, M]/M*¤mmDeltaŽt, FoldList[prop, v0, Range[0,M-1]/M*¤mmDeltaŽt]}]]
\end{mathematica}
\begin{itemize}
	\item The definition of \mm{propApprox} now needs a time-dependent potential \mm{Wt[x,t]} that it can evaluate as the propagation proceeds. This potential must be specified as a pure function with two arguments, as in the example below.
	\item The exponentials for the potential propagation, calculated once-and-for-all on line~\ref*{mathline:potexp} of \mm{\ref{math:splitstep}}, are now re-calculated in each call of the \mm{propPot2} function.
	\item In the Trotter propagation step of \autoref{eq:splitstep} we evaluate the potential twice in each propagation interval $[t,t+\ddd{\Delta}{t}/M]$: once at $t+\frac14\ddd{\Delta}{t}/M$ for the first half-step with the potential operator $\op{V}$, and once at $t+\frac34\ddd{\Delta}{t}/M$ for the second half-step.
	\item On line~\ref*{mathline:nestfold} of \mm{\ref{math:splitstepT}} we have replaced \mm{NestList} by \mm{FoldList}, which is more flexible: for example, \mm{FoldList[f,x,\{a,b,c\}]} calculates the list $\{x,f(x,a),f(f(x,a),b),f(f(f(x,a),b),c)\}$. By giving the list of propagation interval starting times as the last argument of \mm{FoldList}, the \mm{prop} function is called repeatedly, with the current interval starting time as the second argument.
\end{itemize}
As an example, we calculate the time-dependent density profile under the same conditions as above, except that the gravitational acceleration is modulated periodically: $W(x,t)=W(x)\cdot[1+A\cdot\sin(\omega t)]$. The oscillation frequency $\omega=\pi/t_1$ is chosen to drive the bouncing particle resonantly and enhance its amplitude. This time-dependent potential is passed as the first argument to \mm{propApprox}:
\begin{mathematica}
	¤mathin With[{A = 0.1, ¤mmomega = ¤mmpi/t1, ¤mmDeltaŽt = 50, M = 1000},
	¤mathnl   Wt[x_, t_] = W[x]*(1 + A*Sin[¤mmomega*t]);
	¤mathnl   ¤mmrho = ArrayPad[Abs[propApprox[Wt,¤mmDeltaŽt,M,vv][[All,2]]]^2/¤mmDelta, {{0, 0}, {1, 1}}];
	¤mathnl   ArrayPlot[Reverse[Transpose[¤mmrho]], DataRange -> {{0, ¤mmDeltaŽt}, {0, a}}]
\end{mathematica}
\begin{center}
\includegraphics[width=0.6\textwidth]{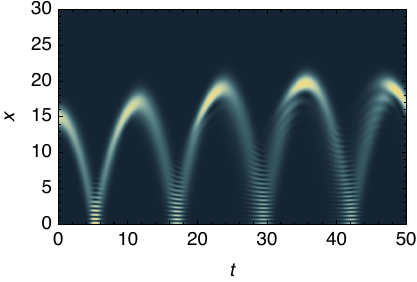}
\end{center}
The increase in bouncing amplitude can be seen clearly in this density plot.

\subsubsection{exercises}

\begin{questions}
	\item\label{Q:Trotter} Convince yourself that the Trotter expansion of \autoref{eq:Trotter} is really necessary, \ie, that $e^{X+Y} \neq e^X e^Y$ if $X$ and $Y$ do not commute.\index{Trotter expansion} \emph{Hint:} use two concrete non-commuting objects $X$ and $Y$, for example two random \num{2 x 2} matrices as generated with $\mm{\text{RandomReal}[\{0,1\},\{2,2\}]}$.
\pagenote[\ref{Q:Trotter}]{
\begin{mathematica}
	¤protect¤mathin¤ X¤ =¤ RandomReal[{0,1},¤ {2,2}]
	¤protect¤mathout¤ {{0.580888,¤ 0.80848},¤ {0.218175,¤ 0.979598}}
	¤protect¤mathin¤ Y¤ =¤ RandomReal[{0,1},¤ {2,2}]
	¤protect¤mathout¤ {{0.448364,¤ 0.774595},¤ {0.490198,¤ 0.310169}}
	¤protect¤mathin¤ X.Y¤ -¤ Y.X
	¤protect¤mathout¤ {{0.227318,¤ -0.420567},¤ {0.225597,¤ -0.227318}}
	¤protect¤mathin¤ MatrixExp[X¤ +¤ Y]
	¤protect¤mathout¤ {{4.68326,¤ 6.06108},¤ {2.71213,¤ 5.68068}}
	¤protect¤mathin¤ MatrixExp[X].MatrixExp[Y]
	¤protect¤mathout¤ {{5.0593,¤ 5.38209},¤ {3.10936,¤ 5.31705}}
\end{mathematica}}
	\item\label{Q:Hscaling} Given a particle moving in the range $x\in[0,a]$ with the Hamiltonian
	\begin{equation}
		\Ham = -\frac{\hbar^2}{2m} \frac{\dd^2}{\dd[x]^2} + W_0\sin(10\pi x/a),
	\end{equation}
	compute its time-dependent wavefunction starting from a ``moving Gaussian'' $\psi(t=0)\propto e^{-\frac{(x-a/2)^2}{4\sigma^2}}e^{\ii k x}$ with $\sigma=0.05a$ and $k=100/a$. Study $\avg{\op{x}}(t)$ using first $W_0=0$ and then $W_0=\num{5000}\frac{\hbar^2}{ma^2}$. \emph{Hint:} use natural units such that \mm{a=m=\mmhbar=1} for simplicity.
\pagenote[\ref{Q:Hscaling}]{We use the split-step propagation code of \autoref{sec:dynamics1D} with the potential
\begin{mathematica}
	¤protect¤mathin¤ a¤ =¤ m¤ =¤ ¤mmhbar¤ =¤ 1;
	¤protect¤mathin¤ With[{W0¤ =¤ 0¤ *¤ ¤mmhbar^2/(m*a^2)},
	¤protect¤mathnl¤ ¤ ¤ W[x_]¤ =¤ W0*Sin[10*¤mmpi*x/a];]
\end{mathematica}
	and the initial wavefunction
\begin{mathematica}
	¤protect¤mathin¤labelŽmath:initialgaussian¤ With[{x0=a/2,¤ ¤mmsigma=0.05*a,¤ k=100/a},
	¤protect¤mathnl¤ ¤ ¤ v0=Normalize[Function[x,¤ E^(-((x-x0)^2/(4*¤mmsigma^2)))*E^(I*k*x)]¤ /@¤ xgrid];]
\end{mathematica}
	For $W_0=0$ the Gaussian wavepacket bounces back and forth between the simulation boundaries and disperses slowly; the self-interference at the reflection points is clearly visible:
\begin{center}
\ifthenelse{\boolean{smallfigures}}%
{\includegraphics[height=0.15\textheight]{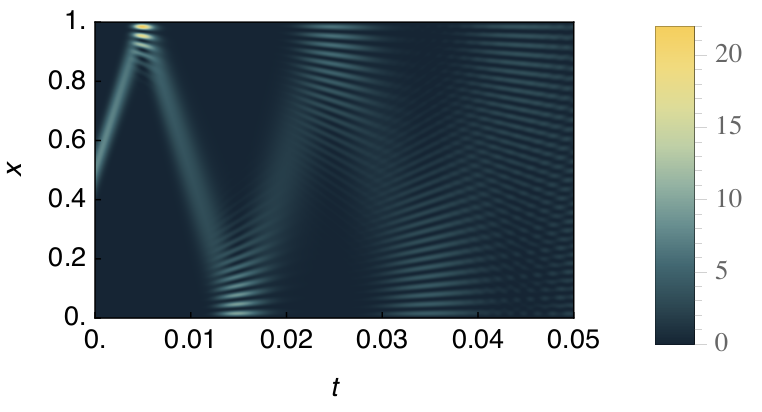}}%
{\includegraphics[height=0.15\textheight]{1Dparticledynamics_sin0}}
\includegraphics[height=0.15\textheight]{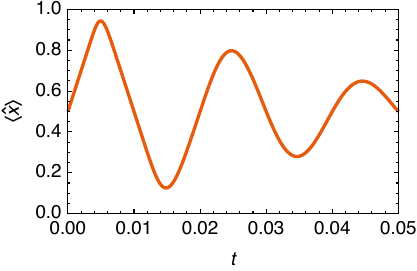}
\end{center}
	For $W_0=\num{5000}\frac{\hbar^2}{ma^2}$ the Gaussian wavepacket remains mostly trapped:
\begin{center}
\ifthenelse{\boolean{smallfigures}}%
{\includegraphics[height=0.15\textheight]{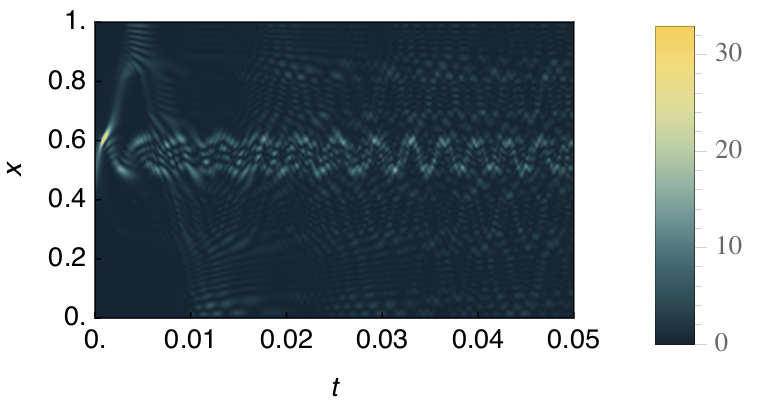}}%
{\includegraphics[height=0.15\textheight]{1Dparticledynamics_sin}}
\includegraphics[height=0.15\textheight]{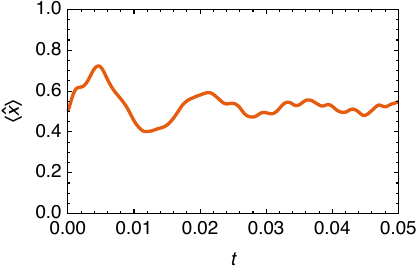}
\end{center}}
\end{questions}

\section[non-linear Schr\"odinger equation]{Many particles in one dimension: dynamics with the non-linear Schr\"odinger equation}\index{Schroedinger equation@Schr\"odinger equation!non-linear}
\label{sec:tdGPE}

The advantage of the split-step evolution of \autoref{eq:splitstep} becomes particularly clear when the system's energy depends on the wavefunction in a more complicated way than in the usual time-independent Schr\"odinger equation. A widely used example is the nonlinear energy functional\footnote{A functional is an operation that calculates a number from a given function. For example, $E[\psi]:L^2\to\dsR$ converts a wavefunction $\psi\in L^2$ into an energy $E\in\dsR$. See \url{https://en.wikipedia.org/wiki/Functional_(mathematics)}.}
\begin{equation}
	\label{eq:nonlinearenergyfunctional}
	E[\psi] = \underbrace{-\frac{\hbar^2}{2m} \int_{-\infty}^{\infty}\dd[x]\psi^*(x)\psi''(x)}_{E\ix{kin}[\psi]}
		+ \underbrace{\int_{-\infty}^{\infty}\dd[x]V(x)\abs{\psi(x)}^2}_{E\ix{pot}[\psi]}
		+ \underbrace{\frac{\kappa}{2}\int_{-\infty}^{\infty}\dd[x]\abs{\psi(x)}^4}_{E\ix{int}[\psi]},
\end{equation}
in which the last term describes the \emph{mean-field interactions}\index{mean-field interaction} between $N$ particles that are all in wavefunction $\psi(x)$ (normalized to $\int_{-\infty}^{\infty}\dd[x]\abs{\psi(x)}^2=1$), and which are therefore in a joint product wavefunction $\psi(x)^{\otimes N}$ (see \autoref{eq:productwf}). Each particle sees a potential $V\ix{int}(x)=\frac{\kappa}{2}\abs{\psi(x)}^2$ generated by the average density $(N-1)\abs{\psi(x)}^2$ of other particles with the same wavefunction, usually through collisional interactions. 
In three dimensions, the coefficient $\kappa=(N-1)\times 4\pi \hbar^2a_s/m$ approximates the mean-field $s$-wave scattering between a particle and the $(N-1)$ other particles, with $s$-wave scattering length $a_s$\index{s-wave scattering@$s$-wave scattering} (see \autoref{sec:BEC}); in the present one-dimensional example, no such identification is made.

In order to find the ground state (energy minimum) of \autoref{eq:nonlinearenergyfunctional} under the constraint of wavefunction normalization $\int_{-\infty}^{\infty}\dd[x]\abs{\psi(x)}^2=1$, we use the Lagrange multiplier\footnote{See \url{https://en.wikipedia.org/wiki/Lagrange_multiplier}.}\index{Lagrange multiplier} method: using the Lagrange multiplier
$\mu$ called the \emph{chemical potential}\index{chemical potential}, we conditionally minimize the energy with respect to the wavefunction by setting its functional derivative\footnote{See \url{https://en.wikipedia.org/wiki/Functional_derivative}.}
\begin{equation}
	\frac{\delta}{\ddd{\delta}{\psi}^*(x)}\left(E[\psi]-\mu\int_{-\infty}^{\infty}\dd[x]\abs{\psi(x)}^2\right)=
	-\frac{\hbar^2}{m} \psi''(x)
	+ 2V(x)\psi(x)
	+ 2\kappa\abs{\psi(x)}^2\psi(x)
	-2\mu\psi(x)
	=0
\end{equation}
to zero. This yields the non-linear Schr\"odinger equation\index{Schroedinger equation@Schr\"odinger equation!non-linear}
\begin{equation}
	\label{eq:TIGPE}
	\Bigg[ -\frac{\hbar^2}{2m} \frac{\dd^2}{\dd[x]^2}
	+ \underbrace{V(x) + \kappa\abs{\psi(x)}^2}_{V\ix{eff}(x)} \Bigg]\psi(x)
	=\mu \psi(x),
\end{equation}
also called the \emph{Gross--Pitaevskii equation} for the description of dilute Bose--Einstein condensates.\index{Bose-Einstein condensate@Bose--Einstein condensate}
By analogy to the linear Schr\"odinger equation, it also has a time-dependent form for the description of Bose--Einstein condensate dynamics,
\begin{equation}
	\label{eq:TDGPE}
	\ii\hbar\frac{\ddd{\partial}{\psi(x,t)}}{\ddd{\partial}{t}} = \Bigg[
		-\frac{\hbar^2}{2m}\frac{\partial^2}{\ddd{\partial}{x}^2}
		+ \underbrace{V(x,t) + \kappa\abs{\psi(x,t)}^2}_{V\ix{eff}(x,t)}
	\Bigg]\psi(x,t).
\end{equation}
For any $\kappa\neq0$ there is no solution of the form of \autoref{eq:tdsesol}. But the split-step method of \autoref{eq:splitstep} can still be used to simulate \autoref{eq:TDGPE} because the wavefunction-dependent effective potential $V\ix{eff}(x,t)$ is still diagonal in the position representation. We extend the Mathematica code of \mm{\ref{math:splitstepT}} by modifying the \mm{propPot2} method to include a non-linear term with prefactor \mm{\mmkappa} (added as an additional argument to the \mm{propApprox} function), and do not forget that the wavefunction at grid point $x_j$ is $\psi(x_j)=v_j/\sqrt{\Delta}$:\index{split-step method}
\begin{mathematica}
	¤mathin¤labelŽmath:NLSE propApprox[Wt_, ¤mmkappa_?NumericQ, ¤mmDeltaŽt_?NumericQ, M_Integer /; M >= 1,
	¤mathnl            v0_ /; VectorQ[v0, NumericQ]] :=
\end{mathematica}
and
\begin{mathematica}[firstnumber=\getrefnumber{mathline:1Dproppot2}]
	¤mathnl     propPot2[t_, v_] := Exp[¤mmlambda/2*((Wt[#,t]&/@xgrid) + ¤mmkappa*Abs[v]^2/¤mmDelta)]*v;
\end{mathematica}
As an example, we plot the time-dependent density for the time-independent gravitational well $W(x,t)=m g x$ and $\kappa=-3\cdot(g\hbar^4/m)^{1/3}$ (attractive interaction), $\kappa=0$ (no interaction), $\kappa=+3\cdot(g\hbar^4/m)^{1/3}$ (repulsive interaction):
\begin{mathematica}
	¤mathin With[{¤mmkappa = -3 * (g*¤mmhbar^4/m)^(1/3), ¤mmDeltaŽt = 50, M = 10^3},
	¤mathnl   ¤mmrho = ArrayPad[Abs[propApprox[W[#1]&,¤mmkappa,¤mmDeltaŽt,M,vv][[All,2]]]^2/¤mmDelta,{{0,0},{1,1}}];
	¤mathnl   ArrayPlot[Reverse[Transpose[¤mmrho]], DataRange -> {{0, ¤mmDeltaŽt}, {0, a}}]
\end{mathematica}
\begin{center}
\includegraphics[width=0.3\textwidth]{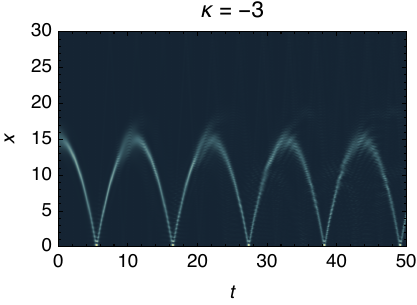}
\includegraphics[width=0.3\textwidth]{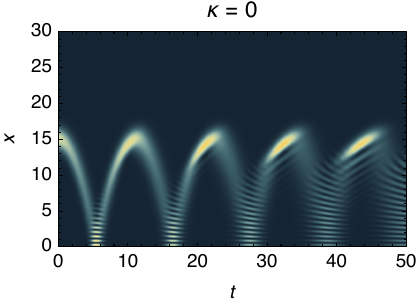}
\includegraphics[width=0.3\textwidth]{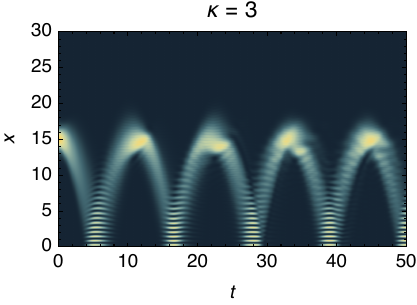}
\end{center}
Observations:
\begin{itemize}
	\item The noninteractive case ($\kappa=0$) shows a slow broadening and decoherence of the wavepacket.
	\item Attractive interactions ($\kappa<0$) make the wavepacket collapse to a tight spot and bounce almost like a classical particle.
	\item Repulsive interactions ($\kappa>0$) make the wavepacket broader, which slows down its decoherence.
\end{itemize}

\subsubsection{exercises}

\begin{questions}
	\item\label{Q:doublewell} Dimensionless problem (\mm{a=m=\mmhbar=1}): Given a particle moving in the range $x\in[0,1]$ with the non-linear Hamiltonian
	\begin{equation}
		\Ham = -\frac12 \frac{\dd^2}{\dd[x]^2} + \underbrace{\Omega\left[\left(\frac{x-\frac12}{\delta}\right)^2-1\right]^2}_{W(x)} + \kappa\abs{\psi(x)}^2,
	\end{equation}
	do the following calculations:
	\begin{enumerate}
		\item Plot the potential $W(x)$ for $\Omega=1$ and $\delta=\frac14$ (use $\kappa=0$). What are the main characteristics of this potential? \emph{Hint:} compute $W(\frac12)$, $W'(\frac12)$, $W(\frac12\pm\delta)$, $W'(\frac12\pm\delta)$.
		\item Calculate and plot the time-dependent density $\abs{\psi(x,t)}^2$ for $\Omega=250$, $\delta=\frac14$, and $\kappa=0$, starting from $\psi_0(x)\propto\exp\left[-\left(\frac{x-x_0}{2\sigma}\right)^2\right]$ with $x_0=\num{0,2694}$ and $\sigma=\num{0,0554}$. Calculate the probabilities for finding the particle in the left half ($x<\frac12$) and in the right half ($x>\frac12$) up to $t=20$. What do you observe?
		\item What do you observe for $\kappa=0.5$? Why?
	\end{enumerate}
\pagenote[\ref{Q:doublewell}]{
	\begin{enumerate}
		\item $W(x)$ is a double-well potential with minima at $x=\frac12\pm\delta$ and a barrier height of $\Omega$:
\begin{mathematica}
	¤protect¤mathin¤ W[{¤mmOmega_,¤ ¤mmdelta_},¤ x_]¤ =¤ ¤mmOmega*(((x-1/2)/¤mmdelta)^2-1)^2;
	¤protect¤mathin¤ Table[W[{¤mmOmega,¤mmdelta},x],¤ {x,1/2-¤mmdelta,1/2+¤mmdelta,¤mmdelta}]
	¤protect¤mathout¤ {0,¤ ¤mmOmega,¤ 0}
	¤protect¤mathin¤ Table[D[W[{¤mmOmega,¤mmdelta},y],y]¤ /.¤ y->x,¤ {x,1/2-¤mmdelta,1/2+¤mmdelta,¤mmdelta}]
	¤protect¤mathout¤ {0,¤ 0,¤ 0}
	¤protect¤mathin¤ Plot[W[{1,¤ 1/4},¤ x],¤ {x,¤ 0,¤ 1}]
\end{mathematica}
\begin{center}
\includegraphics[height=0.15\textheight]{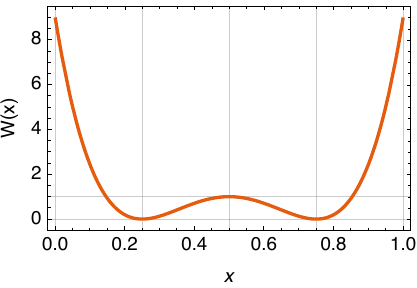}
\end{center}
		\item As in \ref{Q:Hscaling}, we use the split-step propagation code of \autoref{sec:dynamics1D} with the potential
\begin{mathematica}
	¤protect¤mathin¤ With[{¤mmOmega¤ =¤ 250,¤ ¤mmdelta¤ =¤ 1/4},
	¤protect¤mathnl¤ ¤ ¤ W[x_]¤ =¤ W[{¤mmOmega,¤ ¤mmdelta},¤ x];]
\end{mathematica}		
		With the initial state from \mm{\ref{math:initialgaussian}} (\ref{Q:Hscaling}) with \mm{x0=0.2694}, \mm{\mmsigma=0.0554}, \mm{k=0}, the time-dependent density is seen to oscillate between the wells:
\begin{mathematica}
	¤protect¤mathin¤ With[{¤mmDeltaŽt¤ =¤ 20,¤ M¤ =¤ 10^4},
	¤protect¤mathnl¤ ¤ ¤ V¤ =¤ propApprox[¤mmDeltaŽt,¤ M,¤ v0];]
	¤protect¤mathin¤ ¤mmrho¤ =¤ ArrayPad[(nmax+1)*Abs[#[[2]]]^2&¤ /@¤ V,¤ {{0,0},{1,1}}];
	¤protect¤mathin¤ ArrayPlot[Reverse[Transpose[¤mmrho]]]
\end{mathematica}
\begin{center}
\ifthenelse{\boolean{smallfigures}}%
{\includegraphics[height=0.15\textheight]{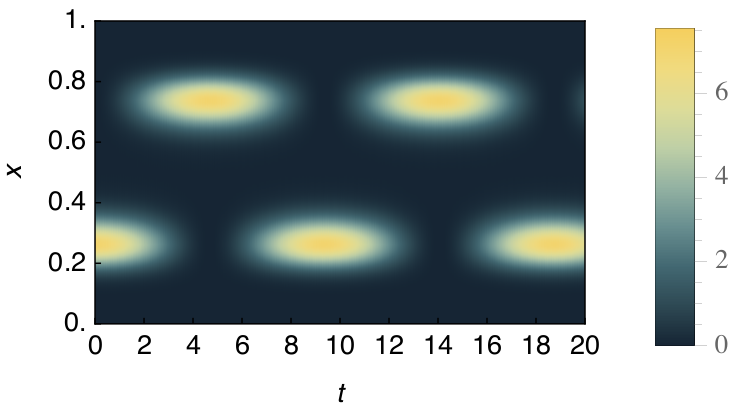}}%
{\includegraphics[height=0.15\textheight]{1Dparticledynamics_doublewell}}
\end{center}
		This oscillation is apparent in the left/right probabilities:
\begin{mathematica}
	¤protect¤mathin¤ ListLinePlot[{{#[[1]],Norm[#[[2,;;(nmax/2)]]]^2}&¤ /@¤ V,
	¤protect¤mathnl¤ ¤ ¤ ¤ ¤ ¤ ¤ ¤ ¤ ¤ ¤ ¤ ¤ ¤ ¤ {#[[1]],Norm[#[[2,nmax/2+1;;]]]^2}&¤ /@¤ V}]
\end{mathematica}
\begin{center}
\includegraphics[height=0.15\textheight]{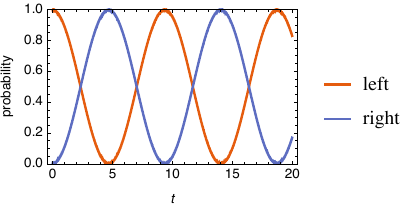}
\end{center}
		\item Now we use \mm{\ref{math:NLSE}} and observe that the attractive interactions prevent the particle from tunneling between the wells:
\begin{mathematica}
	¤protect¤mathin¤ With[{¤mmkappa¤ =¤ 0.5,¤ ¤mmDeltaŽt¤ =¤ 20,¤ M¤ =¤ 10^4},
	¤protect¤mathnl¤ ¤ ¤ V¤ =¤ propApprox[W[#1]&,¤ ¤mmkappa,¤ ¤mmDeltaŽt,¤ M,¤ v0];]
	¤protect¤mathin¤ ¤mmrho¤ =¤ ArrayPad[(nmax+1)*Abs[#[[2]]]^2&¤ /@¤ V,¤ {{0,0},{1,1}}];
	¤protect¤mathin¤ ArrayPlot[Reverse[Transpose[¤mmrho]]]
\end{mathematica}
\begin{center}
\ifthenelse{\boolean{smallfigures}}%
{\includegraphics[height=0.15\textheight]{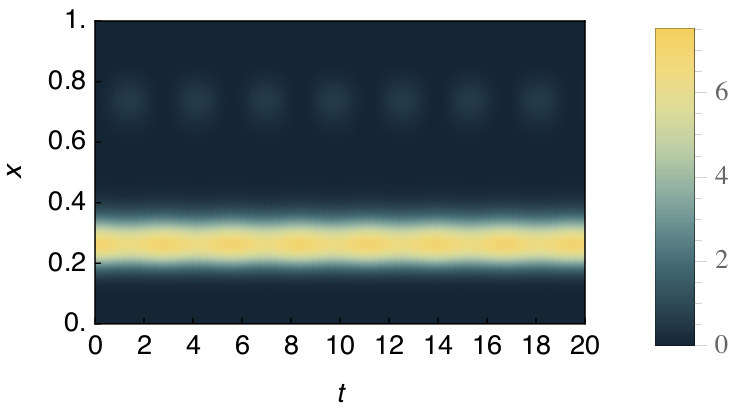}}%
{\includegraphics[height=0.15\textheight]{1Dparticledynamics_doublewell_interacting}}
\end{center}
\begin{mathematica}
	¤protect¤mathin¤ ListLinePlot[{{#[[1]],Norm[#[[2,;;(nmax/2)]]]^2}&¤ /@¤ V,
	¤protect¤mathnl¤ ¤ ¤ ¤ ¤ ¤ ¤ ¤ ¤ ¤ ¤ ¤ ¤ ¤ ¤ {#[[1]],Norm[#[[2,nmax/2+1;;]]]^2}&¤ /@¤ V}]
\end{mathematica}
\begin{center}
\includegraphics[height=0.15\textheight]{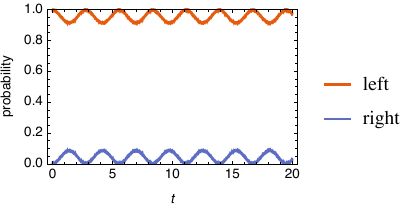}
\end{center}
	\end{enumerate}}
\end{questions}

\subsection[ground state of the non-linear Schr\"odinger equation]{\label{sec:itpGP}imaginary-time propagation for finding the ground state of the non-linear Schr\"odinger equation\hspace{\stretch{1}}\attachcode{GroundState}{non-linear Schr\"odinger equation and imaginary-time propagation in 1D}}\index{imaginary-time propagation}

In the previous section we have looked at the dynamical evolution of a Bose--Einstein condensate with the time-dependent nonlinear \hyperref[eq:TDGPE]{Schr\"odinger equation~\ref*{eq:TDGPE}}, which could be performed with minimal modifications to previous Mathematica code. The time-independent nonlinear \hyperref[eq:TIGPE]{Schr\"odinger equation~\ref*{eq:TIGPE}}, on the other hand, seems at first sight inaccessible to our established methods: it is an operator eigenvalue equation, with the operator acting non-linearly on the wavefunction and thus invalidating the matrix diagonalization method of \autoref{sec:TimeIndepSchr}. How can we determine the ground state of \autoref{eq:TIGPE}?

You may remember from statistical mechanics that at temperature $T$, the density operator of a system governed by a Hamiltonian $\Ham$ is
\begin{equation}
	\op{\rho}(\beta) = \frac{e^{-\beta\Ham}}{Z(\beta)}
\end{equation}
with $\beta=1/(k\ix{B}T)$ the reciprocal temperature in terms of the Boltzmann constant\index{Boltzmann constant} $k\ix{B}=\SI{1,3806488(13)e-23}{J/K}$. The partition function $Z(\beta)=\Tr e^{-\beta\Ham}$ ensures that the density operator has the correct norm, $\Tr\op{\rho}(\beta)=1$.

We know that at zero temperature the system will be in its ground state $\ket{\gamma}$,\footnote{For simplicity we assume here that the ground state is non-degenerate.}
\begin{equation}
	\lim_{\beta\to\infty} \op{\rho}(\beta) = \ket{\gamma}\bra{\gamma}.
\end{equation}
If we multiply this equation by an arbitrary state $\ket{\psi}$ from the right,  we find
\begin{equation}
	\lim_{\beta\to\infty} \op{\rho}(\beta)\ket{\psi} = \ket{\gamma}\scp{\gamma}{\psi}.
\end{equation}
Assuming that $\scp{\gamma}{\psi}\neq0$ (which is true for almost all states $\ket{\psi}$), the ground state is therefore
\begin{equation}
	\label{eq:itpgs}
	\ket{\gamma} = \frac{\lim_{\beta\to\infty} \op{\rho}(\beta)\ket{\psi}}{\scp{\gamma}{\psi}}
	= \frac{1}{\scp{\gamma}{\psi}} \lim_{\beta\to\infty} \frac{1}{Z(\beta)} \times e^{-\beta\Ham}\ket{\psi}.
\end{equation}
This means that if we take almost any state $\ket{\psi}$ and calculate $\lim_{\beta\to\infty} \frac{1}{Z(\beta)} e^{-\beta\Ham}\ket{\psi}$, we find a state that is proportional to the ground state (the prefactors $\frac{1}{\scp{\gamma}{\psi}}$ and $\frac{1}{Z(\beta)}$ are merely scalar prefactors). But we already know how to do this: the wavefunction $e^{-\beta\Ham}\ket{\psi}$ is calculated from $\ket{\psi}$ by \emph{imaginary-time propagation}. In fact the split-step algorithm of \autoref{sec:dynamics1D} remains valid if we replace $\ii(t-t_0)/\hbar \mapsto \beta$.
The advantage of \autoref{eq:itpgs} over the matrix method of \autoref{sec:TimeIndepSchr} is that the former can be implemented even if the Hamiltonian depends on the wavefunction, as in \autoref{eq:TIGPE}.
The only caveat is that, while regular time propagation (\autoref{sec:dynamics1D}) is unitary, imaginary-time propagation is not. The wavefunction must therefore be re-normalized after each imaginary-time evolution step (with the \mm{Normalize} function\index{Mathematica!vector!normalize}).

To implement this method of calculating the ground state by imaginary-time propagation, we set $\beta=M\cdot\ddd{\delta}{\beta}$ and modify \autoref{eq:itpgs} to
\begin{equation}
	\ket{\gamma} \propto \lim_{M\cdot\ddd{\delta}{\beta}\to\infty} e^{-M\,\ddd{\delta}{\beta}\,\Ham}\ket{\psi}
	= \lim_{M\cdot\ddd{\delta}{\beta}\to\infty} \left[ e^{-\ddd{\delta}{\beta}\,\Ham}\right]^M \ket{\psi}
	\stackrel{\stackrel{\text{Trotter \autoref{eq:Trotter}}}{\downarrow}}{=}
	\lim_{\ddd{\delta}{\beta}\to0} \lim_{M\cdot\ddd{\delta}{\beta}\to\infty} \left[ e^{-\frac{\ddd{\delta}{\beta}}{2}\op{V}}e^{-\ddd{\delta}{\beta}\,\op{T}}e^{-\frac{\ddd{\delta}{\beta}}{2}\op{V}}\right]^M \ket{\psi}.
\end{equation}
In practice we choose a small but finite ``imaginary-time'' step $\ddd{\delta}{\beta}$, and keep multiplying the wavefunction by $e^{-\frac{\ddd{\delta}{\beta}}{2}\op{V}}e^{-\ddd{\delta}{\beta}\,\op{T}}e^{-\frac{\ddd{\delta}{\beta}}{2}\op{V}}$ until the normalized wavefunction no longer changes and the infinite-$\beta$ limit ($M\cdot\ddd{\delta}{\beta}\to\infty$) has effectively been reached.
\begin{mathematica}
	¤mathin¤labelŽmath:gsnl groundstate[g_?NumericQ, ¤mmdelta¤mmbeta_?NumericQ, tolerance_: 10^-10] := 
	¤labelŽmathline:groundstatevars¤mathnl Module[{Ke, propKin, propPot2, v0, ¤mmgamma},
	¤mathnlc   (* compute the diagonal elements of exp[-¤mmdelta¤mmbeta*T] *)
	¤mathnl   Ke = Exp[-¤mmdelta¤mmbeta*Range[nmax]^2*¤mmpi^2/2] //N;
	¤mathnlc   (* propagate by a full imaginary-time-step with T *)
	¤mathnl   propKin[v_] := Normalize[FourierDST[Ke*FourierDST[v,1],1]];
	¤mathnlc   (* propagate by a half imaginary-time-step with V *)
	¤mathnl   propPot2[v_] := Normalize[Exp[-¤mmdelta¤mmbeta/2*(Wgrid + g*(nmax+1)*Abs[v]^2)]*v];
	¤mathnlc   (* propagate by a full imaginary-time-step by *)
	¤mathnlc   (* H=T+V using the Trotter approximation *)
	¤mathnl   prop[v_] := propPot2[propKin[propPot2[v]]];
	¤mathnlc   (* random starting point *)
	¤mathnl   v0 = Normalize@RandomComplex[{-1-I,1+I}, nmax];
	¤mathnlc   (* propagation to the ground state *)
	¤mathnl   ¤mmgamma = FixedPoint[prop,v0,SameTest->Function[{v1,v2},Norm[v1-v2]<tolerance]];
	¤labelŽmathline:groundstatelast¤mathnlc   (* return the ground-state coefficients *)
	¤mathnl   ¤mmgamma]
\end{mathematica}
The last argument, \mm{tolerance}, is optional and is given the default value $10^{-10}$ if not specified (see \autoref{sec:optionalargument}).
The \mm{FixedPoint} function\index{Mathematica!fixed point of a map} is used to apply the imaginary-time propagation until the result no longer changes (two consecutive results are considered equal if the function given as \mm{SameTest} returns true when applied to these two results).

Multiplying \autoref{eq:TIGPE} by $\psi^*(x)$ and integrating over $x$ gives
\begin{equation}
	\label{eq:chempot}
	\mu = \underbrace{-\frac{\hbar^2}{2m} \int_{-\infty}^{\infty}\dd[x]\psi^*(x)\psi''(x)}_{E\ix{kin}[\psi]}
		+ \underbrace{\int_{-\infty}^{\infty}\dd[x]V(x)\abs{\psi(x)}^2}_{E\ix{pot}[\psi]}
		+ \underbrace{g\int_{-\infty}^{\infty}\dd[x]\abs{\psi(x)}^4}_{2E\ix{int}[\psi]},
\end{equation}
which is very similar to \autoref{eq:nonlinearenergyfunctional} apart from a factor of two for $E\ix{int}$.
We use this to calculate the total energy and the chemical potential in \mm{\ref{math:gsnl}} by replacing lines~\ref*{mathline:groundstatelast}ff with
\begin{mathematica}[firstnumber=\getrefnumber{mathline:groundstatelast}]
	¤mathnlc   (* energy components *)
	¤mathnl   Ekin = ¤mmpi^2/2*Range[nmax]^2.Abs[FourierDST[¤mmgamma,1]]^2;
	¤mathnl   Epot = Wgrid.Abs[¤mmgamma]^2;
	¤mathnl   Eint = (g/2)(nmax+1)*Total[Abs[¤mmgamma]^4];
	¤mathnlc   (* total energy *)
	¤mathnl   Etot = Ekin + Epot + Eint;
	¤mathnlc   (* chemical potential *)
	¤mathnl   ¤mmmu = Ekin + Epot + 2*Eint;
	¤mathnlc   (* return energy, chemical potential, coefficients *)
	¤mathnl   {Etot, ¤mmmu, ¤mmgamma}]
\end{mathematica}
and adding the local variables \mm{Ekin}, \mm{Epot}, \mm{Eint}, \mm{Etot}, and \mm{\mmmu} on line~\ref*{mathline:groundstatevars}.

As an example we calculate the ground-state density for the gravity well of \autoref{sec:gravitywell} with three different values of the interaction strength $\kappa$ [in units of $(g\hbar^4/m)^{1/3}$]:
\begin{mathematica}
	¤mathin With[{¤mmkappa = 3 * (g*¤mmhbar^4/m)^(1/3), ¤mmdelta¤mmbeta = 10^-4},
	¤mathnl   {Etot, ¤mmmu, ¤mmgamma} = groundstate[¤mmdelta¤mmbeta, ¤mmkappa];
	¤mathnl   ListLinePlot[Join[{{0, 0}}, Transpose[{xgrid,Abs[¤mmgamma]^2/¤mmDelta}], {{a, 0}}],
	¤mathnl     PlotRange -> All, PlotLabel -> {Etot, ¤mmmu}]]
\end{mathematica}
\begin{center}
\includegraphics[width=\textwidth]{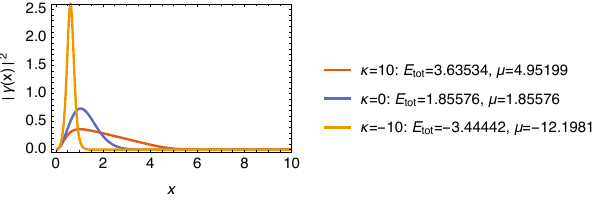}
\end{center}
Note that for $\kappa=0$ the Gross--Pitaevskii equation is the Schr\"odinger equation, and the chemical potential is equal to the total energy, matching the exact result of \mm{\ref{math:gravitywellgse}}.

\subsubsection{exercises}

\begin{questions}
	\item\label{Q:harmonicnl} Dimensionless problem (\mm{a=m=\mmhbar=1}): Given a particle moving in the range $x\in[0,1]$ with the non-linear Hamiltonian
	\begin{equation}
		\Ham = -\frac12 \frac{\dd^2}{\dd[x]^2} + 2500\left(x-\frac12\right)^2 + \kappa\abs{\psi(x)}^2,
	\end{equation}
	do the following calculations:
	\begin{enumerate}
		\item For $\kappa=0$ calculate the exact ground state $\ket{\zeta}$ (assuming that the particle can move in the whole domain $x\in\dsR$) and its energy eigenvalue. \emph{Hint:} assume $\zeta(x)=\exp\left[-\left(\frac{x-\frac12}{2\sigma}\right)^2\right]/\sqrt{\sigma\sqrt{2\pi}}$ and find the value of $\sigma$ that minimizes $\me{\zeta}{\Ham}{\zeta}$.
		\item Calculate the ground state $\lim_{\beta\to\infty} e^{-\beta\Ham}\ket{\zeta}$ and its chemical potential by imaginary-time propagation (with normalization of the wavefunction after each propagation step), using the code given above.
		\item Plot the ground-state density for different values of $\kappa$.
		\item Plot the total energy and the chemical potential as functions of $\kappa$.
	\end{enumerate}
\pagenote[\ref{Q:harmonicnl}]{
	\begin{enumerate}
		\item We do this calculation in Mathematica, for the more general potential $W(x)=\frac12 k(x-\frac12)^2$:
\begin{mathematica}
	¤protect¤mathin¤ ¤mmzeta[x_]¤ =¤ E^(-((x-1/2)/(2*¤mmsigma))^2)/Sqrt[¤mmsigma*Sqrt[2*¤mmpi]];
	¤protect¤mathin¤ Assuming[¤mmsigma>0,¤ Integrate[¤mmzeta[x]^2,¤ {x,¤ -¤mminfty,¤ ¤mminfty}]]
	¤protect¤mathout¤ 1
	¤protect¤mathin¤ Assuming[¤mmsigma>0,
	¤protect¤mathnl¤ ¤ ¤ e¤ =¤ Integrate[¤mmzeta[x]*(-1/2*¤mmzeta¤textquotesingle¤textquotesingleŽ[x]¤ +¤ 1/2*k*(x-1/2)^2*¤mmzeta[x]),¤ {x,¤ -¤mminfty,¤ ¤mminfty}]]
	¤protect¤mathout¤ (1+4*k*¤mmsigma^4)/(8*¤mmsigma^2)
	¤protect¤mathin¤ Solve[D[e,¤ ¤mmsigma]¤ ==¤ 0,¤ ¤mmsigma]
	¤protect¤mathout¤ {{¤mmsigma¤ ->¤ -1/(Sqrt[2]*k^(1/4))},¤ {¤mmsigma¤ ->¤ -I/(Sqrt[2]*k^(1/4))},
	¤protect¤mathnl¤ ¤ {¤mmsigma¤ ->¤ I/(Sqrt[2]*k^(1/4))},¤ {¤mmsigma¤ ->¤ 1/(Sqrt[2]*k^(1/4))}}
\end{mathematica}
		Of these four solutions, we choose $\sigma=(4k)^{-1/4}$ because it is real and positive:
\begin{mathematica}
	¤protect¤mathin¤ e¤ /.¤ ¤mmsigma¤ ->¤ (4*k)^(-1/4)
	¤protect¤mathout¤ Sqrt[k]/2
\end{mathematica}
		For $k=\num{5000}$, the ground state is therefore $\zeta(x)$ with $\sigma=\num{20000}^{-1/4}\approx\num{0,0840896}$ and energy $E=\sqrt{\num{5000}}/2\approx\num{35,3553}$.
		\item We use the same code as in \autoref{sec:itpGP} but with the potential
\begin{mathematica}
	¤protect¤mathin¤ With[{k¤ =¤ 5000},
	¤protect¤mathnl¤ ¤ ¤ W[x_]¤ =¤ 1/2*k*(x-1/2)^2;]
\end{mathematica}
	Further, we use \mm{nmax=1000} to describe the wavefunction with strongly attractive interactions better.
	The result matches the Gaussian approximation: both the energy and the chemical potential are approximately $\sqrt{k}/2$,
\begin{mathematica}
	¤protect¤mathin¤ groundstate[10^-4,¤ 0][[;;2]]
	¤protect¤mathout¤ {35.3553,¤ 35.3553}
\end{mathematica}
		\item Ground-state density for repulsive interactions:
\begin{mathematica}
	¤protect¤mathin¤ With[{¤mmkappa¤ =¤ 100,¤ ¤mmdelta¤mmbeta¤ =¤ 10^-4},
	¤protect¤mathnl¤ ¤ ¤ {Etot,¤ ¤mmmu,¤ ¤mmgamma}¤ =¤ groundstate[¤mmdelta¤mmbeta,¤ ¤mmkappa];
	¤protect¤mathnl¤ ¤ ¤ ListLinePlot[Join[{{0,¤ 0}},¤ Transpose[{xgrid,Abs[¤mmgamma]^2/¤mmDelta}],¤ {{a,¤ 0}}],
	¤protect¤mathnl¤ ¤ ¤ ¤ ¤ PlotRange¤ ->¤ All,¤ PlotLabel¤ ->¤ {Etot,¤ ¤mmmu}]]
\end{mathematica}
\begin{center}
\includegraphics[height=0.15\textheight]{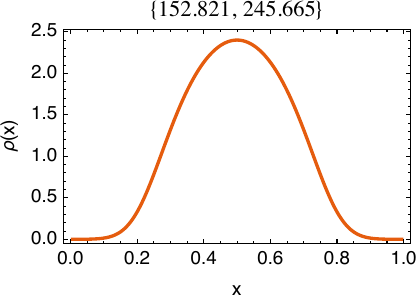}
\end{center}
		Ground-state density for no interactions:
\begin{mathematica}
	¤protect¤mathin¤ With[{¤mmkappa¤ =¤ 0,¤ ¤mmdelta¤mmbeta¤ =¤ 10^-4},
	¤protect¤mathnl¤ ¤ ¤ {Etot,¤ ¤mmmu,¤ ¤mmgamma}¤ =¤ groundstate[¤mmdelta¤mmbeta,¤ ¤mmkappa];
	¤protect¤mathnl¤ ¤ ¤ ListLinePlot[Join[{{0,¤ 0}},¤ Transpose[{xgrid,Abs[¤mmgamma]^2/¤mmDelta}],¤ {{a,¤ 0}}],
	¤protect¤mathnl¤ ¤ ¤ ¤ ¤ PlotRange¤ ->¤ All,¤ PlotLabel¤ ->¤ {Etot,¤ ¤mmmu}]]
\end{mathematica}
\begin{center}
\includegraphics[height=0.15\textheight]{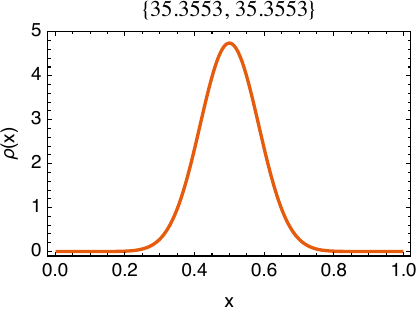}
\end{center}
		Ground-state density for attractive interactions:
\begin{mathematica}
	¤protect¤mathin¤ With[{¤mmkappa¤ =¤ -100,¤ ¤mmdelta¤mmbeta¤ =¤ 10^-4},
	¤protect¤mathnl¤ ¤ ¤ {Etot,¤ ¤mmmu,¤ ¤mmgamma}¤ =¤ groundstate[¤mmdelta¤mmbeta,¤ ¤mmkappa];
	¤protect¤mathnl¤ ¤ ¤ ListLinePlot[Join[{{0,¤ 0}},¤ Transpose[{xgrid,Abs[¤mmgamma]^2/¤mmDelta}],¤ {{a,¤ 0}}],
	¤protect¤mathnl¤ ¤ ¤ ¤ ¤ PlotRange¤ ->¤ All,¤ PlotLabel¤ ->¤ {Etot,¤ ¤mmmu}]]
\end{mathematica}
\begin{center}
\includegraphics[height=0.15\textheight]{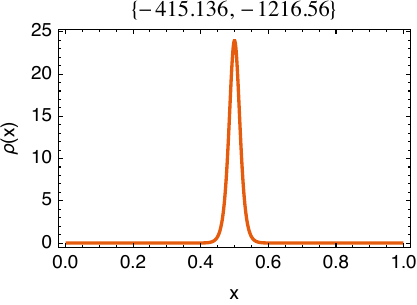}
\end{center}
		\item The energy and chemical potential differ for $\kappa\neq0$:
\begin{mathematica}
	¤protect¤mathin¤ With[{¤mmdelta¤mmbeta¤ =¤ 10^-4},
	¤protect¤mathnl¤ ¤ ¤ ListLinePlot[Transpose[Table[{{¤mmkappa,groundstate[¤mmdelta¤mmbeta,¤mmkappa][[1]]},
	¤protect¤mathnl¤ ¤ ¤ ¤ ¤ {¤mmkappa,groundstate[¤mmdelta¤mmbeta,¤mmkappa][[2]]}},¤ {¤mmkappa,¤ -100,¤ 100,¤ 10}]]]]
\end{mathematica}
\begin{center}
\includegraphics[height=0.15\textheight]{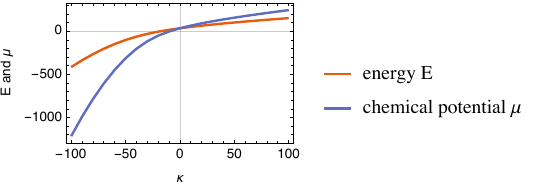}
\end{center}
	\end{enumerate}}
\end{questions}

\section{several particles in one dimension: interactions}\index{interaction}

In \autoref{sec:tdGPE} we have studied a simple mean-field description of many-particle systems, with the advantage of simplicity and the disadvantage of not describing inter-particle correlations. Here we use a different approach that captures the full quantum mechanics of many-particle systems (including correlations), with the disadvantage of much increased calculation size.

We have seen in \autoref{sec:coupledDOF} how to describe quantum-mechanical systems with more than one degree of freedom. This method can be used  for describing several particles moving in one dimension. In the following we look at two examples of interacting particles.

When more than one particle is present in a system, we must distinguish between bosons\index{boson} and fermions\index{fermion}. Whenever the Hamiltonian is symmetric under particle exchange (which is the case in this section), each one of its eigenstates can be associated with an irreducible representation of the particle permutation group. For two particles, the only available choices are the symmetric and the antisymmetric irreducible representations, and therefore every numerically calculated eigenstate can be labeled as either bosonic (symmetric) or fermionic (antisymmetric). For more particles, however, other irreducible representations exist,\footnote{See \url{https://en.wikipedia.org/wiki/Irreducible_representation}.} meaning that some numerically calculated eigenstates of the Hamiltonian may not be physical at all because they are neither bosonic (fully symmetric) nor fermionic (fully antisymmetric).

\subsection[two identical particles in one dimension with contact interaction]{\label{sec:contactinteraction}two identical particles in one dimension with contact interaction\hspace{\stretch{1}}\attachcode{ContactInteraction}{two particles in 1D with contact interaction}}\index{contact interaction}

We first look at two identical particles moving in a one-dimensional square well of width $a$ and interacting through a contact potential $V\ix{int}(x_1,x_2)=\kappa\times\delta(x_1-x_2)$. Such potentials are a good approximation of the $s$-wave scattering interactions taking place in cold dilute gases. The Hamiltonian of this system is\footnote{Notice that we write this Hamiltonian in an abbreviated form. The full operator form, with terms similar to \autoref{eq:potop} but containing double integrals over space, is cumbersome to write (see \autoref{eq:Hintfull}).}
\begin{equation}
	\label{eq:Ham2contact}
	\Ham = \underbrace{-\frac{\hbar^2}{2m} \left[ \frac{\partial^2}{\ddd{\partial}{x}_1^2} + \frac{\partial^2}{\ddd{\partial}{x}_2^2} \right]}_{\op{T}}
		+ \underbrace{V(x_1) + V(x_2)}_{\op{V}} + \underbrace{\kappa \delta(x_1-x_2)}_{\Ham\ix{int}},
\end{equation}
where $V(x)$ is the single-particle potential (as in \autoref{sec:1part1dim}) and $\kappa$ is the interaction strength, often related to the $s$-wave scattering length $a_s$.\index{s-wave scattering@$s$-wave scattering} For the time being we do not need to specify whether the particles are bosons or fermions.

We describe this system with the tensor-product basis constructed from two finite-resolution position basis sets:\index{tensor!product}
\begin{equation}
	\label{eq:basis2D}
	\ket{j_1, j_2} = \ket{j_1} \otimes \ket{j_2} \qquad \text{for $j_1,j_2 \in\{1,2,3,\dots,n\ix{max}\}$}.
\end{equation}
Most of the matrix representations of the terms in \autoref{eq:Ham2contact} are constructed as tensor products of the matrix representations of the corresponding single-particle representations since $\op{T}=\op{T}_1\otimes\one+\one\otimes\op{T}_2$ and $\op{V}=\op{V}_1\otimes\one+\one\otimes\op{V}_2$. The only new element is the interaction Hamiltonian $\Ham\ix{int}$. Remembering that its formal operator definition is
\begin{equation}
	\label{eq:Hintfull}
	\Ham\ix{int} = \kappa \int_{-\infty}^{\infty} \dd[x_1]\dd[x_2] \Big[\ket{x_1} \otimes \ket{x_2}\Big] \delta(x_1-x_2) \Big[\bra{x_1} \otimes \bra{x_2}\Big]
\end{equation}
(while \autoref{eq:Ham2contact} is merely a shorthand notation), we calculate its matrix elements in the finite-precision position basis with
\begin{equation}
	\label{eq:interactionme1}
	\me{j_1,j_2}{\Ham\ix{int}}{j_1',j_2'} = \kappa \int_{-\infty}^{\infty} \dd[x_1]\dd[x_2] \scp{j_1}{x_1}\scp{j_2}{x_2} \delta(x_1-x_2) \scp{x_1}{j_1'} \scp{x_2}{j_2'}
	= \kappa \int_0^a \dd[x] \vartheta_{j_1}(x)\vartheta_{j_2}(x)\vartheta_{j_1'}(x)\vartheta_{j_2'}(x).
\end{equation}
These quartic overlap integrals can be calculated by a four-dimensional type-I discrete sine transform (see \autoref{eq:Xnj} and \ref{Q:quarticoverlaps}),
\begin{multline}
	\label{eq:overlap4}
	\int_0^a \dd[x] \vartheta_{j_1}(x)\vartheta_{j_2}(x)\vartheta_{j_3}(x)\vartheta_{j_4}(x) =\\
	\frac{1}{2a} \sum_{n_1,n_2,n_3,n_4=1}^{n\ix{max}}
	X_{n_1 j_1} X_{n_2 j_2} X_{n_3 j_3} X_{n_4 j_4}
	 \Big[
		\delta_{n_1+n_2,n_3+n_4}+\delta_{n_1+n_3,n_2+n_4}+\delta_{n_1+n_4,n_2+n_3}\\
		-\delta_{n_1,n_2+n_3+n_4}-\delta_{n_2,n_1+n_3+n_4}-\delta_{n_3,n_1+n_2+n_4}-\delta_{n_4,n_1+n_2+n_3} \Big],
\end{multline}
which we evaluate in Mathematica very efficiently and all at once with
\begin{mathematica}
	¤mathin¤labelŽmath:quarticoverlaps overlap4 = FourierDST[Table[KroneckerDelta[n1+n2,n3+n4]
	¤mathnl  +KroneckerDelta[n1+n3,n2+n4]+KroneckerDelta[n1+n4,n2+n3]
	¤mathnl  -KroneckerDelta[n1,n2+n3+n4]-KroneckerDelta[n2,n1+n3+n4]
	¤mathnl  -KroneckerDelta[n3,n1+n2+n4]-KroneckerDelta[n4,n1+n2+n3],
	¤mathnl  {n1,nmax},{n2,nmax},{n3,nmax},{n4,nmax}],1]/(2*a);
\end{mathematica}

\paragraph{Mathematica code}
As before, we assume that the quantities $a$, $m$, and $\hbar$ are expressed in a suitable set of units (see \autoref{sec:1Dunits}).
First we define the grid size and the unit operator \mm{id} acting on a single particle:
\begin{mathematica}
	¤mathin m = ¤mmhbar = 1;   (* for example *)
	¤mathin a = 1;       (* for example *)
	¤mathin nmax = 50;   (* for example *)
	¤mathin ¤mmDelta = a/(nmax+1);
	¤mathin xgrid = Range[nmax]*¤mmDelta;
	¤mathin id = IdentityMatrix[nmax, SparseArray];
\end{mathematica}
The total kinetic Hamiltonian is assembled via a Kronecker product\index{Mathematica!Kronecker product} (tensor product) of the two single-particle kinetic Hamiltonians:
\begin{mathematica}
	¤mathin T1M = SparseArray[Band[{1,1}]->Range[nmax]^2*¤mmpi^2*¤mmhbar^2/(2*m*a^2)];
	¤mathin T1P = FourierDST[T1M, 1];
	¤mathin TP = KroneckerProduct[T1P, id] + KroneckerProduct[id, T1P];
\end{mathematica}
The same for the potential Hamiltonian (here we assume no potential, that is, a square well; but you may modify this):
\begin{mathematica}
	¤mathin W[x_] = 0;
	¤mathin Wgrid = W /@ xgrid;
	¤mathin V1P = SparseArray[Band[{1,1}]->Wgrid];
	¤mathin VP = KroneckerProduct[V1P, id] + KroneckerProduct[id, V1P];
\end{mathematica}
The interaction Hamiltonian is constructed from \mm{\ref{math:quarticoverlaps}} with the \mm{ArrayFlatten} command, which flattens the combination basis set $\ket{j_1}\otimes\ket{j_2}$ into a single basis set $\ket{j_1,j_2}$, or in other words, which converts the $n\ix{max}\times n\ix{max}\times n\ix{max}\times n\ix{max}$-matrix \mm{overlap4} into a $n\ix{max}^2\times n\ix{max}^2$-matrix:
\begin{mathematica}
	¤mathin HintP = ArrayFlatten[overlap4];
\end{mathematica}
The full Hamiltonian, in which the amplitude of the potential can be adjusted with the prefactor $\Omega$ and the interaction strength with $g$, is
\begin{mathematica}
	¤mathin¤labelŽmath:contactHam HP[¤mmOmega_, ¤mmkappa_] = TP + ¤mmOmega*VP + ¤mmkappa*HintP;
\end{mathematica}
We calculate eigenstates (the ground state, for example) with the methods already described previously. The resulting wavefunctions are in the tensor-product basis of \autoref{eq:basis2D}, and they can be plotted with\index{Mathematica!plotting}
\begin{mathematica}
	¤mathin¤labelŽmath:plot2Dwf plot2Dwf[¤mmpsi_] := Module[{¤mmpsi1,¤mmpsi2},
	¤mathnlc   (* make a square array of wavefunction values *)
	¤mathnl   ¤mmpsi1 = ArrayReshape[¤mmpsi, {nmax,nmax}];
	¤mathnlc   (* add a frame of zeros at the edges *)
	¤mathnlc   (* representing the boundary conditions *)
	¤mathnl   ¤mmpsi2 = ArrayPad[¤mmpsi1, 1];
	¤mathnlc   (* plot *)
	¤mathnl   ArrayPlot[Reverse[Transpose[¤mmpsi2]]]
\end{mathematica}
Assuming that a given wavefunction \mm{\mmpsi} is purely real-valued,\footnote{The eigenvectors of Hermitian operators can always be chosen to have real coefficients. Proof: Suppose that $\matr{H}\cdot\vect{\psi}=E\vect{\psi}$ for a vector $\vect{\psi}$ with complex entries. Complex-conjugate the eigenvalue equation, $\matr{H}\dagg\cdot\vect{\psi}^*=E^*\vect{\psi}^*$; but $\matr{H}\dagg=\matr{H}$ and $E^*=E$, and hence $\vect{\psi}^*$ is also an eigenvector of $\matr{H}$ with eigenvalue $E$. Thus we can introduce two real-valued vectors $\vect{\psi}\ix{r}=\vect{\psi}+\vect{\psi}^*$ and $\vect{\psi}\ix{i}=\ii(\vect{\psi}-\vect{\psi}^*)$, representing the real and imaginary parts of $\vect{\psi}$, respectively, which are both eigenvectors of $\matr{H}$ with eigenvalue $E$. Mathematica (as well as most other matrix diagonalization algorithms) automatically detect Hermitian matrices and return eigenvectors with real coefficients.} we can plot it with
\begin{mathematica}
	¤mathin plot2Dwf[v/¤mmDelta]
\end{mathematica}
Here we plot the four lowest-energy wavefunctions for $\Omega=0$ (no potential, the particles move in a simple infinite square well) and $\kappa=+25$ (repulsive interaction), using $n\ix{max}=50$ grid points, with the title of each panel showing the energy and the symmetry (see below). White corresponds to zero wavefunction, red is positive $\psi(x_1,x_2)>0$, and blue is negative $\psi(x_1,x_2)<0$.
\begin{center}
\includegraphics[width=0.24\textwidth]{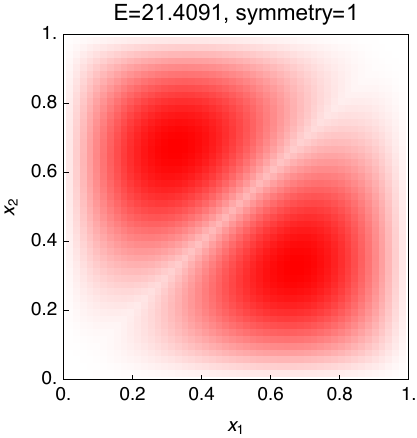}
\includegraphics[width=0.24\textwidth]{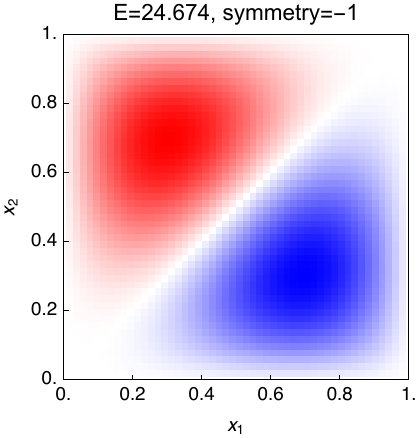}
\includegraphics[width=0.24\textwidth]{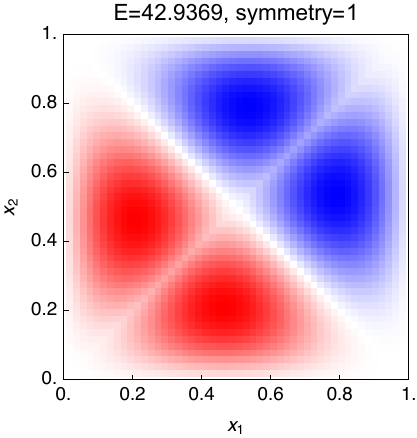}
\includegraphics[width=0.24\textwidth]{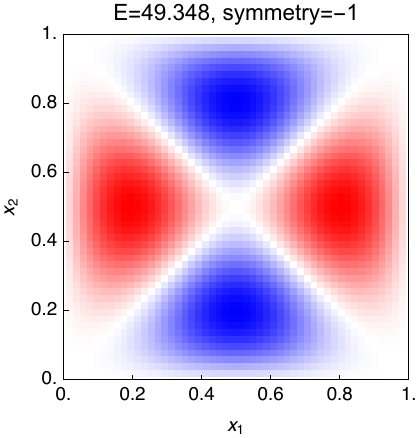}
\end{center}
We can see that in the ground state for $g>0$ the particles avoid each other, \ie, the ground-state wavefunction $\psi(x_1,x_2)$ is reduced whenever $x_1=x_2$.

And here are the lowest four energy eigenstate wavefunctions for $\kappa=-10$:
\begin{center}
\includegraphics[width=0.24\textwidth]{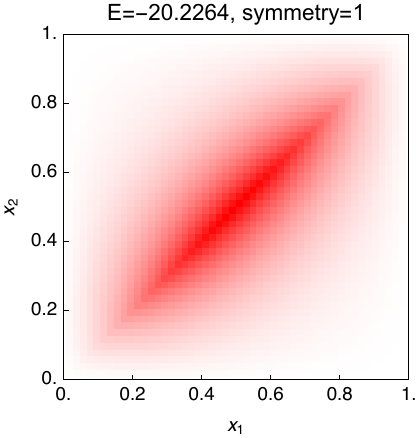}
\includegraphics[width=0.24\textwidth]{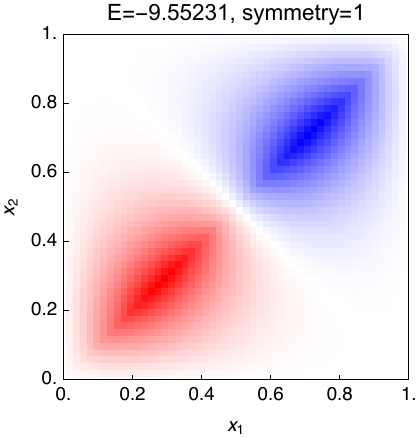}
\includegraphics[width=0.24\textwidth]{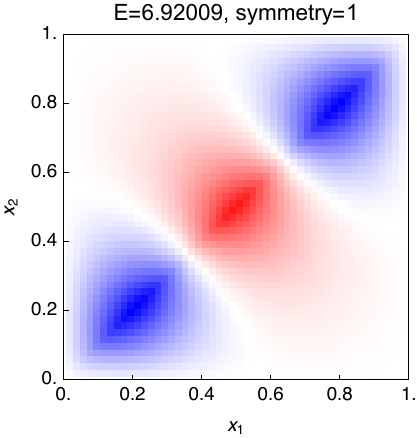}
\includegraphics[width=0.24\textwidth]{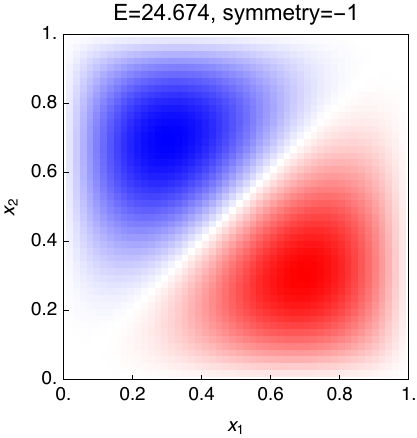}
\end{center}
We can see that in the ground state for $\kappa<0$ the particles attract each other, \ie, the ground-state wavefunction $\psi(x_1,x_2)$ is increased whenever $x_1=x_2$.
We also notice that the second-lowest state for $\kappa=+25$ is exactly equal to the fourth-lowest state for $\kappa=-10$: its wavefunction vanishes whenever $x_1=x_2$ and thus the contact interaction has no influence on this state.

In the above plots we have noted the symmetry of each eigenstate (symmetric or antisymmetric with respect to particle exchange), which is calculated with the integral
\begin{equation}
	\label{eq:symmetry}
	\mathcal{S}[\psi] = \int_0^a \dd[x_1] \dd[x_2] \psi^*(x_1,x_2) \psi(x_2,x_1)
	= \begin{cases}
		+1 & \text{for symmetric states $\psi(x_2,x_1)=\psi(x_1,x_2)$,}\\
		-1 & \text{for antisymmetric states $\psi(x_2,x_1)=-\psi(x_1,x_2)$.}
	\end{cases}
\end{equation}
In Mathematica, the mirrored wavefunction $\psi(x_2,x_1)$ is calculated with the particle interchange operator $\op{\Xi}$ defined as
\begin{mathematica}
	¤mathin¤labelŽmath:interchange ¤mmXi = ArrayFlatten[SparseArray[{i_,j_,j_,i_} -> 1, {nmax, nmax, nmax, nmax}]];
\end{mathematica}
such that $\psi(x_2,x_1)=\scp{x_2,x_1}{\psi}=\me{x_1,x_2}{\op{\Xi}}{\psi}$. The symmetry of a state, defined in \autoref{eq:symmetry}, is therefore the expectation value of the $\op{\Xi}$ operator:
\begin{mathematica}
	¤mathin symmetry[v_] := Re[Conjugate[v].(¤mmXi.v)]
\end{mathematica}
Here we show the numerical energy eigenvalues of the contact interaction Hamiltonian, colored according to their symmetry: red dots indicate symmetric states ($\mathcal{S}=+1$), whereas blue dots indicate antisymmetric states ($\mathcal{S}=-1$).
\begin{center}
\includegraphics[width=0.5\textwidth]{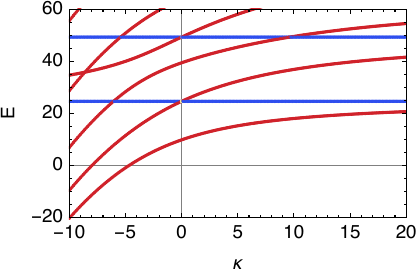}
\end{center}
In this representation it becomes even clearer that antisymmetric states are independent of the contact interaction because their wavefunction vanishes whenever $x_1=x_2$ (see \ref{Q:interchange2}).

\subsubsection{bosons and fermions}
\index{boson}\index{fermion}

The reason why every state in the above calculation is either symmetric or antisymmetric with respect to particle interchange is that the Hamiltonian \mm{\ref{math:contactHam}} commutes with the particle interchange operator \mm{\ref{math:interchange}} (see \ref{Q:interchange1}). As a result, $\Ham$ and $\op{\Xi}$ can be diagonalized simultaneously.

We notice that $\op{\Xi}$ has only eigenvalues $\pm1$:
\begin{mathematica}
	¤mathin ¤mmXi //Eigenvalues //Counts
	¤mathout <| -1 -> 1225, 1 -> 1275 |>
\end{mathematica}
The $n\ix{max}(n\ix{max}+1)/2$ eigenvalues $+1$ correspond to eigenvectors that are symmetric under particle interchange and form a basis of the symmetric subspace of the full Hilbert space (bosonic states); the $n\ix{max}(n\ix{max}-1)/2$ eigenvalues $-1$ correspond to eigenvectors that are antisymmetric under particle interchange and form a basis of the antisymmetric subspace of the full Hilbert space (fermionic states). By constructing a matrix whose rows are the symmetric eigenvectors, we construct an operator $\op{\Pi}\ix{s}$ that projects from the full Hilbert space onto the space of symmetric states,
\begin{mathematica}
	¤mathin ¤mmepsilon = Transpose[Eigensystem[Normal[¤mmXi]]];
	¤mathin ¤mmPiŽs = Select[¤mmepsilon, #[[1]] == 1 &][[All, 2]] //Orthogonalize //SparseArray;
\end{mathematica}
Similarly we construct a projector $\op{\Pi}\ix{a}$ onto the space of antisymmetric states,
\begin{mathematica}
	¤mathin ¤mmPiŽa = Select[¤mmepsilon, #[[1]] == -1 &][[All, 2]] //Orthogonalize //SparseArray;
\end{mathematica}
With the help of these projectors, we define the Hamiltonians of the system restricted to the symmetric or antisymmetric subspace, respectively:
\begin{mathematica}
	¤mathin HPs[¤mmOmega_, ¤mmkappa_] = ¤mmPiŽs.HP[¤mmOmega,¤mmkappa].Transpose[¤mmPiŽs];
	¤mathin¤labelŽmath:HPa HPa[¤mmOmega_, ¤mmkappa_] = ¤mmPiŽa.HP[¤mmOmega,¤mmkappa].Transpose[¤mmPiŽa];
\end{mathematica}
If the two particles in the present problem are indistinguishable bosons, then they can only populate the symmetric states (red dots in the above eigenvalue plot). We calculate the $m$ lowest energy eigenstates of this symmetric subspace with the restricted Hamiltonian \mm{HPs}:
\begin{mathematica}
	¤mathin Clear[sgs];
	¤mathin sgs[¤mmOmega_?NumericQ, ¤mmkappa_?NumericQ, m_Integer /; m >= 1] := sgs[¤mmOmega, ¤mmkappa, m] =
	¤mathnl   {-#[[1]], #[[2]].¤mmPiŽs} &[Eigensystem[-HPs[N[¤mmOmega], N[¤mmkappa]], m,
	¤mathnl   Method -> {"Arnoldi", "Criteria" -> "RealPart", MaxIterations -> 10^6}]]
\end{mathematica}
Notice that we convert the calculated eigenstates back into the full Hilbert space by multiplying the results with \mm{\mmPi s} from the right.

In the same way, if the two particles in the present problem are indistinguishable fermions, then they can only populate the antisymmetric states (blue dots in the above eigenvalue plot). We calculate the $m$ lowest energy eigenstates of this antisymmetric subspace with the restricted Hamiltonian \mm{HPa}:
\begin{mathematica}
	¤mathin Clear[ags];
	¤mathin¤labelŽmath:ags ags[¤mmOmega_?NumericQ, ¤mmkappa_?NumericQ, m_Integer /; m >= 1] := ags[¤mmOmega, ¤mmkappa, m] =
	¤mathnl   {-#[[1]], #[[2]].¤mmPiŽa} &[Eigensystem[-HPa[N[¤mmOmega], N[¤mmkappa]], m,
	¤mathnl   Method -> {"Arnoldi", "Criteria" -> "RealPart", MaxIterations -> 10^6}]]
\end{mathematica}
As an example, here we calculate the six lowest energy eigenvalues of the full Hamiltonian for $\Omega=0$ and $\kappa=5$:
\begin{mathematica}
	¤mathin gs[0, 5, 6][[1]] //Sort
	¤mathout¤labelŽmath:cigs {15.2691, 24.674, 32.3863, 45.4849, 49.348, 58.1333}
\end{mathematica}
The six lowest \emph{symmetric} energy eigenvalues are
\begin{mathematica}
	¤mathin sgs[0, 5, 6][[1]] //Sort
	¤mathout¤labelŽmath:scigs {15.2691, 32.3863, 45.4849, 58.1333, 72.1818, 93.1942}
\end{mathematica}
The six lowest \emph{antisymmetric} energy eigenvalues are
\begin{mathematica}
	¤mathin ags[0, 5, 6][[1]] //Sort
	¤mathout¤labelŽmath:acigs {24.674, 49.348, 64.1524, 83.8916, 98.696, 123.37}
\end{mathematica}
From \mm{\ref{math:scigs}} and \mm{\ref{math:acigs}}  we can see which levels of \mm{\ref{math:cigs}} are symmetric or antisymmetric.

\subsubsection{exercises}

\begin{questions}
	\item\label{Q:quarticoverlaps} Show that \autoref{eq:overlap4} is plausible by setting \mm{nmax=3}, evaluating \mm{\ref{math:quarticoverlaps}}, and then comparing its values to explicit integrals from \autoref{eq:overlap4} for several tuples $(j_1,j_2,j_3,j_4)$. \emph{Hint:} use \mm{a=1} for simplicity.
\pagenote[\ref{Q:quarticoverlaps}]{We do this calculation for $a=1$; the prefactor $a^{-1}$ of the right-hand side of \autoref{eq:overlap4} can be found by a variable substitution $x\mapsto a x'$. From the momentum basis functions
\begin{mathematica}
	¤protect¤mathin¤ ¤mmphi[n_,¤ x_]¤ =¤ Sqrt[2]*Sin[n*¤mmpi*x];
\end{mathematica}
		we define the position basis functions
\begin{mathematica}
	¤protect¤mathin¤ ¤mmtheta[nmax_,¤ j_,¤ x_]¤ :=¤ 1/Sqrt[nmax+1]*Sum[¤mmphi[n,j/(nmax+1)]*¤mmphi[n,x],¤ {n,nmax}]
\end{mathematica}
		The exact overlap integrals are
\begin{mathematica}
	¤protect¤mathin¤ J[nmax_,¤ {j1_,j2_,j3_,j4_}]¤ :=
	¤protect¤mathnl¤ ¤ ¤ Integrate[¤mmtheta[nmax,j1,x]*¤mmtheta[nmax,j2,x]*¤mmtheta[nmax,j3,x]*¤mmtheta[nmax,j4,x],¤ {x,0,1}]
\end{mathematica}
	We make a table of overlap integrals, calculated both exactly and approximately through \mm{\ref{math:quarticoverlaps}}, and show that the difference is zero (up to numerical inaccuracies):
\begin{mathematica}
	¤protect¤mathin¤ With[{nmax¤ =¤ 3},
	¤protect¤mathnl¤ ¤ ¤ A¤ =¤ Table[J[nmax,¤ {j1,j2,j3,j4}],¤ {j1,nmax},{j2,nmax},{j3,nmax},{j4,nmax}];
	¤protect¤mathnl¤ ¤ ¤ B¤ =¤ FourierDST[Table[KroneckerDelta[n1+n2,n3+n4]
	¤protect¤mathnl¤ ¤ ¤ ¤ ¤ +KroneckerDelta[n1+n3,n2+n4]+KroneckerDelta[n1+n4,n2+n3]
	¤protect¤mathnl¤ ¤ ¤ ¤ ¤ -KroneckerDelta[n1,n2+n3+n4]-KroneckerDelta[n2,n1+n3+n4]
	¤protect¤mathnl¤ ¤ ¤ ¤ ¤ -KroneckerDelta[n3,n1+n2+n4]-KroneckerDelta[n4,n1+n2+n3],
	¤protect¤mathnl¤ ¤ ¤ ¤ ¤ {n1,nmax},¤ {n2,nmax},¤ {n3,nmax},¤ {n4,nmax}],1]/2;]
	¤protect¤mathnl¤ A¤ -¤ B¤ //Abs¤ //Max
	¤protect¤mathout¤ 8.88178*10^-16
\end{mathematica}}
	\item\label{Q:distance} In the problem of \autoref{sec:contactinteraction}, calculate the expectation value of the inter-particle distance $\avg{x_1-x_2}$, and its variance $\avg{(x_1-x_2)^2}-\avg{x_1-x_2}^2$, in the ground state as a function of $\kappa$ (still keeping $\Omega=0$). \emph{Hint:} Using \autoref{eq:positionoperator}, the position operators \mm{x1} and \mm{x2} are approximately
\begin{mathematica}
	¤mathin x = SparseArray[Band[{1,1}]->xgrid];
	¤mathin x1 = KroneckerProduct[x, id];
	¤mathin x2 = KroneckerProduct[id, x];
\end{mathematica}
\pagenote[\ref{Q:distance}]{We define memoizing functions that calculate $\avg{\op{x}_1-\op{x}_2}$ and $\avg{(\op{x}_1-\op{x}_2)^2}$ with
\begin{mathematica}
	¤protect¤mathin¤ Clear[¤mmDeltaŽ1,¤mmDeltaŽ2];
	¤protect¤mathin¤ ¤mmDeltaŽ1[¤mmkappa_?NumericQ]¤ :=¤ ¤mmDeltaŽ1[¤mmkappa]¤ =
	¤protect¤mathnl¤ ¤ ¤ With[{¤mmgamma=gs[0,¤mmkappa,1][[2,1]]},¤ Re[Conjugate[¤mmgamma].((x1-x2).¤mmgamma)]]
	¤protect¤mathin¤ ¤mmDeltaŽ2[¤mmkappa_?NumericQ]¤ :=¤ ¤mmDeltaŽ2[¤mmkappa]¤ =
	¤protect¤mathnl¤ ¤ ¤ With[{¤mmgamma=gs[0,¤mmkappa,1][[2,1]]},¤ Re[Conjugate[¤mmgamma].((x1-x2).(x1-x2).¤mmgamma)]]
\end{mathematica}
		The mean distance in the ground state is zero for symmetry reasons: (notice the numerical inaccuracies)
\begin{mathematica}
	¤protect¤mathin¤ ListLinePlot[Table[{¤mmkappa,¤ ¤mmDeltaŽ1[¤mmkappa]},¤ {¤mmkappa,¤ -25,¤ 25,¤ 1}]]
\end{mathematica}
\begin{center}
\includegraphics[width=0.4\textwidth]{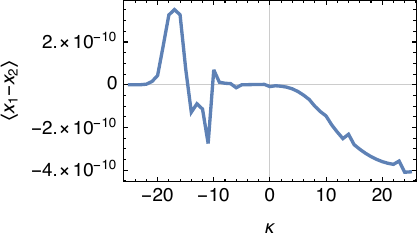}
\end{center}
		The variance of the distance in the ground state increases with $\kappa$:
\begin{mathematica}
	¤protect¤mathin¤ ListLinePlot[Table[{¤mmkappa,¤ ¤mmDeltaŽ2[¤mmkappa]-¤mmDeltaŽ1[¤mmkappa]^2},¤ {¤mmkappa,¤ -25,¤ 25,¤ 1}]]
\end{mathematica}
\begin{center}
\includegraphics[width=0.4\textwidth]{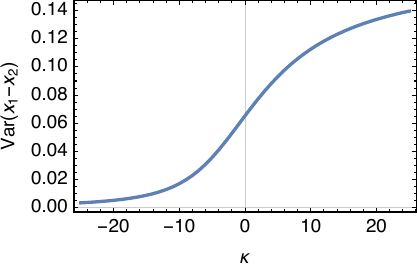}
\end{center}}
	\item\label{Q:interchange1} Show in Mathematica (by explicit calculation) that the Hamiltonian \mm{\ref{math:contactHam}} commutes with the particle interchange operator \mm{\ref{math:interchange}}. \emph{Hint:} use the \mm{Norm} function to calculate the matrix norm of the commutator.
\pagenote[\ref{Q:interchange1}]{We show that all three terms of the Hamiltonian commute with the particle interchange operator:
\begin{mathematica}
	¤protect¤mathin¤ ¤mmXi.TP¤ -¤ TP.¤mmXi¤ //Norm
	¤protect¤mathout¤ 0.
	¤protect¤mathin¤ ¤mmXi.VP¤ -¤ VP.¤mmXi¤ //Norm
	¤protect¤mathout¤ 0
	¤protect¤mathin¤ ¤mmXi.HintP¤ -¤ HintP.¤mmXi¤ //Norm
	¤protect¤mathout¤ 1.15903*10^-14
\end{mathematica}}
	\item\label{Q:interchange2} Show in Mathematica (by explicit calculation) that the antisymmetric Hamiltonian \mm{\ref{math:HPa}} does not depend on $\kappa$.
\pagenote[\ref{Q:interchange2}]{
\begin{mathematica}
	¤protect¤mathin¤ D[Normal[HPa[¤mmOmega,¤ ¤mmkappa]],¤ ¤mmkappa]¤ //Abs¤ //Max
	¤protect¤mathout¤ 1.03528*10^-15
\end{mathematica}}
	\item\label{Q:contactexact} The contact-interaction problem of this section can be solved analytically if $W(x)=0$, which allows us to check the accuracy of the presented numerical calculations. We will study the dimensionless (\mm{a=m=\mmhbar=1}) Hamiltonian $\Ham=-\frac12\left[\frac{\partial^2}{\ddd{\partial}{x}_1^2}+\frac{\partial^2}{\ddd{\partial}{x}_2^2}\right]+\kappa\delta(x_1-x_2)$.
	\begin{enumerate}
		\item The ground-state wavefunction will be of the form
			\small
			\begin{equation}
				\psi(x_1,x_2) = A\times\begin{cases}
					\cos[\alpha (x_1+x_2-1)] \cos[\beta (x_1-x_2+1)]\\
					\quad-\cos[\alpha (x_1-x_2+1)] \cos[\beta (x_1+x_2-1)] & \text{ if $0\le x_1\le x_2\le 1$,}\\
					\cos[\alpha (x_2+x_1-1)] \cos[\beta (x_2-x_1+1)]\\
					\quad-\cos[\alpha (x_2-x_1+1)] \cos[\beta (x_2+x_1-1)] & \text{ if $0\le x_2\le x_1\le 1$.}
				\end{cases}
			\end{equation}
			\normalsize
			Check that this wavefunction satisfies the boundary conditions $\psi(x_1,0)=\psi(x_1,1)=\psi(0,x_2)=\psi(1,x_2)=0$, that it is continuous across the boundary $x_1=x_2$ (\ie, that the two pieces of the wavefunction match up), and that it satisfies the symmetries of the calculation box: $\psi(x_1,x_2)=\psi(x_2,x_1)=\psi(1-x_1,1-x_2)$.
		\item Insert this wavefunction into the time-independent Schr\"odinger equation. Find the energy eigenvalue by assuming $x_1\neq x_2$. You should find $E=\alpha^2+\beta^2$.
		\item Express the Hamiltonian and the wavefunction in terms of the new coordinates $R=(x_1+x_2)/\sqrt{2}$ and $r=(x_1-x_2)/\sqrt{2}$. \emph{Hints:} $\frac{\partial^2}{\ddd{\partial}{x}_1^2}+\frac{\partial^2}{\ddd{\partial}{x}_2^2}=\frac{\partial^2}{\ddd{\partial}{R}^2}+\frac{\partial^2}{\ddd{\partial}{r}^2}$ and $\delta(u x)=u^{-1}\delta(x)$.
		\item Integrate the Schr\"odinger equation, expressed in the $(R,r)$ coordinates, over $r\in[-\epsilon,\epsilon]$ and take the limit $\epsilon\to0^+$. Verify that the resulting expression is satisfied if $\alpha \tan(\alpha)=\beta\tan(\beta)=\kappa/2$. \emph{Hint:} Do the integration analytically and use $\int_a^b\dd[r] \frac{\partial^2}{\ddd{\partial}{r}^2}f(r)=f'(b)-f'(a)$.
		\item The ground state is found by numerically solving $\alpha\tan(\alpha)=\beta\tan(\beta)=\kappa/2$. Out of the many solutions of these equations, we choose the correct ones for the ground state by specifying the starting point of the numerical root solver:
\begin{mathematica}
	¤mathin Clear[a,b];
	¤mathin a[-¤mminfty] = ¤mmpi/2;
	¤mathin a[0] = ¤mmpi;
	¤mathin a[¤mminfty] = 3¤mmpi/2;
	¤mathin a[¤mmkappa_?NumericQ] := a[¤mmkappa] = u /.
	¤mathnl   FindRoot[u*Tan[u]==¤mmkappa/2, {u,¤mmpi+ArcTan[¤mmkappa/(2¤mmpi)]}]
	¤mathin b[-¤mminfty] = I*¤mminfty;
	¤mathin b[0] = 0;
	¤mathin b[¤mminfty] = ¤mmpi/2;
	¤mathin b[¤mmkappa_ /; ¤mmkappa >= 0] := b[¤mmkappa] = u /. FindRoot[u*Tan[u] == ¤mmkappa/2,
	¤mathnl   {u, If[¤mmkappa<¤mmpi, 1, ¤mmpi/2 - ¤mmpi/¤mmkappa + 2¤mmpi/¤mmkappa^2]}]
	¤mathin b[¤mmkappa_ /; ¤mmkappa < 0] := b[¤mmkappa] = I*u /. FindRoot[u*Tanh[u] == -¤mmkappa/2, {u,-¤mmkappa/2}]
\end{mathematica}
		Compare the resulting $\kappa$-dependent ground state energy to the numerically calculated ground-sta\-te energies from \mm{\ref{math:contactHam}}.
	\end{enumerate}
\pagenote[\ref{Q:contactexact}]{We define the wavefunctions $\psi_1(x_1,x_2)$ for $x_1<x_2$ and $\psi_2(x_1,x_2)$ for $x_1>x_2$:
\begin{mathematica}
	¤protect¤mathin¤ ¤mmpsi1[x1_,x2_]¤ =¤ A*(Cos[¤mmalpha*(x1+x2-1)]*Cos[¤mmbeta*(x1-x2+1)]
	¤protect¤mathnl¤ ¤ ¤ ¤ ¤ ¤ ¤ ¤ ¤ ¤ ¤ ¤ ¤ ¤ ¤ ¤ ¤ -Cos[¤mmalpha*(x1-x2+1)]*Cos[¤mmbeta*(x1+x2-1)]);
	¤protect¤mathin¤ ¤mmpsi2[x1_,x2_]¤ =¤ ¤mmpsi1[x2,x1];
\end{mathematica}
	\begin{enumerate}
		\item Check the boundary conditions $\psi(x_1,0)=\psi(x_1,1)=\psi(0,x_2)=\psi(1,x_2)=0$:
\begin{mathematica}
	¤protect¤mathin¤ {¤mmpsi2[x1,0],¤ ¤mmpsi1[x1,1],¤ ¤mmpsi1[0,x2],¤ ¤mmpsi2[1,x2]}¤ //FullSimplify
	¤protect¤mathout¤ {0,¤ 0,¤ 0,¤ 0}
\end{mathematica}
		Check that the two pieces match up for $x_1=x_2$:
\begin{mathematica}
	¤protect¤mathin¤ ¤mmpsi1[x,x]¤ ==¤ ¤mmpsi2[x,x]
	¤protect¤mathout¤ True
\end{mathematica}
		Check the symmetries of the wavefunction:
\begin{mathematica}
	¤protect¤mathin¤ ¤mmpsi1[x1,x2]¤ ==¤ ¤mmpsi2[x2,x1]¤ ==¤ ¤mmpsi2[1-x1,1-x2]¤ //FullSimplify
	¤protect¤mathout¤ True
\end{mathematica}
		\item Check that the two pieces of the wavefunction satisfy the Schr\"odinger equation whenever $x_1\neq x_2$, with energy value $E=\alpha^2+\beta^2$:
\begin{mathematica}
	¤protect¤mathin¤ -1/2*D[¤mmpsi1[x1,x2],{x1,2}]+D[¤mmpsi1[x1,x2],{x2,2}]¤ ==
	¤protect¤mathnl¤ ¤ ¤ (¤mmalpha^2+¤mmbeta^2)*¤mmpsi1[x1,x2]¤ //FullSimplify
	¤protect¤mathout¤ True
	¤protect¤mathin¤ -1/2*D[¤mmpsi2[x1,x2],{x1,2}]+D[¤mmpsi2[x1,x2],{x2,2}]¤ ==
	¤protect¤mathnl¤ ¤ ¤ (¤mmalpha^2+¤mmbeta^2)*¤mmpsi2[x1,x2]¤ //FullSimplify
	¤protect¤mathout¤ True
\end{mathematica}
		\item The transformed Hamiltonian is
\begin{equation}
	\Ham = \left[ -\frac12 \frac{\partial^2}{\ddd{\partial}{R}^2} \right] + \left[ -\frac12 \frac{\partial^2}{\ddd{\partial}{r}^2}
	+ \frac{\kappa}{\sqrt{2}} \delta(r) \right]
\end{equation}
		and the transformed wavefunctions are
\begin{mathematica}
	¤protect¤mathin¤ ¤mmpsi1[(R+r)/Sqrt[2],(R-r)/Sqrt[2]]¤ //FullSimplify
	¤protect¤mathout¤ A*(Cos[¤mmalpha*(R*Sqrt[2]-1)]*Cos[¤mmbeta*(r*Sqrt[2]+1)]
	¤protect¤mathnl¤ ¤ ¤ -Cos[¤mmbeta*(R*Sqrt[2]-1)]*Cos[¤mmalpha*(r*Sqrt[2]+1)])
	¤protect¤mathin¤ ¤mmpsi2[(R+r)/Sqrt[2],(R-r)/Sqrt[2]]¤ //FullSimplify
	¤protect¤mathout¤ A*(Cos[¤mmalpha*(R*Sqrt[2]-1)]*Cos[¤mmbeta*(r*Sqrt[2]-1)]
	¤protect¤mathnl¤ ¤ ¤ -Cos[¤mmbeta*(R*Sqrt[2]-1)]*Cos[¤mmalpha*(r*Sqrt[2]-1)])
\end{mathematica}
		with
\begin{equation}
	\psi(R,r) = \begin{cases}
		\psi_1(R,r) & \text{if $r<0$}\\
		\psi_2(R,r) & \text{if $r>0$}
	\end{cases}
\end{equation}
		\item The Schr\"odinger equation in $(R,r)$ coordinates is
\begin{equation}
	\label{eq:contactexactSchr}
	\left[ -\frac12 \frac{\partial^2}{\ddd{\partial}{R}^2} \right]\psi(R,r) + \left[ -\frac12 \frac{\partial^2}{\ddd{\partial}{r}^2}
	+ \frac{\kappa}{\sqrt{2}} \delta(r) \right] \psi(R,r) = (\alpha^2+\beta^2)\psi(R,r)
\end{equation}
		We integrate \autoref{eq:contactexactSchr} over $r\in[-\epsilon,\epsilon]$:
\begin{equation}
	-\frac12 \int_{-\epsilon}^{\epsilon}\dd[r]\frac{\partial^2\psi(R,r)}{\ddd{\partial}{R}^2}
	-\frac12 \int_{-\epsilon}^{\epsilon}\dd[r]\frac{\partial^2\psi(R,r)}{\ddd{\partial}{r}^2}
	+\frac{\kappa}{\sqrt{2}} \int_{-\epsilon}^{\epsilon}\dd[r] \delta(r)\psi(R,r)
	= (\alpha^2+\beta^2) \int_{-\epsilon}^{\epsilon} \dd[r]\psi(R,r)
\end{equation}
		Using partial integration on the second term of the left-hand side:
\begin{equation}
	-\frac12 \int_{-\epsilon}^{\epsilon}\dd[r]\frac{\partial^2\psi(R,r)}{\ddd{\partial}{R}^2}
	-\frac12 \left[ \left. \frac{\partial\psi(R,r)}{\ddd{\partial}{r}} \right|_{r=\epsilon}-\left. \frac{\partial\psi(R,r)}{\ddd{\partial}{r}} \right|_{r=-\epsilon}\right]
	+\frac{\kappa}{\sqrt{2}} \psi(R,0)
	= (\alpha^2+\beta^2) \int_{-\epsilon}^{\epsilon} \dd[r]\psi(R,r)
\end{equation}
		In the limit $\epsilon\to0^+$ this equation becomes
\begin{equation}
	-\frac12 \left[ \left. \frac{\partial\psi_2(R,r)}{\ddd{\partial}{r}} \right|_{r=0}-\left. \frac{\partial\psi_1(R,r)}{\ddd{\partial}{r}} \right|_{r=0}\right]
	+\frac{\kappa}{\sqrt{2}} \psi(R,0)
	= 0
\end{equation}
	Inserting the definitions of $\psi_1$ and $\psi_2$:
\begin{mathematica}
	¤protect¤mathin¤ -1/2*((D[¤mmpsi2[(R+r)/Sqrt[2],(R-r)/Sqrt[2]],r]/.r->0)
	¤protect¤mathnl¤ ¤ ¤ -¤ (D[¤mmpsi1[(R+r)/Sqrt[2],(R-r)/Sqrt[2]],r]/.r->0))
	¤protect¤mathnl¤ ¤ ¤ +¤ ¤mmkappa/Sqrt[2]*¤mmpsi1[R/Sqrt[2],R/Sqrt[2]]¤ //FullSimplify
	¤protect¤mathout¤ A*(Cos[¤mmbeta*(R*Sqrt[2]-1)]*(2¤mmalpha*Sin[¤mmalpha]-¤mmkappa*Cos[¤mmalpha])
	¤protect¤mathnl¤ ¤ ¤ -Cos[¤mmalpha*(R*Sqrt[2]-1)]*(2¤mmbeta*Sin[¤mmbeta]-¤mmkappa*Cos[¤mmbeta]))/Sqrt[2]
\end{mathematica}
	The only way that this expression can be zero for all values of $R\in[0,\sqrt{2}]$ is if $2\alpha\sin(\alpha)-\kappa \cos(\alpha)=2\beta\sin(\beta)-\kappa \cos(\beta)=0$, and hence if $\alpha\tan(\alpha)=\beta\tan(\beta)=\kappa/2$.
		\item See the attached Mathematica notebook \texttt{ContactInteraction.nb}.
	\end{enumerate}}
\end{questions}

\subsection{two particles in one dimension with arbitrary interaction}
\label{sec:arbitraryinteraction}

Two particles in one dimension interacting via an arbitrary potential have a Hamiltonian very similar to \autoref{eq:Ham2contact}, except that the interaction is now
\begin{equation}
	\Ham\ix{int} = V\ix{int}(x_1,x_2),
\end{equation}
or, more explicitly as an operator in the Dirac position basis,
\begin{equation}
	\Ham\ix{int} = \int_0^a \dd[x_1]\dd[x_2] \ket{x_1}\otimes\ket{x_2} V\ix{int}(x_1,x_2) \bra{x_1}\otimes\bra{x_2}.
\end{equation}
As an example, for the Coulomb interaction\index{Coulomb interaction} we have $V\ix{int}(x_1,x_2)=\frac{Q_1 Q_2}{4\pi\epsilon_0\abs{x_1-x_2}}$ with $Q_1$ and $Q_2$ the electric charges of the two particles. For many realistic potentials $V\ix{int}$ only depends on $\abs{x_1-x_2}$.

In the finite-resolution position basis, the matrix elements of this interaction Hamiltonian can be approximated with a method similar to what we have already seen, for example in \autoref{sec:potentialenergy}:
\begin{multline}
	\label{eq:Hpotmegen}
	\me{j_1,j_2}{\Ham\ix{int}}{j_1',j_2'} = 
	\int_0^a \vartheta_{j_1}(x_1)\vartheta_{j_2}(x_2) V\ix{int}(x_1,x_2) \vartheta_{j_1'}(x_1)\vartheta_{j_2'}(x_2) \dd[x_1] \dd[x_2]\\
	\approx V\ix{int}(x_{j_1},x_{j_2}) \int_0^a \vartheta_{j_1}(x_1)\vartheta_{j_2}(x_2) \vartheta_{j_1'}(x_1)\vartheta_{j_2'}(x_2) \dd[x_1] \dd[x_2]
	= \delta_{j_1,j_1'} \delta_{j_2,j_2'} V\ix{int}(x_{j_1},x_{j_2}).
\end{multline}
This approximation is easy to evaluate without the need for integration over basis functions. But realistic interaction potentials are usually singular for $x_1=x_2$ (consider, for example, the Coulomb potential), and therefore the approximate \autoref{eq:Hpotmegen} fails for the evaluation of the matrix elements $\me{j,j}{\Ham\ix{int}}{j,j}$. This problem cannot be solved in all generality, and we can either resort to more accurate integration (as in \autoref{sec:contactinteraction}) or we can replace the true interaction potential with a less singular version: for the Coulomb potential, we could for example use a truncated singularity for $\abs{x}<\delta$ for some small distance $\delta$:\index{Coulomb interaction!truncated}
\begin{equation}
	\label{eq:truncatedCoulomb}
	V\ix{int}(x) = \frac{Q_1 Q_2}{4\pi\epsilon_0} \times \begin{cases}
		\frac{1}{\abs{x}} & \text{if $\abs{x}\ge\delta$}\\
		\frac{1}{\delta} & \text{if $\abs{x}<\delta$}
	\end{cases}
\end{equation}
As long as the particles move at energies much smaller than $V\ix{int}(\pm\delta)=\frac{Q_1 Q_2}{4\pi\epsilon_0\delta}$ they cannot distinguish the true Coulomb potential from this truncated form.

\subsubsection{exercises}

\begin{questions}
	\item\label{Q:Coulomb} Consider two indistinguishable bosons in an infinite square well, interacting via the truncated Coulomb potential of \autoref{eq:truncatedCoulomb}. Calculate the expectation value of the inter-particle distance, $\avg{x_1-x_2}$, and its variance, $\avg{(x_1-x_2)^2}-\avg{x_1-x_2}^2$, in the ground state as a function of the Coulomb interaction strength (attractive and repulsive). \emph{Hint:} set $\delta=\Delta=a/(n\ix{max}+1)$ in \autoref{eq:truncatedCoulomb}.
\pagenote[\ref{Q:Coulomb}]{We solve this problem with the code of \autoref{sec:contactinteraction}, in the same way as \ref{Q:distance}. The interaction potential is, according to \autoref{eq:truncatedCoulomb},
\begin{mathematica}
	¤protect¤mathin¤ With[{¤mmdelta=¤mmDelta},
	¤protect¤mathnl¤ ¤ ¤ Q[x_]¤ =¤ Piecewise[{{1/Abs[x],Abs[x]>¤mmdelta},¤ {1/¤mmdelta,Abs[x]<=¤mmdelta}}];]
\end{mathematica}
		and the interaction Hamiltonian $\Ham\ix{int}\approx \kappa/\abs{x}$ is approximately
\begin{mathematica}
	¤protect¤mathin¤ HintP¤ =¤ SparseArray[{j1_,j1_,j2_,j2_}¤ :>¤ Q[xgrid[[j1]]-xgrid[[j2]]],
	¤protect¤mathnl¤ ¤ ¤ {nmax,nmax,nmax,nmax}]¤ //ArrayFlatten;
\end{mathematica}
		With these definitions, the energy levels are (with \mm{a=m=\mmhbar=1})
\begin{center}
\includegraphics[width=0.4\textwidth]{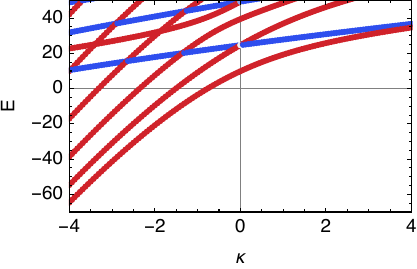}
\end{center}
		We see that the lowest energy level is always symmetric under particle exchange (colored in red); the bosonic ground state is therefore just the lowest energy level. The expectation value $\avg{\op{x}_1-\op{x}_2}$ is zero by symmetry; its variance is
\begin{center}
\includegraphics[width=0.4\textwidth]{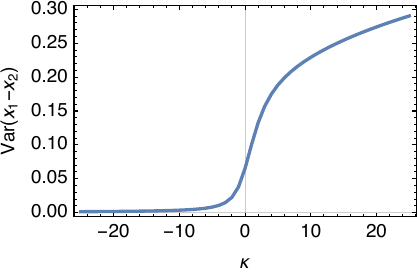}
\end{center}}
	\item\label{Q:CoulombF} Answer \ref{Q:Coulomb} for two indistinguishable fermions.
\pagenote[\ref{Q:CoulombF}]{We see in the answer of \ref{Q:Coulomb} that the lowest fermionic state (blue) depends on the coupling strength $\kappa$. In the spirit of \autoref{sec:contactinteraction} we define the fermionic Hamiltonian with \mm{\ref{math:HPa}} and calculate the fermionic ground state with \mm{\ref{math:ags}}. The expectation values of $\op{x}_1-\op{x}_2$ and $(\op{x}_1-\op{x}_2)^2$ are calculated from the antisymmetric ground state with
\begin{mathematica}
	¤protect¤mathin¤ Clear[F¤mmDeltaŽx,¤ F¤mmDeltaŽx2];
	¤protect¤mathin¤ F¤mmDeltaŽx[¤mmkappa_?NumericQ]¤ :=¤ F¤mmDeltaŽx[¤mmkappa]¤ =
	¤protect¤mathnl¤ ¤ ¤ With[{¤mmgamma=ags[0,¤mmkappa,1][[2,1]]},¤ Re[Conjugate[¤mmgamma].((x1-x2).¤mmgamma)]]
	¤protect¤mathin¤ F¤mmDeltaŽx2[¤mmkappa_?NumericQ]¤ :=¤ F¤mmDeltaŽx2[¤mmkappa]¤ =
	¤protect¤mathnl¤ ¤ ¤ With[{¤mmgamma=ags[0,¤mmkappa,1][[2,1]]},¤ Re[Conjugate[¤mmgamma].((x1-x2).(x1-x2).¤mmgamma)]]
\end{mathematica}
		The expectation value $\avg{\op{x}_1-\op{x}_2}$ is zero by symmetry; its variance is larger than that for bosons:
\begin{center}
\includegraphics[width=0.6\textwidth]{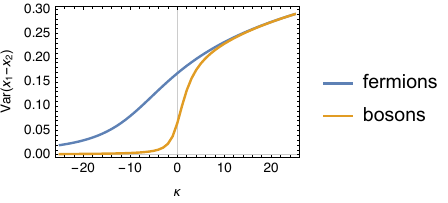}
\end{center}}
\end{questions}

\section[one particle in several dimensions]{\label{sec:BEC}one particle in several dimensions\hspace{\stretch{1}}\attachcode{3DBEC}{Bose--Einstein condensate in 3D}}

An important application of the imaginary-time propagation method of \autoref{sec:itpGP} is the calculation of the shape of a three-dimensional Bose--Einstein condensate\index{Bose-Einstein condensate@Bose--Einstein condensate}. In this section we use such a calculation as an example of how to extend single-particle lattice quantum mechanics to more spatial dimensions.

The non-linear Hamiltonian describing a three-dimensional Bose--Einstein condensate in a harmonic trap (to use a very common case) is
\begin{equation}
	\label{eq:BEC3DHam}
	\Ham = -\frac{\hbar^2}{2m} \left( \frac{\partial^2}{\ddd{\partial}{x}^2} + \frac{\partial^2}{\ddd{\partial}{y}^2} + \frac{\partial^2}{\ddd{\partial}{z}^2} \right)
	+ \frac{m}{2} \left( \omega_x^2 x^2 + \omega_y^2 y^2 + \omega_z^2 z^2 \right)
	+ (N-1)\frac{4\pi\hbar^2 a_s}{m}\abs{\psi(x,y,z)}^2,
\end{equation}
where we have assumed that the single-particle wavefunction $\psi(x,y,z)$ is normalized: $\int \abs{\psi(x,y,z)}^2\dd[x] \dd[y] \dd[z]=1$. As before, the contact interaction is described by the $s$-wave scattering length $a_s$. We will call $\kappa=\frac{4\pi\hbar^2 a_s}{m}$ the interaction constant, as in previous sections.

We perform this calculation in a square box, where $\abs{x}\le\frac{a}{2}$, $\abs{y}\le\frac{a}{2}$, and $\abs{z}\le\frac{a}{2}$; we will need to choose $a$ large enough so that the BEC fits into this box, but small enough so that we do not need an unreasonably large $n\ix{max}$ for the description of its wavefunction. Notice that this box is shifted by $\frac{a}{2}$ compared to the $[0\dots a]$ boxes used so far; this does not influence the calculations in any way. 


The ground state of the non-linear Hamiltonian of \autoref{eq:BEC3DHam} can be found by three-dimensional imaginary-time propagation, starting from (almost) any arbitrary state. Here we assemble a Mathematica function \mm{groundstate} that, given an imaginary time step $\ddd{\delta}{\beta}$, propagates a random initial state until the state is converged to the ground state.

The units of the problem are dealt with as in \autoref{sec:1Dunits}, differing from \mm{\ref{math:coupledunits}}ff in that here we choose the length and time units freely:
\begin{mathematica}
	¤mathin LengthUnit = Quantity["Micrometers"];   (* choose freely *)
	¤mathin TimeUnit = Quantity["Seconds"];         (* choose freely *)
	¤mathin MassUnit = Quantity["ReducedPlanckConstant"]*TimeUnit/LengthUnit^2;
	¤mathin EnergyUnit = Quantity["ReducedPlanckConstant"]/TimeUnit;
	¤mathin ¤mmhbar = Quantity["ReducedPlanckConstant"]/(EnergyUnit*TimeUnit);
\end{mathematica}
We will be considering $N=\num{1000}$ $^{87}$Rb atoms\index{Rubidium-87} in a magnetic trap with trap frequencies $\omega_x=2\pi\times\SI{115}{Hz}$ and $\omega_y=\omega_z=2\pi\times\SI{540}{Hz}$. The $^{87}$Rb atoms are assumed to be in the $\ket{F=1,M_F=-1}$ hyperfine ground state, where their $s$-wave scattering length is $a_s=\num{100,4}a_0$ (with $a_0=\SI{52,9177}{pm}$ the Bohr radius).\index{s-wave scattering@$s$-wave scattering}\index{Planck's constant}
\begin{mathematica}
	¤mathin m = Quantity[86.909187, "AtomicMassUnit"]/MassUnit;
	¤mathin a = Quantity[10, "Micrometers"]/LengthUnit;
	¤mathin ¤mmomegaŽx = 2*¤mmpi*Quantity[115, "Hertz"]*TimeUnit;
	¤mathin ¤mmomegaŽy = 2*¤mmpi*Quantity[540, "Hertz"]*TimeUnit;
	¤mathin ¤mmomegaŽz = 2*¤mmpi*Quantity[540, "Hertz"]*TimeUnit;
	¤mathin as = Quantity[100.4, "BohrRadius"]/LengthUnit;
	¤mathin ¤mmkappa = 4*¤mmpi*¤mmhbar^2*as/m;
\end{mathematica}
Next we define the grid on which the calculations will be done. In each Cartesian direction there are $n\ix{max}$ grid points $x_j=\mm{xgrid[[j]]}$ on the interval $[-a/2,+a/2]$:
\begin{mathematica}
	¤mathin nmax = 50;
	¤mathin ¤mmDelta = a/(nmax+1);
	¤mathin xgrid = a*(Range[nmax]/(nmax+1) - 1/2);
\end{mathematica}
We  define the dimensionless harmonic-trap potential: the potential has its minimum at the center of the calculation box, \ie, at $x=y=z=0$.
\begin{mathematica}
	¤mathin W[x_,y_,z_] = m/2 * (¤mmomegaŽx^2*x^2 + ¤mmomegaŽy^2*y^2 + ¤mmomegaŽz^2*z^2);
\end{mathematica}
We only need the values of this potential on the grid points. To evaluate this, we build a three-dimensional array whose element \mm{Wgrid[[jx,jy,jz]]} is given by the grid-point value \mm{W[xgrid[[jx]],xgrid[[jy]],xgrid[[jz]]]}:
\begin{mathematica}
	¤mathin Wgrid=Table[W[xgrid[[jx]],xgrid[[jy]],xgrid[[jz]]],{jx,nmax},{jy,nmax},{jz,nmax}];
\end{mathematica}
We could also define this more efficiently through functional programming:\index{Mathematica!functional programming}\index{Mathematica!outer product}
\begin{mathematica}
	¤mathin Wgrid = Outer[W, xgrid, xgrid, xgrid];
\end{mathematica}
The structure of the three-dimensional \mm{Wgrid} array of potential values mirrors the structure of the wavefunction that we will be using: any wavefunction \mm{v} will be a $n\ix{max}\times n\ix{max}\times n\ix{max}$ array of coefficients in our finite-resolution position basis:
\begin{equation}
	\psi(x,y,z) = \sum_{j_x,j_y,j_z=1}^{n\ix{max}} \mm{v[[jx,jy,jz]]} \vartheta_{j_x}(x) \vartheta_{j_y}(y) \vartheta_{j_z}(z).
\end{equation}
From \autoref{eq:positionbasislocality} we find that on the three-dimensional grid points the wavefunction takes the values
\begin{equation}
	\psi(x_{j_x},x_{j_y},x_{j_z}) = \mm{v[[jx,jy,jz]]}/\Delta^{3/2}.
\end{equation}
The norm of a wavefunction is
\begin{equation}
	\int_{-\infty}^{\infty}\abs{\psi(x,y,z)}^2\dd[x] \dd[y] \dd[z]
	= \sum_{j_x,j_y,j_z=1}^{n\ix{max}} \abs{\mm{v[[jx,jy,jz]]}}^2
	= \mm{Norm[Flatten[v]]\^{}2},
\end{equation}
from which we define a wavefunction normalization function
\begin{mathematica}
	¤mathin nn[v_] := v/Norm[Flatten[v]]
\end{mathematica}
The ground state calculation then proceeds by imaginary-time propagation, with step size $\ddd{\delta}{\beta}$ corresponding to an evolution $e^{-\ddd{\delta}{\beta}\Ham}$ per step. The calculation is done for $N=\mm{n}$ particles. Remember that the \mm{FourierDST}\index{discrete sine transform} function can do multi-dimensional discrete sine transforms, and therefore the kinetic-energy propagator can still be evaluated very efficiently. The last argument, \mm{tolerance}, is optional and is given the value \num{e-6} if not specified (see \autoref{sec:optionalargument}).
\begin{mathematica}
	¤mathin groundstate[n_?NumericQ, ¤mmdelta¤mmbeta_?NumericQ, tolerance_:10^(-6)] :=
	¤mathnl   Module[{Kn, Ke, propKin, propPot2, prop, v0, ¤mmgamma, Ekin, Epot, Eint, Etot, ¤mmmu},
	¤mathnlc     (* compute the diagonal elements of exp[-¤mmdelta¤mmbeta*T] *)
	¤mathnl     Kn = ¤mmpi^2*¤mmhbar^2/(2*m*a^2)*Table[nx^2+ny^2+nz^2,
	¤mathnl       {nx,nmax}, {ny,nmax}, {nz,nmax}];
	¤mathnl     Ke = Exp[-¤mmdelta¤mmbeta*Kn] //N;
	¤mathnlc     (* propagate by a full imaginary-time-step with T *)
	¤mathnl     propKin[v_] := nn[FourierDST[Ke*FourierDST[v, 1], 1]];
	¤mathnlc     (* propagate by a half imaginary-time-step with V *)
	¤mathnl     propPot2[v_] := nn[Exp[-(¤mmdelta¤mmbeta/2)*(Wgrid+¤mmkappa*(n-1)*Abs[v]^2/¤mmDelta^3)]*v];
	¤mathnlc     (* propagate by a full imaginary-time-step by *)
	¤mathnlc     (* H=T+V using the Trotter approximation *)
	¤mathnl     prop[v_] := propPot2[propKin[propPot2[v]]]
	¤mathnlc     (* random starting point *)
	¤mathnl     v0 = nn @ RandomVariate[NormalDistribution[], {nmax, nmax, nmax}];
	¤mathnlc     (* propagation to the ground state *)
	¤mathnl     ¤mmgamma = FixedPoint[prop, v0,
	¤mathnl       SameTest -> Function[{v1,v2}, Norm[Flatten[v1-v2]]<tolerance]];
	¤mathnlc     (* energy components *)
	¤mathnl     Ekin = Flatten[Kn].Flatten[Abs[FourierDST[¤mmgamma, 1]]^2];
	¤mathnl     Epot = Flatten[Wgrid].Flatten[Abs[¤mmgamma]^2];
	¤mathnl     Eint = (¤mmkappa/2)*(n-1)*Total[Flatten[Abs[¤mmgamma]^4]]/¤mmDelta^3;
	¤mathnlc     (* total energy *)
	¤mathnl     Etot = Ekin + Epot + Eint;
	¤mathnlc     (* chemical potential *)
	¤mathnl     ¤mmmu = Ekin + Epot + 2*Eint;
	¤mathnlc     (* return energy, chemical potential, coefficients *)
	¤mathnl     {Etot, ¤mmmu, ¤mmgamma}]
\end{mathematica}
As an example, we calculate the ground state for $N=\num{1000}$ atoms and a time step of $\ddd{\delta}{\beta}=\num{e-5}$ time units, using the default convergence tolerance:
\begin{mathematica}
	¤mathin {Etot, ¤mmmu, ¤mmgamma} = groundstate[1000, 10^(-4)];
	¤mathin {Etot, ¤mmmu} * UnitConvert[EnergyUnit, "Joules"]
	¤mathout {6.88125*10^-31 J, 8.68181*10^-31 J}
\end{mathematica}
A more common energy unit is the Hertz, arrived at via Planck's constant:
\begin{mathematica}
	¤mathin {Etot, ¤mmmu} * UnitConvert[EnergyUnit/Quantity["PlanckConstant"], "Hertz"]
	¤mathout {1038.51 Hz, 1310.25 Hz}
\end{mathematica}
One way of plotting the ground-state density in 3D is as an iso-density surface. We plot the surface at half the peak density with
\begin{mathematica}
	¤mathin ¤mmrho = Abs[¤mmgamma]^2/¤mmDelta^3;
	¤mathin ListContourPlot3D[¤mmrho,
	¤mathnl   DataRange -> a*(1/(nmax+1)-1/2)*{{-1,1},{-1,1},{-1,1}},
	¤mathnl   Contours -> {Max[¤mmrho]/2}, BoxRatios -> Automatic]
\end{mathematica}
Here we show several such iso-density surfaces:
\begin{center}
\includegraphics[width=\textwidth]{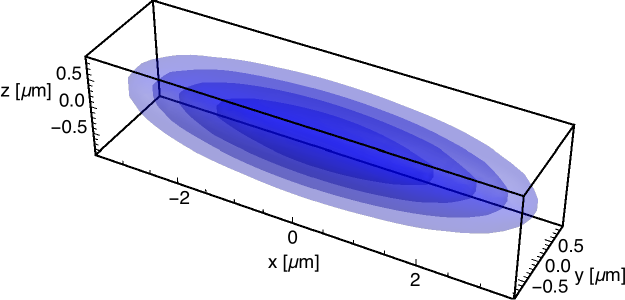}
\end{center}
For more quantitative results we can, for example, calculate the expectation values $\mm{X}=\avg{x}$, $\mm{Y}=\avg{y}$, $\mm{Z}=\avg{z}$, $\mm{XX}=\avg{x^2}$, $\mm{YY}=\avg{y^2}$, $\mm{ZZ}=\avg{z^2}$. We could define coordinate arrays as
\begin{mathematica}
	¤mathin xc = Table[xgrid[[jx]], {jx,nmax}, {jy,nmax}, {jz,nmax}];
	¤mathin yc = Table[xgrid[[jy]], {jx,nmax}, {jy,nmax}, {jz,nmax}];
	¤mathin zc = Table[xgrid[[jz]], {jx,nmax}, {jy,nmax}, {jz,nmax}];
\end{mathematica}
but we define them more efficiently as follows:
\begin{mathematica}
	¤mathin ones = ConstantArray[1, nmax];
	¤mathin xc = Outer[Times, xgrid, ones, ones];
	¤mathin yc = Outer[Times, ones, xgrid, ones];
	¤mathin zc = Outer[Times, ones, ones, xgrid];
\end{mathematica}
The desired expectation values are then computed with
\begin{mathematica}
	¤mathin X = Total[Flatten[xc * ¤mmrho]];
	¤mathin Y = Total[Flatten[yc * ¤mmrho]];
	¤mathin Z = Total[Flatten[zc * ¤mmrho]];
	¤mathin XX = Total[Flatten[xc^2 * ¤mmrho]];
	¤mathin YY = Total[Flatten[yc^2 * ¤mmrho]];
	¤mathin ZZ = Total[Flatten[zc^2 * ¤mmrho]];
\end{mathematica}
The root-mean-square size of the BEC is calculated from these as the standard deviations of the position operators in the three Cartesian directions:
\begin{mathematica}
	¤mathin {Sqrt[XX-X^2], Sqrt[YY-Y^2], Sqrt[ZZ-Z^2]} * LengthUnit
	¤mathout {1.58829 ¤mmmuŽm, 0.417615 ¤mmmuŽm, 0.417615 ¤mmmuŽm}
\end{mathematica}

\subsection{exercises}

\begin{questions}
	\item\label{Q:BECnoint} Take the BEC Hamiltonian of \autoref{eq:BEC3DHam} in the absence of interactions ($a_s=0$) and calculate analytically the expectation values $\avg{x^2}$, $\avg{y^2}$, $\avg{z^2}$ in the ground state.
\pagenote[\ref{Q:BECnoint}]{The expectation values are the usual ones of the harmonic oscillator, given by
\begin{align}
	\label{eq:nivariances}
	\avg{x^2} &= \frac{\hbar}{2m\omega_x}, &
	\avg{y^2} &= \frac{\hbar}{2m\omega_y}, &
	\avg{z^2} &= \frac{\hbar}{2m\omega_z}.
\end{align}
		They are independent in the three Cartesian directions.}
	\item\label{Q:BECTF} Take the BEC Hamiltonian of \autoref{eq:BEC3DHam} in the limit of strong interactions (Thomas--Fermi limit)\index{Thomas-Fermi approximation@Thomas--Fermi approximation}, where the kinetic energy can be neglected. The Gross--Pitaevskii equation is then
\begin{equation}
	\left[ \frac{m}{2} \left( \omega_x^2 x^2 + \omega_y^2 y^2 + \omega_z^2 z^2 \right)
	+ (N-1)\frac{4\pi\hbar^2 a_s}{m}\abs{\psi(x,y,z)}^2 \right] \psi(x,y,z) = \mu\psi(x,y,z),
\end{equation}
	which has two solutions:
\begin{equation}
	\abs{\psi(x,y,z)}^2 = \begin{cases}
		0 & \text{or}\\
		\frac{\mu-\frac{m}{2} \left( \omega_x^2 x^2 + \omega_y^2 y^2 + \omega_z^2 z^2 \right)}{(N-1)\frac{4\pi\hbar^2 a_s}{m}}.
	\end{cases}
\end{equation}
	Together with the conditions that $\abs{\psi(x,y,z)}^2\ge0$, that $\psi(x,y,z)$ should be continuous, and that $\int \abs{\psi(x,y,z)}^2\dd[x] \dd[y] \dd[z] = 1$, this gives us the Thomas--Fermi ``inverted parabola'' density
\begin{equation}
	\label{eq:TFdensity}
	\abs{\psi(x,y,z)}^2 = \begin{cases}
		\rho_0\left[1-\left(\frac{x}{R_x}\right)^2-\left(\frac{y}{R_y}\right)^2-\left(\frac{z}{R_z}\right)^2\right] & \text{if $\left(\frac{x}{R_x}\right)^2+\left(\frac{y}{R_y}\right)^2+\left(\frac{z}{R_z}\right)^2 \le 1$,}\\
		0 & \text{if not},
	\end{cases}
\end{equation}
	which is nonzero only inside an ellipsoid with Thomas--Fermi radii
\begin{subequations}
\label{eq:TFradii}
\begin{align}
	\label{eq:TFradiusX}R_x &= \left[ \frac{15\hbar^2a_s(N-1)\omega_y\omega_z}{m^2\omega_x^4}\right]^{\frac15}=\left[\frac{15\kappa(N-1)\omega_y\omega_z}{4\pi m\omega_x^4}\right]^{\frac15},\\
	R_y &= \left[ \frac{15\hbar^2a_s(N-1)\omega_z\omega_x}{m^2\omega_y^4}\right]^{\frac15}=\left[\frac{15\kappa(N-1)\omega_x\omega_z}{4\pi m\omega_y^4}\right]^{\frac15},\\
	R_z &= \left[ \frac{15\hbar^2a_s(N-1)\omega_x\omega_y}{m^2\omega_z^4}\right]^{\frac15}=\left[\frac{15\kappa(N-1)\omega_x\omega_y}{4\pi m\omega_z^4}\right]^{\frac15}.
\end{align}
\end{subequations}
	The density at the origin of the ellipsoid is
\begin{equation}
	\label{eq:TFcentraldensity}
	\rho_0 = \frac{1}{8\pi} \left[ \frac{225m^6\omega_x^2\omega_y^2\omega_z^2}{\hbar^6a_s^3(N-1)^3}\right]^{\frac15}
	= \left[\frac{225m^3\omega_x^2\omega_y^2\omega_z^2}{512\pi^2\kappa^3(N-1)^3}\right]^{\frac15}
\end{equation}
	and the chemical potential is
\begin{equation}
	\mu = \frac12 \left[ 225m\hbar^4a_s^2(N-1)^2\omega_x^2\omega_y^2\omega_z^2\right]^{\frac15}
	= \left[\frac{225}{512\pi^2}m^3\kappa^2(N-1)^2\omega_x^2\omega_y^2\omega_z^2\right]^{\frac15}.
\end{equation}
	Using this Thomas--Fermi density profile, calculate the expectation values $\avg{x^2}$, $\avg{y^2}$, $\avg{z^2}$ in the ground state of the Thomas--Fermi approximation. \emph{Hints:} Calculate $\avg{x^2}$ using \autoref{eq:TFdensity} without substituting \hyperref[eq:TFradii]{Equations~\ref*{eq:TFradii}} and \autoref{eq:TFcentraldensity}; do these substitutions only after having found the result. You can find $\avg{y^2}$ and $\avg{z^2}$ by analogy, without repeating the calculation.
\pagenote[\ref{Q:BECTF}]{We calculate the integral over the density in Cartesian coordinates by integrating only over the ellipsoid in which the density is nonzero:
\begin{mathematica}
	¤protect¤mathin¤ A¤ =¤ Assuming[Rx>0¤ &&¤ Ry>0¤ &&¤ Rz>0,
	¤protect¤mathnl¤ ¤ ¤ Integrate[¤mmrho0*(1-(x/Rx)^2-(y/Ry)^2-(z/Rz)^2),
	¤protect¤mathnl¤ ¤ ¤ ¤ ¤ {x,¤ -Rx,¤ Rx},
	¤protect¤mathnl¤ ¤ ¤ ¤ ¤ {y,¤ -Ry*Sqrt[1-(x/Rx)^2],¤ Ry*Sqrt[1-(x/Rx)^2]},
	¤protect¤mathnl¤ ¤ ¤ ¤ ¤ {z,¤ -Rz*Sqrt[1-(x/Rx)^2-(y/Ry)^2],¤ Rz*Sqrt[1-(x/Rx)^2-(y/Ry)^2]}]]
	¤protect¤mathout¤ 8/15*¤mmpi*Rx*Ry*Rz*¤mmrho0
\end{mathematica}
		Similarly, we calculate the integral of the density times $x^2$ with
\begin{mathematica}
	¤protect¤mathin¤ B¤ =¤ Assuming[Rx>0¤ &&¤ Ry>0¤ &&¤ Rz>0,
	¤protect¤mathnl¤ ¤ ¤ Integrate[x^2¤ *¤ ¤mmrho0*(1-(x/Rx)^2-(y/Ry)^2-(z/Rz)^2),
	¤protect¤mathnl¤ ¤ ¤ ¤ ¤ {x,¤ -Rx,¤ Rx},
	¤protect¤mathnl¤ ¤ ¤ ¤ ¤ {y,¤ -Ry*Sqrt[1-(x/Rx)^2],¤ Ry*Sqrt[1-(x/Rx)^2]},
	¤protect¤mathnl¤ ¤ ¤ ¤ ¤ {z,¤ -Rz*Sqrt[1-(x/Rx)^2-(y/Ry)^2],¤ Rz*Sqrt[1-(x/Rx)^2-(y/Ry)^2]}]]
	¤protect¤mathout¤ 8/105*¤mmpi*Rx^3*Ry*Rz*¤mmrho0
\end{mathematica}
		The expectation value $\avg{x^2}$ is the ratio of these two integrals,
\begin{mathematica}
	¤protect¤mathin¤ B/A
	¤protect¤mathout¤ Rx^2/7
\end{mathematica}
		With the value of $R_x$ given in \autoref{eq:TFradiusX}, this becomes
\begin{subequations}
\begin{align}
	\label{eq:TFvariances}
	\avg{x^2} &= \frac17 \left[ \frac{15\hbar^2a_s(N-1)\omega_y\omega_z}{m^2\omega_x^4}\right]^{\frac25} = \frac17 \left[ \frac{15\kappa(N-1)\omega_y\omega_z}{4\pi m\omega_x^4}\right]^{\frac25},\\
	\avg{y^2} &= \frac17 \left[ \frac{15\hbar^2a_s(N-1)\omega_x\omega_z}{m^2\omega_y^4}\right]^{\frac25} = \frac17 \left[ \frac{15\kappa(N-1)\omega_x\omega_z}{4\pi m\omega_y^4}\right]^{\frac25},\\
	\avg{z^2} &= \frac17 \left[ \frac{15\hbar^2a_s(N-1)\omega_x\omega_y}{m^2\omega_z^4}\right]^{\frac25} = \frac17 \left[ \frac{15\kappa(N-1)\omega_x\omega_y}{4\pi m\omega_z^4}\right]^{\frac25}.
\end{align}
\end{subequations}
		We see that, in contrast to \ref{Q:BECnoint}, the expectation values of the three Cartesian directions are not independent of each other's trapping frequencies.}
	\item\label{Q:BECcompareexpectationvalues} Compare the numerical expectation values $\avg{x^2}$, $\avg{y^2}$, $\avg{z^2}$ of our Mathematica code to the analytic results of \ref{Q:BECnoint} and \ref{Q:BECTF}. What is the maximum $^{87}$Rb atom number $N$ which allows a reasonably good description (in this specific trap) with the non-interacting solution? What is the minimum atom number which allows a reasonably good description with the Thomas--Fermi solution?
\pagenote[\ref{Q:BECcompareexpectationvalues}]{We plot the second moments of \autoref{eq:nivariances} and \autoref{eq:TFvariances} as functions of the particle number $N$:
\begin{center}
	\includegraphics[width=0.6\textwidth]{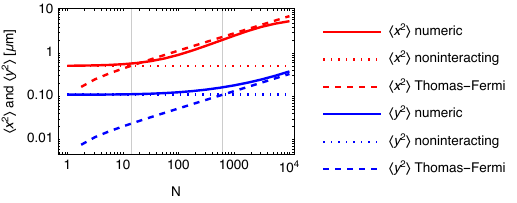}
\end{center}
		The values of $\avg{z^2}$ are equal to those of $\avg{y^2}$ because of the cylindrical symmetry of the problem.
		The crossover point where the Thomas--Fermi second moment is equal to the noninteracting second moment is at
\begin{align}
	\bar{N}_x &= \frac{49}{60a_s\omega_y\omega_z}\sqrt{\frac{7\hbar\omega_x^3}{2m}}+1, &
	\bar{N}_y &= \frac{49}{60a_s\omega_x\omega_z}\sqrt{\frac{7\hbar\omega_y^3}{2m}}+1, &
	\bar{N}_z &= \frac{49}{60a_s\omega_x\omega_y}\sqrt{\frac{7\hbar\omega_z^3}{2m}}+1,
\end{align}
		indicated with vertical lines in the above plot.
		The noninteracting limit, \autoref{eq:nivariances}, is good for $N\lesssim10$. The Thomas--Fermi limit, \autoref{eq:TFvariances}, is good for $N\gtrsim\num{5000}$. Notice that for $N\gtrsim\num{3000}$ the numeric value of $\avg{x^2}$ deviates from the Thomas--Fermi limit because of the finite size of the calculation box.}
\end{questions}

%
%

%
%
%
%
%
%
%

\chapter{combining spatial motion and spin}
\label{chap:spacespin}
\restartlist{questions}
\chapterpicture{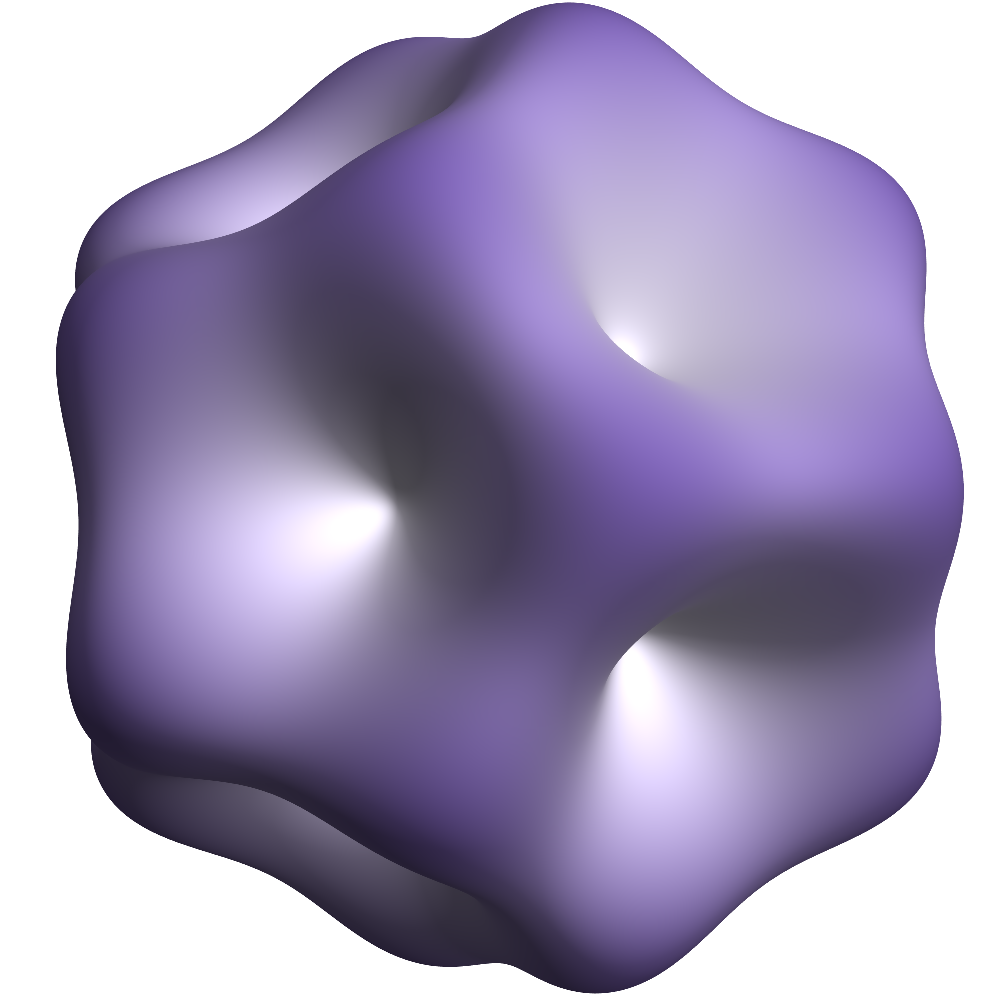}
In this chapter we put together all the techniques studied so far: internal-spin\index{spin} degrees of freedom (\autoref{chap:spin}) and spatial (motional) degrees of freedom\index{real-space dynamics} (\autoref{chap:1D}) are combined with the tensor-product formalism\index{tensor!product} (\autoref{chap:basis}). We arrive at a complete numerical description of interacting spin-ful particles moving through space. To showcase these powerful tools, we study Rashba coupling as well as the Jaynes--Cummings model.

\clearpage
\section{one particle in 1D with spin}
\label{sec:1part1dimwithspin}

\subsection{separable Hamiltonian}
\label{sec:spacespinsep}

The simplest problem combining a spatial and a spin degree of freedom in a meaningful way consists of a single spin-$1/2$ particle moving in one dimension in a state-selective potential:
\begin{equation}
	\label{eq:spacespinham1}
	\Ham = -\frac{\hbar^2}{2m} \frac{\dd^2}{\dd[x]^2} + V_0(x) + V_z(x) \op{S}_z,
\end{equation}
where $\op{S}_z=\frac12\op{\sigma}_z$ is given by the Pauli matrix.
As was said before, \autoref{eq:spacespinham1} is a short-hand notation of the full Hamiltonian
\begin{equation}
	\label{eq:spacespinham1full}
	\Ham = -\frac{\hbar^2}{2m} \int_{-\infty}^{\infty}\dd[x]  \ket{x} \frac{\dd^2}{\dd[x]^2}\bra{x} \otimes \one
	+ \int_{-\infty}^{\infty}\dd[x] \ket{x}V_0(x)\bra{x} \otimes \one
	+ \int_{-\infty}^{\infty}\dd[x] \ket{x}V_z(x)\bra{x} \otimes \op{S}_z,
\end{equation}
where it is more evident that the first two terms act only on the spatial part of the wavefunction, while the third term couples the two degrees of freedom.

The Hilbert space\index{Hilbert space} of this particle consists of a one-dimensional degree of freedom $x$, which we had described in \autoref{chap:1D} with a basis built from square-well eigenstates, and a spin-$1/2$ degree of freedom $\opvect{S}=\frac12\opvect{\sigma}$ described in the Dicke basis (\autoref{chap:spin}). This tensor-product structure of the Hilbert space allows us to simplify the matrix elements of the Hamiltonian by factoring out the spin degree of freedom,
\scriptsize
\begin{align}
	\me{\phi,{\uparrow}}{\Ham}{\psi,{\uparrow}} &=
	-\frac{\hbar^2}{2m} \int_{-\infty}^{\infty} \phi^*(x)\psi''(x)\dd[x] \scp{{\uparrow}}{{\uparrow}}
	+\int_{-\infty}^{\infty} \phi^*(x)V_0(x) \psi(x)\dd[x] \scp{{\downarrow}}{{\downarrow}}
	+\frac12\int_{-\infty}^{\infty} \phi^*(x)V_z(x) \psi(x)\dd[x] \me{{\uparrow}}{\op{\sigma}_z}{{\uparrow}}\nonumber\\
	&= -\frac{\hbar^2}{2m} \int_{-\infty}^{\infty} \phi^*(x)\psi''(x)\dd[x]
	+\int_{-\infty}^{\infty} \phi^*(x)V_0(x) \psi(x)\dd[x]
	+\frac12 \int_{-\infty}^{\infty} \phi^*(x)V_z(x) \psi(x)\dd[x]\nonumber\\
	\me{\phi,{\uparrow}}{\Ham}{\psi,{\downarrow}} &=
	-\frac{\hbar^2}{2m} \int_{-\infty}^{\infty} \phi^*(x)\psi''(x)\dd[x] \scp{{\uparrow}}{{\downarrow}}
	+\int_{-\infty}^{\infty} \phi^*(x)V_0(x) \psi(x)\dd[x] \scp{{\uparrow}}{{\downarrow}}
	+\frac12\int_{-\infty}^{\infty} \phi^*(x)V_z(x) \psi(x)\dd[x] \me{{\uparrow}}{\op{\sigma}_z}{{\downarrow}}\nonumber\\
	&= 0\nonumber\\
	\me{\phi,{\downarrow}}{\Ham}{\psi,{\uparrow}} &=
	-\frac{\hbar^2}{2m} \int_{-\infty}^{\infty} \phi^*(x)\psi''(x)\dd[x] \scp{{\downarrow}}{{\uparrow}}
	+\int_{-\infty}^{\infty} \phi^*(x)V_0(x) \psi(x)\dd[x] \scp{{\downarrow}}{{\uparrow}}
	+\frac12\int_{-\infty}^{\infty} \phi^*(x)V_z(x) \psi(x)\dd[x] \me{{\downarrow}}{\op{\sigma}_z}{{\uparrow}}\nonumber\\
	&= 0\nonumber\\
	\me{\phi,{\downarrow}}{\Ham}{\psi,{\downarrow}} &=
	-\frac{\hbar^2}{2m} \int_{-\infty}^{\infty} \phi^*(x)\psi''(x)\dd[x] \scp{{\downarrow}}{{\downarrow}}
	+\int_{-\infty}^{\infty} \phi^*(x)V_0(x) \psi(x)\dd[x] \scp{{\downarrow}}{{\downarrow}}
	+\frac12\int_{-\infty}^{\infty} \phi^*(x)V_z(x) \psi(x)\dd[x] \me{{\downarrow}}{\op{\sigma}_z}{{\downarrow}}\nonumber\\
	&= -\frac{\hbar^2}{2m} \int_{-\infty}^{\infty} \phi^*(x)\psi''(x)\dd[x]
	+\int_{-\infty}^{\infty} \phi^*(x)V_0(x) \psi(x)\dd[x]
	-\frac12 \int_{-\infty}^{\infty} \phi^*(x)V_z(x) \psi(x)\dd[x].
\end{align}
\normalsize
We see that this Hamiltonian does not mix states with different spin states (since all matrix elements where the spin state differs between the left and right side are equal to zero). We can therefore solve the two disconnected problems of finding the particle's behavior with spin up or with spin down, with effective Hamiltonians
\begin{align}
	\label{eq:sepham}
	\Ham_{{\uparrow}} &= -\frac{\hbar^2}{2m} \frac{\dd^2}{\dd[x]^2} + V_0(x) + \frac12 V_z(x), &
	\Ham_{{\downarrow}} &= -\frac{\hbar^2}{2m} \frac{\dd^2}{\dd[x]^2} + V_0(x) - \frac12 V_z(x).
\end{align}
These Hamiltonians now only describe the spatial degree of freedom, and the methods of \autoref{chap:1D} can be used without further modifications.

\subsection{non-separable Hamiltonian}

A more interesting situation arises when the Hamiltonian is not separable as in \autoref{sec:spacespinsep}. Take, for example, the Hamiltonian of \autoref{eq:spacespinham1} in the presence of a uniform transverse magnetic field $B_x$,
\begin{equation}
	\label{eq:spacespinham2}
	\Ham = -\frac{\hbar^2}{2m} \frac{\dd^2}{\dd[x]^2} + V_0(x) + V_z(x) \op{S}_z + B_x\op{S}_x.
\end{equation}
The interaction Hamiltonian with the magnetic field is not separable:
\begin{align}
	\me{\phi,{\uparrow}}{B_x\op{S}_x}{\psi,{\uparrow}} &= \frac12B_x \int_{-\infty}^{\infty} \phi^*(x)\psi(x)\dd[x]\me{{\uparrow}}{\op{\sigma}_x}{{\uparrow}} = 0\nonumber\\
	\me{\phi,{\uparrow}}{B_x\op{S}_x}{\psi,{\downarrow}} &= \frac12B_x \int_{-\infty}^{\infty} \phi^*(x)\psi(x)\dd[x]\me{{\uparrow}}{\op{\sigma}_x}{{\downarrow}} = \frac12 B_x \int_{-\infty}^{\infty} \phi^*(x)\psi(x)\dd[x]\nonumber\\
	\me{\phi,{\downarrow}}{B_x\op{S}_x}{\psi,{\uparrow}} &= \frac12B_x \int_{-\infty}^{\infty} \phi^*(x)\psi(x)\dd[x]\me{{\downarrow}}{\op{\sigma}_x}{{\uparrow}} = \frac12 B_x \int_{-\infty}^{\infty} \phi^*(x)\psi(x)\dd[x]\nonumber\\
	\me{\phi,{\downarrow}}{B_x\op{S}_x}{\psi,{\downarrow}} &= \frac12B_x \int_{-\infty}^{\infty} \phi^*(x)\psi(x)\dd[x]\me{{\downarrow}}{\op{\sigma}_x}{{\downarrow}} = 0.
\end{align}
Therefore we can no longer study separate Hamiltonians as in \autoref{eq:sepham}, and we must instead study the joint system of spatial motion and spin. In what follows we study a simple example of such a Hamiltonian, both analytically and numerically. We take the trapping potential to be harmonic,
\begin{equation}
	\label{eq:harmonicpotential}
	V_0(x) = \frac12 m \omega^2 x^2
\end{equation}
and the state-selective potential as a homogeneous force,
\begin{equation}
	\label{eq:force}
	V_z(x) = -F x.
\end{equation}

\subsubsection{ground state for $B_x=0$}

For $B_x=0$ we know that the ground states of the two spin sectors are the ground states of the effective Hamiltonians of \autoref{eq:sepham}, which are Gaussians:
\begin{align}
	\label{eq:Bx0gs}
	\scp{x}{\gamma_{{\uparrow}}} &= \frac{e^{-\left(\frac{x-\mu}{2\sigma}\right)^2}}{\sqrt{\sigma\sqrt{2\pi}}} \otimes\ket{\uparrow}
	& \scp{x}{\gamma_{{\downarrow}}} &= \frac{e^{-\left(\frac{x+\mu}{2\sigma}\right)^2}}{\sqrt{\sigma\sqrt{2\pi}}} \otimes\ket{\downarrow}
\end{align}
with $\mu = \frac{F}{2m\omega^2}$ and $\sigma=\sqrt{\frac{\hbar}{2m\omega}}$. These two ground states are degenerate, with energy $E=\frac12\hbar\omega-\frac{F^2}{8m\omega^2}$. In both of these ground states the spatial and spin degrees of freedom are entangled: the particle is more likely to be detected in the $\ket{\uparrow}$ state on the right side ($x>0$), and more likely to be detected in the $\ket{\downarrow}$ state on the left side ($x<0$) of the trap. This results in a positive expectation value of the operator $\op{x}\otimes\op{S}_z$:
\begin{equation}
	\me{\gamma_{{\uparrow}}}{\op{x}\otimes\op{S}_z}{\gamma_{{\uparrow}}}
	= \me{\gamma_{{\downarrow}}}{\op{x}\otimes\op{S}_z}{\gamma_{{\downarrow}}}
	= \frac{\mu}{2} = \frac{F}{4m\omega^2}.
\end{equation}

\subsubsection{perturbative ground state for $B_x> 0$}

For small $|B_x|$ the ground state can be described by a linear combination of the states in \autoref{eq:Bx0gs}. If we set
\begin{equation}
	\ket{\gamma\ix{p}} = \alpha \times \ket{\gamma_{{\uparrow}}} + \beta \times \ket{\gamma_{{\downarrow}}}
\end{equation}
with $|\alpha|^2+|\beta|^2=1$, we find that the expectation value of the energy is
\begin{multline}
	\me{\gamma\ix{p}}{\Ham}{\gamma\ix{p}} = |\alpha|^2 \me{\gamma_{{\uparrow}}}{\Ham}{\gamma_{{\uparrow}}}
	+ \alpha^*\beta \me{\gamma_{{\uparrow}}}{\Ham}{\gamma_{{\downarrow}}}
	+ \beta^*\alpha \me{\gamma_{{\downarrow}}}{\Ham}{\gamma_{{\uparrow}}}
	+ |\beta|^2 \me{\gamma_{{\downarrow}}}{\Ham}{\gamma_{{\downarrow}}}\\
	= \frac12\hbar\omega-\frac{F^2}{8m\omega^2}
	+ \frac12 B_x (\alpha^*\beta+\beta^*\alpha) e^{-\frac{F^2}{4m\hbar\omega^3}}
\end{multline}
For $B_x>0$ this energy is minimized for $\alpha=1/\sqrt{2}$ and $\beta=-1/\sqrt{2}$, and the perturbative ground state is therefore the anti-symmetric combination of the states in \autoref{eq:Bx0gs}
\begin{equation}
	\label{eq:BxgsP}
	\scp{x}{\gamma\ix{p}} = \frac{e^{-\left(\frac{x-\mu}{2\sigma}\right)^2}}{\sqrt{2\sigma\sqrt{2\pi}}} \otimes\ket{\uparrow}
	- \frac{e^{-\left(\frac{x+\mu}{2\sigma}\right)^2}}{\sqrt{2\sigma\sqrt{2\pi}}} \otimes\ket{\downarrow}.
\end{equation}
with energy
\begin{equation}
	\me{\gamma\ix{p}}{\Ham}{\gamma\ix{p}} = \frac12\hbar\omega-\frac{F^2}{8m\omega^2}-\frac12 B_x e^{-\frac{F^2}{4m\hbar\omega^3}}.
\end{equation}
The energy splitting between this ground state and the first excited state,
\begin{equation}
	\label{eq:BxesP}
	\scp{x}{\epsilon\ix{p}} = \frac{e^{-\left(\frac{x-\mu}{2\sigma}\right)^2}}{\sqrt{2\sigma\sqrt{2\pi}}} \otimes\ket{\uparrow}
	+ \frac{e^{-\left(\frac{x+\mu}{2\sigma}\right)^2}}{\sqrt{2\sigma\sqrt{2\pi}}} \otimes\ket{\downarrow}.
\end{equation}
is $\Delta E = \me{\epsilon\ix{p}}{\Ham}{\epsilon\ix{p}}-\me{\gamma\ix{p}}{\Ham}{\gamma\ix{p}}=B_x e^{-\frac{F^2}{4m\hbar\omega^3}}$, which can be very small for large exponents $\frac{F^2}{4m\hbar\omega^3}$.

\subsubsection[numerical calculation of the ground state]{\label{sec:1Dspacespingroundstate}numerical calculation of the ground state\hspace{\stretch{1}}\attachcode{ParticleMotionWithSpin}{one particle in 1D with spin}}

For a numerical description of this particle we use dimensionless units such that \mm{a=m=\mmhbar=1}; other units can be used in the same was as presented in \autoref{sec:1Dunits}.
We describe the spatial degree of freedom with the finite-resolution position basis of \autoref{sec:positionbasis1D}, centered at $x=0$ as in \autoref{sec:BEC}:
\begin{mathematica}
	¤mathin a = m = ¤mmhbar = 1;
	¤mathin nmax = 100;
	¤mathin ¤mmDelta = a/(nmax+1);
	¤mathin xgrid = a*(Range[nmax]/(nmax+1)-1/2);
\end{mathematica}
The operator $\op{x}$ is approximately diagonal in this representation (see \autoref{eq:positionoperator}):
\begin{mathematica}
	¤mathin xop = SparseArray[Band[{1,1}] -> xgrid];
\end{mathematica}
The identity operator on the spatial degree of freedom is
\begin{mathematica}
	¤mathin idx = IdentityMatrix[nmax, SparseArray];
\end{mathematica}
The identity and Pauli operators for the spin degree of freedom are
\begin{mathematica}
	¤mathin ids = IdentityMatrix[2, SparseArray];
	¤mathin {sx,sy,sz}=Table[SparseArray[PauliMatrix[i]/2], {i,3}];
\end{mathematica}
The kinetic energy operator is constructed via a discrete sine transform, as before:
\begin{mathematica}
	¤mathin TM = SparseArray[Band[{1,1}]->Range[nmax]^2*¤mmpi^2*¤mmhbar^2/(2*m*a^2)];
	¤mathin TP = FourierDST[TM, 1];
\end{mathematica}
From these we assemble the Hamiltonian, assuming that $F$ and $B_x$ are expressed in matching units:
\begin{mathematica}
	¤mathin H[¤mmomega_, F_, Bx_] = 
	¤mathnl   KroneckerProduct[TP, ids]
	¤mathnl   + m*¤mmomega^2/2 * KroneckerProduct[xop.xop, ids]
	¤mathnl   - F * KroneckerProduct[xop, sz]
	¤mathnl   + Bx * KroneckerProduct[idx, sx];
\end{mathematica}
We compute the ground state of this Hamiltonian with
\begin{mathematica}
	¤mathin Clear[gs];
	¤mathin gs[¤mmomega_?NumericQ, F_?NumericQ, Bx_?NumericQ] :=
	¤mathnl   gs[¤mmomega, F, Bx] = -Eigensystem[-H[N[¤mmomega],N[F],N[Bx]], 1,
	¤mathnl     Method -> {"Arnoldi", "Criteria" -> "RealPart", MaxIterations -> 10^6}]
\end{mathematica}
Once a ground state $\ket{\gamma}$ has been calculated, for example with
\begin{mathematica}
	¤mathin ¤mmgamma = gs[100, 5000, 500][[2, 1]];
\end{mathematica}
the usual problem arises of how to display and interpret the wavefunction.
Instead of studying the coefficients of \mm{\mmgamma} directly, we calculate several specific properties of the ground state in what follows.
\begin{description}
	\item[Operator expectation values:] The mean spin direction (magnetization) $\avg{\opvect{S}}=\{\avg{\op{S}_x},\avg{\op{S}_y},\avg{\op{S}_z}\}$ is calculated directly from the ground-state coefficients list with
\begin{mathematica}
	¤mathin mx = Re[Conjugate[¤mmgamma].(KroneckerProduct[idx,sx].¤mmgamma)];
	¤mathin my = Re[Conjugate[¤mmgamma].(KroneckerProduct[idx,sy].¤mmgamma)];
	¤mathin mz = Re[Conjugate[¤mmgamma].(KroneckerProduct[idx,sz].¤mmgamma)];
	¤mathin {mx,my,mz}
	¤mathout {-0.233037, 0., -2.08318*10^-12}
\end{mathematica}
		The mean position $\avg{\op{x}}$ and its standard deviation are calculated with
\begin{mathematica}
	¤mathin X = Re[Conjugate[¤mmgamma].(KroneckerProduct[xop,ids].¤mmgamma)];
	¤mathin XX = Re[Conjugate[¤mmgamma].(KroneckerProduct[xop.xop,ids].¤mmgamma)];
	¤mathin {X, Sqrt[XX-X^2]}
	¤mathout {1.2178*10^-11, 0.226209}
\end{mathematica}
		Even though we found $\avg{\op{x}}=0$ and $\avg{\op{S}_z}=0$ above, these coordinates are correlated: calculating $\avg{\op{x}\otimes\op{S}_z}$,
\begin{mathematica}
	¤mathin Xz = Re[Conjugate[¤mmgamma].(KroneckerProduct[xop,sz].¤mmgamma)]
	¤mathout 0.0954168
\end{mathematica}
	\item[Reduced density matrix of the spatial degree of freedom:]\index{partial trace} Using \mm{\ref{math:traceoutpsi2}} we trace out the spin degree of freedom (the last two dimensions) to find the density matrix in the spatial coordinate:
\begin{mathematica}
	¤mathin ¤mmrhoŽx = traceout[¤mmgamma, -2];
	¤mathin ArrayPlot[Reverse[Transpose[ArrayPad[¤mmrhoŽx/¤mmDelta, 1]]]]
\end{mathematica}
\begin{center}
\includegraphics[width=0.5\textwidth]{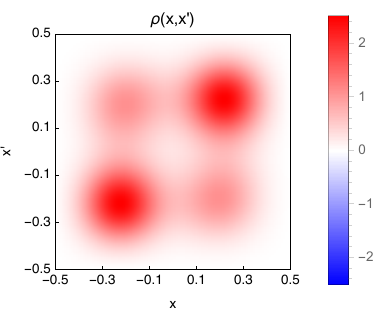}
\end{center}
	\item[Reduced density matrix of the spin degree of freedom:] We can do the same for the reduced matrix of the spin degree of freedom, using \mm{\ref{math:traceoutpsi1}}, and find a \num{2 x 2} spin density matrix:
\begin{mathematica}
	¤mathin ¤mmrhoŽs = traceout[¤mmgamma, nmax]
	¤mathout {{0.5, -0.233037}, {-0.233037, 0.5}}
\end{mathematica}
	\item[Spin-specific spatial densities:] The reduced density matrix of particles in the spin-up state is found by projecting the ground state $\ket{\gamma}$ onto the spin-up sector with the projector $\op{\Pi}_{{\uparrow}}=\ket{\uparrow}\bra{\uparrow}=\frac12\one+\op{S}_z$.\footnote{Remember that $\one=\ket{\uparrow}\bra{\uparrow}+\ket{\downarrow}\bra{\downarrow}$ and $\op{S}_z=\frac12\ket{\uparrow}\bra{\uparrow}-\frac12\ket{\downarrow}\bra{\downarrow}$.} Thus, $\ket{\gamma_{{\uparrow}}}=\op{\Pi}_{{\uparrow}}\ket{\gamma}$ only describes the particles that are in the spin-up state:
\begin{mathematica}
	¤mathin ¤mmgammaŽup = KroneckerProduct[idx, ids/2+sz].¤mmgamma;
	¤mathin ¤mmrhoŽxup = traceout[¤mmgammaŽup, -2];
\end{mathematica}
	In the same way the reduced density matrix of particles in the spin-down state $\ket{\gamma_{{\downarrow}}}=\op{\Pi}_{{\downarrow}}\ket{\gamma}$ is calculated with the down-projector $\op{\Pi}_{{\downarrow}}=\ket{\downarrow}\bra{\downarrow}=\frac12\one-\op{S}_z$:
\begin{mathematica}
	¤mathin ¤mmgammaŽdn = KroneckerProduct[idx, ids/2-sz].¤mmgamma;
	¤mathin ¤mmrhoŽxdn = traceout[¤mmgammaŽdn, -2];
\end{mathematica}
\begin{center}
\includegraphics[width=0.45\textwidth]{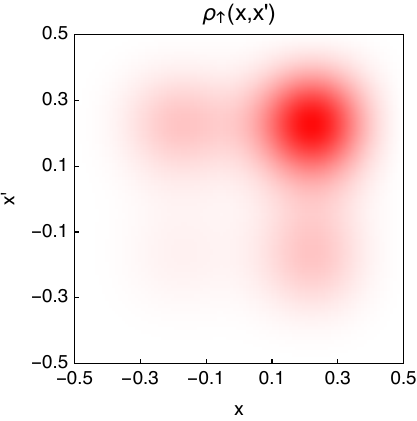}
\includegraphics[width=0.45\textwidth]{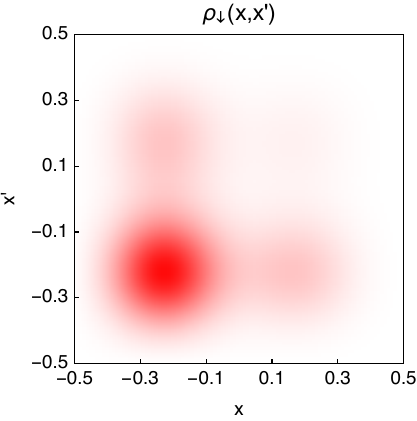}
\end{center}
	The positive correlation between the spin and the mean position, $\avg{\op{x}\otimes\op{S}_z}>0$, is clearly visible in these plots.
	
	Since $\op{\Pi}_{{\uparrow}}+\op{\Pi}_{{\downarrow}}=\one$, these two spin-specific spatial density matrices add up to the total density shown previously. This also means that the spin-specific density matrices do not have unit trace:
\begin{mathematica}
	¤mathin {Tr[¤mmrhoŽxup], Tr[¤mmrhoŽxdn]}
	¤mathout {0.5, 0.5}
\end{mathematica}
	Hence we have 50\% chance of finding the particle in the up or down spin states.
	\item[Space-dependent spin expectation value:] Similarly, we can calculate the reduced density matrix of the spin degree of freedom at a specific point in space by using projection operators $\op{\Pi}_j=\ket{j}\bra{j}$ onto single position-basis states $\ket{j}$:
\begin{mathematica}
	¤mathin ¤mmgammaŽx[j_Integer /; 1 <= j <= nmax] :=
	¤mathnl   KroneckerProduct[SparseArray[{j, j} -> 1, {nmax, nmax}], ids].¤mmgamma
	¤mathin ¤mmrhoŽsx[j_Integer /; 1 <= j <= nmax] := traceout[¤mmgammaŽx[j], nmax]
\end{mathematica}
	We notice that, as before, these spatially-local reduced density matrices do not have unit trace, but their traces sum up to 1:
\begin{mathematica}
	¤mathin Sum[Tr[¤mmrhoŽsx[j]], {j, nmax}]
	¤mathout 1.
\end{mathematica}
	In fact, the traces of these local reduced density matrices give the probability of finding the particle at the given position. We can use this interpretation to calculate the mean spin expectation value of a particle measured at a given grid point:
\begin{mathematica}
	¤mathin meansx[j_Integer /; 1 <= j <= nmax] := Tr[¤mmrhoŽsx[j].sz]/Tr[¤mmrhoŽsx[j]]
	¤mathin ListLinePlot[Transpose[{xgrid, Table[meansx[j], {j, nmax}]}]]
\end{mathematica}
\begin{center}
\includegraphics[width=0.5\textwidth]{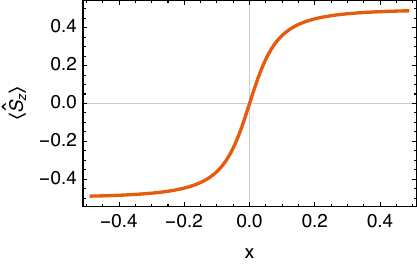}
\end{center}
This graph confirms the observation that particles detected on the left side are more likely to be in the $\ket{\downarrow}$ state, while particles detected on the right side are more likely to be in the $\ket{\uparrow}$ state.
\end{description}

\subsection{exercises}

\begin{questions}
	\item\label{Q:spacespin1} In the problem described by the Hamiltonian of \autoref{eq:spacespinham2}, calculate the following expectation values (numerically) for several parameter sets $\{\omega, F, B_x\}$:
		\begin{enumerate}
			\item $\avg{x}$ for particles detected in the $\ket{\uparrow}$ state
			\item $\avg{x}$ for particles detected in the $\ket{\downarrow}$ state
			\item $\avg{x}$ for particles detected in any spin state
			\item the mean and variance of $\op{x}\otimes\op{S}_z$
		\end{enumerate}
\pagenote[\ref{Q:spacespin1}]{The operators for these expectation values are
	\begin{enumerate}
		\item \mm{A1 = KroneckerProduct[xop,\mmPi$_{\uparrow}$] = KroneckerProduct[xop,ids/2+sz]}
		\item \mm{A2 = KroneckerProduct[xop,\mmPi$_{\downarrow}$] = KroneckerProduct[xop,ids/2-sz]}
		\item \mm{A3 = KroneckerProduct[xop,ids] = A1+A2}
		\item \mm{A4 = KroneckerProduct[xop,sz] = (A1-A2)/2}
	\end{enumerate}
	With these we evaluate the quantities
	\begin{enumerate}
		\item \mm{Re[Conjugate[\mmgamma].(A1.\mmgamma)]}
		\item \mm{Re[Conjugate[\mmgamma].(A2.\mmgamma)]}
		\item \mm{Re[Conjugate[\mmgamma].(A3.\mmgamma)]}
		\item \mm{Re[Conjugate[\mmgamma].(A4.\mmgamma)]} for the mean\\
			\mm{Re[Conjugate[\mmgamma].(A4.A4.\mmgamma)-(Conjugate[\mmgamma].(A4.\mmgamma))\^{}2]} for the variance
	\end{enumerate}}
\end{questions}

\section[one particle in 2D with spin: Rashba coupling]{\label{sec:Rashba}one particle in 2D with spin: Rashba coupling\hspace{\stretch{1}}\attachcode{RashbaCoupling}{one particle in 2D with Rashba coupling}}
\index{Rashba coupling}

A particularly interesting kind of interaction is the Rashba coupling between a particle's momentum and its spin.\footnote{See \url{https://en.wikipedia.org/wiki/Rashba_effect}.} In general, this interaction is proportional to a component of the vector product $\opvect{\kappa}=\opvect{p}\times\opvect{S}$. For a particle moving in two dimensions $(x,y)$, the coupling involves the $z$-component $\op{\kappa}_z=\op{p}_x\otimes\op{S}_y-\op{p}_y\otimes\op{S}_x$.

In this section we study the 2D Rashba Hamiltonian
\begin{equation}
	\label{eq:RashbaHamAbbrev}
	\Ham = \frac{\op{p}_x^2+\op{p}_y^2}{2m} + V(x,y) + \delta \op{S}_z + \alpha(\op{p}_x\otimes\op{S}_y-\op{p}_y\otimes\op{S}_x)
\end{equation}
in a square box where $-\frac{a}{2}\le x,y\le \frac{a}{2}$ as before. With a Hilbert space composed as the tensor product of the $x$, $y$, and spin coordinates, in this order, the full Hamiltonian thus becomes
\begin{multline}
	\label{eq:RashbaHam}
	\Ham
	= \left[ -\frac{\hbar^2}{2m}\int_{-a/2}^{a/2}\dd[x]\ket{x}\frac{\partial^2}{\ddd{\partial}{x}^2}\bra{x} \right] \otimes\one\otimes\one
	+\one\otimes\left[-\frac{\hbar^2}{2m}\int_{-a/2}^{a/2}\dd[x]\ket{y}\frac{\partial^2}{\ddd{\partial}{y}^2}\bra{y}\right]\otimes\one\\
	+ \left[ \int_{-a/2}^{a/2}\dd[x]\dd[y]\ket{x}\ket{y}V(x,y)\bra{x}\bra{y} \right] \otimes\one
	+ \delta(\one\otimes\one\otimes \op{S}_z)
	+\alpha(\op{p}_x\otimes\one\otimes\op{S}_y-\one\otimes\op{p}_y\otimes\op{S}_x).
\end{multline}
For simplicity we will set $V(x,y)=0$; but any nonzero potential can be used with the techniques introduced previously. Further, we use \mm{a=m=\mmhbar=1} to simplify the units; but as usual, any system of units may be used (see \autoref{sec:1Dunits}).

Since both the kinetic and the interaction operator are most easily expressed in the momentum representation, we use the momentum representation (see \autoref{sec:momentumbasis1D}) to express the spatial degrees of freedom of the Hamiltonian. The identity operator is
\begin{mathematica}
	¤mathin nmax = 50;
	¤mathin ¤mmDelta = a/(nmax+1);
	¤mathin idM = IdentityMatrix[nmax, SparseArray];
\end{mathematica}
We use the exact form of the kinetic operator\index{operator!kinetic} from \mm{\ref{math:HkinM}} and the exact form of the momentum operator\index{operator!momentum} from \mm{\ref{math:poperator}}. As discussed previously, these two forms do not exactly satisfy $\op{T}=\op{p}^2/(2m)$. They are, however, the best available low-energy forms.

For the spin degree of freedom, we assume $S=1/2$, giving us the usual spin operators and the identity operator,
\begin{mathematica}
	¤mathin {sx,sy,sz} = Table[SparseArray[PauliMatrix[i]/2], {i, 3}];
	¤mathin idS = IdentityMatrix[2, SparseArray];
\end{mathematica}
With these definitions, we assemble the Rashba Hamiltonian of \autoref{eq:RashbaHam} in the momentum representation with
\begin{mathematica}
	¤mathin¤labelŽmath:RashbaHam HM[¤mmdelta_, ¤mmalpha_] = KroneckerProduct[TM, idM, idS]
	¤mathnl   + KroneckerProduct[idM, TM, idS]
	¤mathnl   + ¤mmdelta*KroneckerProduct[idM, idM, sz]
	¤mathnl   + ¤mmalpha*(KroneckerProduct[pM, idM, sy] - KroneckerProduct[idM, pM, sx]);
\end{mathematica}
Given a state \mm{\mmgamma}, for example the ground state of \mm{\ref{math:RashbaHam}} for specific values of $\delta=1$ and $\alpha=20$, we calculate the mean value $\avg{\op{x}^2}=\avg{\op{x}^2\otimes\one\otimes\one}$ with the position operator \mm{xM} expressed in the momentum basis:
\begin{mathematica}
	¤mathin xgrid = a*(Range[nmax]/(nmax + 1) - 1/2);
	¤mathin xP = SparseArray[Band[{1, 1}] -> xgrid];
	¤mathin xM = FourierDST[xP, 1];
	¤mathin Conjugate[¤mmgamma].(KroneckerProduct[xM.xM, idM, idS].¤mmgamma) //Re
	¤mathout 0.0358875
\end{mathematica}
In the same way, we calculate the mean value $\avg{\op{y}^2}=\avg{\one\otimes\op{y}^2\otimes\one}$:
\begin{mathematica}
	¤mathin Conjugate[¤mmgamma].(KroneckerProduct[idM, xM.xM, idS].¤mmgamma) //Re
	¤mathout 0.0358875
\end{mathematica}
In order to study the spatial variation of the spin (the expectation value of the spin degree of freedom if the particle is detected at a specific spatial location), we calculate the reduced density matrix of the spin degree of freedom at a specific grid point $(x_i,y_j)$ of the position grid.\footnote{Naturally, the following calculations would be simpler if we had represented the ground state in the position basis; however, we use this opportunity to show how to calculate in the momentum basis.}
For this, we first project the ground-state wavefunction \mm{\mmgamma} onto the spatial grid point at $x=x_i$ and $y=y_j$ using the projector $\ket{i}\bra{i}\otimes\ket{j}\bra{j}$ in the momentum representation:
\begin{mathematica}
	¤mathin ¤mmPiŽP[j_] := SparseArray[{j, j} -> 1, {nmax, nmax}]
	¤mathin ¤mmPiŽM[j_] := FourierDST[¤mmPiŽP[j], 1]
	¤mathin gP[i_,j_] := KroneckerProduct[¤mmPiŽM[i], ¤mmPiŽM[j], idS].¤mmgamma
\end{mathematica}
Tracing out the spatial degrees of freedom with the procedure of \autoref{sec:rdm} gives the $2\times2$ spin density matrix at the desired grid point,
\begin{mathematica}
	¤mathin RsP[i_, j_] := traceout[gP[i,j], nmax^2]
\end{mathematica}
The trace \mm{Tr[RsP[i,j]]} of such a reduced density matrix gives the probability of finding the particle at grid point $(x_i,y_j)$:
\begin{center}
\includegraphics[width=0.45\textwidth]{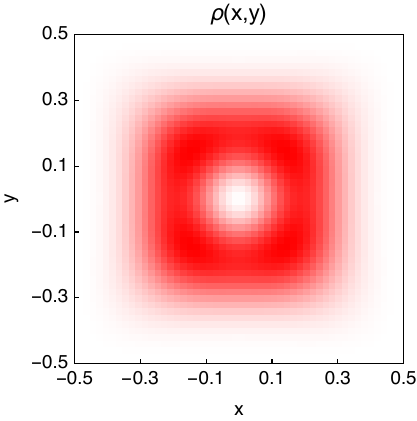}
\end{center}
We can extract more information from these reduced spin density matrices: the magnetization (mean spin direction) at a grid point has the Cartesian components
\begin{mathematica}
	¤mathin mxP[i_, j_] := Re[Tr[RsP[i,j].sx]/Tr[RsP[i,j]]]
	¤mathin myP[i_, j_] := Re[Tr[RsP[i,j].sy]/Tr[RsP[i,j]]]
	¤mathin mzP[i_, j_] := Re[Tr[RsP[i,j].sz]/Tr[RsP[i,j]]]
\end{mathematica}
Plotting these components over the entire grid shows interesting patterns of the mean spin orientation (magnetization) in the ground state:
\begin{center}
\includegraphics[width=0.3\textwidth]{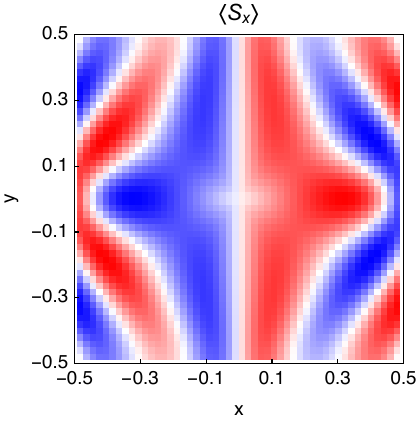}
\includegraphics[width=0.3\textwidth]{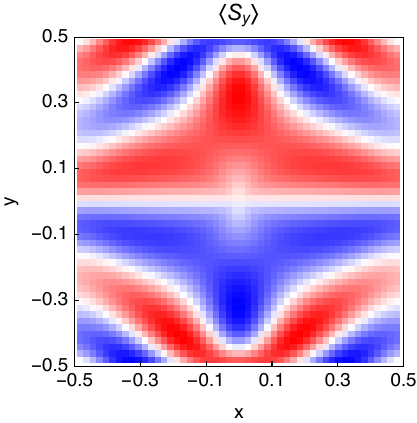}
\includegraphics[width=0.3\textwidth]{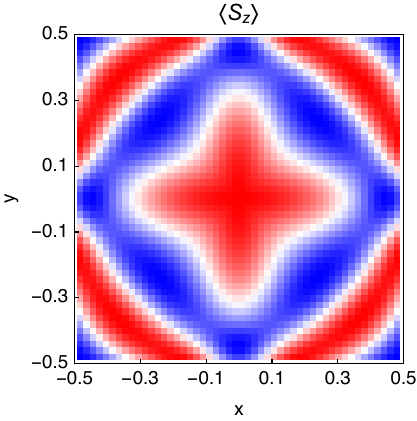}
\includegraphics[width=0.07\textwidth]{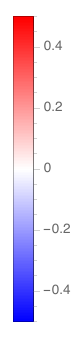}
\end{center}

\subsection{exercises}

\begin{questions}
	\item\label{Q:Rashba} While the Hamiltonian of \autoref{eq:RashbaHamAbbrev} and \autoref{eq:RashbaHam} contains the distinct operators $\op{p}_x$ and $\op{p}_y$, the Mathematica form of the this Hamiltonian assembled in \mm{\ref{math:RashbaHam}} contains the same matrix \mm{pM} representing both $\op{p}_x$ and $\op{p}_y$. Why is this so? What distinguishes the Mathematica representations of these two operators?
\pagenote[\ref{Q:Rashba}]{The ordering of the subspaces of the Hilbert space is what matters here. We have defined the Hilbert space to be a tensor product of the $x$, $y$, and spin degrees of freedom, in this order. In \mm{\ref{math:RashbaHam}} the operators $\op{p}_x$ and $\op{p}_y$ are distinguished by the position in the Kronecker product in which \mm{pM} appears.}
\end{questions}

\section[phase-space dynamics in the Jaynes--Cummings model]{\label{sec:JaynesCummings}phase-space dynamics in the Jaynes--Cummings model\hspace{\stretch{1}}\attachcode{JaynesCummingsModel}{phase-space dynamics in the Jaynes--Cummings model}}
\index{Jaynes-Cummings model@Jaynes--Cummings model}

As a final example, we study the interaction of an atom with the light field in an optical cavity. The atom is assumed to have only two internal states: the ground state $\ket{g}$ and some excited state $\ket{e}$. The atomic state is described as a (pseudo-)spin-1/2 system with the operators (see \ref{Q:pseudospin})
\begin{align}
	\label{eq:pseudospin}
	\op{S}_x &= \frac{\ket{e}\bra{g}+\ket{g}\bra{e}}{2}, &
	\op{S}_y &= \frac{\ket{e}\bra{g}-\ket{g}\bra{e}}{2\ii}, &
	\op{S}_z &= \frac{\ket{e}\bra{e}-\ket{g}\bra{g}}{2},
\end{align}
as well as $\op{S}^{\pm}=\op{S}_x\pm\ii\op{S}_y$ (see \autoref{sec:electronspinB}). The cavity field is assumed to consist of only one mode, described with creation and annihilation operators $\op{a}\dagg$ and $\op{a}$, respectively; all other cavity modes are assumed to be so far off-resonant that they are not coupled to the atom.

The Jaynes--Cummings Hamiltonian\footnote{See \url{https://en.wikipedia.org/wiki/Jaynes-Cummings_model}.} describing the combined system, as well as the coupling between the atom and the cavity field, is
\begin{equation}
	\label{eq:JCHam0}
	\Ham\ix{JC} =
		\underbrace{\hbar\omega\ix{a}\op{S}_z}\ix{atom}
		+ \underbrace{\hbar\omega\ix{c}(\op{a}\dagg\op{a}+\frac12)}\ix{cavity field}
		+ \underbrace{\hbar g(\op{S}^+\op{a}+\op{a}\dagg\op{S}^-)}\ix{coupling}.
\end{equation}
\begin{itemize}
	\item The atomic Hamiltonian describes the energy difference $\hbar\omega\ix{a}$ between the two internal states of the atom.
	\item The cavity field Hamiltonian decribes the energy of $\op{n}=\op{a}\dagg\op{a}$ photons in the cavity mode, each photon carrying an energy $\hbar\omega\ix{c}$.
	\item The coupling term describes the deexcitation of the field $\op{a}$ together with the excitation of the atom $\op{S}^+$, as well as the reverse process of the excitation of the field $\op{a}\dagg$ together with the deexcitation of the atom $\op{S}^-$ (see \ref{Q:Splusminus}).
\end{itemize}
The cavity mode of the Jaynes--Cummings model is usually studied in the Fock basis\index{Fock basis} of definite photon number, using harmonic-oscillator eigenfunctions as basis states. Here we take an alternative approach and look at the $X-P$ phase space spanned by the dimensionless quadrature operators $\op{X}$ and $\op{P}$,\footnote{In a harmonic oscillator of mass $m$ and angular frequency $\omega$, we usually introduce the position operator $\op{x}=\sqrt{\frac{\hbar}{m\omega}}\op{X}$ and the momentum operator $\op{p}=\sqrt{\hbar m \omega}\op{P}$. Here we restrict our attention to the dimensionless quadratures $\op{X}$ and $\op{P}$.} which are related to the creation and annihilation operators $\op{a}\dagg$ and $\op{a}$ via
\begin{align}
	\label{eq:XPvsaad}
	\op{X} &= \frac{\op{a}+\op{a}\dagg}{\sqrt{2}} & \op{a} &= \frac{\op{X}+\ii\op{P}}{\sqrt{2}} = \frac{\op{X}+\frac{\partial}{\ddd{\partial}\op{X}}}{\sqrt{2}}\nonumber\\
	\op{P} &= -\ii\frac{\partial}{\ddd{\partial}\op{X}}=\frac{\op{a}-\op{a}\dagg}{\ii\sqrt{2}} & \op{a}\dagg &= \frac{\op{X}-\ii\op{P}}{\sqrt{2}} = \frac{\op{X}-\frac{\partial}{\ddd{\partial}\op{X}}}{\sqrt{2}}
\end{align}
with the commutators $[\op{a},\op{a}\dagg]=1$ and $[\op{X},\op{P}]=\ii$ (see \ref{Q:XPcommutator}).
We note that the quadrature $\op{X}$ is the amplitude of the electromagnetic field of the cavity mode, and $\op{P}$ its conjugate momentum; there is no motion in real space in this problem, only in amplitude space.
Using these quadrature operators, we write the Jaynes--Cummings Hamiltonian as (see \ref{Q:JCHamTransform})
\begin{align}
	\label{eq:JCHam}
	\Ham\ix{JC} &=
		\hbar\omega\ix{a}\op{S}_z
		+ \hbar\omega\ix{c}(\frac12\op{P}^2+\frac12\op{X}^2)
		+ \sqrt{2}\hbar g(\op{X}\op{S}_x-\op{P}\op{S}_y)\nonumber\\
		&= \hbar\omega\ix{a} \one \otimes \op{S}_z
		+ \hbar\omega\ix{c}(\frac12\op{P}^2+\frac12\op{X}^2) \otimes \one
		+ \sqrt{2}\hbar g(\op{X}\otimes\op{S}_x-\op{P}\otimes\op{S}_y),
\end{align}
where we have made its tensor-product structure explicit in the second line.
To assemble this Hamiltonian in Mathematica, we define the Hilbert space to be the tensor product of the $X-P$ phase space and the spin-1/2 space, in this order.

The phase space is defined as before (see \autoref{chap:1D}) in a calculation box $X\in[-\frac{a}{2},\frac{a}{2}]$ divided into a grid of $n\ix{max}+1$ intervals. We choose $a$ such that the state fits well into the box (considering that the ground state of the cavity field has a size $\avg{\op{X}^2}^{1/2}=\avg{\op{P}^2}^{1/2}=1/\sqrt{2}$), and we choose $n\ix{max}$ such that the Wigner quasi-probability distribution plots have equal ranges in $X$ and $P$. Naturally, any other values of $a$ and $n\ix{max}$ can be chosen.
\begin{mathematica}
	¤mathin ¤mmhbar = 1;   (* natural units *)
	¤mathin a = 10;
	¤mathin nmax = Round[a^2/¤mmpi]
	¤mathout 32
	¤mathin ¤mmDelta = a/(nmax+1);
\end{mathematica}
We represent the phase space in the position basis. The $\op{X}$\index{operator!position} quadrature operator is defined as in \autoref{eq:positionoperator},
\begin{mathematica}
	¤mathin xgrid = a*(Range[nmax]/(nmax+1) - 1/2);
	¤mathin X = SparseArray[Band[{1, 1}] -> xgrid];
\end{mathematica}
The definition of the $\op{P}$\index{operator!momentum} quadrature operator follows \mm{\ref{math:poperator}}, with $\op{P}^2$ defined directly through \mm{\ref{math:HkinM}} for better accuracy at finite $n\ix{max}$:
\begin{mathematica}
	¤mathin P = FourierDST[SparseArray[{n1_, n2_} /; OddQ[n1-n2] ->
	¤mathnl   4*I*n1*n2)/(a*(n2^2-n1^2)), {nmax, nmax}], 1];
	¤mathin P2 = FourierDST[SparseArray[Band[{1, 1}] -> Range[nmax]^2*¤mmpi^2/a^2], 1];
\end{mathematica}
Finally, the phase-space identity operator is
\begin{mathematica}
	¤mathin idX = IdentityMatrix[nmax, SparseArray];
\end{mathematica}
The operators on the pseudo-spin degree of freedom are defined directly from the Pauli matrices instead of using the general definitions of \autoref{eq:spinopme1} and \autoref{eq:spinopme2}:
\begin{mathematica}
	¤mathin {Sx, Sy, Sz} = Table[SparseArray[PauliMatrix[i]/2], {i, 3}];
	¤mathin idS = IdentityMatrix[2, SparseArray];
\end{mathematica}
The Hamiltonian of \autoref{eq:JCHam} is assembled from three parts:
\begin{mathematica}
	¤mathin Ha = KroneckerProduct[idX, Sz];
	¤mathin Hc = KroneckerProduct[X.X/2 + P2/2, idS];
	¤mathin¤labelŽmath:HintJC Hint = Sqrt[2]*(KroneckerProduct[X,Sx]-Re[KroneckerProduct[P,Sy]]);
	¤mathin HP[¤mmomegaŽa_, ¤mmomegaŽc_, g_] = ¤mmhbar*¤mmomegaŽa*Ha + ¤mmhbar*¤mmomegaŽc*Hc + ¤mmhbar*g*Hint;
\end{mathematica}
Remember that we use \mm{P2} instead of \mm{P.P} for the operator $\op{P}^2$ for better accuracy. We use the \mm{Re} operator in \mm{\ref{math:HintJC}} to eliminate the imaginary parts, which are zero by construction but render the expression \mm{Complex}-valued nonetheless.

In the Mathematica notebook attached to this section, the dynamics induced by this time-independent Hamiltonian is studied in the weak and strong coupling regimes, using the technique of \autoref{sec:timeindependentH} to propagate the initial wavefunction.

Given a calculated space$\otimes$spin wavefunction $\mm{\mmpsi}$ (a vector of $2n\ix{max}$ complex numbers), we calculate the $n\ix{max}\times n\ix{max}$ reduced density matrix of the phase-space degree of freedom (cavity field) with \mm{\ref{math:traceoutpsi2}}, tracing out the spin degree of freedom (the last 2 dimensions):
\begin{mathematica}
	¤mathin¤labelŽmath:rX ¤mmrhoŽX = traceout[¤mmpsi, -2];
\end{mathematica}
Similarly, we calculate the \num{2 x 2} reduced density matrix of the spin degree of freedom (atomic state) with \mm{\ref{math:traceoutpsi1}}, tracing out the phase-space degree of freedom (the first $n\ix{max}$ dimensions):
\begin{mathematica}
	¤mathin ¤mmrhoŽS = traceout[¤mmpsi, nmax];
\end{mathematica}
Expectation values in the field or spin degrees of freedom are then easily calculated from these reduced density matrices.

To illustrate these techniques, we calculate the time-dependent wavefunction in the resonant weak-coupling regime ($\omega\ix{a}=\omega\ix{c}=1$, $g=0.1$; initial state: coherent field state at $\avg{\op{X}}=\sqrt{2}$ and $\avg{\op{P}}=0$, spin down). First we show the time-dependence of the atomic spin expectation values, calculated from a reduced spin density matrix with
\begin{mathematica}
	¤mathin {Tr[¤mmrhoŽS.Sx], Tr[¤mmrhoŽS.Sy], Tr[¤mmrhoŽS.Sz]}
\end{mathematica}
\begin{center}
\includegraphics[width=0.5\textwidth]{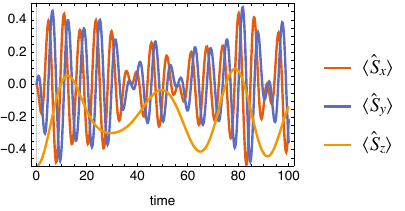}
\end{center}
Observations:
\begin{itemize}
	\item At $t=0$ we recognize the initial spin-down state: $\avg{\op{S}_x}=\avg{\op{S}_y}=0$ and $\avg{\op{S}_z}=-\frac12$.
	\item The $S_x$ and $S_y$ spin components rotate rapidly due to the pseudo-spin excitation energy $\hbar\omega\ix{a}$ (phase factor $e^{-\ii\omega\ix{a}t}$). They are \ang{90} out of phase.
	\item The $S_z$ spin component has a complicated time dependence. Since the atomic energy is $\hbar\omega\ix{a}\avg{\op{S}_z}$, this curve shows the energy flowing between the atom and the cavity light field. 
\end{itemize}
The phase-space Wigner quasi-probability distribution\index{Wigner distribution} of the cavity field, calculated using \mm{\ref{math:WignerR}} from the reduced phase space density matrix of \mm{\ref{math:rX}}, using the same weak-coupling conditions as above, is plotted here at two evolution times:
\begin{mathematica}
	¤mathin WignerDistributionPlot[¤mmrhoŽX, {-a/2, a/2}]
\end{mathematica}
\begin{center}
\includegraphics[width=0.45\textwidth]{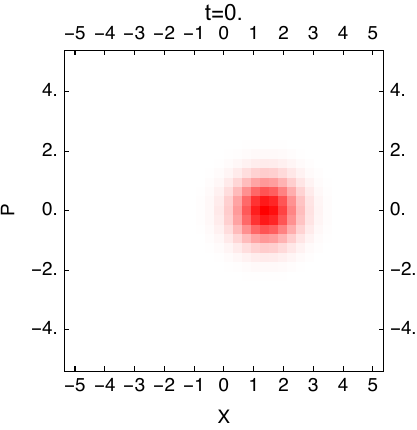}
\includegraphics[width=0.45\textwidth]{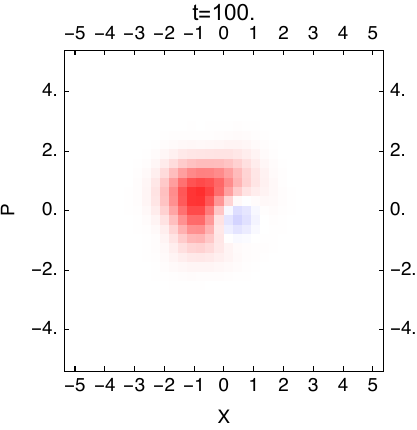}
\includegraphics[width=0.07\textwidth]{WignerDistribution_legend}
\end{center}
Observations:
\begin{itemize}
	\item At $t=0$ we recognize the initial state: a coherent state (circular Gaussian of minimal area $\avg{\op{X}^2}=\avg{\op{P}^2}=\frac12$) displaced by $\delta=\sqrt{2}$ in the $X$-direction, implying $\delta^2/2=1$ photon present initially.
	\item At $t=100$ the structure of the Wigner distribution has taken on a qualitatively different shape, including a significant negative-valued region. Such negative regions are forbidden in classical phase-space distributions and hence indicate an essentially quantum-mechanical state.
\end{itemize}

\subsection{exercises}

\begin{questions}
	\item\label{Q:pseudospin} Show that the operators of \autoref{eq:pseudospin} represent a pseudo-spin-1/2,\index{pseudospin} like in \ref{Q:spinverify}.
\pagenote[\ref{Q:pseudospin}]{We do the first two checks of \ref{Q:spinverify}:
	\begin{align}
		1.&& [\op{S}_x,\op{S}_y]&=[\frac{\ket{e}\bra{g}+\ket{g}\bra{e}}{2}, \frac{\ket{e}\bra{g}-\ket{g}\bra{e}}{2\ii}]\nonumber\\
		&&&=\frac{(\ket{e}\bra{g}+\ket{g}\bra{e})(\ket{e}\bra{g}-\ket{g}\bra{e})-(\ket{e}\bra{g}-\ket{g}\bra{e})(\ket{e}\bra{g}+\ket{g}\bra{e})}{4\ii}\nonumber\\
		&&&=\frac{(-\ket{e}\bra{e}+\ket{g}\bra{g})-(\ket{e}\bra{e}-\ket{g}\bra{g})}{4\ii}\nonumber\\
		&&&=\frac{\ket{g}\bra{g}-\ket{e}\bra{e}}{2\ii}
		=\ii\frac{\ket{e}\bra{e}-\ket{g}\bra{g}}{2}
		=\ii\op{S}_z\\
		&& [\op{S}_y,\op{S}_z]&=[\frac{\ket{e}\bra{g}-\ket{g}\bra{e}}{2\ii}, \frac{\ket{e}\bra{e}-\ket{g}\bra{g}}{2}]\nonumber\\
		&&&=\frac{(\ket{e}\bra{g}-\ket{g}\bra{e})(\ket{e}\bra{e}-\ket{g}\bra{g})-(\ket{e}\bra{e}-\ket{g}\bra{g})(\ket{e}\bra{g}-\ket{g}\bra{e})}{4\ii}\nonumber\\
		&&&=\frac{(-\ket{e}\bra{g}-\ket{g}\bra{e})-(\ket{e}\bra{g}+\ket{g}\bra{e})}{4\ii}\nonumber\\
		&&&=\frac{-\ket{e}\bra{g}-\ket{g}\bra{e}}{2\ii}=\ii\frac{\ket{e}\bra{g}+\ket{g}\bra{e}}{2}=\ii\op{S}_x\\
		&& [\op{S}_z,\op{S}_x]&=[\frac{\ket{e}\bra{e}-\ket{g}\bra{g}}{2},\frac{\ket{e}\bra{g}+\ket{g}\bra{e}}{2}]\nonumber\\
		&&&=\frac{(\ket{e}\bra{e}-\ket{g}\bra{g})(\ket{e}\bra{g}+\ket{g}\bra{e})-(\ket{e}\bra{g}+\ket{g}\bra{e})(\ket{e}\bra{e}-\ket{g}\bra{g})}{4}\nonumber\\
		&&&=\frac{(\ket{e}\bra{g}-\ket{g}\bra{e})-(-\ket{e}\bra{g}+\ket{g}\bra{e})}{4}\nonumber\\
		&&&=\frac{\ket{e}\bra{g}-\ket{g}\bra{e}}{2}
		=\ii\frac{\ket{e}\bra{g}-\ket{g}\bra{e}}{2\ii}=\ii\op{S}_y\\
		2.&&\op{S}_x^2+\op{S}_y^2+\op{S}_z^2 &=
		\left(\frac{\ket{e}\bra{g}+\ket{g}\bra{e}}{2}\right)^2
		+\left(\frac{\ket{e}\bra{g}-\ket{g}\bra{e}}{2\ii}\right)^2
		+\left(\frac{\ket{e}\bra{e}-\ket{g}\bra{g}}{2}\right)^2\nonumber\\
		&&&= \frac{\ket{e}\bra{e}+\ket{g}\bra{g}}{4}+\frac{\ket{e}\bra{e}+\ket{g}\bra{g}}{4}+\frac{\ket{e}\bra{e}+\ket{g}\bra{g}}{4}\nonumber\\
		&&&= \frac34(\ket{g}\bra{g}+\ket{e}\bra{e})=\frac34\one \text{ and hence $S=1/2$.}
	\end{align}}
	\item\label{Q:Splusminus} Express $\op{S}^{\pm}$ in terms of $\ket{g}$, $\ket{e}$, $\bra{g}$, and $\bra{e}$.
\pagenote[\ref{Q:Splusminus}]{$\op{S}^+=\op{S}_x+\ii\op{S}_y=\frac{\ket{e}\bra{g}+\ket{g}\bra{e}}{2}+\ii\frac{\ket{e}\bra{g}-\ket{g}\bra{e}}{2\ii}=\frac{\ket{e}\bra{g}+\ket{g}\bra{e}}{2}+\frac{\ket{e}\bra{g}-\ket{g}\bra{e}}{2}=\ket{e}\bra{g}$. $\op{S}^-=\op{S}_x-\ii\op{S}_y=\frac{\ket{e}\bra{g}+\ket{g}\bra{e}}{2}-\ii\frac{\ket{e}\bra{g}-\ket{g}\bra{e}}{2\ii}=\frac{\ket{e}\bra{g}+\ket{g}\bra{e}}{2}-\frac{\ket{e}\bra{g}-\ket{g}\bra{e}}{2}=\ket{g}\bra{e}$. We can see that $\op{S}^+$ is the operator that excites the atom ($\op{S}^+\ket{g}=\ket{e}$) and $\op{S}^-$ is the operator that deexcites the atom ($\op{S}^+\ket{e}=\ket{g}$).}
	\item\label{Q:XPcommutator} Show that $[\op{X},\op{P}]=\ii$ using \autoref{eq:XPvsaad} and assuming that $[\op{a},\op{a}\dagg]=1$.
\pagenote[\ref{Q:XPcommutator}]{$[\op{X},\op{P}]=\op{X}\op{P}-\op{P}\op{X}=\frac{\op{a}+\op{a}\dagg}{\sqrt{2}}\frac{\op{a}-\op{a}\dagg}{\ii\sqrt{2}} - \frac{\op{a}-\op{a}\dagg}{\ii\sqrt{2}}\frac{\op{a}+\op{a}\dagg}{\sqrt{2}}=\frac{(\op{a}\op{a}-\op{a}\op{a}\dagg+\op{a}\dagg\op{a}-\op{a}\dagg\op{a}\dagg)-(\op{a}\op{a}+\op{a}\op{a}\dagg-\op{a}\dagg\op{a}-\op{a}\dagg\op{a}\dagg)}{2\ii}=\ii[\op{a},\op{a}\dagg]=\ii$.}
	\item\label{Q:JCHamTransform} Show that \autoref{eq:JCHam} follows from \autoref{eq:JCHam0} using \autoref{eq:XPvsaad}.
\pagenote[\ref{Q:JCHamTransform}]{Cavity field: $\op{a}\dagg\op{a}=\frac{\op{X}-\ii\op{P}}{\sqrt{2}}\frac{\op{X}+\ii\op{P}}{\sqrt{2}}=\frac{\op{X}^2+\op{P}^2+\ii[\op{X},\op{P}]}{2}=\frac{\op{X}^2+\op{P}^2-1}{2}$ and hence $\op{a}\dagg\op{a}+\frac12=\frac12\op{P}^2+\frac12\op{X}^2$. Coupling: $\op{S}^+\op{a}+\op{a}\dagg\op{S}^-=(\op{S}_x+\ii\op{S}_y)\frac{\op{X}+\ii\op{P}}{\sqrt{2}}+\frac{\op{X}-\ii\op{P}}{\sqrt{2}}(\op{S}_x-\ii\op{S}_y)=\frac{\op{S}_x\op{X}+\ii\op{S}_x\op{P}+\ii\op{S}_y\op{X}-\op{S}_y\op{P}+\op{X}\op{S}_x-\ii\op{X}\op{S}_y-\ii\op{P}\op{S}_x-\op{P}\op{S}_y}{\sqrt{2}}$. Since the operators on the field and atom degrees of freedom commute (for example, $[\op{X},\op{S}_x]=[\op{X}\otimes\one,\one\otimes\op{S}_x]=0$), this becomes $\op{S}^+\op{a}+\op{a}\dagg\op{S}^-=\frac{\op{X}\op{S}_x+\ii\op{P}\op{S}_x+\ii\op{X}\op{S}_y-\op{P}\op{S}_y+\op{X}\op{S}_x-\ii\op{X}\op{S}_y-\ii\op{P}\op{S}_x-\op{P}\op{S}_y}{\sqrt{2}}=\sqrt{2}(\op{X}\op{S}_x-\op{P}\op{S}_y)$.}
\end{questions}

\index{Mathematica!recursion|seealso{recursion}}
\index{Mathematica!units|see{physical units}}
\index{Gross-Pitaevskii equation|see{Schr\"odinger equation, non-linear}}
\index{reduced density matrix|see{partial trace}}
\index{Mathematica!procedure|see{function}}
\index{momentum operator|see{operator, momentum}}
\index{kinetic energy|see{operator, kinetic}}
\index{potential energy|see{operator, potential}}
\index{Heisenberg principle|see{uncertainty principle}}

\backmatter

\refstepcounter{dummy}\label{sec:attached}
\listofattachments

\vspace{1cm}
\noindent
If you cannot see the hyperlinks to the embedded Mathematica notebooks (white space instead of clickable links between the \embeddedcodecolorname\ square brackets on the target pages), please use the \AdobeReaderLink\ to view this document.

\renewcommand\indexname{index}
\printindex


\ifthenelse{\boolean{includesolutions}}{
\stepcounter{chapter}%
\appendix%
\renewcommand\notesname{solutions to exercises}%
\printpagenotes}{}

\end{document}